%% file: SuperReviewArXiv.tex
\documentclass[12pt]{article}
\usepackage{graphicx}
\usepackage{pict2e}
\usepackage{epsfig}
\usepackage{amssymb}
\usepackage{amsmath}
\usepackage{amsfonts, latexsym}
 \usepackage[sort&compress]{natbib}
 \numberwithin{equation}{section}

\bibpunct{[}{]}{,}{n}{}{;}
\usepackage{fullpage}
\begin{document}
\newcommand{\chapter}{\section}
\newcommand{\eref}[1]{(\ref{#1})}

\newcommand{\be}{\begin{equation}}
\newcommand{\ee}{\end{equation}}
\newcommand{\ba}{\begin{array}}
\newcommand{\ea}{\end{array}}
\newcommand{\bea}{\begin{eqnarray}}
\newcommand{\eea}{\end{eqnarray}}
\newcommand{\sinc}{{\rm sinc}\ }
\newcommand{\Bop}{\bf P}
\newcommand{\Boi}{\bf I}
\newcommand{\Bot}{\bf T}
\newcommand{\Boa}{\bf A}
\newcommand{\Boe}{\bf E}
\newcommand{\Bos}{\bf S}
\newcommand{\Bol}{\bf L}
\newcommand{\Bor}{\bf R}
\newcommand{\fl}{}
\newcommand{\h}{\hbar}
\def\R{{\mathbb R}}
\def\C{{\mathbb C}} 
\def\F{{\mathbb F}} 
\newcommand{\eo}{ \omega}
\newcommand{\dta}{\delta}
\newcommand{\psip}{{ \psi}}
\newcommand{\boa}{\bf a}
\newcommand{\boj}{\bf j}
\newcommand{\boe}{\bf e}
\newcommand{\bof}{\bf f}
\newcommand{\bog}{\bf g}
\newcommand{\both}{\theta}
\newcommand{\fis}{{\cal F}}
\newcommand{\mis}{{\cal M}}
\newcommand{\his}{{\cal H}}
\newcommand{\ris}{{\cal R}}
\newcommand{\nos}{{\not \!\! C}}
\newcommand{\raw}{\rightarrow}
\newcommand{\varp}{{ \varphi}}
\newtheorem{definition}{Definition}
\newtheorem{theorem}{Theorem}
\newtheorem{lemma}{Lemma}
\newtheorem{corollary}{Corollary}
\newtheorem{example}{Example}
\newtheorem{examples}{Examples}
\newtheorem{exercise}{Exercise}
\newtheorem{remark}{Remark}
\newtheorem{remarks}{Remarks}
\newtheorem{assumptions}{Assumptions}
\newcommand{\JMP}{J. Math Phys.}
\newcommand{\JPA}{J. Phys. A}

  \newcommand{\lab}{\label}                
\newcommand{\zbar}{\overline{z}}

\newcommand {\ch} {\mathcal{H}}

\newcommand {\cn} {\mathcal{N}}
\newcommand {\cm} {\mathcal{M}}
\newcommand {\cl} {\mathcal{L}}
\newcommand{\cS}{\mathcal{S}}
\newcommand{\nn}{\nonumber\\ }

%\newtheorem{comment}{Comment}

%%%%%%%%%%%%%%%%%%%%%%%%%%%%%%%%%%%%%
%\textwidth6.5in
%\textheight8.5in
%\oddsidemargin0in
%\evensidemargin0in
\pagenumbering{arabic}
%\frontmatter
\title{Classical and Quantum Superintegrability with Applications}
\author{W. Miller, Jr$^1$., S. Post$^2$ and P. Winternitz$^3$\\
$^1$ School of Mathematics, University of Minnesota,\\
 Minneapolis, Minnesota,
55455, U.S.A.\\$^2$ Department of Mathematics,
University of Hawai`i\\
Honolulu HI 96822, U.S.A.\\
$^3$ Centre de Recherches Math\'ematiques et\\ D\'epartement de Math\'ematiques   
et de Statistique,\\
 Universit\'e de Montr\'eal, C.P. 6128,\\ Montr\'eal, Qu\'ebec H3C 3J7, Canada.\\
miller@ima.umn.edu, sarah@math.hawaii.edu, wintern@crm.umontreal.ca}
\date{}
\maketitle

\begin{abstract}
A superintegrable system is, roughly speaking, a system that allows more integrals of motion than degrees of freedom. This review is devoted to finite dimensional classical and quantum superintegrable systems with scalar potentials and integrals of motion that are polynomials in the momenta. We present a classification of second-order superintegrable systems in two-dimensional Riemannian and pseudo-Riemannian spaces. It is based on the study of the quadratic algebras of the integrals of motion and on the equivalence of different systems under coupling constant metamorphosis. The determining equations for the existence of integrals of motion of arbitrary order in real Euclidean space $E_2$ are presented and partially solved for the case of third-order integrals. A systematic exposition is given of systems in two and higher dimensional space that allow integrals of arbitrary order. The algebras of integrals of motions are not necessarily quadratic but close polynomially or rationally. The relation between superintegrability and the classification of orthogonal polynomials is analyzed.

 \end{abstract}

%\maketitle

%\vspace{0.3cm}
\tableofcontents

% \markboth
\listoffigures
\newpage
%\chapter{Preface}
\section{Introduction} 
A standard way to gain insight into the behavior of a physical system is to construct a mathematical model of the system, analyze the model, use it to make 
physical  predictions that follow from the model and compare the results with experiment. The models provided by classical  and quantum mechanics 
have been and continue to be spectacularly successful in this regard. However, the systems of ordinary and partial differential equations provided by these
models can be very complicated. Usually they cannot be solved analytically and solutions can only be 
approximated numerically. A relatively few systems, however, can be solved exactly with explicit analytic expressions that predict future behavior, and 
with adjustable parameters, such as mass or initial position, so that one can can determine the effect on the system of changing these parameters. 
These are classical and quantum integrable Hamiltonian systems. A special subclass of these systems, called superintegrable, is extremely important for developing 
insight into physical principles, for they can be solved algebraically as well as analytically,  and many of the simpler systems are featured prominently in 
textbooks.  What distinguishes these systems is their symmetry, but often of a much subtler kind than just group symmetry. The symmetries in their totality form
 quadratic,  cubic and other higher order algebras, not necessarily Lie algebras, and are sometimes referred to as `hidden symmetries'. Famous examples are the 
classical harmonic anisotropic oscillator (Lissajous patterns)  and Kepler systems (planetary orbits), the quantum Coulomb system (energy levels of the hydrogen atom, 
leading to the periodic table of the elements) and the quantum isotropic oscillator.   The Hohmann transfer, a fundamental procedure for the positioning of 
satellites and orbital maneuvering of interplanetary spacecraft is based on the superintegrability of the Kepler system.

Superintegrable systems admit the maximum possible symmetry and this forces analytic and algebraic solvability. The special functions of mathematical physics and 
their properties are closely related to their origin and use in providing explicit solutions for superintegrable systems, for instance via separation of variables 
in the partial differential equations of mathematical physics.  These systems appear in a wide variety of modern physical and mathematical theories, 
from semiconductors to supersymmetric field theories. As soon as a system is discovered it tends to be implemented as a model, due to the fact that it can be
solved explicitly. Perturbations of superintegrable systems are frequently used to  study the behavior of more complex systems, e.g., 
the periodic table is based on perturbations of the 
superintegrable hydrogen atom system.
\textbf{•}
The principal research activity in this area  involves the discovery, classification and  solution of superintegrable  systems, and elucidation of 
their structure, particularly the underlying symmetry algebra structure, as well as application of the results in a wide variety of fields. 
Superintegrability has deep historical roots, but the modern theory was inaugurated by Smorodinsky, Winternitz and collaborators in 1965 \cite{FMSUW, FSUW, MSVW} who explored 
multiseparability in 2 and 3 dimensional Euclidean spaces. Wojciechowski seems to have coined the term `superintegrable' and applied it about 1983 \cite{WOJ}. The earlier terminology was ``systems with
accidental degeneracy", going back to Fock and Bargmann \cite{Fock, Bargmann}. Other terms used in
this context were ``higher symmetries", or ``dynamical symmetries" \cite{FMSUW, FSUW, MSVW}. Some explicit solutions of the n-body problem by 
Calogero \cite{CALO, calogero1971solution, calogero1975exactly, ruhl1995exact}, dating from the late 1960s were crucial examples for the theory. The technique of coupling constant metamorphosis to map between integrable and 
superintegrable systems was introduced in the mid-1980s \cite{HGDR, BKM}.  Interest increased greatly due to papers by Evans about 1990 \cite{EVAN, EVA, Evans1991}
which contained many examples that generalized fundamental solvable quantum mechanical systems in 3 dimensions. About 1995 researchers such as Letourneau and Vinet 
 \cite{VILE} recognized the very close relationship between Quasi-Exact Solvability (QES) \cite{turbiner1987spectral,Turbiner1988,gonzalez1994quasi} for quantum systems in one dimension and second order superintegrable
 systems in two and higher dimensions. Beginning about 2000 the structure theory and classification of second order superintegrable systems has 
been largely worked out, with explicit theorems that provide a concrete foundation for the observations made in explicit examples  (Daskaloyannis,
 Kalnins, Kress, Miller, Pogosyan, etc.). The quadratic algebras of symmetries of the second order superintegrable systems, and their representation 
theory has been studied since about 1992 with results by Zhedanov, Daskaloyannis, Kalnins, Kress, Marquette,  Miller, Pogosyan, Post,  Vinet,  Winternitz, etc. New applications of 
the theory to other branches of physics are appearing, e.g. work by Quesne on variable mass Hamiltonians \cite{Quesne2007}. More recently,
 important examples of
 physically interesting third and fourth order quantum superintegrable systems were announced by Evans and Verrier \cite{Evans2008a}, and by Rodriguez, Tempesta,  and Winternitz
 \cite{RTW2008, RTW2009}. 
  
Superintegrable systems of second-order, i.e., classical systems where the defining symmetries are second-order in the momenta and quantum systems where
 the symmetries are second-order partial differential operators, have been well studied and there is now a developing  structure and classification theory.
 The classification theory for third-order systems that separate in orthogonal coordinate systems, i.e. that also admit a second-order integral, has begun 
and many new systems have recently been found, including quantum systems with no classical analog and systems with potentials associated with Painlev\' e
transcendents \cite{Gravel, GW, MW2008, marquette2009painleve, TW20101}. These are quantum systems that could not be obtained by quantizing classical ones. Their quantum limits are sometimes free motion. In other cases, the $\hbar$ going to 0 limit are  singular, in the sense that the quantum potential satisfies partial differential equations in which the leading terms vanish for $\hbar$ going to 0. 
However for nonseparable third-order  and general higher-order superintegrable systems much less is known. In particular until very recently there were few examples
 and  almost no structure theory and classification theory. 
  
This situation has changed dramatically with the publication of the 2009 paper ``An infinite family of solvable and integrable quantum systems on a plane'' by
 F.\ Tremblay, V.A. \ Turbiner and P.\ Winternitz \cite{TTW, TTW2}. The authors'  paper  had an immediate effect on the active field of classical and quantum superintegrable systems.
  Their examples and conjectures have led rapidly to new classes of higher order superintegrable systems, thereby reinvigorating research activity and publications
 in the subject. The authors introduced a 
family of both classical and quantum mechanical potentials in the plane, parametrized by the constant $k$,  conjectured and gave evidence that these systems 
were both classically and quantum superintegrable for all rational $k$, with integrals of  arbitrarily large order. It has now been verified  that that the conjectures
 were correct (Gonera, Kalnins, Kress, Miller, Quesne, Pogosyan, \cite{CQ10, KMPog10, KKM10a, KKM2011Recurr, CG}), Higher order superintegrable systems had been thought to be uncommon, 
 but  are now seen to be 
 ubiquitous with  a clear path to construct families of other candidates at will \cite{ballesteros2013anisotropic, chanu2011three, KKM10, KKM2011Recurr, marquette2013new, Marquette20103, marquette2013quartic, ranada2013higher,  PostVinet2012, PW2010, chanu2012generalizations}.  Tools are being developed for the verification of classical and quantum 
superintegrability of higher order that can be applied to a variety of Hamiltonian systems. A structure theory for these systems, classification results
 and applications are following.

This review is focused on the structure and classification of maximal superintegrable systems and their symmetry algebras, classical and quantum. Earlier reviews exist, including those on the group theory of the hydrogen atom and Coulomb problem \cite{kalnins1976group, englefield1972group, BANITZ}, oscillators \cite{louck1973canonical1, louck1973canonical2, MoshSmir} and accidental degeneracy or symmetry in general \cite{moshinsky1975canonical, McIntosh}.

 There are other interesting approaches to the theory that we don't address here. In particular there is a geometrical approach to the classical
 theory, based on foliations, e.g. \cite{Fasso, Mischenko, nekhoroshev1972, tempesta2012generalized} and those using the methods of differential Galois theory \cite{MPY2010, MPT} or invariant theory of Killing tensors \cite{adlam2007orbit, adlam2008geometric}. Many authors approach classical and quantum superintegrable systems from an external point of view. They use elegant 
techniques such as R-matrix theory and coalgebra symmetries to produce superintegrable systems with generators that are embedded in a larger associative algebra with simple structure, such as a Lie enveloping algebra, e.g.
 \cite{ADS2006, brink1998hidden, BH, HER1, Sasaki2000, Sasaki2001, BEHRR, tsiganov2008maximally, riglioni2013classical, Ran2012}. Here we take an internal point of view. The fundamental object for us is the symmetry algebra generated  by the system. 
A useful analogy is differential geometry where a Riemannian space can be considered either as
 embedded in Euclidean  space or as  defined intrinsically via a metric.
 
 Let us just list some of the reasons why superintegrable systems are interesting
both in
classical and quantum physics.
\begin{enumerate} 
\item  In classical mechanics, superintegrability restricts trajectories to an $n - k$
dimensional
subspace of phase space ($0<k<n$). For $k = n - 1$ (maximal superintegrability), this
implies that all finite trajectories are closed and motion is periodic \cite{nekhoroshev1972}.
\item  At least in principle, the trajectories can be calculated without any calculus.
\item Bertrand's theorem \cite{Bertrand} states that the only spherically symmetric potentials $V(r)$ for
which all bounded trajectories are closed are the Coulomb-Kepler system and the
harmonic
oscillator, hence no other superintegrable systems are spherically symmetric.
\item  The algebra of integrals of motion  is a non-Abelian and interesting
one. Usually it is a finitely generated polynomial algebra, only exceptionally a finite
dimensional Lie algebra or Kac-Moody algebra \cite{daboul1993hydrogen}. 
\item  In the special case of quadratic superintegrability (all integrals of motion are
at most quadratic polynomials in the moments), integrability is related to
separation of variables in the Hamilton-Jacobi equation, or
Schr\"odinger equation, respectively.
\item In quantum mechanics,
superintegrability leads to an additional degeneracy of energy levels, sometimes
called "accidental degeneracy". The term was coined by Fock\cite{Fock} and used by
Moshinsky and collaborators \cite{moshinsky1972canonical, louck1973canonical1, moshinsky1983, moshinsky1983b}, though the point of their studies was to show that
this degeneracy is certainly no accident.
\item A conjecture, born out by all known examples, is that all maximally superintegrable
systems are exactly solvable \cite{TempTW}. If the conjecture is true, then the energy levels can
be calculated algebraically. The wave functions are polynomials (in appropriately
chosen variables) multiplied by some gauge factor.
\item  The non-Abelian polynomial algebra of integrals of motion provides energy spectra
and information on wave functions. Interesting relations exist between
superintegrability and supersymmetry in quantum mechanics
\end{enumerate} 
As a comment, let us mention that superintegrability has also been called non-Abelian
integrability. From this point of view, infinite dimensional integrable systems (soliton
systems) described e.g. by the Korteweg-de-Vries equation, the nonlinear Schrdinger
equation, the Kadomtsev-Petviashvili equation, etc. are actually superintegrable.
Indeed, the generalized symmetries of these equations form infinite dimensional
non-Abelian algebras (the Orlov-Shulman symmetries) with infinite dimensional
Abelian subalgebras of commuting flows\cite{orlov1985additional, orlov1986additional, orlov1997algebra}

Before we delve into the specifics of superintegrability theory, we give a simplified version of the requisite mathematics and physics governing Hamiltonian 
dynamical systems in Section \ref{Chapter0}. Then in Section \ref{Chapter1} we study, as examples, the 2D Kepler system and the 2D hydrogen atom in detail,
 both in Euclidean space and on the 2-sphere, as well as the hydrogen atom in 3D Euclidean space with its $O(4)$ symmetry. The examples illustrate basic features of superintegrability: complete solvability of the systems 
via the symmetry algebra, important applications to physics, and relation of superintegrable systems via contraction.  These well known systems have been 
studied literally for centuries, but the pure superintegrability approach has novel features. In Section \ref{Chapter2ndorder} we sketch the structure 
and classification theory for second-order superintegrable systems, the most tractable class of such systems. Section \ref{thirdorderchapter} is devoted 
to the classification of higher-order systems in 2D Euclidean space, where the quantization problem first becomes serious.   In Sections
 \ref{higherorderclassicalchapter} and \ref{higherorderquantum} we present examples of  higher order classical and quantum systems and tools 
for studying their structure. Here great strides have been made but, as yet, there is no classification theory. Sections \ref{Stackeltransform} 
is devoted to the generalized St\"ackel transform, an invertible structure preserving transformation 
of one superintegrable system to another that is basic to the classification theory. Since superintegrability is a concept that distinguishes 
completely solvable  physical systems it should be no surprise that there are profound relations to the theory of special functions. 
In Sections \ref{Askeyscheme} we make that especially clear by showing that the Askey Scheme for orthogonal hypergeometric polynomials  
can be derived from  contractions of 2D second-order superintegrable systems.

% \mainmatter
 
%\vspace{0.3cm}
\include{Chapter0} 
\include{Chapter1}
\include{Secondorderchapter}

\include{Thirdorderchapter}

\include{higherorderclassicalchapter}
\include{higherorderquantumchapter}

\include{Stackeltransformchapter}
\include{Askeyschemechapter}
\include{Conclusions}
%\backmatter
%\clearpage
%\addcontentsline{toc}{chapter}{Bibliography}

\include{bib}

    %Bibliography
  
\end{document}

%% file: Chapter0.tex
\section{Background and Definitions} \label{Chapter0}

\subsection{Classical mechanics}

The Hamiltonian formalism describes  dynamics of a physical system in $n$ dimensions
by relating the time derivatives of the position coordinates and the momenta to a single
function on the phase space, the Hamiltonian ${\cal H}$. A physical system describing the position of
a particle at time $t$ involves $n$ position coordinates $q_j(t)$, and $n$ momentum coordinates,
$p_j(t)$. The phase space of a physical system is
described by points $(p_j, q_j)\in  F^{2n}$, where $F$ is the base field, usually $\R$ or $\C$.  
In its simplest form, the Hamiltonian can be interpreted as the total energy of the system:
$ {\cal H} = T + V$,
where $T$ and $V$ are kinetic and potential energy, respectively. Explicitly, 
\be\label{relham} {\cal H} = \frac{1}{2m}\sum_{j,k}g^{jk}({\bf q}) p_jp_k+ V ({\bf q}) \ee
where $g^{jk}$ is a contravariant metric tensor on some real or complex Riemannian  manifold. 
That is $g^{-1}=\det(g^{jk})\ne 0$, $g^{jk}=g^{kj}$  and the metric on the manifold is given by $ds^2=\sum_{j,k=1}^ng_{jk}dq^j\ dq^k$ where $(g_{jk})$ is 
the covariant metric tensor, the matrix inverse to $(g^{jk})$. Under a local  transformation $q'_j=f_j({\bf q})$ 
the contravariant tensor and  momenta transform according to 
\be\label{covariantvector}(g')^{\ell h}=\sum_{j,k} \frac{\partial q'_\ell}{\partial q_j}\frac{\partial q'_h}{\partial q_k}g^{jk},\quad
 p'_\ell =\sum_{j=1}^n \frac{\partial q_j}{\partial q'_\ell} p_j.\ee
 Here $m$ is a scaling parameter that can  be interpreted as the mass of the particle.
For the Hamiltonian (\ref{relham}) the relation  between the momenta and the velocities is 
$ p_j=m\sum_{\ell=1}^ng_{j\ell}\dot{q}_\ell$,
so that 
$T=\frac{1}{2m}\sum_{j,k}g^{jk}({\bf q}) p_jp_k=\frac{m}{2}\sum_{\ell,h=1}^n g_{\ell h}({\bf q})\dot{q}_\ell\dot{q}_h$.
Once the velocities are given, the momenta are scaled linearly in $m$. In mechanics the exact value of $m$ may be important
but for mathematical structure calculations it  can be scaled to any nonzero value. To make direct contact with mechanics we may set $m=1$; 
for  structure calculations we will usually set $m=1/2$. The above formulas
show how to rescale for  differing values of $m$.

The dynamics of the system are given by 
Hamilton's equations, \cite{Goldstein, Arnold}
\be\label{hameqns} \frac{d q_j}{dt}  =\frac{\partial {\cal H}}{\partial p_j},\qquad  \frac{dp_j}{dt}=-\frac{\partial {\cal H}}{\partial q_j},\quad j=1,\ldots,n.\ee 
Solutions of these  equations give the trajectories of the  system.
\begin{definition} The {\bf Poisson bracket} of two functions ${\cal R}({\bf p}; {\bf q})$, ${\cal S}({\bf p}, {\bf  q})$  on the phase space is the function
 \be\label{poissonbracket} \{{\cal R},{\cal S}\}({\bf p},{\bf q})=\sum_{j=1}^n\left(\frac{\partial{\cal R}}{\partial p_j}\frac{\partial{\cal S}}{\partial q_j}
-\frac{\partial{\cal R}}{\partial q_j}\frac{\partial{\cal S}}{\partial p_j}\right).\ee
\end{definition}
The Poisson bracket obeys the following properties, for ${\cal  R}, {\cal  S}, {\cal T}$
functions on the phase space and $a, b$ constants.
\be\label{antisymmetry} \{{\cal R},{\cal S}\}=-\{{\cal S},{\cal R}\},\quad {\rm anti-symmetry}\ee
\be\label{bilinearity} \{{\cal R}, a{\cal S}+ b{\cal T}\}= a\{{\cal R}, {\cal S}\} + b\{{\cal R},{\cal T }\},\quad  {\rm bilinearity}\ee 
\be\label{jacobi} \{{\cal R},\{ {\cal S},{\cal  T}\}\}  + \{{\cal S}, \{{\cal T}, {\cal  R}\}\}+\{{\cal T}, \{{\cal R},{\cal  S}\}\} = 0,\quad{\rm  Jacobi\ identity}\ee
\be\label{leibniz} \{{\cal R}, {\cal ST}\}= \{{\cal R},{\cal  S}\}{\cal T}+{\cal  S}\{{\cal R},{\cal  T}\},\quad   {\rm Leibniz\ rule}\ee 
\be\label{chainrule} \{f({\cal R}),{\cal S}\}=f'({\cal R})\{{\cal R},{\cal S}\},\quad {\rm chain\ rule.}\ee
With $\delta_{jk}$  the Kronecker delta, coordinates $ ({\bf q}, {\bf  p})$ satisfy canonical relations
\be\label{canrel} \{p_j, p_k\} = \{q_j, q_k\} = 0,\quad  \{p_j, q_k\} = \delta_{jk}.\ee

\begin{definition} A set of $2n$ coordinate functions ${\bf Q}({\bf q},{\bf p}), {\bf P}({\bf q},{\bf p})$ is called {\bf canonical} if the functions 
satisfy the canonical relations
$$  \{P_j, P_k\} = \{Q_j, Q_k\} = 0,\quad  \{P_j, Q_k\} = \delta_{jk},\ 1\le j,k\le n.$$
\end{definition}
Canonical coordinates  are true coordinates on the phase space, i.e., they can be inverted locally to express ${\bf q},{\bf p}$
 as functions of ${\bf Q},{\bf P}$. Furthermore, under this change of coordinates the Poisson bracket (\ref{poissonbracket})  maintains its form, i.e., 
$$\{{\cal R},{\cal S}\}({\bf P},{\bf Q})=\sum_{j=1}^n\left(\frac{\partial{\cal R}}{\partial P_j}\frac{\partial{\cal S}}{\partial Q_j}
-\frac{\partial{\cal R}}{\partial Q_j}\frac{\partial{\cal S}}{\partial P_j}\right).$$
Any coordinate change of the form  $q'_j({\bf q}), p'_k({\bf q},{\bf p})$, where ${\bf q}'$ depends only on
 $\bf q$ and  ${\bf p}'$ is defined by (\ref{covariantvector}), is always canonical.

In terms of the Poisson bracket, we can rewrite Hamilton's equations as
\be \label{hameqns2} \frac{dq_j}{dt} = \{{\cal H},q_j\}, \quad  \frac{dp_k}{dt} = \{{\cal H}, p_k\}.\ee
For any function  ${\cal R}({\bf q},{\bf  p})$, its dynamics
along a trajectory ${\bf q}(t), {\bf p}(t)$ is
\be\label{timedependence} \frac{d{\cal R}}{dt}=\{{\cal H}, {\cal R}\}.\ee
Thus  ${\cal R}({\bf q}, {\bf p})$  will be constant along a trajectory if and only $\{{\cal H},{\cal R}\}=0$.
\begin{definition}  If $\{{\cal H},{\cal S}\}=0$, then ${\cal S}({\bf q},{\bf  p})$  is a {\bf constant of the
motion}.
\end{definition}
\begin{definition}
Let ${\cal F}=(f_1({\bf q},{\bf p}), \ldots, f_N({\bf q},{\bf p}))$ be a set of $N$ functions defined and locally analytic in some region of 
 a $2n$-dimensional phase space. We say ${\cal F}$ is {\bf functionally independent}  if the $N\times 2n$ matrix
$\left( \frac{\partial f_\ell}{\partial  q_j},\frac{\partial f_\ell}{\partial p_k}\right)$
has rank $N$ throughout the region. The set is {\bf functionally dependent} if the rank is strictly less than $N$ on the region.
\end{definition}
If a set is functionally dependent, then locally there exists a nonzero  analytic function $F$ of $N$ variables such that $F(f_1,\ldots,f_n)=0$ identically on the region. Conversely, if $F$ 
exists then the rank of the matrix is $<N$. Clearly,  a set of $N>2n$ functions will be functionally dependent.

\begin{definition} A system with Hamiltonian ${\cal H}$ is {\bf integrable} if it admits $n$ constants of
the motion
$ {\cal P}_1 = {\cal H},{\cal P}_2,\ldots,{\cal P}_n$
that are in involution:
\be\label{involution}\{{\cal P}_j,{\cal P}_k\}=0, \quad 1\le j,k\le n,\ee
and are functionally independent.  
\end{definition}
Now suppose $\cal H$ is integrable with associated constants of the motion ${\cal P}_j$. Up to a canonical change of variables it is possible to assume that  $\det(\frac{\partial {\cal P}_j}{\partial p_k})\ne 0$. Then by the inverse function theorem we can 
solve the $n$ equations ${\cal P}_j({\bf q}, {\bf p}) = c_j$ for the momenta
to obtain $p_k = p_k({\bf q}, {\bf c})$, $ k = 1,\ldots, n$, where ${\bf c} = (c_1,\ldots, c_n)$ is a vector of constants. 
For an integrable system, if
a particle with position $\bf q$  lies on the common intersection of the hypersurfaces ${\cal P}_j = c_j$
for constants $c_j$, then its momentum $\bf p$ is completely determined.
Also, if a particle following a trajectory of an integrable system lies
on the common intersection of the hypersurfaces ${\cal P}_j = c_j$ at  time $t_0$, where the ${\cal P}_j$ are constants of the motion, then it 
lies on the same common intersection for all $t$ near $t_0$.
Considering $p_j({\bf q},{\bf  c})$ and using the conditions (\ref{involution}) and the chain rule it is straightforward
to verify $\frac{\partial p_j}{\partial q_k}=\frac{\partial p_k}{\partial q_j}$. 
Therefore, there exists a function
$u({\bf q},{\bf  c})$ such that $p_\ell= \frac{\partial u}{\partial q_\ell}$, $\ell=1,\ldots,n$. Note that
$ {\cal P}_j({\bf q},\frac{\partial u}{\partial {\bf q}})=c_j,\quad j=1,\ldots,n$,
and, in particular, $u$ satisfies the {\bf Hamilton-Jacobi equation}
\be\label{hamjaceqn} {\cal H}\left({\bf q},\frac{\partial u}{\partial {\bf q}}\right)=E,\ee
where $E = c_1$. By construction $\det(\frac{\partial u}{\partial q_j\partial c_k})\ne 0,$  
 and such a solution of the Hamilton-Jacobi equation depending nontrivially on $n$ parameters $\bf c$ is called a {\bf complete integral}. This argument is reversible: 
a complete integral of  (\ref{hamjaceqn}) determines $n$ constants of
the motion in involution, ${\cal P}_1,\ldots, {\cal P}_n$.
\begin{theorem} \label{hjtheorem} A system is integrable $\iff$ (\ref{hamjaceqn})  admits
a complete integral.
\end{theorem}
 A powerful method for demonstrating that a system is integrable is
to  exhibit a complete integral by using  additive separation of
variables. 

It is a standard result in classical mechanics that  for an integrable system one can
 integrate Hamilton's equations and obtain the trajectories,  \cite{Goldstein, Arnold}.  
The Hamiltonian
formalism is  is well suited to exploiting symmetries of the system and  an important tool in laying the
framework for quantum mechanics.

\subsection{Quantum mechanics}
We give a brief  introduction to  basic principles necessary to understand
quantum superintegrable systems. We ignore such issues as the domains of unbounded operators 
and continuous spectra, and proceed formally. In any case, we will be mainly interested in bound states and their discrete spectra. 

In quantum mechanics,  physical states are represented as one dimensional subspaces
in a complex, projective Hilbert space: two states are  equivalent if
they differ by a constant multiplicative factor. 
 A standard Euclidean space model for a quantum mechanical bound state system with $n$ degrees of freedom is the one where the state vectors are complex square integrable functions 
$\Phi({\bf x},t)$ on $\R^n$ and the transition amplitude between two states $\Phi,\Psi$ is the inner product
$\langle \Psi,\Phi\rangle =\int_{\R^n}\Psi({\bf x},t)\overline{\Phi({\bf x},t)}\ d{\bf x}$.
Usually,  states are normalized: $||\Psi||^2=\langle \Psi,\Psi\rangle=1$.
If $A$ is a self-adjoint operator (observable)  then
$$\langle \Psi,A\Phi\rangle = \int_{\R^n}\Psi({\bf x},t)\overline{A\Phi({\bf x},t)}\ d{\bf x}=\int_{\R^n}(A\Psi({\bf x}))\overline{\Phi({\bf x})}\ d{\bf x}=\langle A\Psi,\Phi\rangle .$$
In this model $q_j\to X_j=x_j$, i.e., multiplication by  Cartesian variable $x_j$,  $p_j\to P_j=-i\hbar\partial_{x_j}$. 
In analogy to  ${\cal H}=\frac{1}{2m}\sum_{j=1}^n p_j^2+V({\bf q})$, we have the  quantum Hamiltonian on n-dimensional Euclidean space,
\be\label{quantumh1}H = -\frac{\hbar^2}{2m}\sum_{j=1}^n\frac{\partial^2}{\partial x_j^2}+V({\bf x}).\ee

Observables correspond to quantities that can be measured. 
The probability of an arbitrary state, $\Phi$ being measured with a given
eigenvalue $\lambda$ is determined by computing its transition amplitude, $|\langle \Psi_\lambda,\Phi\rangle |$, where $\Psi_\lambda$ is an eigenvector
of an operator $A$ with eigenvalue $\lambda$.
As in classical mechanics, we create most of our observables out of the quantities
position and momentum. However, in
quantum mechanics these quantities correspond to operators.  That is, the self-adjoint operator $X_j$ gives the
value of the coordinate $x_j$ while the self-adjoint operator $P_j$ gives the $j$th momentum.
We can define a bilinear product on the operators by the commutator $[A,B] =
AB- BA$. This  product satisfies the same relations (\ref{antisymmetry}-\ref{leibniz}) as the Poisson bracket.
The  position and momentum operators satisfy the 
commutation relations,
\be\label{quantumcomrel} [X_j ,X_k] = [P_j , P_k] = 0,\quad  [X_j , P_k] = i\hbar \delta_{j k},\ee
where $\hbar$ is the reduced Planck constant and $i=\sqrt{-1}$. Equations (\ref{quantumcomrel}) define a Heisenberg algebra $H_n$  in a n-dimensional space.

The time evolution of quantum mechanical states is determined by  $H$: 
The dynamics is given by the {\bf time-dependent Schr\"odinger Equation},
\be\label{timedepschr} i\hbar \frac{d}{dt}\Phi(t)=H\Phi(t).\ee
Using the Hamiltonian, we can define the time evolution operator $ U(t) = \exp(iHt/\hbar)$
as a one parameter group of unitary operators determined by the self-adjoint operator $H$. A state
at an arbitrary time $t$ will be given in terms of the state at time $0$ by $\Phi(t) = U(t)\Phi(0)$.   If a state at time $0$ is an eigenvector for
the Hamiltonian, say $H\Phi_E(0) =E\Phi_E(0)$,  then the time evolution operator acts on
an eigenvector by a (time dependent) scalar function,
$ \Phi_E(t)=U(t)\Phi_E(0)=e^{iEt/\hbar}\Phi_E(0)$.
The eigenvalue problem
\be\label{timeindepschr} H\Phi=E\Phi\ee
is called the {\bf time-independent  Schr\"odinger Equation}.
If we can find a basis of eigenvectors for $H$, we can expand an
arbitrary vector in terms of this energy  basis to get  the time evolution of the state.
Finding eigenvalues and eigenvectors for Hamiltonians is a
fundamental task in quantum mechanics.

 The expected value of an observable $A$ in a state $\Phi(t)$ obeys the  relation
\be \label{expectedvalueevolution} \frac{d}{dt}\langle \Phi,A\Phi\rangle=\frac{i}{\hbar}\langle \Phi,[H,A]\Phi\rangle.\ee
This is in analogy with the classical relation (\ref{timedependence}).
If  $[H,A]=0$, then we can choose a basis for the Hilbert space
which is a  set of simultaneous eigenfunctions of $A$ and $H$, \cite{JVN}. It is important to determine a complete set of commuting observables to specify  the system. This idea leads naturally to quantum integrability.

\begin{definition} A quantum mechanical system in $n$ dimensions is {\bf integrable} if there
exist $n$ integrals of motion, $L_j$ $j=1,\ldots, n$,  that satisfy the following conditions.
\begin{itemize}
\item They are well defined Hermitian operators in the enveloping algebra of the
Heisenberg algebra $H_n$ or convergent series in the basis vectors $X_j, \, 
P_j$, $j=1, \ldots,  n.$
\item They are algebraically independent in the sense that no Jordan polynomials formed entirely out of anti-commutators in $L_j$ vanishes identically
\item the integrals $L_ j$ commute pair-wise. 
\end{itemize}
\end{definition}
The extension of these ideas from classical mechanics to quantum mechanics is straightforward, except for  functional independence. There is no 
agreed-upon operator equivalence for this concept. In this review we use  {\bf algebraic independence} of a set of  operators 
$S_1,\ldots, S_h$. We say the operators are algebraically independent if there is no nonzero Jordan polynomial that vanishes identically. That is, there is no symmetrized polynomial $P$ in $h$ non-commuting  variables such that 
$P(S_1,\ldots,S_h)\equiv 0$. By symmetrized we mean that if the monomial  $\alpha S_jS_k$ appears in $P$ it occurs in the symmetric form via the anti-commutator $\alpha( S_jS_k+ S_kS_j)$.
Similarly third order products are symmetrized sums of 6 monomials, etc.

If the classical Hamiltonian admits a constant of the motion ${\cal S}({\bf q},{\bf p})$  we also want to find a corresponding operator $S$, 
expressible in terms of operators $X_j,P_j$,
that commutes with $H$. However, in quantum mechanics the position and momentum operators do not
commute so the analogy with the function ${\cal S}$ isn't clear. We usually take symmetrized products of any terms with mixed position
and momentum, though there are other conventions.
The quantization problem of determining  suitable  operator counterparts to  classical Hamiltonians
 and constants of the motion  is 
very difficult, sometimes impossible to solve  but, for many of the simpler 
superintegrable systems  the solution of the quantization problem is unique and straightforward.

More generally we can consider quantum systems on a Riemannian manifold, in analogy with the classical systems (\ref{relham}):
\be  H = -\frac{\hbar^2}{2m\sqrt{g}}\sum_{j,k=1}^n\frac{\partial}{\partial x_j}\left(\sqrt{g} g^{jk}\frac{\partial}{\partial x_k}\right) + V ({\bf x})=
\label{laplacebeltrami} -\frac{\hbar^2}{2m}\Delta_n+V\ee
where $\Delta_n$ is the {\bf Laplace-Beltrami operator} on the manifold \cite{CourantHilbert},(Volume I). Inner products are computed using the volume maeasure $\sqrt{g} \ d{\bf x}$.
In the special case of Cartesian coordinates in Euclidean space, expression (\ref{laplacebeltrami}) reduces to (\ref{quantumh1}).
%It is a standard result in differential geometry that the Laplace-Beltrami operator  and hence  $ H$ is defined independent of local coordinates.

We  often consider the time-independent Schr\"odinger equation $H\Psi=E\Psi$ for functions $\Psi({\bf x})$ on 
{\bf complex} Riemannian manifolds with complex coordinates $x_j$. In these cases we are going beyond quantum mechanics and Hilbert
spaces and considering formal eigenvalue problems. However many concepts  carry over, such as symmetry operators, integrability,  superintegrability,
relations to special functions, etc.
While for direct physical application it is important to specify the mass $m$ and to retain the Planck constant $\hbar$,  for many mathematical
computations  we can rescale these constants, without loss of generality.  We  rewrite the
Schr\"odinger eigenvalue equation as
\be\label{scaledseqn} \Delta_n \Psi({\bf x})-\frac{2m}{\hbar^2}V({\bf x})\Psi({\bf x})= -\frac{2m}{\hbar^2}E\Psi({\bf x})\ee
or $(\Delta_n+{\tilde V})\Psi={\tilde E}\Psi$ where ${\tilde V},{\tilde E}$ are rescaled potential and energy eigenvalue, respectively.
This is  appropriate because many of the superintegrable systems we consider have potentials of the form $V=\alpha V_0$ where $\alpha$
 is an arbitrary parameter. Thus,  we will make the most convenient choice of $-\hbar^2/2m$, usually $+1$ or $-1/2$, with the understanding
that the result can be scaled to reinsert $\hbar$ and  obtain any desired $m$. In cases where $V$ doesn't admit an arbitrary multiplicative factor or 
the quantization problem can't be solved we will have  to retain the original formulation with $H$ given by (\ref{laplacebeltrami}). We shall see in Section 5 that in quantum mechanics integrable and
superintegrable systems exist that have no classical counterparts. The
limit $\hbar \rightarrow0$ corresponds to free motion, or this limit may be singular. This
occurs when integrals of motion of order three or higher are involved.
Thus, for the case of third and higher-order integrals of motion, it is best to keep $\hbar$ explicitly in all formulas in order to be able to pass to the classical limit.

\subsection{Superintegrability}
\label{superintegrability}
Let us first focus on the explicit solvability  properties of  classical systems  ${\cal H}=\sum g^{jk}p_jp_k+V$ and operator  systems $H=\Delta_n+V$. For this review, we will be focusing mainly on classical integrable and superintegrable systems that are polynomial in the momenta both for the purposes of quantization as well as simply for the fact that these are the most well studied. We thus define {\it  polynomial} integrability and superintegrability for classical systems. The quantum analogy of these systems will then be integrals which are {\it finite-order} differential operators and hence we define quantum integrability and superintegrability of finite-order. For most of the review will will omit the the adjective ``polynomial" or ``of finite-order" and just speak of integrability and superintegrability. 

Similarly,  we define superintegrability  as having more than $n$ integrals of motion. There is thus a notion of minimal and maximal superintegrability. We will focus in this review on maximal superintegrability and often omit the adjective maximal. Maximal superintegrability requires that there exists $2n-1$ integrals of motion, one of which (the Hamiltonian) commutes with all of the others. In this sense, maximal superintegrability coincides with non-commutative integrability \cite{nekhoroshev1972, Mischenko} which requires $2n-r$ integrals, $r$ of which commute with all of the integrals.

\subsubsection{Classical superintegrability and the order of integrable systems}

\begin{definition}
A Hamiltonian system  is {\bf (polynomially)  integrable} if it is integrable and the  constants of the motion are 
each polynomials in the momenta  globally defined (except possibly for singularities on lower dimensional manifolds).
\end{definition}

Systems with symmetry beyond polynomial integrability  are  (polynomially) superintegrable; 
those with maximum possible symmetry are maximally (polynomially) superintegrable. 
\begin{definition}
A classical Hamiltonian system in $n$ dimensions is {\bf  (polynomially) superintegrable} if it admits $n+k$ with $k=1,\ldots, n-1$ functionally independent constants  of the motion that are polynomial in the momenta and are globally defined except possibly for singularities on a lower dimensional manifold. It is {\bf minimally (polynomially) superintegrable} if $k=1$  and {\bf maximally (polynomially) superintegrable} if $k=n-1$.
\end{definition}
Every constant of the motion  $\cal S$, polynomial or not, is  a solution of the equation $\{{\cal H},{\cal S}\}=0$ where $\cal H$ 
is the  Hamiltonian. This is a linear  homogeneous first order partial differential equation for $\cal S$ in $2n$ variables. 
It is a well-known result  that every solution of such equations can  be expressed as a function $F(f_1,\ldots,f_{2n-1})$ of $2n-1$ 
functionally independent solutions, \cite{CourantHilbert}. Thus there always exist $2n-1$ independent functions, locally defined, 
in involution with the Hamiltonian, the  largest possible number. However, it is rare to 
find $2n-1$ such functions that are globally defined and polynomial in the momenta. Thus superintegrable systems are very special.

Most authors require a superintegrable system to be  integrable. We have not done so here because we  know of no proof (or counter example) that every superintegrable system 
in our sense is necessarily integrable. 
 It is a theorem 
that at most $n$ functionally independent constants of the motion can be in mutual involution, \cite{Arnold}.
 However, several distinct $n$-subsets of the $2n-1$ polynomial constants of the motion 
for a superintegrable system could be in involution. In that case the system is {\bf multi-integrable}.

 Another  feature of 
superintegrable systems is that the classical orbits traced out by the trajectories  can be determined algebraically, without the need for integration. 
Along any trajectory each of the symmetries  is constant: ${\cal L}_s=c_s$, $s=1,\ldots,2n-1$. Each equation ${\cal L}_s({\bf q},{\bf p})=c_s$ determines a 
$2n-1$ dimensional hypersurface in the $2n$ dimensional phase space, and the trajectory must lie in that hypersurface. Thus the trajectory lies in the
 common intersection of $2n-1$ 
independent hypersurfaces; hence it  must be a curve.
An important  property of real superintegrable systems is loosely stated as ``all bounded trajectories are periodic" \cite{SCQS}. The formal proof  is in \cite{nekhoroshev1972}.  

The polynomial constants of motion for a system with Hamiltonian H are
elements of the (polynomial) Poisson algebra of the system. Stated as a
lemma we have:
\begin{lemma} Let $\cal H$  be a Hamiltonian with constants of the motion ${\cal L}, {\cal K}$. Then $\alpha {\cal L}
+\beta {\cal K}$, ${\cal L}{\cal K}$ and $\{{\cal L},{\cal K}\}$ are also constants of the motion.\end{lemma}
More generally, any set $F_k$ of $k$ polynomial constants of the motion ${\cal L}_1,\ldots,{\cal L}_k$ will generate a symmetry algebra $S_{F_k}$, 
a subalgebra of $S_{\cal H}$, simply by taking all possible finite combinations of scalar multiples, sums, products and Poission brackets of the generators. 
(Since $\cal H$ is always a constant of the motion, we will always require  that ${\cal H}$ must belong to the symmetry algebra generated by $ F_k$.) We will be particularly interested in finding sets of generators for which $S_{F_k}=S_{\cal H}$.  The $n$ defining constants of the motion of a polynomially integrable system do not generate a very interesting symmetry algebra, because all Poisson brackets of the generators vanish. However, for
the $2n-1$ generators of a polynomial superintegrable system the brackets cannot all vanish and the symmetry algebra has nontrivial structure.

The {\bf order} $O({\cal L})$ of a polynomial constant of the motion $\cal L$ is its order as a polynomial in the momenta.  (Note that the order is an intrinsic property of a symmetry: It doesn't change under a transformation from position coordinates $\bf q$ to coordinates ${\bf q}'$.) Here ${\cal H}$ has order 2. The {\bf order} $O(F_k)$  of a set of generators $F_k=\{ {\cal L}_1,\ldots,{\cal L}_k\}$, is the maximum order of the generators, excluding $\cal H$.

Let $S=S_{F_k}$ be a symmetry algebra of a Hamiltonian system generated by the set $F_k$. Clearly, many different sets $F'_{k'}$ can generate the same symmetry algebra. Among all these there will be at least one set of generators $F^0_{k_0}$ for which $\ell=O(F^0_{k_0})$ is a minimum. Here $\ell$ is unique, although $F^0_{k_0}$ is not. We define the {\bf order} of $S$ to be $\ell$.

\subsubsection{Extension to quantum systems}

The extension of superintegrability to quantum systems is relatively straightforward. We state our definitions for systems determined by Schr\"odinger operators of the form
(\ref{laplacebeltrami}) in $n$ dimensions.

\begin{definition}
A quantum  system is  {\bf  integrable}  (of finite-order) if it is integrable and the integrals of motion $L_k$ are finite-order differential operators. 
\end{definition}
\begin{definition}
A quantum system in $n$ dimensions is {\bf  superintegrable} (of finite-order) if it admits $n+k$ $k=1,\ldots,n$ algebraically independent  finite-order partial differential operators 
 ${ L}_1={ H},\ldots,{ L}_{n+k}$ in the variables ${\bf x}$ 
globally defined (except for  singularities on lower dimensional manifolds),  such that $[H,{ L}_j]$=0.
It is {\bf minimally superintegrable  (of finite-order)} if $k=1$  and {\bf maximally  superintegrable} (of finite-order) if $k=n-1$.
\end{definition}
Note that, unlike in the case of classical superintegrability, there is no proof that $2n-1$ is indeed the maximal number of possible symmetry operators. However, there are no counterexamples known to the authors, that 
the maximum possible number of algebraically independent 
symmetry operators for a Hamiltonian of form (\ref{laplacebeltrami}) is $2n-1$.

In analogy with the classical case the  symmetry operators  for quantum Hamiltonian $ H$ form the  {\bf symmetry algebra} $S_{ H}$ of the quantum system, closed under scalar multiplication, multiplication and the commutator:
\begin{lemma} Let $ H$  be a Hamiltonian with symmetries ${ L}, { K}$, and $\alpha,\beta$ be scalars. Then $\alpha { L}
+\beta { K}$, ${L}{ K}$ and $[{L},{K}]$ are also symmetries.\end{lemma}
We will use the term symmetries and integrals of motion interchangeably. 

Any set $F_k$ of $k$ symmetry operators ${ L}_1,\ldots,{ L}_k$ will generate a symmetry algebra $S_{F_k}$, a subalgebra of $S_{ H}$, simply by taking all possible finite combinations of  scalar multiples, sums,  products and commutators of the generators. (Since $ H$ is always a symmetry, we will always require that  ${ H}$ belongs to the symmetry algebra generated by $F_k$.) 
We will be particularly interested in finding sets of generators for which $S_{F_k}=S_{ H}$.   However, for
the $2n-1$ generators of a superintegrable system the commutators cannot all vanish and the symmetry algebra has non-abelian structure, \cite{JVN}.

The {\bf order} $O(L)$ of a symmetry $\cal L$ is its order as a linear  differential operator.  
 The {\bf order} $O(F_k)$  of a set of generators $F_k=\{ { L}_1={ H},\ldots,{ L}_k\}$, is the maximum order of the generators,  excluding $H$. 
Let $S=S_{F_k}$ be a symmetry algebra of a Hamiltonian system generated by the set $F_k$. Many different sets $F'_{k'}$ of symmetry operators
 can generate the same symmetry algebra. Among all these there will be a set  $F^0_{k_0}$ for which $\ell=O(F^0_{k_0})$ is a minimum.  We define the 
{\bf order} of $S$ to be $\ell$.

%% file: Chapter1.tex
\section{Important  Examples}\label{Chapter1}

We present some simple but important examples of superintegrable systems and show, for classical systems, how the trajectories  can be 
determined geometrically (without solving Newton's or Hamilton's equations) and, for quantum systems, how the energy spectrum can be determined algebraically 
(without solving the Schr\"odinger equation), simply by exploiting the structure of the symmetry 
algebra.  We  treat  two-dimensional versions of the Kepler and hydrogen atom systems in flat space and in positive constant curvature space and the hydrogen atom in three-dimensional Euclidean space.

\subsection{The classical Kepler system}
The Kepler problem is a specific case of the two body problem for which one of the bodies is stationary relative to the other and the bodies interact
 according to an inverse square law. The motion of two isolated bodies satisfies this condition to good approximation if one is significantly more massive than the other. 
Kepler specifically investigated the motion of the planets around the sun and stated three laws of planetary motion: 1)
Planetary orbits are planar ellipses with the Sun positioned at a focus. 2) 
A planetary orbit sweeps out equal areas in equal time. 3) The square of the period of an orbit is  proportional to the cube of 
the length of the semi-major axis of the ellipse.

We will see that the precise mathematical statements of these laws are recovered simply via  superintegrability analysis. Since planetary orbits lie in a plane we 
can write the Hamiltonian system in two dimensional Euclidean space with Cartesian coordinates: 
\begin{equation}\label{keplerhamiltonian}{\cal H}={\cal L}_1=\frac12 (p_1^2+p_2^2)-\frac{\alpha}{\sqrt{q_1^2+q_2^2}}, \quad \alpha>0.\end{equation}
 Hamilton's equations of motion are $$ {\dot q_1}=p_1,
\ {\dot q_2}=p_2,\ {\dot p_1}=\frac{\alpha q_1}{(q_1^2+q_2^2)^{3/2}},\ {\dot p_2}=\frac{\alpha q_2}{(q_1^2+q_2^2)^{3/2}},$$
leading to Newton's equations
$ {\ddot {\bf q}}=\frac{\alpha {\bf q}}{({\bf q}\cdot{\bf q})^{3/2}}$.
We shall not solve these equations to obtain  time dependence of the trajectories, but rather show how superintegrability alone implies Kepler's 
laws and determines the orbits.
Here $n=2$ and $2n-1=3$ so three constants of the motion are required for superintegrability. Kepler's second law is a statement of the 
conservation of angular momentum.  The conserved quantity is 
${\cal L}_2=q_1p_2-q_2p_1$
which can be verified by checking that  $\{{\cal H},{\cal L}_2\}=0$. Angular momentum is conserved in any Hamiltonian system with potential that depends only on the 
radial distance $r=\sqrt{q_1^2+q_2^2}$. However, the gravitational potential (along with the isotropic oscillator potential) 
 is special in that it admits a third constant 
of the motion, see the Bertrand Theorem, \cite{Bertrand, Goldstein}.  Consider the  2-vector 
\begin{equation}\label{lrl}\mathbf{e}=\left({\cal L}_3,{\cal L}_4\right)=
\left({\cal L}_2p_2-\frac{\alpha q_1}{\sqrt{q_1^2+q_2^2}},-{\cal L}_2p_1-\frac{\alpha q_2}{\sqrt{q_1^2+q_2^2}}\right)
\end{equation}
in the $q_1-q_2$ plane. One can check that $\{{\cal H},{\cal L}_3\}=\{{\cal H},{\cal L}_4\}=0$, so  the components of $\mathbf{e}$ are constants 
of the motion and the classical Kepler system is superintegrable. 
The quantity $\mathbf{e}$ is called the Laplace-Runge-Lenz vector. We will see that for an elliptical orbit or a hyperbolic trajectory 
 the Laplace-Runge-Lenz vector is directed along the axis formed by the origin  $(q_1,q_2)=(0,0)$  and the perihelion (point of closest approach) of the trajectory 
to the origin.  The perihelion is time-invariant in the Kepler problem, so the direction in which $\mathbf{e}$ points must be a constant of the
 motion. The length squared of the Laplace vector is determined by the energy and angular momentum:
\begin{equation}\label{keplercasimir}\mathbf{e}\cdot\mathbf{e}\equiv{\cal L}_3^2+{\cal L}_4^2=2{\cal L}_2^2{\cal H}+\alpha^2.\end{equation}
We have found 4 constants of the motion and only $2n-1=3$ can be functionally independent. The functional dependence is given by 
(\ref{keplercasimir}). We can use the symmetries, ${\cal L}_1,\cdots,{\cal L}_4$ to generate the symmetry algebra. The remaining nonzero Poisson brackets are 
\be\label{keplerstructure}
\{{\cal L}_2,{\cal L}_3\}=-{\cal L}_4,\quad \{{\cal L}_2,{\cal L}_4\}={\cal L}_3,
\quad \{{\cal L}_3,{\cal L}_4\}=2{\cal L}_2{\cal H}.\ee
(The first  equations show that $\bf e$ transforms as a 2-vector under rotations about the origin.)  The structure equations {\it do not}
define a Lie algebra, due to the quadratic term term ${\cal L}_2{\cal H}$. They, together with the Casimir (\ref{keplercasimir}), define a quadratic algebra, 
a Lie algebra 
only if ${\cal H}$ is restricted to a constant energy. An alternative is to consider ${\cal H}$ as a ``loop parameter" and then the operators ${\cal L}_j$ generate a twisted Kac-Moody algebra \cite{daboul1993hydrogen, daboul2001time}.

\begin{figure}\
\begin{centering}
\includegraphics[scale=1]{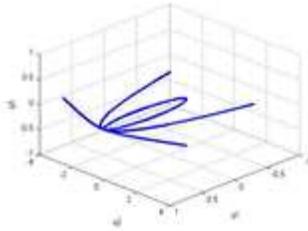}

\caption[Possible Kepler trajectories]{Possible Kepler trajectories: an ellipse, parabola, and hyperbola }\label{fig3}
\end{centering}
\end{figure}

Now suppose we have a specific solution of Hamilton's equations with angular momentum ${\cal L}_2=\ell$ and energy ${\cal H}=E$.
  Because of the radial symmetry  we are free to choose the coordinate axes of the Cartesian coordinates centered at $(0,0)$ 
in any orientation we wish. 
 We choose {\bf peripatetic  coordinates} such that the 
Laplace vector corresponding to the solution  is pointed in the direction of the 
positive $q_1$-axis. In these coordinates we have ${\cal L}_4=0$, ${\cal L}_3=e_1 >0$ and $e_1^2= 2\ell^2E+\alpha^2$.   
Then the first two structure equations simplify to 
$ p_1=-\frac{\alpha q_2}{\ell \sqrt{q_1^2+q_2^2}}$,
$p_2=\frac{e_1}{\ell}+\frac{\alpha q_1}{\ell\sqrt{q_1^2+q_2^2}}$.
The original expression for ${\cal L}_2$  allows us to write:
\[\ell=\frac{q_1 e_1}{\ell }+\frac{\alpha q_1^2}{\ell \sqrt{q_1^2+q_2^2}}+\frac{\alpha q_2^2}{\ell\sqrt{q_1^2+q_2^2}}=
\frac{q_1 e_1}{\ell}+\frac{\alpha \sqrt{q_1^2+q_2^2}}{\ell},\]
which after rearrangement and squaring becomes
\begin{equation}\label{keplerconics}\left(1-\frac{e_1^2}{\alpha^2}\right)q_1^2+\frac{2\ell^2 e_1}{\alpha^2}q_1+q_2^2=\frac{\ell^4}{\alpha^2}.\end{equation}
As is well known from second year calculus, these are conic sections in the $q_1-q_2$ plane; our trajectories are ellipses, parabolas, and hyperbolas, depending on 
the discriminant  of the equation, i.e., depending on the constants of the motion, $e_1,\ell, E$. There are special cases of circles, stationary points, 
and straight lines. The three general cases are presented in Figure \ref{fig3}.

Returning to Kepler's laws, the first is now obvious. The only closed trajectories, or orbits, are elliptical. Kepler's second law is
 a statement of conservation of angular momentum. Indeed,  introduce polar coordinates such that $q_1=r\cos\phi$, $q_2=r\sin\phi$ 
 note  that along the trajectory
$${\ell}={\cal L}_2=q_1(t)p_2(t)-q_2(t)p_1(t)= q_1\frac{dq_2}{dt}-q_2\frac{dq_1}{dt}=r^2\frac{d\phi}{dt}.$$
The area traced out from time $0$ to time $t$ is  
$A(t)=\frac12\int_{\phi(0)}^{\phi(t)}\!r^2(\phi)\,\mathrm{d}\phi$.
Differentiating with respect to time:
$\frac{d}{dt}A(t)=\frac{1}{2}r^2\frac{d\phi(t)}{dt}=\frac{\ell}{2}$,
so the rate is  constant. 
Note that Kepler's third law is only valid for closed trajectories: ellipses.  We may write the period $T$  of such an orbit in terms of the constants 
of the motion. Explicit evaluation for  $\phi(0)=0$, $\phi(T)=2\pi$, yields $A(T)=\frac{\ell T}{2}$ as the area of the ellipse. 
Kepler's third law follows easily from equation (\ref{keplerconics}) and the simple calculus expression for the area of an ellipse.

%%%%%%%%%%%%%%%%%%%%%%%%%%%%%%%%%%%%%%%%%%%%%%%%%%%%%%%%%%%%%%%%%%%%%%%%%%%%%%%%
\subsection{A Kepler analogue on the 2-sphere}
\label{kepler_analogues}
%%%%%%%%%%%%%%%%%%%%%%%%%%%%%%%%%%%%%%%%%%%%%%%%%%%%%%%%%%%%%%%%%%%%%%%%%%%%%%%%

There are analogs of the Kepler problem on spaces of nonzero constant curvature  
that are also superintegrable.  We consider a 2-sphere analog and show 
that superintegrability yields information about  the trajectories, and  that in a particular 
limit we recover the Euclidean space problem. 
 It is convenient 
to consider the  2-sphere as a two dimensional surface embedded in Euclidean 3-space.  Let $s_1,s_2,s_3$ be  standard Cartesian coordinates. Then the equation 
\begin{equation}\label{sphere}s_1^2+s_2^2+s_3^2=1\end{equation}
defines the unit sphere. The embedding  phase space is now six dimensional with conjugate momenta $p_1,p_2,p_3$. The phase space for motion on the 2-sphere will be a four 
dimensional submanifold of this Euclidean phase space. One of the constraints is (\ref{sphere}). Since the tangent vector to any trajectory constrained to
 the sphere is orthogonal to the normal vector, we have the additional phase 
space constraint  $s_1p_1+s_2p_2+s_3p_3=0$.  
The Hamiltonian  is 
\begin{equation}\label{hamiltoniansphere}{\cal H}={\cal J}_1^2+{\cal J}_2^2+{\cal J}_3^2+\frac{\alpha s_3}{\sqrt{ s_1^2+s_2^2}}\end{equation}
with $\alpha<0$ and ${\cal J}_1=s_2p_3-s_3p_2$, where ${\cal J}_2$, ${\cal J}_3$ are obtained by cyclic permutations of $1,2,3$.  If the universe has constant positive curvature, this 
 would be a possible model for planetary motion about the Sun, \cite{Segal}. Note that the ${\cal J}_k$ are 
angular momentum generators, although ${\cal J}_1,{\cal J}_2$ are not constants of the motion. 
Due to the embedding, we have 
$${\cal H}'=p_1^2+p_2^2+p_3^2+\frac{\alpha s_3}{(s_1^2+s_2^2+s_3^2)\sqrt{s_1^2+s_2^2}}=\frac{{\cal H} +(s_1p_1+s_2p_2+s_3p_3)^2}{s_1^2+s_2^2+s_3^2},$$ 
so we can use the usual  Euclidean Poisson bracket
$\{{\cal F}, {\cal G}\}=\sum_{i=1}^3(-\partial_{s_i}{\cal F}\partial_{p_i}{\cal G}+\partial_{p_i}{\cal F}\partial_{s_i}{\cal G})$
for our computations if at the end we restrict to the unit  sphere.  
(Note that we have here normalized the parameters $m/2=1$.) The Hamilton equations for the trajectories $s_j(t), p_j(t)$ in phase space are
$ \frac{ds_j}{dt}=\{ {\cal H},s_j\}$, $ \frac{dp_j}{dt}=\{ {\cal H},p_j\}$, $j=1,2,3$.
 The classical basis for the constants  is
\be\label{constants6} {\cal L}_1=2{\cal J}_1 {\cal J}_3-\frac{\alpha s_1}{\sqrt{s_1^2+s_2^2}},\  {\cal L}_2=2{\cal J}_2{\cal J}_3-\frac{\alpha s_2}{\sqrt{s_1^2+s_2^2}},\ 
{\cal X}={\cal J}_3.\ee
The structure  and Casimir relations  are
\be\label{structure6}\{{\cal X},{\cal L}_1\}=-{\cal L}_2,\  \{{\cal X},{\cal L}_2\}={\cal L}_1,\
 \{{\cal L}_1,{\cal L}_2\}=4({\cal H}-2{\cal X}^2){\cal X},
\ee
\be\label{classcasimir6}{\cal L}_1^2 + {\cal L}_2^2+4 {\cal X}^4 -4{\cal H} {\cal X}^2-\alpha^2=0.
\ee
Here, $({{\cal L}_1},{{\cal L}_2})$ transforms as a vector with respect to rotations about the $s_3$-axis.
 The Casimir relation expresses the square of the length of this vector in terms of the other constants of the motion:
${\cal L}_1^2+{\cal L}_2^2=\kappa^2$, $\kappa^2=\alpha^2+4{\cal H}{\cal X}^2-4 {\cal X}^4$,
where $\kappa\ge 0$. We choose the $s_1,s_2,s_3$ coordinate system so that the vector points in the direction of the positive $s_1$ axis:
 $({\cal L}_1,{\cal L}_2)=(\kappa,0)$. Then 
\be\label{symmetries}{\cal J}_1{\cal J}_3=\frac{{\alpha}s_1}{2\sqrt{s_1^2+s_2^2}}+\frac{\kappa}{2},\quad
{\cal J}_2{\cal J}_3=\frac{{\alpha}s_2}{2\sqrt{s_1^2+s_2^2}},\ee
\[{\cal J}_3={\cal X},\quad
{\cal J}_1^2+{\cal J}_2^2+{\cal J}_3^2={\cal H}-\frac{{\alpha}s_3}{\sqrt{s_1^2+s_2^2}}.\]
Substituting the first three equations into the fourth, we obtain:
\be\label{cone2}\left({\cal H}{\cal X}^2-(\frac{\alpha^2}{4}+\frac{\kappa^2}{4})-{\cal X}^4\right)^2(s_1^2+s_2^2)-\alpha^2(\frac{\kappa s_1}{2}+{\cal X}^2s_3)^2=0\ee
For fixed values of the constants of the motion equation (\ref{cone2}) describes a cone. Thus the orbit lies on the 
intersection of this cone with the unit sphere $s_1^2+s_2^2+s_3^2=1$,  a conic section. This is the spherical geometry analog of Kepler's first law.
A convenient way to view the trajectories is to project them  onto the $(s_1,s_2)$-plane:  $(s_1,s_2,s_3)\to (s_1,s_2)$. The projected points describe a curve in the unit disc $s_1^2+s_2^2\le 1$. This curve is defined by 
\be\label{projectedtraj}[{\cal H}{\cal X}^2-(\frac{\alpha^2}{4}+\frac{\kappa^2}{4})-{\cal X}^4]^2(s_1^2+s_2^2)-\alpha^2(\frac{\kappa s_1}{2}\pm{\cal X}^2\sqrt{1-s_1^2-s_2^2})^2=0.\ee
The plus sign corresponds to the projection of the trajectory from the northern hemisphere, the minus sign to  projection from the southern hemisphere.

Notice that the potential has an attractive singularity at the north pole: $s_1=s_2=0,s_3=1$ and a repulsive singularity at the south pole.
In polar coordinates $s_1=r\cos\phi$, $s_2=r\sin\phi$ the equation of the  projected orbit on the $(s_1,s_2)$-plane is given by
\be\label{polarorbit} r^2=\frac{4{\cal X}^4}{4{\cal X}^4+(-\alpha+\kappa\cos\phi)^2},\quad 0\le\phi<2\pi.\ee

\begin{theorem}
For nonzero angular momentum, the projection sweeps out area   $A(t)$ in the plane at a constant rate $\frac{dA}{dt}={\cal X}$ with respect to the origin (0,0).
\end{theorem}
\begin{theorem}
For nonzero angular momentum, the period of the orbit is \be\label{polarperiod}T=\frac{4{\cal X}^3\pi \left(\frac{1}{4{\cal X}^4+(\kappa-\alpha)^2}\sqrt{\frac{4{\cal X}^4+(\kappa-\alpha)^2}{4{\cal X}^4+(\kappa+\alpha)^2}}
+\frac{1}{{\cal X}^4+(\kappa+\alpha)^2}\right)}{\sqrt{2\sqrt{\frac{4{\cal X}^4+(\kappa-\alpha)^2}{4{\cal X}^4+(\kappa+\alpha)^2}}+2\frac{4{\cal X}^4+\alpha^2-\kappa^2}{4{\cal X}^4+(\alpha+\kappa)^2}}}.\ee
\end{theorem}

\subsubsection{Contraction to the Euclidean space Kepler problem} \label{limit}

The sphere model can be considered as describing a 2-dimensional bounded ``universe'' of radius $1$. 
Suppose an observer is situated  ``near''  the attractive north pole. This observer uses a system of units with unit 
length $\epsilon$ where $0<\epsilon<<1$. The observer is unable to detect that she is on a 2-sphere; to her the universe appears flat. In her units, the coordinates are
\be \label{contractioncoords} \ba{l}s_1=\epsilon x,\quad  s_2=\epsilon y,\quad   s_3= 1+O(\epsilon^2),\\
 p_1=\frac{p_x}{\epsilon},\quad  p_2=\frac{p_y}{\epsilon},\quad  p_3= -(xp_x+yp_y)+O(\epsilon^2).\ea\ee
 Here, $\epsilon^2$ is so small that to the observer, it appears that the universe is the plane $s_3=1$ with local Cartesian coordinates $(x,y)$. 
We define new constants $\beta,{ h}$ by 
\be \alpha=\beta/\epsilon, \qquad {\cal H}-{\cal X}^2=h/\epsilon^2. \label{contractionparam}\ee
Substituting into (\ref{hamiltoniansphere}), we find
$\frac{1}{\epsilon^2}(p_x^2+p_y^2)+\frac{\beta}{\epsilon^2\sqrt{x^2+y^2}}=\frac{ h}{\epsilon^2}$.
Multiplying both sides of this equation by $\epsilon^2$  we obtain the Hamiltonian for the Euclidean  Kepler system:
\be\label{KeplerHam} p_x^2+p_y^2+\frac{\beta}{\sqrt{x^2+y^2}}={ h}.\ee
Using the same procedure we find that the constants of the motion become
$ {\cal X}=xp_y-yp_x$, $  {\cal L}_1=\frac{e_1}{\epsilon}$,$ {\cal L}_2=\frac{e_2}{\epsilon}$,
where $(e_1,e_2)$ is the Laplace-Runge-Lenz vector for the Kepler system:
$ e_1=-2{\cal X}p_y-\frac{\beta x}{\sqrt{x^2+y^2}}$,$ e_2=2{\cal X}p_x-\frac{\beta y}{\sqrt{x^2+y^2}}$.
The same procedure applied to the structure equations yields
\[
\{{\cal X},{e}_1\}=-{e}_2,\  \{{\cal X},{e}_2\}={e}_1,\
 \{{e}_1,{e}_2\}=4{h}{\cal X},\ 
{e}_1^2 + {e}_2^2 -4{h} {\cal X}^2-\beta^2=0.
\]
Thus the length of the Laplace vector is $k=\kappa/\epsilon$ where $k^2=4h{\cal X}^2+\beta^2$. To the observer, the trajectories lie in the plane $s_3=1$.
and  equation (\ref{projectedtraj}) for  the paths of the trajectories is
\be\label{Keplertraj}[{h}{\cal X}^2-(\frac{\beta^2}{4}+\frac{k^2}{4})]^2(x^2+y^2)-\beta^2(\frac{kx}{2}+{\cal X}^2)^2=0.\ee
The solutions of the Kepler trajectory equations (\ref{Keplertraj}) are the usual conic sections: intersections of a plane and a cone.

\medskip
\subsubsection{Trajectory determination} 

To obtain a plot of a  trajectory, defined by its constants of the motion and  parametrized by time, we can integrate Hamilton's equations
 numerically. To do this it is necessary to identify a point in six-dimensional phase space that lies on the trajectory, to 
serve as an intial point for integration. Thus for each set of constants of the motion we must find a distinguished point.

First we take the case ${\cal X}\ne 0$. We project the orbits onto the ${s}_1-{s}_2$ unit disk.
From (\ref{polarorbit}), we see that ${r}^2$,  is minimized when ${\phi}=0$. This is the perihelion of the orbit, which implies that
 ${q}_1={r}({0})=\sqrt{\frac{4{\cal X}^4}{4{\cal X}^4+(-{\alpha}+{\kappa})^2}}$, ${q}_2=0$, ${p}_1=0$, 
 ${p}_2=\frac{\cal X}{{q}_1}$, ${q}_3=\frac{({\alpha}+{\kappa})}{\sqrt{4{\cal X}^4+(-{\alpha}+{\kappa})^2}}$ and ${p}_3=0$.
 Note that aphelion occurs at ${q}_1=-{r}({\pi})=\sqrt{\frac{4{\cal X}^4}{4{\cal X}^4+(-{\alpha}-{\kappa})^2}}$, ${q}_2=0$, ${p}_1=0$,  ${p}_2=-\frac{\cal X}{{q}_1}$.

The Casimir relation  implies $\kappa^2=\alpha^2+4({\cal H}-{\cal X}^2){\cal X}^2>0$. If ${\cal H}={\cal X}^2$, then $\alpha^2=\kappa^2$ and $-\alpha=\kappa$, so the aphelion is precisely unity, while perihelion is less than unity.
If ${\cal H}<{\cal X}^2$, we have $-\alpha<\kappa$, which implies aphelion occurs in the northern hemisphere.
If ${\cal H}>{\cal X}^2$, we have $-\alpha>\kappa$, which implies aphelion occurs in the southern hemisphere.

In the case of zero angular momentum, ${\cal X}=0$,  (\ref{symmetries}) implies ${s}_2=0$, and ${\alpha}={\kappa}$, 
so ${\cal H}={{\cal J}_2}^2+\frac{{\alpha}{s}_3}{{s}_1}$. The projection  onto the disk
is a segment of the  line $s_2=0$, so the motion occurs on a segment of a great circle on the 2-sphere. 
All trajectories crash into the north pole in finite time.

\medskip
{\bf Classification of trajectories}

{\bf Case 1:} The trajectory is contained within the northern hemisphere. 
We have ${r}^2={4{\cal X}^2}/{4{\cal X}^2+(-{\alpha}+{\kappa}\cos{\phi})^2}<1$ for all ${\phi},$  and ${\cal X}^2> {\cal H}$.

{\bf Case 2:} The projection has one point of tangency with the  circle; the trajectory is contained in the closure of a  hemisphere,
 $\frac{\alpha}{\kappa}=\pm1$ so ${\cal H}={\cal X}^2$.

{\bf Case 3:} The projection has two points of tangency with the unit circle; the trajectory has points in both hemispheres. 
We have  ${\cal H}>{\cal X}^2$.

\medskip

We correlate these cases  with the original Kepler orbits, using  (\ref{Keplertraj}). Dividing both sides of the equation by ${\beta}^2{k}^2$, we write the 
discriminant as
 ${\Delta}=\left(1-\frac{{\beta}^2}{{k}^2}\right)\left(\frac{{\beta}^2}{{k}^2}\right)=\left(1-\frac{{\alpha}^2}{{\kappa}^2}\right)\left(\frac{{\alpha}^2}
{{\kappa}^2}\right)$. When $\frac{{\alpha}^2}{{\kappa}^2}>1$, ${\Delta}<0$; when $\frac{{\alpha}^2}{{\kappa}^2}=1$, ${\Delta}=0$; 
and when $\frac{{\alpha}^2}{{\kappa}^2}<1$, ${\Delta}>0$. Thus cases 1, 2, and 3 are, respectively, analogies of ellipses, parabolas, 
and hyperbolas in the original Kepler problem.

\begin{figure}[h] \begin{centering}
\includegraphics[scale=1]{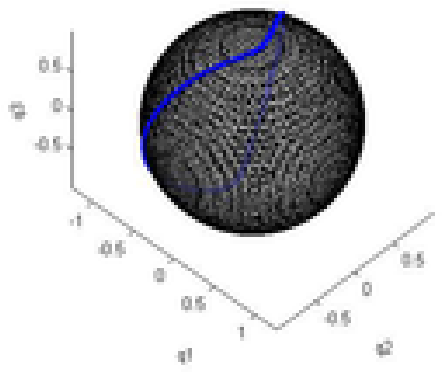}\includegraphics[scale=1]{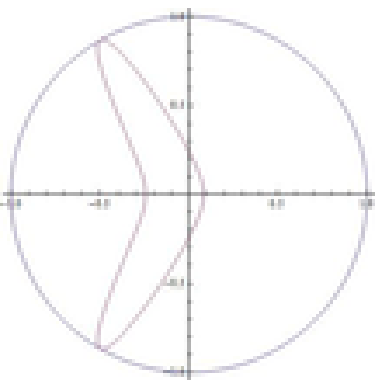}
\caption[A hyperbolic type orbit on the sphere]{A hyperbolic type orbit determined by the parameters $q_1=2/\sqrt{580}$, $q_2=0$, $q_3=211/\sqrt{580}$, $p_1=0$, $p_2=\sqrt{580}/2$. $p_3=0$, $\alpha=-8$, $\kappa=16$, ${\cal X}=1$. To the left, the orbit on the sphere and the right, the projection of the orbit.  Note that the orbit crosses into the southern hemisphere.}\label{hyperbola1.eps}
\end{centering}
\end{figure}

%\begin{figure}[p] \begin{centering}
%\caption{The projection of the hyperbolic orbit, \label{hyperbolaproj.eps}
%\end{centering}
\begin{figure}[h] \begin{centering}
\includegraphics[scale=1]{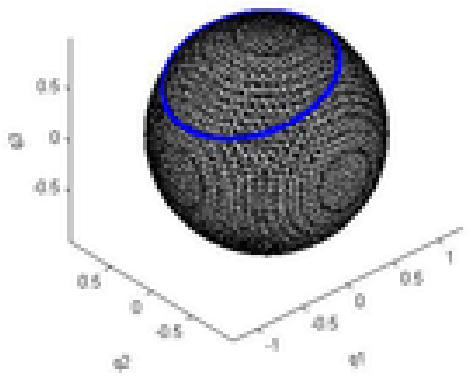} \includegraphics[scale=1]{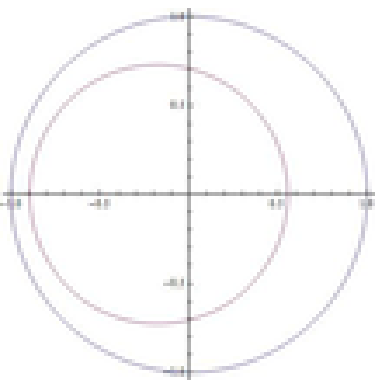}
\caption[An elliptic type orbit on the sphere]{An elliptic  type orbit determined by the parameters $q_1=2/\sqrt{580}$, $q_2=0$, $q_3=211/\sqrt{580}$, $p_1=0$, $p_2=\sqrt{580}/2$. $p_3=0$, $\alpha=-8$, $\kappa=16$, ${\cal X}=1$. To the left, the orbit on the sphere and the right, the projection of the orbit. Note that the orbit is restricted to the northern hemisphere.}\label{ellipse.eps}
\end{centering}
\end{figure} 
%\begin{figure}[p] \begin{centering}
%\caption{The projection of the elliptic orbit, }\label{ellipseproj.eps}
%\end{centering}
%\end{figure} 
%\begin{figure}[p]
%\centerline{\includegraphics[width=6cm]{3.eps}}
%\caption{A circular  type orbit. The projection of this orbit onto the unit disk is a circle. It is restricted to the northern hemisphere.}\label{3.eps}
%\end{figure} 
\begin{figure}[h] \begin{centering}
\includegraphics[scale=1]{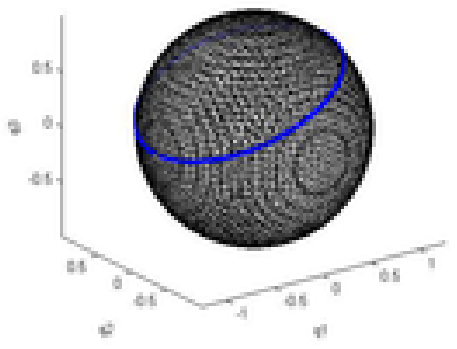}\includegraphics[scale=1]{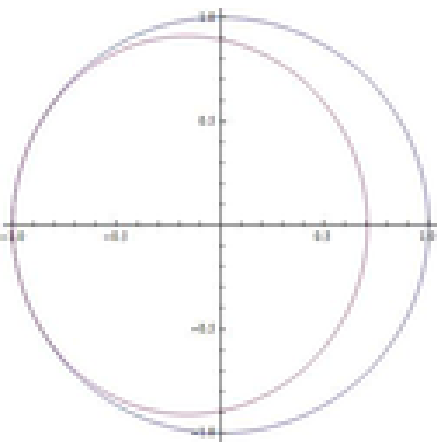}
\caption[A parabolic type orbit on the sphere]{A parabolic  type orbit determined by the parameters $q_1=3/\sqrt{13}$, $q_2=0$, $q_3=3/\sqrt{13}$, $p_1=0$, $p_2=\sqrt{13}/2$. $p_3=0$, $\alpha=-2$, $\kappa=1$, ${\cal X}=1$. To the left, the orbit on the sphere and the right, the projection of the orbit. It is restricted to the northern hemisphere but tangent to the equator.}\label{parabola.eps}
\end{centering}
\end{figure} 
%\begin{figure}[p] \begin{centering}
%\caption{The projection of the parabolic orbit,  }\label{parabolaproj.eps}
%\end{centering}
%\end{figure} 
%\begin{figure}[p]
%\centerline{\includegraphics[width=6cm]{4.eps}}
%\caption{A  great circle trajectory with negative energy. The trajectory is restricted to the northern hemisphere.}\label{4.eps}
%\end{figure} 
%\begin{figure}[p]
%\centerline{\includegraphics[width=6cm]{5.eps}}
%\caption{A great circle trajectory with positive energy. The trajectory crosses into the southern hemisphere.}\label{5.eps}
%\end{figure} 

\subsubsection{The Hohmann transfer on the 2-sphere}
The Hohmann transfer is a rocket maneuver used to place satellites in geocentric orbits about the earth and for interplanetary  navigation, \cite{Curtis}.  It
is based on the superintegrability of the classical Kepler system and 
 uses  impulse maneuvers to change trajectories. For an impulse maneuver the rocket
engine is turned on for a very short time but with a powerful 
thrust. At the instant $t_0$ of
firing, the rocket which  was on a  trajectory  with present linear momentum ${\bf p}^{(0)}$ and position ${\bf s}^{(0)}$ 
 immediately  follows a new trajectory with
initial position still ${\bf s}^{(0)}$ but new linear momentum ${\bf p}^{(1)}$ at
time $t_0$. The
change in momentum  is $\Delta {\bf p}={\bf p}^{(1)}-{\bf p}^{(0)}$, and it
completely determines the new trajectory of the rocket. This is an
idealization, but for very brief rocket firing during a long mission it
can be quite accurate and energy efficient. The mass of the rocket will decrease due to the
engine firing: however,  the mass of the rocket factors out of the
orbit equations, so  will not  effect our calculations.

\begin{figure}
\begin{centering}

\includegraphics[scale=1]{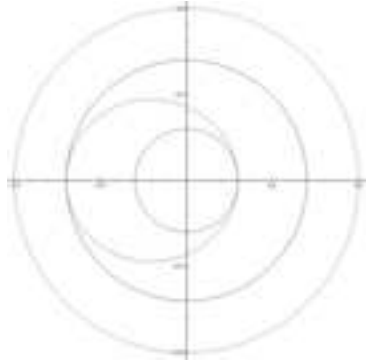}

\caption[Hohman transfer trajectories on the sphere]{The Hohman transfer trajectories projected onto the unit disk. Here $\alpha=-1$, $r_p=.3$, 
$r_a=.7$, $\kappa=.05142$ and ${\cal X}=.0567$.}\label{Hohmann1}
\end{centering}
\end{figure}

The Hohmann transfer, proposed by Hohmann in 1925 \cite{Curtis}, uses impulse maneuvers to take a rocket from a near-Earth circular orbit 
to a higher circular orbit with an expenditure of  minimum $\Delta({\bf p})$. 
To explain the basic idea we apply it in the  setting of the 2-sphere universe. Suppose we have a rocket with engines turned off that is in a 
counter-clockwise circular orbit around the  north pole.  Thus the Laplace vector is zero, i.e.,
 $\kappa^{(0)}=0$, the angular momentum is
${\cal X}^{(0)}>0$ and the energy ${\cal H}^{(0)}$. The projection  in the $(s_1,s_2)$-plane is the circle with radius 
 $r^{(0)}=2({\cal X}^{(0)})^2/\sqrt{4({\cal X}^{(0)})^4+\alpha^2}$.
We use impulse maneuvers to put the rocket in a new counter-clockwise circular orbit  farther from the pole. 
The constants of the motion for the target orbit 
are given as  $\kappa^{(1)}=0$, ${\cal X}^{(1)}>0$ and the energy ${\cal H}^{(1)}$. The projection of the target orbit in the $(s_1,s_2)$-plane is the circle with radius 
 $r^{(1)}=2({\cal X}^{(1)})^2/\sqrt{4({\cal X}^{(1)})^4+\alpha^2}$
where $r^{(1)}>r^{(0)}$. We choose  periaptic coordinates so that at time $t=0$ the projection of the rocket orbit is crossing the positive $s_1$-axis at the point
$s_1=r^{(0)}, s_2=0$. The linear momentum is $p_1=0, p_2={\cal X}^{(0)}/ r^{(0)}$. (On the 2-sphere we have $s_3=\alpha/r^{(0)}, p_3=0$.)  
At  $t=0$ we fire the rocket motor briefly in a direction tangent to the orbit,
to boost the momentum  from $p_2$ to $p_2+\Delta_1 (p_2)$. (Here, the phase space coordinates $p_1, p_3,s_1,s_2,s_3$ remain unchanged.
If $\Delta_1 (p_2)$ is appropriately chosen  the rocket  will now 
follow an elliptical type orbit in the $(s_1,s_2)$-plane with equation
$r^2=4{\cal X}^4/(4{\cal X}^4+(-\alpha+\kappa\cos\phi)^2)$
where ${\cal X},\kappa$ are new constants of the motion. We  design this orbit such that the perigee radius
$(\phi=0)$ is $r_p=r^{(0)}$ and
the apogee radius $(\phi=\pi)$ is $r_a=r^{(1)}$, so that the ellipse will be tangent to the
circular destination orbit at apogee.  Thus we require 
$$ (r^{(0)})^2=r_p^2=\frac{4{\cal X}^4}{4{\cal X}^4+(-\alpha+\kappa)^2} ,\quad (r^{(1)})^2=r_a^2 =\frac{4{\cal X}^4}{4{\cal X}^4+(\alpha+\kappa)^2} $$
Solving for $\cal X$ and $\kappa$ in these two equations we find
$$\kappa=-\frac{\alpha}{r_a^2-r_p^2}(r_a\sqrt{1-r_p^2}-r_p\sqrt{1-r_a^2})^2,\quad {\cal X}^4=-\frac{\kappa \alpha r_a^2r_p^2}{r_a^2-r_p^2}.$$
Note that $0<r_p<r_a\le 1$. At perigee on the elliptical-type orbit we must have $p_2={\cal X}/r^{(0)}$. Thus $\Delta(p_2)=({\cal X}-{\cal X}^{(0)})/r^{(0)}$.
The time for the satellite to travel from perigee to apogee is half a period
$T/2$ where $T$ is given by (\ref{polarperiod}). 
At apogee we again fire the rocket engine, briefly, to put the satellite in the higher circular orbit. We orient the engine so it fires in direction of 
 the tangent vector of the trajectory. At the instant of firing, $t=T/2$, the linear momentum at  apogee is $p_1=0, p_2=-{\cal X}/r^{(1)}, p_3=0$ and the 
momentum vector is tangent to the trajectory. 
At the instant just after firing the position is the same but the new momentum in the $s_2$-direction is $
p_2+\Delta_2 (p_2)$. We require that $\Delta_2 (p_2)$
is exactly the change in momentum required to put the satellite in the
higher circular orbit, i.e., to make $\kappa^{(1)}=0$ and the angular momentum equal to ${\cal X}^{(1)}$.
 Thus  $p_2+\Delta_2 (p_2)=-{\cal X}^{(1)}/r^{(1)}$, so $\Delta_2(p_2)= ({\cal X}-{\cal X}^{(1)})/r^{(1)}$.
The total delta-$p$ is $\Delta_1(p_2)+\Delta_2(p_2)$ which can be shown to be the minimum required to
move to the higher circular orbit. 
An example of the transfer is  in Figure \ref{Hohmann1}.
Similar Hohmann  transfers can be designed to change to and from  elliptic type and hyperbolic type trajectories.

%\begin{figure}
%\begin{centering}

%\includegraphics[scale=0.5]{hohman2.eps}

%\caption{A second example of  Hohman transfer trajectories projected onto the unit disk. Here $\alpha=-1$, $r_p=.1$, 
%$r_a=.9$, $\kappa=.9072$ and ${\cal X}=.00831$.}\label{Hohmann2}
%\end{centering} 
%\end{figure}

\subsection{Quantum Kepler-Coulomb on the 2-sphere}
\label{quantum_kepler}
Our last example is an analog of the hydrogen atom 
on the 2-sphere. This system is superintegrable and, just as in Section \ref{limit}, we  show that the Euclidean quantum Kepler-Coulomb system can be obtained as a limit 
of the 2-sphere system. 
As in the classical case we embed the 2-sphere in Euclidean 3-space. We define the  Hamiltonian operator for the quantum Kepler system  as 
\begin{equation}\label{qhamiltoniansphere} H=\sum\limits_{i=1}^{3} J_i^2+\frac{\alpha s_3}{\sqrt{s_1^2+s_2^2}}\end{equation}
Where the operator $J_1=s_2\partial_3-s_3\partial_2$ and $J_2$, $J_3$ are obtained by cyclic permutations of this expression. 
Our immediate goal is to determine the bound state eigenvalues 
 of the Hamiltonian operator according to the equation
$H\Psi=E\Psi$
for functions  $\Psi({\bf s})$ square integrable on the 2-sphere $s_1^2+s_2^2+s_3^2=1$. This system was first treated in \cite{schroedinger1940} and \cite{stevenson1941}, by different methods. We compute in Euclidean 3-space with Cartesian 
coordinates $s_1,s_2,s_3$, but at the end we restrict our solutions to the 2-sphere. 
A basis for the symmetry operators is
\[{ L}_1={J}_1 { J}_3+J_3J_1-\frac{\alpha s_1}{\sqrt{s_1^2+s_2^2}},\
{ L}_1={J}_1 { J}_3+J_3J_1-\frac{\alpha s_1}{\sqrt{s_1^2+s_2^2}},\
{ X}={ J}_3,\]
along with $H$. They satisfy the structure relations
\[[ X,{ L}_1]=-{ L}_2,\
[{ X},{L}_2]={ L}_1,\
[{L}_1,{ L}_2]=4{ H}X-8{ X}^3+X,\]
\begin{equation}\label{casimirq}{L}_1^2 + { L}_2^2+4 { X}^4 -4{ H} { X}^2+H-5X^2-\alpha^2=0.\end{equation}
We define a new basis of symmetry operators
$
L_0=iX$,
$L^+=L_1-iL_2$,
$L^-=L_1+iL_2$.
Here $L^\pm$ are raising and lowering operators for the eigenvalues of $L_0$, just as in Lie algebra theory. 
The structure and Casimir relations become
\begin{equation}[L_0,{ L}^\pm]=\pm L^\pm,\ 
[{L}^+,L^-]=8HL_0+16L_0^3+2L_0,\end{equation}
\begin{equation}\label{qc1}L^+L^--(4HL_0+8L_0^3+L_0)=-4L_0^4-4HL_0^2-H-5L_0^2+\alpha^2.\end{equation}
Equivalently (\ref{qc1}) may be written as
\begin{equation}\label{qc2}L^-L^++(4HL_0+8L_0^3+L_0)=-4L_0^4-4HL_0^2-H-5L_0^2+\alpha^2\end{equation}

Now let $V_E$ be the finite dimensional eigenspace of vectors $\Psi$ with eigenvalue $E$. Thus $V_E$ is the vector space of all solutions $\Psi$ of 
the Schr\"odinger equation $H\Psi=E\Psi$.  Since all of the symmetry 
operators commute with $H$, each maps the space $V_E$ into itself. For example, since $[H,L_0]=0$, if $\Psi\in V_E$ then $H(L_0\Psi) =L_0H\Psi=EL_0\Psi$, 
so $L_0\Psi\in V_E$. Suppose $V_E$ is $m$-dimensional. Then we can find a basis of $m$ vectors for $V_E$ with respect to which the action of the 
symmetry operators restricted to the eigenspace is given by $m\times m$  matrices. The matrix corresponding to $H$ is just $E\ {\rm I}$ where 
$\rm I$ is the $m\times m$ identity matrix. The structure equations are now satisfied in the algebra of matrix products and linear combinations.
 We have a representation of the symmetry algebra.
 Quantum superintegrable systems  have 
 degenerate eigenvalues and the reason for this is closely related to the representations.
 We can determine the degeneracies and compute 
the possible eigenvalues $E$ through analysis of the representations.

Here $L_0$ is formally self-adjoint and $L^\pm$ are mutual adjoints.
 We will find an orthogonal   basis for $V_E$ consisting of eigenvectors of $L_0$. For this basis,
 the matrix corresponding to $L_0$ will be diagonal.  The following property is critical. 
Let $\Psi\in V_E$ be an eigenvector of $L_0$ with eigenvalue $\lambda$.
\begin{lemma}\label{actionLp}  Either $L^+\Psi=\theta$ (the zero vector) or $L^+\Psi$ is an eigenvector of $L_0$ with eigenvalue $\lambda+1$. Furthermore, either $L^-\Psi=\theta$  or $L^-\Psi$ is an eigenvector of $L_0$ with eigenvalue $\lambda-1$.
\end{lemma}
Given the eigenvector $\Psi\in V_E$ we define the repeated action of the  symmetry operators as $L_0^k\Psi$ for $k=0,1,2\ldots$. Since $V_E$ is  finite-dimensional, there exist integers $k$ such that $(L^+)^{k+1}\Psi =\theta$. Let $k_0$ be the smallest such integer and define  $\Phi_0=(L^+)^{k_0}\Psi.$ By Lemma \ref{actionLp}, $\Phi_0$ is an eigenvector of $L_0$,with $L_0\Phi_0=\mu\Phi_0$ where $\mu=\lambda+k_0$.  We call $\Phi_0$ a highest weight vector. 
 By repeatedly applying  $L^-$ to $\Phi_0$ we can generate  eigenvectors $\Phi_j=
(L^-)^{j}\Phi_0$.  There will be a smallest  integer $j=n$ such that  $L^-\Phi_n=\theta$. Thus
$$ L_0\Phi_j=(\mu-j)\Phi_j,\quad j=0,1,\ldots,n, \quad L^-\Phi_j=\Phi_{j+1}, \ j=0,\ldots,n-1.$$
Now we apply both sides of the Casimir equation (\ref{qc1}) to $\Phi_j$:
$ L^+\Phi_{j+1}=\left((4E+1)(\mu-j)+8(\mu-j)^3-4(\mu-j)^4
-(4E+5)(\mu-j)^2-E+\alpha^2\right)\Phi_j$.\break  This shows that the raising operator moves us back up the ladder of eigenvectors, one rung at a time.
In the special case $j=n$ we obtain 
\be\label{bottom}(4E+1)(\mu-n)+8(\mu-n)^3-4(\mu-n)^4
-(4E+5)(\mu-n)^2-E+\alpha^2 =0.\ee
Next we apply both sides of  (\ref{qc2}) to $\Phi_0$
to get
$-(4E+1)\mu-8\mu^3-4\mu^4-(4E+5)\mu^2-E+\alpha^2=0$. 
Solving  for $E$ we find
\be\label{Eeqn1} E=-\frac14(2\mu+1)^2+\frac14+\frac{\alpha^2}{(2\mu+1)^2},\ee
expressing $E$ in terms of the highest weight $\mu$. Substituting into (\ref{bottom}) we can obtain
a equation that is polynomial in $\mu$ and which factors as
\be\label{mueqn} 
 (n-2\mu)(n+\sqrt{(n+1)^2+2i\alpha}-2\mu)(n+\sqrt{(n+1)^2-2i\alpha}-2\mu )\ee
 $$\times (n-\sqrt{(n+1)^2+2i\alpha}-2\mu) (n-\sqrt{(n+1)^2-2i\alpha}-2\mu) =0.$$    
 The first factor gives the solution $\mu=n/2$. The other solutions are  complex and eigenvalues of a self-adjoint operator must be real.
 Thus the only solution is $\mu=n/2$. It follows that for every  $n$ there is a  $V_{E_n}$ with dimension $n$, energy
 \be\label{Eeqn2} E_n=-\frac14(n+1)^2+\frac14+\frac{\alpha^2}{(n+1)^2}.\ee
 If we choose a complex solution, say $2\mu=n+\sqrt{(n+1)^2+2i\alpha}$ then we still have  an
  eigenspace with energy  (\ref{Eeqn1}), but $\mu$, $E_n$ are complex, so rejected on physical grounds.
 However,  if $\alpha=-i a$,  $a$  real, then $2\mu_n=n+\sqrt{(n+1)^2+2a}$  and 
 $E_n=-\frac14(2\mu_n+1)^2+\frac14-\frac{a^2}{(2\mu_n+1)^2}$,
 is real. In traditional quantum mechanics this is rejected because  $H$ now has a complex potential function, so is not self-adjoint. 
This is an example of a PT-symmetric quantum mechanical system, \cite{ BB, BBM, znojil1999pt}, a field whose study is now under intensive development.

\subsubsection{Contraction to Euclidean space}

 As  for the classical case, we consider a system of
 units with unit length  $\epsilon$ near the point ${s_1}=0$, ${s_2}=0$, ${s_3}=1$ on the 2-sphere via  \eref{contractioncoords} and \eref{contractionparam}. As in the classical case, the Hamiltonian $H$ on the two-sphere contracts to a Hamiltonian on Euclidean space \begin{equation}h=({{\partial}_x}^2+{{\partial_y}^2})+\frac{\beta}{\sqrt{x^2+y^2}}.\end{equation}
A similar procedure applied to the symmetry operators yields 
\be \ba{l}  {\epsilon}{L_1}={\ell}_1=-{\partial}_y(x{\partial}_y-y{\partial}_x)-(x{\partial}_y-y{\partial}_x){\partial}_y-\frac{{\beta}x}{\sqrt{x^2+y^2}},\\
{\epsilon}{L_2}={\ell}_2=-{\partial}_x(x{\partial}_y-y{\partial}_x)-(x{\partial}_y-y{\partial}_x){\partial}_x-\frac{{\beta}y}{\sqrt{x^2+y^2}},\\
{\epsilon}{ X}={\chi}=x{\partial}_y-y{\partial}_x.\ea\ee The  structure relations contract to 
\be [h,{\ell_1}]=[h,{\ell_2}]=[h,{\chi}]=0,\
[{\chi},{\ell}_1]=-{\ell}_2,\  [{\chi},{\ell}_2]={\ell}_1,\ee
\[ [{\ell}_1,{\ell}_2]=4h{\chi}+{\chi},\
{\ell}_1^2+{\ell}_2^2+h-5{\chi}^2-{\beta}^2=0.\] Setting ${\ell}_0=i{\chi}$, ${\ell}^+={\ell}_1-i{\ell}_2$, and  ${\ell}^-={\ell}_1+i{\ell}_2$, we obtain:
\be [h,{\ell}_0]=[h,{\ell}^+]=[h,{\ell}^-]=0,\ 
[{\ell}_0,{\ell}^{\pm}]=\pm{\ell}^{\pm},\
 [{\ell}^+,{\ell}^-]=2(4h+1){\ell}_0\label{su2}.\ee
 
With the rescaling \eref{contractionparam}, the energy spectrum (\ref{Eeqn2}) becomes
$E_n=-\frac{{\epsilon}^2}4(n+1)^2+\frac{{\epsilon}^2}4-\frac{\beta^2}{(n+1)^2}$.
 Taking the 
limit as ${\epsilon}\rightarrow0$, we recover the spectrum of the hydrogen atom confined to a plane,  
\be E_n=-\frac{\beta^2}{(n+1)^2}.\ee

\subsection{Quantum Coulomb problem in Euclidean space $E_3$}
The Schr\"odinger equation for the hydrogen atom is solved in detail in every textbook on quantum mechanics. Here we shall briefly review its superintegrability \cite{Pauli, Bargmann, Fock}. The quantum Hamiltonian 
\be \label{HCoul3} H=(\vec{p})^2-\frac{\alpha}{r}, \qquad  r=\sqrt{x_1^2+x_2^2+x_3^2}, \quad \alpha>0,\ee
commutes with the angular momentum operators 
\be \vec{L}= \vec{r}\times \vec{p}\ee
and the Laplace-Runge-Lenz vector 
\be \vec{A}=\vec{p}\times \vec{L}-\vec{L}\times \vec{p}-\frac{\alpha}{r}\vec{r}.\ee
The commutation relations between these operators are 
\bea [\vec{L}, H]=[\vec{A}, H]=0, \\
     \fl   [L_j, L_k]=i\varepsilon_{jk\ell}L_{\ell}, \qquad [L_j, A_k]=i\varepsilon_{jk\ell}A_{\ell},\qquad
       [A_j, A_k]=-i\varepsilon_{jk\ell}L_{\ell}H\label{hydrogenalgebra},\eea where $\varepsilon_{jk\ell}$ is the completely antisymmetric tensor.
Thus the angular momentum components $L_j$ generate an $o(3)$ algebra and the Laplace-Runge-Lenz vector transforms like a vector under rotations. However, \eref{hydrogenalgebra} shows that the integrals of motion $\{L_j, A_k\}$ do not strictly speaking form a finite dimensional Lie algebra. The standard procedure \cite{Pauli, Fock, Bargmann} is to restrict to the case of a fixed energy, i.e. put  $H=E$ where $E$ is a fixed constant. For $E<0$ (bound states) the relations \eref{hydrogenalgebra} correspond to the Lie algebra $o(4)$. For $E>0$ (scattering states), we obtain the $o(3,1)$ Lie algebra. For $E=0$, also in the continuous spectrum, we have the Euclidean algebra $e(3).$ A different possibility, the one adopted in this review, is to view $\{\vec{L}, \vec{A}, H\}$ as the generators of a quadratic algebra. Finally, we mention the possibility of considering $H$ as a ``loop parameter" as mentioned above.  The relation \eref{hydrogenalgebra} then leads to an infinite dimensional Lie algebra, namely a twisted centerless Kac-Moody algebra \cite{daboul1993hydrogen, daboul2001time}. In any case, the representations of each of these algebras lead to the same formula for the spectrum of the hydrogen atom, namely the famous Balmer formula, 
\be E=-\frac{2me^4}{\hbar^2}{\frac{1}n}.\label{Balm}\ee 
An interesting historical footnote is that W. Pauli in his famous 1926 paper \cite{Pauli} uses precisely the formulas \eref{hydrogenalgebra}  to derive the Balmer formula \eref{Balm}. He calls $\vec{A}$ the Lenz vector and does not mention the group $O(4)$ nor the algebra $o(4)$ explicitly. That was done about 16 years later by Fock and Bargmann \cite{Fock, Bargmann}. Pauli obtained this result before the Schr\"odinger equation was known \cite{Schro} using only the algebra, no calculus. 

%% file: Secondorderchapter.tex
\section{Second-order systems}\label{Chapter2ndorder}
The first papers to consider  systematically what would now be called superintegrability  began with a search for systems that admitted ``dynamical symmetries" of order 2, connected with separation of variables \cite{FMSUW, FSUW, MSVW}. The structure and classification of  second-order superintegrable systems is the best understood area of the theory. In this chapter, we discuss the classification of second-order superintegrable systems in 2D and 3D and give an indication in the final section how such an analysis could be extended into nD.

\subsection{Second-order superintegrability in 2D}

\subsubsection{Determining equations, classical/quantum isomorphism}
We start with the determining equations (Killing and Bertrand-Darboux) for second-order integrals of motion for Hamiltonians in two-dimensions. We show that there exists an isomorphism between classical and quantum superintegrable systems.

Consider a general Hamiltonian with a scalar potential defined on a manifold written in conformal coordinates so that $ds^2=\lambda(x,y)(dx^2+dy^2).$ (Since all 2D manifolds are conformally flat there always exist such Cartesian-like coordinates $(x,y)$ and it is convenient to work with these.) The corresponding classical Hamiltonian takes the form 
\be  {\cal H}=\frac{1}{\lambda(x,y)}(p_x^2+p_y^2)+V(x,y),\qquad
(x,y)=(x_1,x_2).\label{Hclass}\ee
Since $\cal H$ is an even function of the momenta, any function on phase space which Poisson commutes with it  can always be
chosen to be an even or odd function of the momenta (hence no first
order terms). Thus, we can express any  second-order integral of  the motion as
\be 
{\cal L}=\sum_{i,j=1,2} a^{ji}(x,y)p_jp_i +W(x,y).\label{Lclass}\ee
The conditions that $\{ \cl , \ch \}=0$ are equivalent to the  {\bf Killing equations}
\begin{eqnarray} \label{Killingeqns}a^{ii}_i&=&
-\frac{\lambda_1}{\lambda}a^{i1} -\frac{\lambda_2}{\lambda}a^{i2},\quad i=1,2\\
2a^{ij}_i+a^{ii}_j&=&-\frac{\lambda_1}{\lambda}a^{j1}
-\frac{\lambda_2}{\lambda}a^{j2},\quad i,j=1,2,\ i\ne j.\nonumber
\end{eqnarray}
The subscripts denote partial derivatives. 
The determining equations for the functions $W_j$ are 
\bea W_j=\sum_{k=1}^2\lambda(x,y)a^{jk} V_k  \label{Wi},\eea
with  compatibility condition,  $\partial_1W_2=\partial_2W_1$, given by the  {\bf Bertrand-Darboux equation} 
\be \fl \label{BD} (V_{22}-V_{11})a^{12}+V_{12}(a^{11}-a^{22})= \left[\frac{(\lambda
a^{12})_1-(\lambda a^{11})_2}{\lambda}\right]V_1 + \left[\frac{(\lambda
a^{22})_1-(\lambda a^{12})_2}{\lambda}\right]V_2.
\ee

Now, consider the determining equations for a second-order integral of a quantum Hamiltonian $H$
\be H=\frac{1}{\lambda(x,y)}(p_x^2+p_y^2)+V(x,y),\label{Hquant} \ee
where in this case $p_i$ are the conjugate momenta represented as  $p_i=-i\hbar \partial_i.$
Analogously  the most general form for a formally self-adjoint second-order differential operator which commutes with $H$ is given by a sum of even terms, properly symmetrized 
\be L=\sum_{i,j=1,2} \frac 12 \left\{ a^{ji}(x,y), p_jp_i\right\} +W(x,y)\label{Lquant} .\ee
It is a direct calculation to show that the commutation relation $[H, L]=0$  holds whenever the Killling equations (\ref{Killingeqns}) and the Bertrand-Darboux equations (\ref{BD}) hold. Thus, we have proven the following theorem.  
\begin{theorem}
If $\cl$ (\ref{Lclass}) is a second-order integral for a classical Hamiltonian $\ch$ (\ref{Hclass}) then the second-order differential operator $L$  (\ref{Lquant}) commutes with the quantum Hamiltonian $H$ (\ref{Hquant}). In particular, if $\ch$ (\ref{Hclass})  is superintegrable for a fixed potential $V$, then $H$ (\ref{Hquant}) will also be superintegrable for the same potential. 
\end{theorem}

If the Hamiltonian is superintegrable there exist  two second-order constants in involution with $\mathcal{H}$ and we get two B-D  equations. These two equations can be used to obtain  fundamental PDEs
for the potential \cite{KKM20041}:
\be\label{2Dcanon} V_{22}-V_{11}=A^{22}({\bf x})V_1+B^{22}({\bf x})V_2,\qquad
V_{12}=A^{12}({\bf x})V_1+B^{12}({\bf x})V_2.\ee
If 
the integrability conditions for the PDEs (\ref{2Dcanon}) are satisfied identically, we say
that the potential
is {\bf nondegenerate}.
That means, at each regular point  ${\bf x}_0$ where the $A^{ij},B^{ij}$ are
defined and analytic, we can prescribe the values of
$V,\ V_1,\  V_2$ and $V_{11}$ arbitrarily and there will exist a unique
potential  $V({\bf x})$ with these values at ${\bf x}_0$.

Nondegenerate potentials depend on these 3 parameters, not including the trivial
additive parameter. Degenerate potentials depend on less than 3 parameters. It has been shown that all second-order superintegrable degenerate potentials in 2D are restrictions of nondegenerate ones, although they may admit an additional, first-order integral associated with group symmetry. In this case, the minimal generating set changes along with the structure of the algebra generated by the integrals \cite{KKMP2009}.

\subsubsection{Classification of nondegenerate systems, proof of the existence of quadratic algebras}
The structure theory for nondegenerate systems has been worked out  by Kalnins et al  \cite{KKM20041, KKM20042, KMP2, KMP3}.  
Assume that the basis  symmetries take the form 
${\cal L}_i=\sum_{j,k=1,2} a^{jk}_{(i)}({\bf x})p_jp_k+W^i({\bf x})$. 

\begin{definition} A set of symmetries $\{{\cal L}_i, i=1,\dots, \ell\}$ is said to be {\bf functionally
linearly dependent} if there exist $c_i({\bf x}), $ not identically $0$,  such that 
$$\sum _{i=1}^\ell c_i({\bf x})\sum_{j,k=1}^2a^{jk}_{(i)}({\bf x})p_jp_k=0.$$
The set is  {\bf functionally linearly independent} if it is  not functionally linearly dependent. \end{definition}

This functional linear independence criterion splits
superintegrable systems of all orders into two classes with different
properties. In two-dimensions, there
 is  only one functionally linearly dependent
superintegrable system, namely
 ${\cal H} =p_zp_{\bar z}+V(z)$,
where $V(z)$ is an arbitrary function of $z$ alone. This system
separates in only one set of coordinates $(z,{\bar z})=(x+iy,x-iy)$. For functionally
linearly independent 2D systems the theory is much more interesting. These are exactly the systems which separate in more than one set of orthogonal coordinates.
From \cite{KKM20041} we have the following results.
\begin{theorem}  Let $\cal H$ be the Hamiltonian of a 2d superintegrable
functionally linearly independent system with nondegenerate potential.
\begin{itemize}
\item The space of second-order constants of the motion is
3-dimensional.
\item  The space of  third-order constants of the motion  is
1-dimensional.
\item  The space of fourth-order constants of the motion is
6-dimensional.
\item The space of sixth-order constants is
10-dimensional.
\end{itemize}
\end{theorem}
These results follow from a study of the integrability conditions for constants of the motion that gives these numbers as the upper bounds of the dimensions. To prove that the bounds are achieved we use the  result:
\begin{theorem} Let $\cal K$ be a third-order constant of the motion for a
  superintegrable system with nondegenerate potential $V$:
$$
{\cal K}=\sum ^2_{k,j,i=1}a^{kji}(x,y)p_kp_jp_i+\sum
^2_{\ell=1}b^\ell(x,y)p_\ell.
$$
Then
$b^\ell(x,y)=\sum_{j=1}^2f^{\ell,j}(x,y)\frac{\partial V}{\partial x_j}(x,y)
$
with
$f^{\ell,j}+f^{j,\ell}=0,\quad  1\le \ell,j\le 2$.
The $a^{ijk},b^\ell$ are uniquely determined by the
quantity
$f^{1,2}(x_0,y_0)$
at any regular point $(x_0,y_0)$ of $V$.
\end{theorem}
This follows from a direct computation of the Poisson bracket $\lbrace \mathcal{H,K} \rbrace$ using the B-D relations for the $a^{ik}$'s.
Since $\lbrace \cl_1,\cl_2 \rbrace$ is a non-zero third-order constant of the motion we can see that the number (1) is achieved and we can solve for the coefficients $a^{ijk}(x,y)$ and $b^{\ell}$.
This result enables us to choose standard bases for second- and higher-order
symmetries. Indeed, given any $2\times 2$ symmetric matrix ${\cal A}_0$, and
any regular  point $(x_0,y_0)$ there exists one and only one second-order
symmetry (or constant of the motion) ${\cal L}=a^{kj}(x,y)p_kp_j+W(x,y)$ such that the matrix given by 
$a^{kj}(x_0,y_0)$ is equal to ${\cal A}_0$ and $W(x_0,y_0)=0$. Further, if
$ {\cal L}_\ell=\sum a^{kj}_{(\ell)}p_kp_j+W_{(\ell)},\ \ell=1,2$ are second-order constants of the the motion  and  ${\cal
A}_{(i)}(x,y)=\{a^{kj}_{(i)}(x,y)\}$, $i=1,2$
are $2\times 2$ symmetric  matrix functions, then the Poisson bracket of these
symmetries is given by
$$
\{{\cal L}_1,{\cal L}_2\}=\sum
^2_{k,j,i=1}a^{kji}(x,y)p_kp_jp_i+b^\ell(x,y)p_\ell.
$$
Using this relation we can solve for $f^{k,\ell}$:
$
f^{k,\ell}=2\lambda\sum_j(a^{kj}_{(2)}a^{j\ell}_{(1)}-a^{kj}_{(1)}a^{j\ell}_{(2)})$.
 Thus
$\{{\cal L}_1,{\cal L}_2\}$
is uniquely determined by the skew-symmetric matrix
$$[{\cal A}_{(2)},{\cal A}_{(1)}]\equiv {\cal A}_{(2)}
{\cal A}_{(1)}-{\cal A}_{(1)}{\cal A}_{(2)},
$$
hence by the constant matrix $[{\cal A}_{(2)}(x_0,y_0),{\cal
  A}_{(1)}(x_0,y_0)]$ evaluated at a regular point,  by the Taylor series generated from the Killing equations (\ref{Killingeqns}) and the Betrand-Darboux equations (\ref{BD}).
This allows a standard structure by exploiting identification of the space of  second-order constants of the
motion with the space of $2\times 2$ symmetric matrices and
identification of  third-order constants of the
motion with the space of $2\times 2$ skew-symmetric matrices.

Indeed given a basis for the $3$-dimensional space of symmetric
matrices, $\lbrace A^{ij}\rbrace$, we can define a standard set of basis symmetries ${\cal S}_{(k\ell)}=\sum
a^{ij}({\bf x})p_ip_j+W_{(k\ell)}({\bf x})$ corresponding to a regular point
${\bf x}_0.$ The third-order symmetry  is defined by the commutator of the basis elements; these are necessarily skew-symmetric and hence the identification with $\mathcal{K}$ with the skew-symmetric matrix. The operators $ \mathcal{S}_{(ij)}$ form an alternative basis to basis $\mathcal{L}_k$ for second-order integrals. An intuitive choice of basis for the $3$-dimensional space of symmetric
matrices is 
\be \label{2dbasis}
\lbrace A^{ij} \rbrace\in \left \{ \left [\ba{cc} 1& 0\\0&0\ea \right], \left[\ba{cc} 0& 0\\0&1\ea\right], \left[\ba{cc} 0& 1\\1&0\ea\right]\right \} \ee 
corresponding to  the basis of symmetries $\lbrace \cS_{11}, \cS_{22}, \cS_{12} \rbrace$. Notice, the identity element 
\[\mathcal{I}=\left[\ba{cc} 1& 0\\0&1\ea \right] \]
corresponds to the Hamiltonian.

The following theorems show the existence of the quadratic algebra \cite{KKM20041}.
 \begin{theorem} The 6 distinct monomials
$$({\cal S}_{(11)})^2,\  ({\cal S}_{(22)})^2,\ ({\cal S}_{(12)})^2,\  {\cal
S}_{(11)}{\cal
    S}_{(22)},\ {\cal S}_{(11)}{\cal
    S}_{(12)},\ {\cal S}_{(12)}{\cal
   S}_{(22)},\
$$ form a basis for the space of fourth-order
    symmetries.
\end{theorem}
We note that since $\mathcal{R}=\lbrace \cl_1, \cl_2 \rbrace$ is a basis for the third-order symmetries,  $\lbrace \cl_i,R \rbrace$ are of fourth-order, so  must be polynomials in the second-order symmetries.
\begin{theorem}
The 10 distinct monomials
$$({\cal S}_{(ii)})^3,\ ({\cal S}_{(ij)})^3,\ ({\cal S}_{(ii)})^2{\cal
S}_{(jj)},\ ({\cal S}_{(ii)})^2{\cal S}_{(ij)},\  ({\cal S}_{(ij)})^2
{\cal  S}_{(ii)},\
{\cal S}_{(11)}{\cal S}_{(12)}{\cal
    S}_{(22)},$$
for $i,j=1,2,\ i\ne j$ form a basis for the space of sixth-order
    symmetries.
\end{theorem}
Again, since $\mathcal{R}^2$ is a sixth-order constant we have immediately the existence of the Casimir relation which forces the algebra to close at order 6. This is a remarkable property of superintegrable systems and gives us finitely generated quadratic algebras. 
The analogous results for fifth-order symmetries follow directly from the Jacobi identity.

\subsubsection{St\"ackel transform: Proof of constant curvature}\label{4.1.3}
The St\"ackel transform of a superintegrable system takes it to a new superintegrable system on a different space. The basic observation is, if we have a Hamiltonian $\cal H$ and symmetry $\cal L$, which separates as
$${\cal H}={\cal H}_0+\alpha V_0 +\tilde{\alpha}\qquad {\cal L}={\cal L}_0+\alpha W_0$$
with $\lbrace {\cal H}_0, {\cal L}_0 \rbrace=\lbrace {\cal H}, {\cal L} \rbrace=0$
then we have  $\lbrace \tilde{\cal H}, \tilde {\cal L} \rbrace=0$ where
$$\tilde {\cal H}=\frac{{\cal H}_0}{V_0} \qquad \tilde {\cal L}={\cal L}_0-W_0\tilde{\cal H}.$$
After the transformation, the energy of the Hamiltonian $\tilde {\cal H}$ is $-\alpha$ and the original Hamiltonian is restricted to $\tilde{\alpha}$. In this sense, the coupling constant $\alpha$ has been exchanged with the energy, which can be represented by $-\tilde{\alpha},$ and the St\"ackel transform coincides with coupling constant metamorphosis \cite{HGDR} for second-order superintegrable systems.  See \cite{Post20111} for more information on the two transformations.

We explain the process in greater detail in the case of a 2D classical system; for notational purposes, we denote $\mathcal{H}_0=\frac1\lambda(p_x^2+p_y^2)$ and $\mathcal{L}_0=\sum a^{ij}p_ip_j$.
Suppose we have a non-degenerate superintegrable system
$$\mathcal{H}=\frac1{\lambda(x,y)}(p_x^2+p_y^2)+V(x,y)=\mathcal{H}_0+V'+\alpha U+\tilde{\alpha}$$
where the potential $V(x,y)$ as well as the separated  potential piece $U$ satisfy the differential equations
(\ref{2Dcanon}). 
We  transform to another superintegrable system $\tilde{\cal H}$ via 
\bea \tilde \lambda =\lambda U, &\qquad \qquad \tilde V=V/U. \eea
We then separate a complementary part of a symmetry  $\mathcal{L}$ as $\mathcal{L}_U=\mathcal{L}-W_0=\mathcal{L}_0+W_U$ where
$\lbrace \mathcal{L}_U, \mathcal{H}_0+U\rbrace=0$.
Under this transformation, $\tilde {\cal L}={\cal L}-\frac{W_U}U{\cal H}$ is in involution with $\tilde {\cal H}$. Notice that our new superintegrable system is defined on a new manifold with metric $\tilde \lambda=\lambda U$.
Although the new superintegrable system is on a (possibly) different manifold, the quadratic algebra remains unchanged except for the ordering of parameters. Indeed for a quadratic algebra as
$$ \mathcal{R}^2=F(\cl_1, \cl_2, \ch, \alpha,\beta, \gamma),\ 
\lbrace \cl_i, \mathcal{R}\rbrace =G_i(\cl_1, \cl_2,\ch, \alpha,\beta, \gamma) $$
we have the transformed algebra
$$ \tilde {\cal R}^2=F(\tilde\cl_1,\tilde \cl_2,\tilde{\alpha} ,-\tilde \ch,  \beta, \gamma),\
\lbrace \tilde \cl_i,\tilde {\cal R}\rbrace =G_i(\tilde \cl_1,\tilde  \cl_2,\tilde{\alpha},-\tilde \ch, \beta, \gamma). $$
Thus, it seems natural to view the quadratic algebra as the defining characteristic of a superintegrable system. Since the St\"ackel transform is invertible, we can use it to classify equivalence classes whose defining characteristic will be the form of the quadratic algebra.

In 2D every superintegrable system is St\"ackel equivalent to one on a constant curvature manifold. This was shown in detail in \cite{KKM20042} whose results we summarize here. A basic fact  is that given  a Hamiltonian $\mathcal{H}$ with $\lambda(x,y)$ the metric, the differential equations for the leading order terms of the integral $\mathcal{L}$ (\ref{Killingeqns}) are compatibile whenever  \be \label{fundintcond}
(\lambda_{22}-\lambda_{11})a^{12}-\lambda_{12}(a^{22}-a^{11})=3\lambda_1a^{12}_1-3\lambda_2a^{12}_2+(a^{12}_{11}-a^{12}_{22})\lambda.\ee
Since the St\''ackel transform of ${\cal L}$ does not change the function $a^{12}$ nor the difference $a^{22}-a^{11}$, any Hamiltonian which is St\"ackel equivalent to a Hamiltonian $\mathcal{H}$ with metric  $\lambda$, has a metric $\mu$ which also satisfies the (\ref{fundintcond}). Thus, a study of the possible solutions for the metrics gives all the possible spaces which admit superintegrable systems. These are classified by the following theorem.

\begin{theorem}
\label{Liouville}
If $ds^2=\lambda(dx^2+dy^2)$ is the metric of a nondegenerate
superintegrable system (expressed in coordinates $x,y$ such that
$\lambda_{12}=0$) then $\lambda=\mu$ is a solution of the system
\begin{equation}
\label{th1eqn}
\mu_{12}=0,\qquad \mu_{22}-\mu_{11}=3\mu_1(\ln
a^{12})_1-3\mu_2(\ln
a^{12})_2+(\frac{a^{12}_{11}-a^{12}_{22}}{a^{12}})\mu,
\end{equation}
where either
$$ I)\quad a^{12}=X(x)Y(y), \quad
 X''=\alpha^2 X,\quad Y''=-\alpha^2 Y,$$
or
$$ II)\quad a^{12}=\frac{2X'(x)Y'(y)}{C(X(x)+Y(y))^2},
$$
$$
(X')^2=F(X), \quad X''=\frac12 F'(X),\quad (Y')^2=G(Y),\quad Y''=\frac12
G'(Y),$$
$$ F(X)=
\frac{\alpha}{24}X^4+\frac{\gamma_1}{6}X^3+\frac{\gamma_2}{2}X^2+\gamma_3X+\gamma_4,$$
$$ G(Y)=-
\frac{\alpha}{24}Y^4+\frac{\gamma_1}{6}Y^3-\frac{\gamma_2}{2}Y^2+\gamma_3Y-\gamma_4.$$
Conversely, every solution $\lambda$ of one of these systems defines a
nondegenerate superintegrable system. If $\lambda$ is a solution then
the remaining solutions $\mu$ are exactly the nondegenerate
superintegrable systems that are St\"ackel equivalent to $\lambda$. Furthermore, if $V$ is the potential for the superintegrable system with metric $\lambda$, then $\tilde{ V}=\lambda V /\mu$ is the potential for the St\"ackel equivalent system with metric $\mu.$
\end{theorem}

We note that these spaces are exactly the spaces classified by Koenigs in \cite{Koenigs} where he identified the spaces which admit more that one second-order Killing tensor. As in his paper we write the requirements on $\mu$ and $a^{ij}$ as \be \fl  \label{mueqns1} a_{11}^{12}+a_{22}^{12}=0,\quad \mu_{12}=0,\quad
a^{12}(\mu_{11}-\mu_{22})+3\mu_1a^{12}_1-3\mu_2a^{12}_2+(a^{12}_{11}-a^{12}_{22})\mu=0,\ee
We can then see by direct computation the following theorem.
\begin{lemma} Suppose $\mu= \lambda(x,y), \ a^{12}=a(x,y)$ satisfy
  (\ref{mueqns1}). Then $\mu= {\tilde a}(x,y), \ a^{12}={\tilde
 \lambda}(x,y)$ also satisfy  (\ref{mueqns1}) where

 $${\tilde a}(x,y)=a(x+y,ix-iy),\quad {\tilde
   \lambda}(x,y)=\lambda(x-iy,y-ix).$$ This transformation is invertible.
\end{lemma}
Using this lemma, we obtain the following theorem.
\begin{theorem}\label{constcurve} System (\ref{mueqns1}) characterizes a
nondegenerate
  superintegrable system if and only if the metric ${\tilde a^{12}}(x,y)$
  is of constant curvature. Equivalently, the system (\ref{mueqns1})
characterizes a nondegenerate
  superintegrable system if and only if the symmetry $a^{12}$ is the
  image
$a^{12}={\tilde \lambda}$ where the metric $\lambda$
  is of constant curvature  (i.e. $\lambda_{12}=0$).
\end{theorem}

\medskip\noindent PROOF: System (\ref{mueqns1}) characterizes a nondegenerate
  superintegrable system if and only if the symmetry  $a^{12}$
  satisfies the Liouville equation $(\ln a^{12})_{12}=Ca^{12}$ for
  some constant $C$. (If $C=0$ we have Case I, and if $C\ne 0$ we have
  Case II.) It is straightforward to check that this means that
$$ \frac{{\tilde a^{12}}_{11}+{\tilde a^{12}}_{22}}{({\tilde
   a^{12}})^2}-\frac{{(\tilde a^{12}}_1)^2+({\tilde a^{12}}_2)^2}{({\tilde
    a^{12}})^3}=4iC,
$$
so the scalar curvature of metric ${\tilde a^{12}(dx^2+dy^2)}$ is
constant.
Similarly, if $\lambda$ is of constant curvature then $\tilde\lambda$
satisfies Liouville's equation. $\Box$

\begin{theorem} Every nondegenerate superintegrable 2D system is
  St\"ackel equivalent to a superintegrable system on a
  constant curvature space.
\end{theorem}

This  theorem  greatly simplifies the task of finding superintegrable systems in 2D; we can restrict to the complex plane and  the 2-sphere. Furthermore, we can use  the symmetries for these spaces to find the symmetries for the superintegrable potential. We give examples  in the following section.
\subsubsection{The list}\label{thelist}

We fix some notation. Let $s_1^2+s_2^2+s_3^2=1$ be  the embedding of the  2-sphere in 3D  Euclidean space and $z=x+iy$, $\overline{z}=x-iy$. Define  $J_3=s_1\partial_{s_2}-s_2\partial_{s_1}$ to be the generator of rotation about the $s_3$ axis, with $J_2,J_3$  obtained by cyclic permutation. On the Euclidean plane, we shall also use $J_3$ to denote the generator of rotations in the $x,y$ plane. 
Always  $R=[L_1, L_2]$. We use the notation for the constant curvature superintegrable systems  found in \cite{KKMP}: $Ek$ is the $k$th complex Euclidean system on the list and $Sk$ is the $k$th system on the complex sphere.

All of these systems have the remarkable property that the symmetry algebras generated by $H,L_1,L_2$ for nondegenerate potentials close under commutation. 
Define the third-order commutation $R$ by $R=[L_1,L_2]$. Then the fourth-order operators $[R,L_1],[R,L_2]$ are contained in the associative algebra of 
symmetrized products of the generators:\hfill\break
\[[L_j,R]=\sum_{0\leq e_1+e_2+e_3\leq 2} M^{(j)}_{e_1,e_2,e_3} \{ L_1^{e_1}, L_2^{e_2}\} H^{e_3}, \qquad e_k\geq 0,\]
where  $\{L_1,L_2\}=L_1L_2+L_2L_1$ is the anti-commutator. 
The sixth-order operator $R^2$ is contained in the algebra of symmetrized products  to third-order:
\be\label{Casimir1} R^2 -\sum_{0\leq e_1+e_2+e_3\leq 3} N_{e_1,e_2,e_3} \{ L_1^{e_1}, L_2^{e_2}\} H^{e_3} = 0.\ee
In both equations the constants $M^{(j)}_{e_1, e_2, e_3}$ and $N_{e_1, e_2, e_3}$ are polynomials in the parameters $a_1, a_2, a_3$ of degree $2-e_1-e_2-e_3$ and $3-e_1-e_2-e_3$ respectively.

For systems with one-parameter potentials the situation is different \cite{KKMP2009}.
 There are 4 generators: one first-order $X$ and 3 second-order $H,L_1,L_2$. The commutators $[X,L_1],[X,L_2]$ are second-order and can be expressed as
 \be\label{structure2}[X,L_j]=\sum_{0\leq e_1+e_2+2e_3+e_4\leq 1} P^{(j)}_{e_1,e_2,e_3, e_4} \{L_1^{e_1},L_2^{e_2},X^{2e_3}\} H^{e_4} ,\quad j=1,2,\ee
 where $\{L_1^{e_1},L_2^{e_2}, X^{2e_3+1}\}$ is the symmetrizer of three operators and has 6 terms and $ X^0=H^0=I.$
The commutator $[L_1,L_2]$ is third-order skew-adjoint and can be expressed as a polynomial in the generators via
\be[L_1,L_2]=\sum_{0\leq e_1+e_2+2e_3+e_4\leq 1} Q_{e_1,e_2, e_3, e_4}\{ L_1^{e_1},L_2^{e_2},X^{2e_3+1}\}H^{e_4}. \ee
Finally, since there are at most 3 algebraically independent generators, there will be a polynomial identity satisfied by the 4 generators. It is of fourth-order:
\be\label{Casimir2}\sum_{0\leq e_1+e_2+e_3+e_4\leq 2} S_{e_1, e_2, e_3, e_4}\{L_1^{e_1},L_2^{e_2}, X^{2e_4}\}H^{e_3}=0.\ee
. The constants 
$P^{(j)}_{e_1, e_2, e_3, e_4}$, $Q_{e_1, e_2, e_3, e_4}$, $S_{e_1, e_2, e_3, e_4}$ are polynomials in  $a_1$ of degrees  $1-e_1-e_2-e_3-e_4,$  $1-e_1-e_2-e_3-e_4$  and $2-e_1-e_2-e_3-e_4$ respectively.

Below is a list of all quantum superintegrable systems in 2D constant curvature space. The corresponding classical systems and their structure equations are similar, but may differ with respect to the non-leading-order terms.  The classification of St\"ackel equivalent systems and notation comes from \cite{Kress2007}. 
There are approximately 35 multiparameter families of superintegrable systems on Darboux and Koenig spaces, but all are St\"ackel equivalent to these. 
First we list  non-degenerate systems; they depend on 3 parameters $a_1, a_2, a_3$ as well as a trivial additive parameter which we have dropped. 

\medskip\noindent
{\bf 1)}\ {\bf  Quantum S9:} 
This quantum superintegrable system is defined by 
\bea \ba{rl} H=&J_1^2+J_2^2+J_3^2+\frac{a_1}{s_1^2}+\frac{a_2}{s_2^2}+\frac{a_3}{s_3^2},\\
        L_1=& J_1^2+\frac{a_3 s_2^2}{s_3^2}+\frac{a_2 s_3^2}{s_2^2},\quad
        L_2= J_2^2+\frac{a_1 s_3^2}{s_1^2}+\frac{a_3s_1^2}{s_3^2}.\ea \label{S9Operators}\eea
The algebra is 
\bea \fl \ba{rl} [L_i,R]=&\!\!\!\!4\{L_i,L_k\}-4\{L_i,L_j\}- (8+16a_j)L_j + (8+16a_k)L_k+ 8(a_j-a_k),\\
 R^2=&\!\!\!\!\frac83\{L_1,L_2,L_3\} -(16a_1+12)L_1^2 -(16a_2+12)L_2^2  -(16a_3+12)L_3^2\\
&\!\!\!\!+\frac{52}{3}(\{L_1,L_2\}+\{L_2,L_3\}+\{L_3,L_1\})+ \frac13(16+176a_1)L_1\\
&\!\!\!\!+\frac13(16+176a_2)L_2 + \frac13(16+176a_3)L_3 +\frac{32}{3}(a_1+a_2+a_3)\\ 
&\!\!\!\!+48(a_1a_2+a_2a_3+a_3a_1)+64a_1a_2a_3.\ea \label{S9Structure}\eea
To simplify the expressions, we have used  $L_3=H-L_1-L_2-a_1-a_2-a_3$ and the indices $\{i,j,k\}$ chosen as  a cyclic permutation of $\{1,2,3\}$

This system is St\"ackel equivalent to: 
\bea \fl \ba{lrl}S7:& H=&J_1^2+J_2^2+J_3^2+\frac{a_1s_1}{\sqrt{s_2^2+s_3^2}}+\frac{a_2s_2}{s_3^2\sqrt{s_2^2+s_3^2}}+\frac{a_3}{s_3^2},\\
S8:& H=&J_1^2+J_2^2+J_3^2+a_1\frac{s_1}{\sqrt{s_2^2+s_3^2}}+a_2\frac{s_1+is_2-s_3}{\sqrt{(s_1+is_2)(s_3-is_2)}}+a_3\frac{s_1+is_2+s_3}{\sqrt{(s_1+is_2)(s_3+is_2)}},\ea.\eea

\medskip\noindent
{\bf 2)}\ {\bf  Quantum E1:} 
The quantum system (Smorodinsky-Winternitz I \cite{FMSUW})is defined by 
\bea \ba{rl} H=&\partial_x^2+\partial_y^2+a_1(x^2+y^2)+\frac{a_2}{x^2}+\frac{a_3}{y^2},\\
         L_1=&\partial_y^2+a_1y^2+\frac{a_3}{y^2},\quad
         L_2= (x\partial_y-y\partial_x)^2 +\left(\frac{a_2y^2}{x^2}+\frac{a_3x^2}{y^2}\right).\ea  
         \label{E1Operators}\eea 
 The algebra relations are 
\bea \fl \ba{rl}\left[ L_1,R\right]=&8L_1H-8L_1^2+16a_1L_2-8a_1(1+2a_2+2a_3),\label{E1Structure}\\
 \left[ L_2, R\right]=&8\{L_1, L_2\} -4L_2H-16(1+a_2+a_3)L_1+8(1+2a_3)H, \\
 R^2=&\frac{8}{3}\left(\{L_1,L_2, H\}-\{L_1, L_1,L_2\}\right)-(16a_3+12)H^2\\
 & -\left(\frac{176}{3}+16a_2+16a_3\right)L_1^2 -16a_1L_2^2 +\left(\frac{176}3+32a_3\right)L_1H\\ & +\frac{176a_1}{3}L_2-\frac{16a_1}3\left(12a_2a_3+9a_2+9a_3+2\right).
\ea \eea
This system is St\"ackel equivalent to: 
\bea\fl  \ba{lrl} E16:&H=&\partial_x^2+\partial_y^2+\frac{1}{\sqrt{x^2+y^2}}\left(a_1+a_2\frac{1}{x+\sqrt{x^2+y^2}}+a_3\frac{1}{x-\sqrt{x^2+y^2}}\right),\\
S2:& H=&J_1^2+J_2^2+J_3^2+a_1\frac{1}{(s_1-is_2)^2}+a_2\frac{s_1+is_2}{(s_1-is_2)^2}+\frac{a_3}{s_3^2},\\
S4:& H=&J_1^2+J_2^2+J_3^2+a_1\frac{1}{(s_1-is_2)^2}+a_2\frac{s_1+is_2}{\sqrt{s_1^2+s_2^2}}+a_2\frac{1}{(s_1-is_2)\sqrt{s_1^2+s_2^2}}.\ea\eea
The system $E16$ is also referred to as Smorodinsky-Winternitz III \cite{FMSUW}.

\medskip\noindent
{\bf 3)} {\bf  Quantum E2:} The system (Smorodinsky-Winternitz II \cite{FMSUW})  is given by
\bea \ba{rl}
 H=&\partial_x^2+\partial_y^2+a_1(4x^2+y^2)+a_2 x+\frac{a_3}{y^2},\\
{ L}_1=&\partial_y^2+a_1 y^2+\frac{a_3}{y^2},\\
{ L}_2=&\frac12\{(x\partial_y-y\partial_x), \partial_y\}-y^2(a_1 x+\frac{a_2}{4})+\frac{a_3x}{y^2}.\ea\label{E2Operators}\eea
\bea\fl \ba{rl} [L_1,R]=&-2a_2L_1-16a_1L_2,\\
 \left[ L_2,R\right]=&6L_1^2-4L_1H+2a_2L_2-a_1(8a_3+6),\\
{ R}^2=&4HL_1^2-4L_1^3-16a_1L_2^2-2a_2\{L_1,L_2\}\\
&-4a_1(4a_3+3)H+4a_1(4a_3+11)L_1-\frac{a_2^2}{4}(4a_3+3).\ea\label{E2Structure}\eea
This system is St\"ackel equivalent to: 
\bea\fl  \ba{lrl} S16:&H=&J_1^2+J_2^2+J_3^2+a_1\frac{1}{(s_1-is_2)^2}+a_2\frac{s_3}{(s_1-is_2)^3}+a_3\frac{1-4s_3^2}{(s_1-is_2)^4}.\ea\eea

\medskip\noindent
{\bf 4)}\ {\bf  Quantum E$3'$:} 
The quantum system is defined by 
\bea \ba{rl}
H=&\partial_x^2+\partial_y^2+a_1(x^2+y^2)+a_2x+a_3y,\\
L_1=&\partial_y^2+a_1 y^2+a_3y,\quad
L_2=\partial_x\partial_y+a_1xy+\frac{a_2y}2+\frac{a_3x}{2},\ea\eea
\be\fl \ba{rl} \left[L_1, R\right]=&-4a_1L_2-a_2a_3, \quad
 \left[L_2, R\right]=4a_1L_1-2a_1H-\frac{(a_2+a_3)(a_2-a_3)}{2},\\
R^2=& 4a_1L_1(H-L_1)-4a_1L_2^2+(a_2^2-a_3^2)L_1-2a_2a_3L_2+a_3^2H-4a_1^2.\ea \label{E3'Structure}\ee
This system is St\"ackel equivalent to:
\bea\fl  \ba{lrl} E11:&H=&\partial_x^2+\partial_y^2+a_1 z+a_2\frac{z}{\sqrt{\overline{z}}}+a_3\frac{1}{\sqrt{\overline{z}}},\\
E20:&H=&\partial_x^2+\partial_y^2+\frac{1}{\sqrt{x^2+y^2}}\left(a_1+a_2\left(x+\sqrt{x^2+y^2}\right)+a_3\left(x-\sqrt{x^2+y^2}\right)\right). \ea\eea
The system $E20$ is (Smorodinsky-Winternitz IV \cite{FMSUW}) while $E3'$ is equivalent by a translation to the simple harmonic oscillator. However, it should be emphasized that without the two additional constants incorporated by the translation, the system would not be St\"ackel equivalent to the other two.

\medskip\noindent
{\bf 5)}\ {\bf  Quantum E10:} 
The quantum system is defined by $(\partial=\partial_z)$:
\be \ba{rl} H=&4\partial \overline{\partial}+a_1\left(z\overline{z}-\frac12\overline{z}^3\right)+a_2\left(z-\frac32
\overline{z}^2\right)+a_3\overline{z}, \\
 L_1=&-\partial^2-\frac{a_1\overline{z}^2}4-\frac{a_2\overline{z}}2+\frac{a_3}{12},\nonumber \\
L_2=&\{z\partial-\overline{z}\overline{\partial}, \partial\}-\overline{\partial}^2-(2z+\overline{z}^2)\left(\frac{a_1(2z-3\overline{z}^2)}{16} -\frac{a_2\overline{z}}{2}+\frac{a_3}{4}\right),\ea \label{E10Operators}
\ee
\bea\ba{rl}
\left[ L_1, R\right]=&2a_1L_1-\frac{a_2^2}{2}-\frac{a_1a_3}{6},\\
\left[ L_2, R \right]= &24L_1^2+4a_3L_1-2a_1L_2+a_2H,\\
R^2=&-16L_1^3-\frac{a_1}{4}H^2+2a_1\{L_1, L_2\}-2a_2L_1H-4a_3L_1^2 \\&-(a_2^2+\frac{a_1a_3}{3})L_2-\frac{a_2a_3}{3}H-a_1^2+\frac{a_3^3}{27}.\ea \label{E10Structure}
\eea
This system is St\"ackel equivalent to:
\bea \ba{lrl} E9:&H=&\partial_x^2+\partial_y^2+a_1\frac{1}{\overline{z}}+a_2(z+\overline{z})+a_3\frac{2\overline{z}+z}{\sqrt{\overline{z}}}. \ea\eea

\medskip\noindent
{\bf 6)}\ {\bf  Quantum E8:} 
The quantum system is defined by $(\partial=\partial_z)$:
\bea \ba{rl}H=&4\partial\overline{\partial}+a_1z\overline{z}+\frac{a_2z}{\overline{z}^3}+\frac{a_3}{\overline{z}^2},\\
L_1=& -\partial^2-\frac{a_1}{4}\overline{z}^2+\frac{a_2}{2\overline{z}^2},\quad
L_2=  -\left(z\partial -\overline{z}\overline{\partial}\right)^2+\frac{a_2z^2}{\overline{z}^2}+\frac{a_3z}{\overline{z}},\ea\label{E8Operators}\eea
\bea\label{E8Structure} \ba{l}
\left[ L_1, R\right]= -8L_1^2+2a_1a_2,\quad
        \left[L_2, R\right]= 8\{L_1, L_2\} -16L_1-2a_3H,\\
         R^2= 8\{L_1^2, L_2\} -\frac{176}{3}L_1^2-a_3L_1H+a_2H^2-4a_1a_2L_2-\frac{a_1(3a_3^2-4a_2)}{3}.\ea
\eea 
This system is St\"ackel equivalent to:
\bea \ba{lrl} E7:&H=&\partial_x^2+\partial_y^2+a_1 \frac{\overline{z}}{\sqrt{\overline{z}^2-c^2}}+a_2\frac{z}{\sqrt{\overline{z}^2-c^2}(\overline{z}+\sqrt{\overline{z}^2-c^2})}+a_3z\overline{z}, \qquad c\in \mathbb{C}, \\
E17:&H=&\partial_x^2+\partial_y^2+a_1\frac{1}{\sqrt{z\overline{z}}}+a_2\frac{1}{z^2}+a_2\frac{1}{z\sqrt{z\overline{z}}},\\
E19:&H=&\partial_x^2+\partial_y^2+a_1 \frac{\overline{z}}{\sqrt{\overline{z}^2-4^2}}+a_2\frac{1}{\sqrt{z(\overline{z}+2)}}+a_3\frac{1}{\sqrt{z(\overline{z}-2)}}. \ea\eea

The following are the degenerate systems; they depend on only 1 (other than the trivial additive) parameter and admit a Killing vector. 

\medskip\noindent
{\bf 7)}\ {\bf  Quantum  S3 (Higgs oscillator):}\label{S3algebra}  The system is the same as $S9$ with $a_1=a_2=0$ $a_3=a.$ The symmetry algebra is generated by 
\bea\label{S3Operators} X=J_3, \qquad L_1=J_1^2 + \frac{a s_2^2}{s_3^2}, \qquad 
L_2= \frac12(J_1J_2+J_2J_1)-\frac{a s_1s_2}{s_3^2}\nonumber .\eea
The structure relations for the algebra are given by 
\bea \label{S3Structure}
\ba{l}
\left[ L_1,X \right]=2L_2,\qquad [L_2,X]= -X^2-2L_1+H-a,\\
\left[ L_1,L_2 \right]=-(L_1X+XL_1)-(\frac12+2a)X,\\
0=\{L_1, X^2\}+2L_1^2+2L_2^2-2L_1H+\frac{5+4a}{2}X^2-2aL_1-a.\ea\eea
St\"ackel equivalence to the following potential on the sphere
\bea \ba{lrl} S6:&H=&J_1^2+J_2^2+J_3^2 +a\frac{s_3}{\sqrt{s_1^2+s_2^2}}.\ea\eea

\medskip\noindent
{\bf 8)}\  {\bf  Quantum  E14:}
The system is defined by 
\be \ba{lr} H=\partial_x^2+\partial_y^2  +\frac{a}{{\bar z}^2},   &X=\partial _z,\\
L_1=\frac{i}{2}\lbrace z\partial_z +{\bar z}\partial_{\bar z},\partial_z \rbrace+\frac{a}{\bar z},& 
L_2=\left(z\partial _z+{\bar z}\partial_{\bar z}\right)^2+\frac{az}{\bar z},\ea\label{E14Operators}\ee
\be \ba{lr} [L_1,L_2]=-\lbrace X,L_2 \rbrace -\frac{1}{2}X,&[X, L_1]=-X^2, \quad
\left[X, L_2\right]=2L_1,\\
L_1^2+XL_2X-b H-\frac{1}{4}X^2=0.\ea\label{E14Structure}\ee
This system is St\"ackel equivalent to 
\be  E12: H=\partial_x^2+\partial_y^2 +\frac{a\bar z}{\sqrt{{\bar z}^2+c^2}}.\ee

\medskip\noindent
{\bf 9)}\ {\bf   Quantum  E6:}
The system is defined by 
 \be \ba{lr} H=\partial_x^2+\partial_y^2 + \frac{a}{x^2}, &X=\partial_{y},\\
    L_1=\frac12 \lbrace x\partial_y-y\partial_x,\partial_{x} \rbrace -\frac{ay}{x^2}, &L_2=(x\partial_y-y\partial_x)^2+\frac{ay^2}{x^2},\ea \label{E6Structure}\ee
with symmetry algebra
\be \fl  \ba{l} [L_1,L_2]=\{X,L_2\}+(2a+\frac12)X, \quad [L_1,X]=H-X^2,\quad \left[L_2,X\right]=2L_1,\\
  L_1^2+\frac14\{L_2,X^2\}+\frac12 XL_2X-L_2H+(a+\frac34)X^2=0.\ea\ee
  This system is St\"ackel equivalent to 
  \bea \ba{lrl} S5:&H=&J_1^2+J_2^2+J_3^2 +\frac{a}{(s_1-is_2)^2}.\ea\eea

\medskip \noindent
{\bf 10)}\ {\bf   Quantum  E5:}
The system is defined by
\be \ba {lr}  H=\partial_x^2+\partial_y^2 +ax, &X=\partial_y,\\
L_1=\partial_{xy}+\frac1{2} ay, &L_2=\frac1{2}\lbrace x\partial_y-y\partial_x,\partial_y \rbrace -\frac1{4}ay^2, \ea\label{E10Structure1}\ee
\be \ba{l} [L_1,L_2]=2X^3-HX,\qquad  [L_1,X]=-\frac{a}{2}, \qquad  [L_2,X]=L_1,\\
X^4-HX^2+L_1^2+a L_2=0.\ea \label{E5Structure}\ee
This system is not St\"ackel equivalent to another constant curvature system. 

\medskip\noindent
{\bf 11)}\ {\bf   Quantum  E4:}
The system is defined by 
\be \ba{lr}
H =\partial_x^2+\partial_y^2+ a(x+iy), & X=\partial_x+i\partial_y,\\
L_1=\partial_x^2+a x,& L_2=\frac{i}{2}\lbrace x\partial_y-y\partial_x, X \rbrace -\frac{a}{4}(x+iy)^2,\ea\ee
 \be \ba{c}[L_1,X]=a, \quad [L_2,X]=X^2,\quad 
 [L_1,L_2]=X^3+HX-\left \lbrace L_1,X\right \rbrace,\\
 X^4-2 \left\lbrace L_1,X^2\right\rbrace +2HX^2+H^2+4a L_2=0.\ea\label{E4Structure}\ee
 This system is St\"ackel equivalent to 
 \bea \ba{lrl} E13:&H=&\partial_x^2+\partial_y^2 +\frac{a}{\sqrt{\overline{z}}}.\ea\eea

\medskip \noindent
{\bf 12)}\ {\bf  Quantum E3 (harmonic oscillator):} 
The system is determined by 
 \be \ba{c} H=\partial_x^2+\partial_y^2+a(x^2+y^2),\quad X=x\partial_y-y\partial_x,\nonumber\\
      L_1=\partial_y^2-\omega^2 y^2, \quad L_2= \partial_{xy}+a x y,\ea\label{E3Structure1}\ee 
\be \ba{c}[L_1,X] =2L_2, \quad  [L_2,X]=H-2L_1, \quad [L_1,L_2]=-2a X,\\
L_1^2+L_2^2-L_1H+aX^2-a=0.\ea\label{E3Structure}\ee
This system is St\"ackel equivalent to the Kepler-Coulomb system, 
\bea \ba{lrl} E18:&H=&\partial_x^2+\partial_y^2 +\frac{a}{\sqrt{x^2+y^2}}.\ea\eea
In this case, the St\"ackel transform coincides with the Kustaanheimo-Stiefel transformation \cite{KStransform}.

\subsection{Second-order superintegrability in nD}
\subsubsection{Classification in 3D conformally flat spaces}
In this section, we review the structure and classification of second-order superintegrable systems in 3D conformally flat space. For more detailed proof and further analysis,  see the original papers \cite{KKM3,KKM20052,KKM20061}.
 
Consider a Hamiltonian in  3D conformally flat space, with metric $ds^2=\lambda(x_1, x_2, x_3)(dx_1^2+dx_2^2+dx_3^2),$  given by, in Cartesian-like coordinates,
\be\label{H3D} \mathcal{H}=\frac{1}{\lambda(x_1, x_2, x_3)}\left(p_1^2+p_2^2+p_3^2\right)+V(x_1, x_2, x_3).\ee
The second-order integrals of the motion are determined by the 3D version of the Killing equations, 
\begin{eqnarray} \label{Killingeqns3D}a^{ii}_i&=&\sum_{j=1}^3
-\frac{\lambda_j}{\lambda}a^{ij} ,\quad i=1,2\\
2a^{ij}_i+a^{ii}_j&=&\sum_{k=1}^3-\frac{\lambda_k}{\lambda}a^{jk},\quad i,j=1,2,\ i\ne j,\nonumber
\end{eqnarray}
and the determining equations for the function $W$,
\bea W_j=\sum_{k=1}^3\lambda(x,y)a^{jk} V_k  \label{Wi3D}.\eea
Thus, the requirements for the existence of a second-order integral of motion in 3D can be expressed in terms of the integrability conditions for these equations. 

\begin{theorem}\label{Super3D} A  second-order function on phase space
\be \label{S3D} \mathcal{S} =\sum_{1\leq i,j\leq 3}a^{ij}(x_1,x_2, x_3)p_ip_j+W(x_1,x_2, x_3)\ee
is an integral of the motion for $\mathcal{H}$ (\ref{H3D}) if and only if the following hold:
\begin{enumerate}
\item The leading order term $\mathcal{S}_0=\sum_{1\leq i,j\leq 3}a^{ij}(x_1,x_2, x_3)p_ip_j$ is a conformal symmetry of the free Hamiltonian on flat space $\mathcal{H}_0=p_1^2+p_2^2+p_3^2$; i.e.  $\{\mathcal{H}_0, \mathcal{S}_0\}=f\mathcal{H}_0$ for some function $f=f(x_1,x_2,x_3).$
\item The functions $a^{ij}$ and $\lambda$ satisfy the  integrability conditions for the Killing equations (\ref{Killingeqns3D}), 
\be \label{3DKilling}\left(\lambda_2a^{12}+\lambda_3a^{13}\right)_2=\left(\lambda_1a^{12}+\left[(a^{22}-a^{11})\lambda\right]_2+\lambda_3a^{23}\right)_1\ee
\[ \left(\lambda_2a^{12}+\lambda_3a^{13}\right)_3= \left(\lambda_1a^{13}+\lambda_2a^{23}+\left[(a^{33}-a^{11})\lambda\right]_3\right)_1\]
\[ \left(\lambda_1a^{12}+\left[(a^{22}-a^{11})\lambda\right]_2+\lambda_3a^{23}\right)_2=\left(\lambda_1a^{13}+\lambda_2a^{23}+\left[(a^{33}-a^{11})\lambda\right]_3\right)_3.\]
\item The potential satisfies the Bertrand-Darboux equations
\be\label{3DBD}\sum_{k=1}^3\left[V_{jk}\lambda a^{k\ell}+V_{k\ell}\lambda a^{jk} +V_{k}\left((\lambda a^{k\ell}_j+(\lambda a^{jk}_\ell\right)\right]=0,\ j=1,2,3, \ee
the integrability conditions  for the function $W$ (\ref{Wi3D}). 
\end{enumerate}
\end{theorem}

The classification of the Hamiltonians which satisfy these requirements breaks into two general classes, the degenerate and nondegenerate, but the full structure theory has been worked out only for the nondegenerate case and for systems with bases that are functionally independent.
\begin{definition} The Hamiltonian (\ref{H3D}) is  {\bf nondegenerate} if the Bertrand-Darboux equations (\ref{3DBD}) are satisfied identically  as equations for the potential $V$.
\end{definition}
It can be shown that nondegenerate systems depend on exactly 4 parameters,  not including the trivial additive parameter. 

We have the following results for nondegenerate 3D superintegrable systems in conformally flat space; many of the proofs follow a similar structure as those for the 2D case. 
\begin{theorem}
Let $\mathcal{H}$ be a classical, nondegenerate  second-order superintegrable system in a 3D conformally flat space. Then
\begin{enumerate}
\item the classical Hamiltonian can be extended to a unique covariant quantum second-order superintegrable system. 
The potential of the quantum system differs from that of the classical system by an additive term depending linearly on the scalar curvature. 
\item  the corresponding Hamilton-Jacobi or Schr\"odinger  equations allow separation of variables in multiple coordinate systems,
\item the system is St\"ackel equivalent to either a system on flat space or the sphere. There are exactly 10 systems in flat space and  6 on the sphere (though there is no published proof as yet in the latter case).
\end{enumerate}
\end{theorem}
\begin{theorem}[5$\Rightarrow$ 6] \cite{KKM3}
Let $\mathcal{H}$ be a nondegenerate superintegrable system in a 3D conformally flat space, i.e.  $\mathcal{H}$ admits 5 functionally independent second-order integrals satisfying the requirements of Theorem \ref{Super3D}. Then there exists an additional second-order integral such that the 6 are functionally linearly independent. 
\end{theorem}
With the addition of this 6th integral, the algebra closes to a quadratic algebra. 
\begin{theorem} Let $\mathcal{H}$ be a nondegenerate superintegrable system in 3D conformally flat space which admits 5 (and hence 6) second-order integrals of the motion.  
Then 
\begin{enumerate}
\item  The space of third-order integrals is of dimension 4 and spanned by the Poisson brackets of the 6 generators. 
\item  The space of fourth-order integrals is of dimension 21 and spanned by quadratic polynomials in the generators. 
\item The space of sixth-order integrals is of dimension 56 and spanned by cubic polynomials in the generators. 
\end{enumerate}
Thus, the quadratic algebra closes at sixth-order. Additionally, there is an eighth-order functional relation between the 6, second-order integrals. Furthermore, the theorem holds for quantum systems with Poisson brackets replaced by commutators and polynomials in the generators replaced by their appropriately symmetrized counterparts.
\end{theorem}

Let us consider an example of a quadratic superintegrable system which is nondegenerate, the singular isotropic oscillator:
\be H=\partial_1^2+\partial_2^2+\partial_3^2+a^2(x_1^2+x_2^2+x_3^2)
+\frac{b_1}{x_1^2}+\frac{b_2}{x_2^2}+\frac{b_3}{x_3^2}\quad
\partial_i\equiv\partial_{x_i}.\ee
 A basis for the second-order constants of the motion is (with $H =M_1+M_2+M_3.$)
\be\label{q2ordsym}
\ba{l}M_\ell=\partial_\ell^2+a^2x_\ell^2+\frac{b_\ell}{x_\ell^2},\quad \ell=1,2,3,\\
L_i=(x_j\partial_k-x_k\partial_j)^2+\frac{b_jx_k^2}{x_j^2}+\frac{b_kx_j^2}{x_k^2},\ea\ee
where $i,j,k $ are pairwise distinct and run from 1 to 3. There are 4 linearly independent commutators of the second-order
symmetries: 
\be\label{q3ordsym} S_1=[L_1,M_2]=[M_3,L_1],\
S_2=-[M_3,L_2]=[M_1,L_2],\ee
$$
S_3=-[M_1,L_3]=[M_2,L_3],\ R=[L_1,L_2]=[L_2,L_3]=[L_3,L_1],
$$
$$ [M_i,M_j]=[M_i,L_i]=0,\quad 1\le i,j\le 3.
$$
 The fourth-order structure equations are 
 $[M_i,S_i]=0,\ 1=1,2,3$, and 
\be\fl \label{q4ordsym}\epsilon_{ijk}[M_i,S_j]=8M_iM_k-16a^2L_j+8a^2,\
\epsilon_{ijk}[M_i,R]=8(M_jL_j-M_kL_k)+4(M_k-M_j),\ee
$$\epsilon_{ijk}[S_i,L_j]=8M_iL_i-8M_kL_k+4(M_k-M_i),$$
$$
\epsilon_{ijk}[L_i,S_i]=4\{L_i,M_k-M_j\}+16b_jM_k-16b_kM_j+8(M_k-M_j),
$$
$$\epsilon_{ijk}[L_i,R]=4\{L_i,L_k-L_j\}-16b_jL_j+16b_kL_k+8(L_k-L_j+b_j-b_k).
$$
Here, $\{F,G\}=FG+GF$ and $\epsilon_{ijk}$ is the completely antisymmetric tensor.
The fifth-order structure equations are obtainable directly from the
fourth-order equations and the Jacobi identity. An example of the sixth-order  equations is 
\be\fl  \ba{r}
S_i^2-\frac83\{L_j,M_j,M_k\}+16a^2L_i^2+(16b_k+12)M_j^2+(16b_j+12)M_k^2-\frac{104}{3}M_jM_k\qquad \\
-\frac{176}{3}a^2L_i
-\frac{16}{3}a^2(2+9b_j+9b_k+12b_jb_k)=0.\ea\label{q6ordsym}\ee
The remainder of the structure equations can be found in \cite{KMPost10}, where representations for this quadratic algebra are obtained.

On the other hand, for true 3 parameter potentials (i.e. those which are not restrictions of a 4 parameter potential) the $5 \Rightarrow 6$ theorem no longer holds, the space of third-order integrals is of dimension 3 and the there are sixth-order integrals which can not be expressed as a polynomial in the generators. For example, consider the 
following extension of the Kepler-Coulomb system, here given as a classical system 
\bea \mathcal{H}&=&p_1^2+p_2^2+p_3^2+\frac{\alpha}{\sqrt{x_1^2+x_2^2+x_3}}+\frac{\beta}{x_1^2}+\frac{\gamma}{x_2^2},\nonumber\\
 \mathcal{L}_1&=&J_1^2+\frac{\gamma(x_2^2+x_3^2)}{x_2^2},\ 
\mathcal{L}_2= J_2^2+\frac{\beta(x_1^2+x_3^2)}{x_1^2},\nonumber\\
 \mathcal{L}_3&=& J_3^2+(x_1^2+x_2^2)\left(\frac{\beta}{x_1^2}+\frac{\gamma}{x_2^2}\right),\nonumber\\
 \mathcal{L}_4&=&p_1J_2-p_2J_1+x_3\left(\frac{\alpha}{2\sqrt{x_1^2+x_2^2+x_3}}+\frac{\beta}{x_1^2}+\frac{\gamma}{x_2^2}\right).\eea
The third-order integrals are spanned by 
\bea  \mathcal{R}_1&=&\{ \mathcal{L}_1, \mathcal{L}_2\}=\{\mathcal{L}_2,\mathcal{L}_3\}=\{\mathcal{L}_3,\mathcal{L}_1\}=-4J_1J_2J_3+\ldots\nonumber\\
        \mathcal{R}_2&=&\{\mathcal{L}_1, \mathcal{L}_4\}=2p_1J_1J_3-2p_2J_1^2+\ldots\nonumber\\
         \mathcal{R}_3&=&\{\mathcal{L}_2, \mathcal{L}_5\}=2p_2J_2J_3-2p_3J_2^2+\ldots\nonumber\\
        0&=&\{\mathcal{L}_3, \mathcal{L}_5\},\nonumber\eea
unlike in the non-degenerate case where the dimension was 4. Also unlike the non-degenerate case, there is a sixth-order integral which cannot be expressed as a polynomial in the generators $\mathcal{H}, \mathcal{L}_1, \ldots \mathcal{L}_5. $ Indeed, it is clear that the operator $\mathcal{R}_2\mathcal{R}_3$ will have a leading order term of the form $p_3p_2J_2^3J_3$ which cannot be obtained from any polynomial in the generators $\mathcal{H}, \mathcal{L}_1, \ldots \mathcal{L}_5. $

This example demonstrates the difficulties which are encountered when trying to extend the analysis to ND, while at the same time restricting to second-order superintegrability. Indeed the structure of the previous system can be better understood by considering additional higher-order integrals \cite{DASK2011, KKM2012, KM2012b}. This topic will be taken up in the following chapters. 

The most powerful methods we know for construction of second-order systems in higher dimensional constant curvature spaces are based on orthogonal separation of variables for the zero-potential  Helmholtz and associated Hamilton-Jacobi equations, particularly separation in ``generic"  Jacobi ellipsoidal coordinates, \cite{KALNINS, KKWMPOG,KMP06}. We give an example for $n=3$ and then generalize. 
Here, a ``natural'' basis
for first-order symmetries is given by
$ p_1\equiv p_x$, $ p_2\equiv p_y$, $  p_3\equiv p_z$, $
 J_1=yp_z-zp_y$, $ J_2=zp_x-xp_z$, $J_3=xp_y-yp_x$.
Among the separable systems for the Hamilton-Jacobi equation ${\cal H}=p_x^2+p_y^2+p_z^2=E$  there are  $7$ ``generic'' Euclidean  systems,  depending on a
 scaling parameter $c$ and up to three parameters $e_1,e_2,e_3$. For
 each such set of coordinates  there is exactly one nondegenerate
 superintegrable system that admits separation in these coordinates
 {\it simultaneously for all values of the parameters $c,e_j$}. An example is the system $<23>$ in 
B\^ocher's notation \cite{Bocher},
$$ <23> \quad
x-iy=\frac12 c( \frac{u^2+v^2+w^2}{uvw}- \frac12
\frac{u^2v^2+u^2w^2+v^2w^2}{ u^3v^3w^3}),$$ 
$$z=\frac12 c(\frac{uv}{ w} +\frac {uw}{ v} +\frac {vw}{u}),\quad
x+iy=cuvw.
$$
$${\cal L}_1=J^2_1+J^2_2+J^2_3+2c^2(p_1+ip_2)p_3,\quad 
{\cal L}_2=-2J_3(J_1+iJ_2)+c^2(p_1+ip_2)^2.$$
The symmetries ${\cal L}_1,{\cal L}_2$ determine the separation in the variables $u,v,w$.
 If a nondegenerate
superintegrable system, with potential,  separates in these coordinates for {\it all} values of
the parameter $c$, then the space of second-order symmetries must
contain the $5$ symmetries 
$$ {\cal H}=p_x^2+p_y^2+p_z^2+V,\quad {\cal S}_1=J_1^2+J_2^2+J_3^2+f_1,\quad
{\cal S}_2=J_3(J_1+iJ_2)+f_2,$$
$${\cal S}_3=(p_x+ip_y)^2+f_3,\quad {\cal S}_4=p_z(p_x+ip_y)+f_4.$$
 Solving the Bertrand-Darboux
equations for the potential we find  the unique solution
$$
V({\bf x}):=\alpha(x^2+y^2+z^2)+\frac{\beta}{(x+iy)^2}+\frac{\gamma
  z}{(x+iy)^3}+\frac{\delta (x^2+y^2-3z^2)}{(x+iy)^4}.$$
Finally, we can use the symmetry conditions for this potential to
obtain the full $6$-dimensional space of second-order
symmetries.  The other 6 generic  cases yield corresponding results. All second-order 3D Euclidean systems can be shown to be limits of these generic systems. Further these  generic separable coordinates have analogs for all dimensions $n$ and lead to nondegenerate superintegrable systems. 
The number of distinct generic superintegrable systems for each integer $n\ge 2$ is 
$\sum_{j=0}^np(j),
$
where $p(j)$ is the number of integer partitions of $j$, given by the Euler generating function 
$\frac{1}{\prod_{k=1}^\infty(1-t^k) }=\sum_{j=0}^\infty p(j)t^j$, \cite{KMP06}.
 By taking St\"ackel transforms we can define superintegrable systems on many conformally flat spaces. Similarly, each of the 5 generic separable coordinates on the complex 3-sphere leads to a nondegenerate superintegrable system and this construction generalizes to $n$ dimensions.
The number of distinct generic superintegrable systems for each integer $n\ge 2$ is 
$p(n+1)$
where $p(j)$ is the number of integer partitions of $j$,\cite{KMP06}.

There are many constructions and analyses of second-order systems in $n$ dimensions, e.g., \cite{BH,HER1,Herr2, ballesteros2011quantum, KMWP, KKWMPOG, rodriguez2002quantum} but as yet no general theory. We finish this chapter with a discussion of methods that may give a clear way to obtain general results about second-order superintegrability in $n$ dimensions, those of algebraic geometry

\subsubsection{Classification: Relation with algebraic geometry}
In this section, we review the basic results of \cite{KKM2007a} without the details in order to give a general idea of the method and how it can be extended to higher dimensions. 
For a non-degenerate system in complex Euclidean space, the Bertrand-Darboux equations (\ref{3DBD}) depend on essentially 10 different functions. From the assumption that the integrability conditions  (\ref{3DKilling}) are identically satisfied as well as the integrability conditions for the potential,  the derivatives of these functions can be expressed as quadratic polynomials in the original functions as well as 5 quadratic identities for the 10 functions. 

If we consider the algebraic variety defined by these $5$ polynomial functions in 10D, regular points of this variety will determine a unique superintegrable system as long as the system is closed under differentiation. This adds an additional quadratic function in the points, which, when adjoined to the original set of  equations gives an algebraic variety, closed under differentiation each of  whose regular points determine a non-degenerate superintegrable system in 3D Euclidean space. Two points in the variety correspond to the same system if and only if they lie on the same orbit under the action of the complex Euclidean group  $E(3, \mathbb{C})$. In  \cite{KKM2007c} it is shown that there are exactly 10 orbit families, corresponding to the 10 possible nondegenerate Euclidean systems. In \cite{KKM2007a} a similar analysis has been carried out for 2D Euclidean systems. The corresponding analysis for superintegrable systems on the complex sphere has not yet been carried out. For a start see \cite{KKMP2011}.

The possibility of using methods of algebraic geometry to classify superintegrable systems is very promising and suggests a method to extend the analysis in arbitrary dimension as well as a way to understand the geometry underpinning superintegrable systems.

%% file: Thirdorderchapter.tex
\section{Higher-order determining equations on Euclidean space}\label{thirdorderchapter}
In this section, we move beyond second-order and into higher order superintegrability. We begin the chapter with some general theory about the structure of higher-order integrals for classical and quantum Hamiltonians in 2D real Euclidean space.  A general form for the integrals is given as well as the determining equations. 

In section 5.2, we compare these results to those given in the previous chapter by considering second-order integrals and the classification of the solutions to the determining equations in this case. In section 5.3, we give explicitly the determining equations for third-order integrals. These are  non-linear and very difficult to solve. 
One strategy is to assume separation of variables and this in turn implies the existence of an additional first or second-order integral of motion. All third-order superintegrable systems which admit separation of variables in Cartesian, polar and parabolic coordinates have been classified. The results of these investigations are reviewed in Section 5.3. Section 5.4 discusses the fourth-order case.

\subsection{Determining equations for higher-order integrals}
In order to determine the equations for higher-order integrals, we focus our attention on  2D Hamiltonians with a real, scalar potential
\be \label{Hgen} H=p_1^2+p_2^2+V(x,y).\ee
\begin{theorem}
A classical $N$-th order integral for the Hamiltonian (\ref{Hgen}) has the form 
\be \label{Xclassparity} X=\sum_{\ell=0}^{[\frac{N}{2}]} \sum_{j=0}^{N-2\ell}f_{j,2\ell}p_1^jp_2^{N-j-2\ell},\ee
where $f_{j,k}(x,y)$ are real functions. The integral has the following properties: 
\begin{enumerate}
\item The functions $f_{j, 2\ell}$ and the potential $V(x,y)$ satisfy the determining equations
\be \label{classical deteq}\fl  0=2\frac{\partial f_{j-1,2\ell}}{\partial x}+2\frac{\partial f_{j, 2\ell}}{\partial y }-(j+1)\frac{\partial V}{\partial x} f_{j+1, 2\ell-2}-(N-2\ell+2-j)\frac{\partial V}{\partial y} f_{j, 2\ell-2},\ee
\be f_{j,k}=0, \qquad j<0,\, k<0, \, j+k>N, \, k=2\ell+1.\ee
\item As indicated in (\ref{Xclassparity}), all terms in the polynomial $X$ have the same parity.
\item The leading terms in (\ref{Xclassparity})  (of order N obtained for $\ell=0$) are polynomials of order $N$ in the enveloping algebra of the Euclidean (Poisson) Lie algebra $E(2)$ with basis $\{p_1, p_2, L_3\}.$ 

\end{enumerate}
\end{theorem}
There are analogous results for quantum integrals although the proofs are more involved in the quantum case, see \cite{PostWintNth}. The results are as follows. Note that for higher-order integrals, there is a substantial difference between the classical and quantum case. In order to keep track of this difference, we will continue to normalize the mass to 2 but will leave the dependence of $\hbar$ in the quantum system. 
\begin{theorem}
A quantum $N$-th order integral for the Hamiltonian (\ref{Hgen}) has the form 
\be \label{Xquantparity} X=\frac12\sum_\ell\sum_j (-i\hbar)^{N-2\ell}\lbrace f_{j,2\ell}, \partial_x^j\partial_y^{N-2\ell-j}\rbrace.\ee
\be f_{j,k}=0, \qquad j<0,\, k<0, \, j+k>N, \, k=2\ell+1.\ee
where $f_{j,k}(x,y)$ are real functions. The integral has the following properties: 
\begin{enumerate}
\item  The functions $f_{j, 2\ell}$ and the potential $V(x,y)$ satisfy the determining equations for $j=0\ldots N+1-2\ell$, $\ell=0\ldots \lfloor{(N+2)/2}\rfloor$
\bea \fl \!\!\!\! 0\!&=\!&2\frac{\partial f_{j-1,2\ell}}{\partial x}+2\frac{\partial f_{j, 2\ell}}{\partial y }-\!(j\!+\!1)\frac{\partial V}{\partial x} f_{j+1, 2\ell-2}-\!(N\!-\!2\ell\!+\!2\!-\!j)\frac{\partial V}{\partial y} f_{j, 2\ell-2}-\hbar^2Q_{j,2\ell},\label{quantdeteq}\eea
where $Q_{j,2\ell}$ is a quantum correction term that is polynomial in $\hbar^2$ given below in (\ref{Qjl})
\item As indicated in (\ref{Xquantparity}), the symmetrized integral will have terms which are differential operators of the same parity. 
\item The leading terms in (\ref{Xquantparity})  (of order N obtained for $\ell=0$) are polynomials of order $N$ in the enveloping algebra of the Euclidean Lie algebra $E(2)$ with basis $\{p_1, p_2, L_3\}.$ 
\end{enumerate}
\end{theorem}

Let us just mention that as in the classical case, the observation of the parity constraint reduces the possible functional coefficients by about half. Indeed, were one to express the integral in the standard form with all of the differential operators on the right, the integral would have $N+1$ terms and be given by 
\bea  X&=&\sum_{\ell=0}^{[N/2]}\sum_{j=0}^{N-2\ell}\left(f_{j,2\ell} -\hbar^2\phi_{j,2\ell}\right)\partial_x^{j}\partial_y^{N-2\ell-j}(-i\hbar)^{N-2\ell}\nn
&&-i\hbar\sum_{\ell=0}^{[N-1/2]}\sum_{j=0}^{N-2\ell-1}\phi_{j, 2\ell+1}\partial_x^j\partial_y^{N-2\ell-1-j}(-i\hbar)^{N-2\ell-1},\eea 
where the $\phi_{j,k}$ are defined as
\be \fl \phi_{j,2\ell} =\sum_{b=0}^{\ell-1}\sum_{a=0}^{2b+2}\frac{(-\hbar^2)^b}{2}\left(\ba{c}j+a\\a\ea\right)\left(\ba{c} N-2\ell+2b+2-j-a\\ 2b+2-a\ea \right)\partial_x^a\partial_y^{2b+2-a}f_{j+a, 2\ell-2b-2}\ee
\be \fl \phi_{j, 2\ell+1}=\sum_{b=0}^{\ell}\sum_{a=0}^{1+2b} \frac{(-\hbar^2)^b}{2}\left(\ba{c}j+a\\a\ea\right)\left(\ba{c} N-2\ell+2b-j-a\\ 2b+1-a\ea \right)\partial_x^a\partial_y^{2b+1-a}f_{j+a, 2\ell-2b}.\ee
Note that these are polynomial in $\hbar^2.$
Without prior knowledge of the symmetrized structure, the operator $X$ would seem to depend on $N+1$ functions instead of $\lfloor{(N+1)/2}\rfloor.$

Unlike in the classical case where it was clear that there were only  $\lfloor{(N+2)/2}\rfloor$ sets of determining equations, since $\{H,X\}$ is a degree $N+1$ polynomial with distinct parity, in the quantum case this is not immediately clear. As was shown in \cite{PostWintNth}, every other term vanishes modulo the higher order terms and so  the only independent determining equations in the quantum case are the same as those in the classical case, up to a quantum correction $Q_{j,2\ell}$ given by 
\bea
\label{Qjl} \fl Q_{j,2\ell} &=&\left(2\partial_x\phi_{j-1,2\ell}+2\partial_y\phi_{j,2\ell} +\partial_x^2\phi_{j,2\ell-1}+\partial_y^2\phi_{j,2\ell-1}\right)\nn \fl
&&-\sum_{n=0}^{\ell-2}\sum_{m=0}^{2n+3}(-\hbar^2)^n\left(\ba{c}j+m\\m\ea\right)\left(\ba{c}N-2\ell+2n+4-j-m\\ 2n+3-m\ea\right)(\partial_x^m\partial_y^{2n+3-m}V)f_{j+m,2\ell-2n-4}\nn \fl
&&-\sum_{n=0}^{\ell-2}\sum_{m=0}^{2n+2}(-\hbar^2)^n\left(\ba{c}j+m\\m\ea\right)\left(\ba{c} N-2\ell+2n+3+j-m\\ 2n+2-m\ea\right)(\partial_x^m\partial_y^{2n+2-m}V)\phi_{j+m,2\ell-2n-3}\nn \fl
&&-\sum_{n=0}^{\ell-1}\sum_{m=0}^{2n+1}(-\hbar^2)^{n}\left(\ba{c}j+m\\m\ea\right)\left(\ba{c} N-2\ell+2n+2-j-m\\ 2n+1-m\ea\right)(\partial_x^m\partial_y^{2n+1-m}V)\phi_{j+m,2\ell-2n-2}\nonumber. \eea
In this form, it is clear that the quantum correction terms  are polynomial in $\hbar^2$ and so the quantum determining equations go to the classical ones in the classical limit. In the integrable case, these quantum correct terms have been studied \cite{Hiet1984, Hietarinta1986, hietarinta1998pure, Hiet1989, HG1989, HGDR} and, as will be seen for third- and higher-order systems, there exist quantum integrable and superintegrable systems whose potentials either change or vanish in the classical limit. 

\subsection{Second-order determining equations}
As an example, consider the case of an integral of second-order and compare these determining equations with those analyzed in Section 4.1. There are two sets of determining equations. The equations which require the leading order terms to be in the enveloping algebra of $E_2$,  
\be \label{genkilling} 0=\frac{\partial f_{j-1,0}}{\partial x}+\frac{\partial f_{j, 0}}{\partial y },\qquad j=0, \dots 3,  \ee
these are the Killing equations (\ref{Killingeqns}) from Section 4 on Euclidean space. Note that there are no quantum correction terms for this equation since it corresponds to $\ell=0$, and $\phi_{j,0}$ and hence $Q_{j,0}$ are identically 0.  The next and final set of equations are given by the following equation for $j=0,1$
 \be\label{quantN2} 0=2\frac{\partial f_{j-1,2}}{\partial x}+2\frac{\partial f_{j, 2}}{\partial y }-(j+1)\frac{\partial V}{\partial x} f_{j+1, 0}-(2-j)\frac{\partial V}{\partial y} f_{j, 0}-\hbar^2Q_{j,2},\ee
with quantum correction terms 
\bea Q_{j,2}&=& \left(2\partial_x\phi_{j-1,2\ell}+2\partial_y\phi_{j,2\ell} +\partial_x^2\phi_{j,2\ell-1}+\partial_y^2\phi_{j,2\ell-1}\right)\\
\phi_{j,1}&=&\sum_{a=0}^1\frac12 \left(\ba{c}j+a\\a\ea\right)\left(\ba{c}2-j-a\\
1-a\ea\right)\partial_x^{a}\partial_y^{1-a}f_{j+a, 0},\nonumber \\
\phi_{j,2}&=&\sum_{a=0}^2\frac12 \left(\ba{c}j+a\\a\ea\right)\left(\ba{c}2-j-a\\
2-a\ea\right)\partial_x^{a}\partial_y^{2-a}f_{j+a, 0}.\nonumber \eea

The equations (\ref{quantN2}) are the BD equations (\ref{BD}) with $f_{0,2}=W$ the zeroth-order term of the integral. However, recall from the previous chapter that the BD equations (\ref{BD}) have no quantum correction term. For second-order superintegrable systems the quantum correction term at this level does not vanish identically but only when the higher-order terms satisfy the Killing equations (\ref{genkilling}). 

Suppose now that the second-order terms satisfy these equations (\ref{genkilling}), so that the integral is of the form (without loss of generality we assume $A_{0,0,2}=0$)
\begin{eqnarray}
X&=&A_{2,0,0} L_3^2+A_{1,1,0}(L_3p_1+p_1L_3)+A_{1,0,1}(L_3p_2+p_2L_3)\nn
&&+A_{0,2,0}(p_1^2-p_2^2)
+2A_{0,1,1}p_1p_2+f_{0,2} (x_1,x_2).
\label{inta}
\end{eqnarray}
The remaining determining equations (\ref{quantN2}) for $j=0,1$ become
\bea \label{f02}|
\fl \ba{l} \frac{\partial f_{0,2}}{\partial_x}=-2\!\left(A_{2,0,0}y^2\!+\!2A_{1,1,0}y\!+\!A_{0,2,0}\right)\! V_{x}\!+\!2\!\left(A_{2,0,0}xy\!+\!A_{1,1,0}x\!-\!A_{1,0,1}y\!-\!A_{0,1,1}\right)\!V_{y}  \\
\frac{\partial f_{0,2}}{\partial_y}=-2\!\left(A_{2,0,0} xy\!+\! A_{1,1,0}x\!-\! A_{1,0,1}y\!-\! A_{0,1,1}\right)\! V_{x}  -\!2\left(A_{2,0,0}x^2\!-\! 2A_{1,0,1}x\!-\! A_{0,2,0}\right)\!V_{y}.\ea
\label{phieq1}\eea
The compatibility condition for the system (\ref{f02}) is
\begin{eqnarray}
&&(-A_{2,0,0}xy-A_{1,1,0}x+A_{1,0,1}y+A_{0,1,1})(V_{xx}-V_{yy}) \nonumber\\
&&-\left(A_{2,0,0}(x^2+y^2)+2A_{1,1,0}x+2A_{1,0,1}y+2A_{0,2,0}\right)V_{xy} \nonumber \\
&&-(A_{2,0,0}y+A_{1,0,1})V_{x}+3(A_{2,0,0}x-A_{1,0,1})V_{y}=0.
\label{compatibilityphi}
\end{eqnarray}
Eq. (\ref{compatibilityphi}) is exactly the same equation that we would have obtained if we had required that the potential should allow the separation of variables in the Schr\"odinger equation in one of the coordinate system in which the Helmholtz equation allows separation.

The Hamiltonian (\ref{Hgen}) is form invariant under Euclidean transformations, so we can classify the integrals $X$ into equivalence classes under rotations, translations and linear combinations with $H$. There are two invariants in the space of parameters $A_{j,k,\ell}$, namely
\begin{eqnarray}
\fl I_1=A_{2,0,0},\quad   I_2=(2A_{2,0,0}A_{0,2,0}-A_{1,1,0}^2+A_{1,0,1}^2)^2+4(A_{2,0,0} A_{0,1,1}-A_{1,1,0}A_{1,0,1})^2.
\end{eqnarray}
Solving (\ref{compatibilityphi}) for different values of $I_1$ and $I_2$ we obtain :
\begin{eqnarray}
I_1=I_2=0 & \qquad V_C=f_1(x)+f_2(y)& \/  \nonumber \\
I_1=1,\,I_2=0& \qquad V_R=f(r)+\dfrac{1}{r^2}g(\phi) & \quad x =r \cos \phi ,\,y =r \sin \phi \nonumber \\
I_1=0,\,I_2=1&\qquad V_P=\dfrac{f(\xi)+g(\eta)}{\xi^2+\eta^2} & \quad x=\dfrac{\xi^2-\eta^2}{2} \,,y=\xi \eta \nonumber \\
I_1=1,\,I_2=l^2 \neq 0 & \qquad V_E=\dfrac{f(\sigma)+g(\rho)}{\cos^2\sigma- \cosh^2\rho} &
\begin{array}{rl}
&x=l \cosh \rho \cos \sigma  \\
&y=l \sinh \rho \sin \sigma \\
&0 < l < \infty.
\end{array}
\end{eqnarray}
We see that $V_C,V_R,V_P$ and $V_E$ correspond to separation of variables in Cartesian, polar, parabolic and elliptic coordinates, respectively, and that second order integrability (in $E_2$) implies separation of variables. For second order superintegrability, two integrals of the form (\ref{inta}) exist and the Hamiltonian separates in at least two coordinate systems. 

Four three-parameter families of superintegrable systems exist namely
\begin{eqnarray}
&V_I=\alpha(x^2+y^2)+\dfrac{\beta}{x^2}+\dfrac{\gamma}{y^2},& V_{II}=\alpha(x^2+4y^2)+\dfrac{\beta}{x^2}+\gamma y \nonumber \\
&V_{III}=\dfrac{\alpha}{r}+\dfrac{1}{r^2}(\dfrac{\beta}{\cos^2\frac{\phi}{2}}+\dfrac{\gamma}{\sin^2\frac{\phi}{2}}),& V_{IV}=\dfrac{\alpha}{r}+\dfrac{1}{\sqrt{r}}(\beta \cos \frac{\phi}{2}+\gamma \sin \frac{\phi}{2}).
\end{eqnarray}
The classical trajectories, quantum energy levels and wave functions for all of these systems are known. The potentials $V_I$ and $V_{II}$ are isospectral deformations of the isotropic and an anisotropic harmonic oscillator, respectively, whereas $V_{III}$ and $V_{IV}$ are isospectral deformations of the Kepler-Coulomb potential. They correspond to cases E1, E2, E16, and E18 listed in Section 4.1.4.

While the situation is completely understood for second-order integrals, it is less so for third-order ones. In the next section, we move onto the determining equations for the third-order case and the  classification results that have been obtained.

\subsection{Third-order integrals}
 There are three sets of determining equations corresponding to $\ell=0,1,2$. The highest-order terms are the same as in all dimensions, 
 \be \label{genkilling3} 0=\frac{\partial f_{j-1,0}}{\partial x}+\frac{\partial f_{j, 0}}{\partial y },\qquad j=0, \dots 5   \ee
 and lead to the requirement that the highest-order terms be from the enveloping algebra. They are solved by setting 
 \[ f_{3,0}\equiv -A_{300}y^3+A_{210}y^2-A_{120}y+A_{030},\]
\[ f_{2,0}\equiv 3A_{300}xy^2-2A_{210}xy+A_{201}y^2+A_{120}x-A_{111}y+A_{021},\]
\[ f_{1,0}\equiv -3A_{300}x^2y-2A_{201}xy+A_{210}x^2+A_{111}x-A_{102}y+A_{012},\]
\[ f_{0,0}\equiv A_{300}x^3+A_{201}x^2+A_{102}x+A_{003}.\]
  In this case the quantum integral becomes 
\be \fl X=\sum_{j+k+\ell=3} \frac{A_{j,k,\ell}}2\left\{L_3^j,p_1^kp_2^\ell\right\}  +\frac12\left\{f_{1,2}-\hbar^2A_{2,1,0},p_1\right\} +\frac12\left\{f_{0,2}-\hbar^2A_{2,0,1},p_2\right\}.\ee
 In analogy with other references \cite{GW, Gravel, MW2008,  TW20101}, we set $f_{j,0}=F_{4-j}$ and 
 \[ G_1\equiv f_{1, 2}-\frac32 \hbar^2A_{2,1,0}+5\hbar^2 A_{3,0,0}x , \qquad G_2\equiv f_{0,2}-\frac32 \hbar^2A_{2,0,1}-5\hbar ^2A_{3, 0, 0}y.\]

 The next set of equations, for $\ell=1$ are 
  \be \label{3rdOmid} 0=2\frac{\partial f_{j-1,2}}{\partial x}+2\frac{\partial f_{j, 2}}{\partial y }-(j+1)\frac{\partial V}{\partial x} f_{j+1, 0}-(3-j)\frac{\partial V}{\partial y} f_{j, 0}-\hbar^2Q_{j,2}.\ee
Again, the quantum correction term vanishes on solutions of (\ref{genkilling3}). The equations (\ref{3rdOmid}) become
\bea \label{G1} 2(G_1)_x&=3F_1V_x+F_2V_y, &j=2,\\
2(G_1)_y+2(G_2)_x&=2(F_2V_x+F_3V_y), &j=1,\\
\label{G2}2(G_2)_y&=F_3V_x+3F_4 V_y, &j=0.\eea

Equations (\ref{G1})-(\ref{G2}) satisfy the following linear  compatibility conditions 
\bea 0&=&-F_3V_{xxx}+(2F_2-3F_4)V_{xxy}+(-3F_1+2F_3)V_{xyy}-F_2V_{yyy}\nn 
&+&2(F_{2y}-F_{3x})V_{xx}+2(-3F_{1y}+F_{2x}+F_{3y}-3F_{4x})V_{xy}+2(-F_{2y}+F_{3x})V_{yy}\nn
&+&(-3F_{1yy}+2F_{2xy}-F_{3xx})V_x+(-F_{2yy}+2F_{3xy}-3F_{4xx})V_y.\label{3rdordercomp}\eea
Finally, there is a single $\ell=2$ equation and it is given by 
\bea\label{quantnonlin}\fl  0&=& G_1V_x+G_2V_y\eea
\[\fl \quad  - \frac{\hbar^2}{4}\!\left(F_1V_{xxx}\!+\!F_2V_{xxy}\!+\!F_3V_{xyy}\!+\!F_4V_{yyy}\!+\!(8A_{300}y+2A_{210})V_x\!-\!(8A_{003}x-2A_{201})V_y\right).\]
Again, it is interesting to note that the quantum correction term $Q_{0,4}$ has a nontrivial term depending on $\hbar^4$ but this term vanishes on solutions of (\ref{genkilling3}).  

In general, these equations are difficult to solve and, indeed a full classification of their solutions is still an open question.  Early research on third-order integrals of motion was first performed by  Drach \cite{Drach}. He considered the case of one third-order integral of motion (in addition to the Hamiltonian) in  two-dimensional complex space in classical mechanics. He found 10 different complex potentials which allow a third-order integral. Later it was shown that 7 of them are actually quadratically superintegrable and the third-order integral is reducible, i.e. is the Poisson commutator of two second-order integrals \cite{RAN, Tsig}. 

One approach to solving these systems which has lead to the most complete classification is to assume separation of variables as well as the existence of a third-order integral; hence the systems are superintegrable with integrals of degree 2 and 3. This has been completed in Cartesian coordinate \cite{Gravel}, polar coordinates \cite{TW20101} and parabolic coordinates \cite{PopperPostWint2012}. 

\subsubsection{Cartesian coordinates}
This case was considered in \cite{Gravel}. Here we review these classification results. For uniformity we use the same approach  as was used for the potentials separable in polar \cite{TW20101} and parabolic coordinates \cite{PopperPostWint2012}. The results agree with those of \cite{Gravel}. Also note that the mass here is normalized to $m=2$ while the results in \cite{Gravel} have $m=1.$

Suppose that the Hamiltonian admits separation of variables in Cartesian coordinates as well as a third-order integral. In this case, the potential separates as $V=V_1(x)+V_{2}(y)$ and the linear compatibility condition (\ref{3rdordercomp}) reduces to 
\bea -F_{3}V_{1,xxx}+2(F_{2,y}-F_{3,x})V_{1,xx}-(3F_{1,yy}-2F_{2,xy}+F_{3,xx})V_{1,x}=\nn F_2V_{2,yyy}+(F_{2,y}-F_{3,x})V_{2,yy}+(F_{2,yy}-2F_{3,xy}+3F_{4,xx})V_{2y}.\label{cc}\eea
Differentiating (\ref{cc}) twice with respect to $x$ gives two linear ODEs for $V_1$:
\bea\fl \left( 3\,A_{{300}}{x}^{2}+2\,A_{{201}}x+A
_{{102}} \right)V_1^{(5)}+\left( 36\,A_{{300}}x+12\,A_{{201}} \right) V_{1}^{(4)}+ 84\,A_{{300}}V_{1}^{(3)}
 =0\label{W11},\\ \fl 
 \left( -A_{{111}}x-A_{{210}}{x}^{2}-A_{{012
}} \right) V_1^{(5)} +
 \left( -12\,A_{{210}}x-6\,A_{{111}} \right)V_1^{(4)} -28\,A_{{210}} V_1^{(3)}=0.\label{W12}\eea
Differentiating (\ref{cc}) twice with respect to $y$ gives two linear ODEs for $V_2$:
\bea \fl \left( 3\,A_{{300}}{y}^{2}-2\,A_{{210}}y+A_{{120}} \right) V_2^{(5)}+
 \left( 36\,A_{{300}}y-12\,A_{{210}} \right) V_2^{(4)}+  90\,A_{{300}}V_2^{(3)}=0,\label{W21}\\ \fl 
 \label{W22} \left( A_{{201}}{y}^{2} -A_{{111}}y+A_{{021}}\right) V_{2}^{(5)}+\left( 12\,A_{{201}}y-6\,A_{{111}} \right) V_2^{(4)}+ 30\,A_{{201}} V_2^{(3)} =0.\eea
$V_1$ will satisfy a linear ODE if either (\ref{W11}) or (\ref{W12}) are non-trivially satisfied. For these equations to be  satisfied identically, we must require
\be  A_{300}=A_{111}=A_{210}=A_{201}=A_{012}=A_{102}=0\label{V1nonlinear}.\ee
$V_2$ will also satisfy a linear equation unless equations (\ref{W21}) and (\ref{W22}) are satisfied identically when
\be A_{300}=A_{111}=A_{210}=A_{201}=A_{021}=A_{120}=0\label{V2nonlinear}.\ee
Due to the symmetry in the coordinates $x$ and $y$, there are essentially 3 possibilities for solutions of the determining equations:

\paragraph*{Case 1: Both $V_1$ and $V_2$ satisfy non-linear equations } 
This is the case when all four equations (\ref{W11}-\ref{W22}) are satisfied identically; this means that all the constants of (\ref{V1nonlinear}) and (\ref{V2nonlinear}) are zero. 
In this case, the only non-zero constants are $A_{030}$ and $A_{003}$, equations (\ref{G1}-\ref{G2}) can then be solved and the non-linear equation (\ref{quantnonlin}) separates. Up to translation in $x,y$ and the potential, it remains to only solve the following non-linear ODEs
\bea 12V_1^2-\hbar^2V_1^{(3)}=A_{003}\sigma x,\\ 
     12V_2^2-\hbar^2V_2^{(3)}=A_{030}\sigma y.\eea
If $\sigma=0$, the solutions can be expressed as a sum of Weierstrass elliptic functions in $x$ and $y$, respectively. In this case the system is not superintegrable since the third-order integrals are algebraically dependent on the second-order ones.  Otherwise, the solutions are given in terms of solutions of the first Painlev\'e equation  (${\cal P}_1$) \cite{Ince}
\be \fl V_a=2 \hbar^2\omega_1^2{\cal P}_1(\omega_1 x)+2\hbar^2\omega_2^2{\cal P}_1(\omega_2 x), \quad \omega_1=\left(\frac{A_{003}\sigma}{\hbar^4}\right)^{\frac15}, \, \omega_2=\left(\frac{A_{030}\sigma}{\hbar^4}\right)^{\frac15}.\ee
In the classical case, $\hbar=0,$ the potential becomes
\be V=\pm \sqrt{\beta_1 x}\pm \sqrt{\beta_2 y}.\ee

\paragraph*{Case 2:  $V_1=ax^2+bx+c$ and $V_2$ satisfies a non-linear equation}
This is the case when  only the two  equations (\ref{W11}-\ref{W12}) are satisfied identically; this means that all the constants of (\ref{V1nonlinear}) are zero.

If  $a=b=0$, then the system is translation invariant in $x$; this case was addressed in \cite{GW}. With these restrictions, the determining equations can be solved explicitly and the potential satisfies the   equation, 
\be \hbar^{2}V_{2,yy}=12V^2_2+\sigma_1 V_2+\sigma_2,\label{case2eq}\ee
 for $\sigma_1$ and $\sigma_2$ some constants of integration.
In the classical case $(\hbar=0)$ the leading order term in (\ref{case2eq}) vanishes and the only solution is $V_2=const$ and this corresponds to free motion. In the quantum case, we can integrate (\ref{case2eq}) to obtain a first-order equation of the form
\be \hbar^2(V_{2, y})^2=8(V-c_1)(V-c_2)(V-c_3),\ee
where $c_i$'s are constant (roots of a cubic equation). If all of the $c_i$ are real and different we obtain, depending on the initial conditions we impose, two types of solutions in terms of Jacobi elliptic functions 
\be V_b=2(\hbar \omega)^2k^2 {\rm sn}^2(\omega y, k), \qquad \mbox{ or }\quad  V_c=\frac{2(\hbar \omega)^2}{ {\rm sn}^2(\omega y, k)}.\ee
If the  roots satisfy $c_{1,2}=p+iq$ with $p,q\in \mathbb{R}$, $q>0$, $c_3\in \mathbb{R}$, the solution is 
\be V_d= \frac{2(\hbar \omega)^2}{2 \ {\rm cn}^2(\omega y, k)+1},\ee
(the constants $\omega\in \mathbb{R}$ $0<k<1$ are expressed in terms of $c_1, c_2, c_3)$. The potentials $V_c$ and $V_d$ are singular and all three are periodic. 

If the two roots coincide (i.e. $k=0$ or $k=1$) we get elementary solutions. They yield  superintegrable potentials of the form 
\be V_{e}= \frac{2(\hbar \omega)^2}{ \cosh^2(\omega y, k)}, \qquad V_{f}= \frac{(\hbar \omega)^2}{ \sin^2(\omega y, k)},\quad V_{g}= \frac{2(\hbar \omega)^2}{ \sinh^2(\omega y, k)}.\ee
Here $V_{e}$ corresponds to the well known soliton solution of the Korteweg-de Vries equation, $V_g$ is a ``singular soliton" solution and $V_f$ is periodic and singular.  If all three roots coincide, we re-obtain a known superintegrable potential $2\hbar^2/x^2. $

 There are two cases remaining, namely  $a\ne 0$ or $b\ne 0$ in the potential $V_1$. In the first case, (\ref{G1}-\ref{G2}) can then be solved and (\ref{quantnonlin}) reduces to two non-linear equations. These equations yield trivial solutions (i.e. solutions which satisfy linear equations as in Case 3) unless $A_{030}=0$ and $A_{120}\ne 0$. In this case,(\ref{quantnonlin}) reduced to a single non-linear ODE
\begin{equation}\label{ord4}
0=-\hbar^2V_1^{(4)} -12 a(xV_1)'+12(V_1^2)''-2 a x^2 V_1'' +a^2 x^2.
\end{equation}
As shown in \cite{Gravel}, where the analysis is explained in more detail, the solutions can be expressed in terms of solutions of the fourth Painlev\'e equation ${\cal P}_4$ \cite{Ince}. The potential becomes 
\bea 
\fl V_h&=&a (x^2+y^2)\pm \hbar^2\sqrt{4a} {\cal P}_4'(x,\frac{-4a}{\hbar^2})+4a
{\cal P}_4^2(x,\frac{-4a}{\hbar^2})+4 a x {\cal P}_4(x,\frac{-4
a}{\hbar^2}).
\eea
In the classical limit (as $\hbar\rightarrow 0$) the potential satisfies \begin{equation}\label{implic1}
c x^2- d^2+2 d(V_1-a x^2)(3V_1+ax^2)=(9V_1-ax^2)(V_1-ax^2)^3,
\end{equation}
 where $c$ and $d$ are arbitrary constants. It is interesting to note that it is possible to recover both the simple harmonic oscillator or the anisotropic oscillator with a ratio of $1:3$ as special cases of both the classical and quantum system \cite{Gravel}.

Finally, in the case that $a=0$ and $b\ne 0$, again  (\ref{G1}-\ref{G2}) can  be solved and (\ref{quantnonlin}) reduces to two non-linear equations. Depending on the choices of non-zero $A_{ijk}$ (the leading terms in the third-order integral, the two non-linear equations reduce to 
\begin{equation}\label{painleve1}
\h^2 V_1''=12V_1^2+\lambda x+k, \qquad A_{030}\ne 0
\end{equation}
or 
\begin{equation}\label{painleve2}
0=\left(-3(V_1^2)'+\frac{\h^2}{4} V_1^{(3)}\right)'+b
\left((xV_1')'+2V_1'\right)\qquad A_{030}A_{021}\ne 0.
\end{equation}

The solutions for this quantum system (\ref{painleve1}) are given in terms of solutions to the first Painlev\'e equation ${\cal P}_1,$ ($V_i$ below) and the solutions for  (\ref{painleve2}) are given in terms of solutions to the second Painlev\'e equation ${\cal P}_2,$ ($V_j, V_k$ below)
\bea
&V_i=ay+2\hbar^2\omega^2{\cal P}_1(\omega x),\\
&V_j=b x+a y +2(2\h
b)^{\frac{2}{3}}{\cal P}_2^2\left(\left(\frac{2b}{\h^2}\right)^\frac{1}{3}
x,0\right),\\ 
&V_k=b y +2\left( 2 \hbar^2 a^{2}\right)^\frac{1}{3} \left( {\cal P}_2'(-(4
a \hbar^2)^\frac{1}{3} x, \kappa)+ {\cal P}_2^2(-(4 a \hbar^2)^\frac{1}{3} x,
\kappa)\right).
\eea
In the classical limit the solutions to (\ref{painleve1}) and (\ref{painleve2}) are given, respectively, by 
\bea  V_\ell=ay+b\sqrt{x}, \qquad 
V_m=ay+V_1(x), \qquad d=V_1(V_1-bx)^2.
\eea
Throughout, $a,b,c$ and $d$ are arbitrary constants. We stress that the classical limit of the exotic potentials are always singular and we must take the limit of the determining equations rather than the limit of the solutions.

\paragraph*{Case 3:  $V_1$ and $V_2$ satisfy linear equations}
In the cases where at least one of the equations (\ref{W11}-\ref{W12}) and (\ref{W21}-\ref{W22}) is not satisfied identically, the potential functions satisfy linear equations and in particular are rational functions. In this case, the method of solving the system is to simply solve these linear ODEs for $V_1$ and $V_2$, then solve (\ref{G1}-\ref{G2}) and replace the solutions into (\ref{quantnonlin}). The equation (\ref{quantnonlin}) then becomes a functional equation for the parameters of the system which must hold for all values of $x$ and $y.$ This restricts the possible cases to the following: 
\bea
V=\left\{\ba{l} \frac{\hbar^2}{4\alpha^4}(x^2
+y^2)+\frac{2\hbar^2}{(x-\alpha)^2}+\frac{2\hbar^2}{(x+\alpha)^2},\\
\frac{\hbar^2}{4 \alpha^4}(x^2
+y^2)+\frac{2\hbar^2}{y^2}+\frac{2\hbar^2}{(x+\alpha)^2}+\frac{2\hbar^2}{(x-\alpha)^2},\\
\frac{\hbar^2}{4 \alpha^4}(x^2
+y^2)+\frac{2\hbar^2}{(y-\alpha)^2}+\frac{2\hbar^2}{(x-\alpha)^2}+\frac{2\hbar^2}{(y+\alpha)^2}+\frac{2\hbar^2}{(x+\alpha)^2},\\
a (9 x^2 +y^2), \qquad \mbox{ also classical, 3:1 oscillator},\\
a (9 x^2 +y^2)+\frac{2\hbar^2}{y^2},\\\
\frac{\hbar^2}{4\alpha^4}(9 x^2
+y^2)+\frac{2\hbar^2}{(y+\alpha)^2}+\frac{2\hbar^2}{(y-\alpha)^2}.
\ea
\right.
\eea

Again, this list is taken from \cite{Gravel} although we have omitted the systems which are second-order superintegrable. Recall, that if a system is second-order superintegrable then it will have an integral of third-order obtained from taking the commutator (or Poisson commutator) of the two integrals which are not the Hamiltonian. Also, several of these systems are classically superintegrable, where $\hbar$ is considered a constant not set to 0. For example, the harmonic oscillator with rational frequencies is superintegrable even with the presence of singular terms \cite{EvansVerrier2008, RTW2008}. However the integrals will be higher-order polynomials in the momenta. The specifically quantum potentials here indicate the existence of a factorized form of the integrals for certain, $\hbar$-dependent, choices of constants.

This completes the classification of third-order superintegrable systems which admit separation of variables in Cartesian coordinates.

\subsubsection{Polar coordinates}
Now, let us consider potentials allowing separation of variables in polar coordinates \cite{TW20101} 
 \be V=R(r)+\frac{1}{r^2}S(\theta).\ee
 The linear compatibility condition reduces to 
 \bea\fl \ba{rl} 0&= r^4 F_3R'''+ \left( 2{r}^{4} F_{{3,r}} -2{r}^{2}F_{{2,\theta}}+3r \left( 2
\,{r}^{2}F_{{3}} -F_{{1}}   \right)  \right) R'' \\ &
+\left({r}^{4}F_{{3,rr}}  +2\,{r}^{2}(3F_{{3,r}} +3F_{{
3}}-F
_{{2,r\theta }}) -4\,r F
_{{2, \theta}} +3 F_{{1, \theta\theta}} \right) R'\\ &
+\frac{1}{r^2}F_2S'''-\frac{1}{r^3}\left(2r^3F_{3,r}-2rF_{2,\theta}+6F_1\right)S''\\ &+\frac{1}{r^3}\left(3r^5F_{4,rr}+6r^4F_{4,r}-2r^3F_{3,r\theta}+3r^2F_{2,r}+r(F_{2,\theta\theta}-2F_2)-12F_{1,\theta}\right)S'\\ &-\frac{1}{r^3}\left(2r^4F_{3,rr}-12r^3F_{3,r}+4r^2(3F_{3,r}-F_{2,r\theta})+4rF_{2,\theta}+6F_{1,\theta\theta}+18F_1\right)S,\ea \label{ccP}\eea
where  the $F_i$'s are given by
{
\bea \fl \ba{rl} F_1&=A_1\cos 3\theta +A_2\sin 3\theta +A_3 \cos \theta +A_4\sin\theta,\\ 
F_2&=\frac{-3A_1\sin 3\theta+3A_2\cos 3\theta -A_3\cos \theta +A_4\sin \theta}{r}+B_1\cos 2\theta +B_2\sin 2 \theta+B_0,\\
F_3&= \frac{-3A_1\cos 3\theta-3A_2\sin 3\theta +A_3\cos \theta +A_4\sin \theta}{r^2}+\frac{-2B_1\sin 2\theta +2B_2\cos 2 \theta}{r}+C_1\cos\theta +C_2\sin \theta,\\
F_4&=\frac{A_1\sin 3 \theta -A_2\cos 3\theta -A_3\sin \theta +A_4 \cos \theta}{r^3} -\frac{B_1\cos 2\theta+B_2\sin 2 \theta +B_0}{r^2}-\frac{C_1\sin\theta-C_2\cos\theta}{r}+D_0, \ea \nonumber\eea}
with 
\bea \ba{llll}A_1=\frac{A_{030}-A_{012}}{4}, &A_2=\frac{A_{021}-A_{003}}{4},& A_3=\frac{3A_{030}+A_{012}}{4},&\\
A_4=\frac{3A_{003}+A_{021}}{4}, & B_1=\frac{A_{120}-A_{102}}{2},&B_2=\frac{A_{111}}2, & B_0=\frac{A_{120}+A_{102}}2,\nn
C_1=A_{210}, & C_2=A_{201}, & D_0=A_{300}.& \ea\nonumber\eea

Here we summarize the results of \cite{TW20101}. Beginning with the linear compatibility condition, (\ref{ccP}),  it is possible to obtain by differentiating with respect to $r$ several differential consequences which are ODEs for the function $R(r)$. These are then solved and the possible choices are reduced to 4 cases:  $R(r)=ar^2$, $R(r)=a/r$, $R(r)=0$,  or the equations for $R(r)$ vanish by choice of the $A_{jk\ell}$. 

In the case that $R(r)=ar^2$, the only solution is the second-order superintegrable isotropic  oscillator with singular terms (E1/Smorodinsky-Winternitz I). Similarly, the case where $R(r)=a/r$ leads to a Kepler-Coulomb potential with singular term (E16/Smorodinsky-Winternitz III). 

Finally, consider the cases that the equations for $R$ are satisfied identically. This can occur if $R(r)$ arbitrary and only $A_{300}\ne 0$ or  $R(r)=0$. In the first case, this leads to the potential\cite{TW20101}
\[ V(r, \theta)=R(r)+\frac{2\hbar^2}{r^2}{\cal P}(\theta, t_2, t_3),\]
where $R(r)$ is arbitrary and  ${\cal P}(\theta, t_2, t_3)$ is the Weierstrass elliptic function \cite{ByrdFriedman}. This potential allows a third-order integral, however it is algebraically related to the second-order one and the system is hence not superintegrable. 
However, if $R(r)=0$ then the system admits two third-order integrals one of which is algebraically independent of the two second-order integrals. The remaining cases when $R(r)=0$ include the rational three-body Calogero system (or a special case of the TTW system \cite{TTW})
\[ V=\frac{a}{r^2\cos^2(3\theta)},\]
 which is known to be superintegrable \cite{WOJ}. There are also potentials which satisfy nonlinear equations. These occur when $R(r)=0$ and the linear compatibility condition (\ref{ccP}) for   $S(\theta)$ are satisfied trivially, i.e. when all the constants except $A_{300}$, $A_{210}$ and $A_{201}$ are 0. This leads to a quantum superintegrable potential expressed in terms of the sixth Painlev\'e transcendent\cite{Ince}, ${\cal P}_{VI}(\sin\theta/2).$ The potential is given by 
 \be 
 \fl V(r,\theta)=\frac{2}{r^2}\Bigg(\hbar^2W'(x_{\pm})-\frac{\pm8\hbar^2\cos\theta W(x_{\pm})+4\beta_1+\hbar^2}{4\sin^2\theta}\Bigg),\quad x_{\pm}=\cos^2\theta, \sin^2 \theta, \ee
 with 
 \bea\label{WP6}
\fl  W(x)&=&\frac{x^2(x-1)^2}{4{\cal P}_6({\cal P}_6-1)({\cal  P}_6-x)}\bigg[{\cal P}_6'-\frac{{\cal P}_6({\cal P}_6-1)}{x(x-1)}\bigg]^2+\frac{1}{8}(1-\sqrt{2\gamma_1})^2(1-2{\cal P}_6)\nonumber\\ \fl 
&&-\frac{1}{4}\gamma_2\bigg(1-\frac{2x}{{\cal P}_6}\bigg)-\frac{1}{4}\gamma_3\bigg(1-\frac{2(x-1)}{{\cal P}_6-1}\bigg)+\bigg(\frac{1}{8}-\frac{\gamma_4}{4}\bigg)\bigg(1-\frac{2x({\cal P}_6-1)}{{\cal P}_6-x}\bigg).
\eea

  The classical limit of this system $\hbar=0$ is given by 
  \be V=\frac{\dot{T}(\theta)}{r^2},\label{classicalT}\ee 
  where $T$ satisfies a first order linear ODE,
  \be\fl  3z^2(1+z^2)T'^2+2zTT'-T^2+2(\beta_1z^2-\beta_2z)T'+2\beta_2 T+\frac{K_1}{2}=0, \quad z=\tan \theta\label{implic2} .\ee

 To summarize, for systems which admit separation of variables in polar coordinates as well as a third-order integral, there are the second-order superintegrable systems which separate in polar coordinates, the rational three-body Calogero system, as well as two new quantum systems depending either the Weierstrass elliptic function or the sixth Painlev\'e transcendent. Finally, there is also a new classically superintegrable system with the radial part of the potential satisfying a first-order nonlinear ODE (\ref{classicalT}), (\ref{implic2}).

\subsubsection{Parabolic coordinates}
In the final case considered here, let us assume that the potential separates in parabolic coordinates, 
$ x=\frac12 (\xi^2-\eta^2)$, $y=\xi \eta$,
\bea \label{H} H=-\frac{\hbar^2}{\xi^2+\eta^2}\left(\frac{\partial^2}{\partial \xi^2}+\frac{\partial^2}{\partial \eta^2}\right)+V(\xi,\eta),\\
\label{V} V(\xi, \eta)=\frac{W_1(\xi)+W_2(\eta)}{\xi^2+\eta^2}.\eea
The compatibility condition (\ref{3rdordercomp}) becomes \bea \label{ccPar} \fl \ba{rl}0=& F_3 V_{{\xi \xi \xi}} +(3\,F_{{4}}-2\,F_{{2}})V_{{\xi \xi \eta}} +(3\,F_{{1}}-2\,F_{{3}})V_{{\xi \eta \eta}}+F_{{2}}V_{{\eta \eta \eta}}\\&+
\left(2(F_{{3\,\xi}}-F_{{2\,\eta}}) -{\frac {3\xi F_{{1}} 
 -6\eta F_{{2}}  
+7\,\xi\,F_{{3}} }{  {\xi}^{2}+{\eta}^{2} }}\right)V_{{\xi \xi}}+\left(2(F_{{2\,\eta}}-F_{{3\,\xi}}) -{\frac  {3\,\eta\,F_{{4}}  -6\,\xi\,F_{{3}}+7\eta F_{{2}}   }{  {\xi}^{2}+{\eta}^{2} }}\right)V_{{\eta\eta }}\\&+
\left(2\left(3{F_{{1\,\eta}}}- {F_{{2\,\xi}}
}- {F_{{3\,\eta}}}+3 {F_{{4\,\xi}}} \right)-{\frac {21\,\eta\,F_
{{1}}  -5\,\eta\,F_{{3}} 
  -5\,\xi\,F_{{2}}  +21\,\xi\,F_{{4}}}{ {\xi}^{2}+{\eta}^{2} }}\right)V_{{\xi \eta}}\\&+
 A\,V_\eta + B\,V_\xi ,\ea\eea
  where
 \bea\fl \nonumber A&=& {F_{{2\,\eta\,\eta}}}-2{F_{{3\,
\eta\,\xi}}}+3{F_{{4\,\xi\,\xi}}}
+{\frac {-7\eta\,F_{{2\,\eta}} -\xi\,F_{{2\,\xi}}+6\xi\,F_{{3\,\eta}} +6\eta\,F_{{3\,\xi}}-3\eta\,F_{{4\,\eta}}-21\xi\,F_{{4\,\xi}}  }{{\xi}^{2}+{\eta
}^{2} }}  \nn \fl &&
+2\,{\frac {21\,{\xi}^{2}F_
{{4}}  +F_{{2}}  {\xi}^{
2}+7\,{\eta}^{2}F_{{2}}  -12\,\xi\,\eta\,F_{{3}
}  +3\,F_{{4}}  {\eta}^{
2}}{ \left( {\xi}^{2}+{\eta}^{2} \right) ^{2}}}, \nonumber
\eea
\bea \nonumber  \fl B&=& 3 F_{{1\,\eta\,\eta}}-2 {F_{{2\,
\eta\,\xi}}}+{F_{{3\,\xi,\xi}}}-{\frac {21\eta\,F_{{1\,\eta}}+3\xi\,F_{{1\,\xi}}-6\xi\,F_{{2\,\eta}} -6\eta\,F_{{2\,\xi}}+\eta\,F_{{3\,\eta}}+7\xi\,F_{{3\,\xi}
} }{  {\xi}^{2}+{\eta}^
{2} }}\nn \fl &&+2\,{\frac {F_{{3}}
  {\eta}^{2}+3\,F_{{1}}  {\xi}^{2}+21\,{\eta}^{2}F_{{1}}  -12\,
\xi\,\eta\,F_{{2}}  +7\,{\xi}^{2}F_{{3}}
  }{ \left( {\xi}^{2}+{\eta}^{2} \right) ^{2}}}. \nonumber\eea
As explained in \cite{PopperPostWint2012}, these equations can be solved in a similar manner to the previous cases of Cartesian and polar coordinates. However, unlike the previous cases, all such superintegrable systems are second-order superintegrable. It is not difficult to see that the linear compatibility condition (\ref{ccPar}) vanishes identically only when all of the $A_{jk\ell}$ are zero. Thus, the potential terms $W_1(\xi), W_2(\eta)$ always satisfy some non-trivial linear ODE. All that remains is to show by brute force that the only possible solutions of the determining equations are known second-order superintegrable systems. 

Thus, we have shown that for systems which admit a third-order integral and separation of variables, there are new, truly third-order systems in the case of Cartesian and polar coordinates although not for parabolic. It is unknown what happens in the case of general elliptic coordinates, though we conjecture that, in this case, there will be no non-trivial solutions for the linear compatibility conditions, in analogy with the parabolic cases. Hence, it would seem that the existence of potentials which satisfy non-linear equations exist only when there is separation in subgroup type coordinates. This is still an open question. 

\subsubsection{Summary of the case of third-order integrals}
First of all, let us sum up the cases of third-order superintegrability that have been obtained. In classical mechanics they are given in Table 1. 
\begin{table}[ht]\label{table:classical}
\caption{Classical third-order superintegrable systems}
\[\fl  \nonumber \ba{|l|r|}
\hline
V & \mbox{original reference}\\
\hline
 a (9 x^2 +y^2),& \mbox{ from \cite{Gravel}, } C.4  \\
\pm \sqrt{\beta_1 x}\pm \sqrt{\beta_2 y},& C.5\\
ay^2 +V_{1} \mbox{ where $V_{1}$ satisfies equation \eref{implic1}},& C.6\\
ay+b\sqrt{x},& C.7\\
ay^2 +V_{1} \mbox{ where } (V_{1}-bx)^2 V_{1}=d,& C.8\\
\frac{\alpha}{r^2\sin^23\theta}, & \mbox{ from \cite{TW20101}, } (4.21)  \\
\frac{\dot{T}(\theta)}{r^2} \mbox{ where $T$ satisfies equation (\ref{implic2})},& (5.1).\\\hline
\ea \nonumber\]
\end{table}
We omit those for which the third-order integral is a Poisson commutator of two second-order integrals, since such systems are already second-order superintegrable. The potentials (C.4) and (4.21) were already known (see \cite{JauchHill1940, WOJ}). The others are new. The finite trajectories for all of these potentials have been shown to be periodic (as they must be) \cite{MW2007, MW2008}. 

The results in the quantum case are considerably richer and are reproduced in Table 2.
 \begin{table}[ht]\label{table:quantum}.
\caption{Quantum third-order superintegrable systems}
\[  \fl  \ba{|l|r|}
\hline
V & \mbox{original reference}\\
\hline
 \frac{\hbar^2}{4\alpha^4}(x^2
+y^2)+\frac{2\hbar^2}{(x-\alpha)^2}+\frac{2\hbar^2}{(x+\alpha)^2},& \mbox{ from \cite{Gravel}, } (Q.5)\\
\frac{\hbar^2}{4 \alpha^4}(x^2
+y^2)+\frac{2\hbar^2}{y^2}+\frac{2\hbar^2}{(x+\alpha)^2}+\frac{2\hbar^2}{(x-\alpha)^2},& (Q.6)\\
\frac{\hbar^2}{4 \alpha^4}(x^2
+y^2)+\frac{2\hbar^2}{(y-\alpha)^2}+\frac{2\hbar^2}{(x-\alpha)^2}+\frac{2\hbar^2}{(y+\alpha)^2}+\frac{2\hbar^2}{(x+\alpha)^2},& (Q.7)\\
a (9 x^2 +y^2), & (Q.9)\\
a (9 x^2 +y^2)+\frac{2\hbar^2}{y^2},& (Q.10)\\\
\frac{\hbar^2}{4\alpha^4}(9 x^2
+y^2)+\frac{2\hbar^2}{(y+\alpha)^2}+\frac{2\hbar^2}{(y-\alpha)^2},& (Q.11)\\
2\h^2 \omega_1^2 {\cal P}_1\left(\omega_1 x\right)+2\h^2
\omega_2^2{\cal P}_1\left(\omega_2 x\right),&(Q.17)\\
a (x^2+y^2)\pm \hbar^2\sqrt{4a} {\cal P}_4'(x,\frac{-4a}{\hbar^2})+4a
{\cal P}_4^2(x,\frac{-4a}{\hbar^2})+4 a x {\cal P}_4(x,\frac{-4
a}{\hbar^2}),&(Q.18)\\
a y+2\h^2 \omega^2 {\cal P}_1(\omega x),&(Q.19)\\
b x+a y +2(2\h
b)^{\frac{2}{3}}{\cal P}_2^2\left(\left(\frac{2b}{\h^2}\right)^\frac{1}{3}
x,0\right),&(Q.20)\\
a y +2\left( 2 \h^2 b^{2}\right)^\frac{1}{3} \left(
{\cal P}_2'(-(4 b/\h^2)^\frac{1}{3} x, \kappa)+{\cal  P}_2^2(-(4
b/\h^2)^\frac{1}{3} x, \kappa)\right),& (Q.21)\\
2(\hbar \omega)^2k^2 {\rm sn}^2(\omega y, k),\, \frac{2(\hbar \omega)^2}{ {\rm sn}^2(\omega y, k)},\, \frac{2(\hbar \omega)^2}{2(  {\rm cn}^2(\omega y, k)+1)},& \mbox{ from \cite{GW}, } (4.11)\\
\frac{\alpha}{r^2\sin^23\theta}, & \mbox{ from \cite{TW20101}, } (4.21)  \\
\frac{2}{r^2}\Bigg(\hbar^2W'(x_{\pm})-\frac{\pm8\hbar^2\cos\theta W(x_{\pm})+4\beta_1+\hbar^2}{4\sin^2\theta}\Bigg), \mbox{ where }W \mbox{ satisfies (\ref{WP6})}, &(4.44)\\
 \frac{\hbar^2}{r^2}{\cal P}(\theta, t_2, t_3), &(4.55).\\
\hline
\ea \nonumber\]
\end{table}
The systematic study of superintegrable systems with at least one third-order integral of motion confirms in a striking manner the differences between classical and quantum integrability. In the case of second-order integrability and superintegrability, it suffices to determine all classical superintegrable systems. The quantum systems can be obtained from the classical ones by the usual quantization procedure. The potentials are the same in both cases. The integrals of motion are also the same up to symmetrization of the operators. For third and higher-order integrals the situation is quite different. As we have shown in Section 5.1 and especially in Section 5.3 the determining equations in the quantum case contain terms proportional to $\h^2$ (or powers of $\h^2$). Thus the potentials in the quantum case can be quite different than in the classical one, as can the integrals of motion. 
An extreme case of this occurs when the potential is proportional to $\h^2$ and in the classical limit corresponds to free motion.

 %This is the case for all potentials expressed in terms of elliptic functions and Painlev\'e transcendents.

The Painlev\'e transcendent potentials have interesting polynomial algebras \cite{MW2007, Marquette20103}. Their representations can be used to calculate energy levels. There exist interesting connections with supersymmetric quantum mechanics that make it possible to calculate the wave functions in such potentials \cite{MW2008,marquette2009superintegrability, marquette2009painleve}. This has so far been carried out for the potential expressed in terms of the Painlev\'e transcendent ${\cal P}_4$ \cite{marquette2009painleve}. Interestingly it turns out that this potential is an isospectral deformation of the harmonic oscillator: the spectrum depends linearly on one quantum number with integer values and hence coincides with that of the isotropic 2D harmonic oscillator. The degeneracy of the energy levels is however different and resembles that of an anisotropic harmonic oscillator. An analysis of these questions goes beyond the scope of the present review, though for ${\cal P}_{4}$ they are treated in \cite{marquette2009painleve}. The fact that the Schr\"odinger equation with a Painlev\'e function potential
can be solved algebraically lends further  credence to the conjecture that
all (Euclidean) maximally superintegrable potentials
are exactly solvable \cite{TempTW}.  
 
\subsection{Fourth-order integrals}
We consider the 4th order classical operator defined by 
\be X=\sum_{j=0}^4f_{j,0}p_1^jp_2^{4-j}+\sum_{j=0}^2f_{j,2}p_1^jp_2^{2-j}+f_{0,4}.\ee

Again, the highest order determining equations  require that the term $ f_{j,0}p_1^jp_2^{3-j}$ be in the enveloping algebra of $E(2,\mathbb{R})$ and so we can rewrite it as 
 \[\sum_{j=0}^4 f_{j,0}p_1^jp_2^{4-j}=\sum_{i+j+k=4}A_{ijk}p_1^ip_2^jL_3^k, \qquad L_3=x p_2-yp_1.\]
 In analogy with the third-order case and to draw comparisons with the standard form of the third-order equations in the literature, we make the following definitions:
 \be f_{22}=g_1, \quad f_{12}=g_2, \qquad f_{0,2}=g_3,\ee
 and 
 \bea\fl 
F_1&\equiv& 4(-A_{004}y^4+A_{103}y^3-A_{202}y^2+A_{301}y-A_{400}),\nn\fl 
F_2&\equiv&4A_{004}xy^3+A_{013}y^3-3A_{103}xy^2-A_{112}y^2+2A_{202}xy-A_{211}y-A_{301}x-A_{310},\nn\fl 
F_3&\equiv&-6A_{004}x^2y^2-3A_{013}xy^2+3A_{103}x^2y-A_{002}y^2-A_{202}x^2-A_{211}x+A_{121}y-A_{220},\nn\fl 
F_4&\equiv & 4A_{004}x^3y-A_{103}x^3+3A_{013}x^2y-A_{112}x^2+2A_{022}xy-A_{121}x-A_{103},\nn\fl 
F_5&\equiv&-4(A_{004}x^4+A_{013}x^3+A_{022}x^2+A_{031}x+A_{040}).\eea
In this form, the  determining equations for the classical systems are given by 
\bea\label{cl41} 0=2g_{1,x}+F_1V_x+F_2V_y,\\
\label{cl42}0=2g_{2,x}+2g_{1,y}+3F_2V_x+2F_3V_y,\\
\label{cl43}0=2g_{3,x}+2g_{2,y}+2F_3V_x+3F_4V_y,\\
\label{cl44}0=2g_{3,y}+F_4V_x+F_5V_y,\eea
and 
\bea \label{cl45}f_{04,x}=g_{1}V_x+\frac12 g_{2}V_y,\\
\label{cl46} f_{04,y}=\frac12 g_{2}V_x-g_{3}V_y.\eea

The compatibility equations for \eref{cl41}-\eref{cl44}  are linear and given by 
\[ \fl 0=\partial_{yyy}\left(F_1V_x+F_2V_y \right)-\partial_{xyy}\left(3F_2V_x+2F_3V_y \right) +\partial_{xxy}\left(2F_3V_x+3F_4V_y \right)-\partial_{xxx}\left(F_4V_x+F_5V_y \right).\]
Equations \eref{cl45} and \eref{cl46} have nonlinear compatibility equations.

For the quantum integral, we take the following choice of symmetrization
\be \fl  X=\!\!\!\!\!\!\sum_{j+k+l=4} \!\!\!\frac{A_{jkl}}2 \lbrace L_3^j, p_1^kp_2^l\rbrace+\frac{(-i\hbar)^2}2\left(\lbrace g_1, \partial_x^2\rbrace+\lbrace g_2 \partial_x\partial_y\rbrace+\lbrace g_3, \partial_y^2\rbrace\right)+f_{04}.\ee
Note again that, as shown in \cite{PostWintNth},  a different choice would lead to an $\hbar^2$ correction term in $g_1, g_2$ or $g_3.$
The determining equations are given by the first 4 equations for the classical system given \eref{cl41}, \eref{cl42}, \eref{cl43}, \eref{cl44} and the last two equations with a correction term of order $\hbar^2$ 
\bea \fl \label{cq45} f_{04,x}=&2g_1V_x+& g_{2}V_y\\
\fl &+\frac{\hbar^2}{4}\bigg(&(F_2+F_4)V_{xxy}-4(F_1-F_5)V_{xyy}-(F_2-F_6)V_{yyy}\nn
\fl&&+(F_{2,y}-F_{5,x})V_{xx}-(13F_{1,y}+F_{4,x})V_{xy}-4(F_{2,y}-F_{5,x})V_{yy}\nn
\fl && +2(6A_{400}x^2+62A_{400}y^2+3A_{301}x-29A_{310}y+9A_{220}+3A_{202})V_x\nn
\fl && +2(56A_{400}xy+13A_{310}x-13A_{301}y+3A_{211})V_y\bigg),\nn
\fl\label{cq46} f_{04,y}=&g_{2}V_x+&2g_{3}V_y\\ \fl
&+\frac{\hbar^2}{4}\bigg(&-(F_2+F_4)V_{xxx}+4(F_1-F_5)V_{xxy}+(F_2-F_6)V_{xyy}\nn \fl
 &&+4(F_{1,y}-F_{4,x})V_{xx}-(F_{2,y}+13F_{5,x})V_{xy}-(F_{1,y}-3F_{4,x})V_{yy}\nn \fl
 &&+2(56A_{400}xy-13A_{310}x+13A_{301}y+3A_{211})V_x\nn\fl
 &&+2(62A_{400}x^2+6A_{400}y^2+29A_{A301}x-3A_{310}y+9A_{202}+3A_{220})V_y\bigg).\nonumber \eea
Note that the terms on the second line of each equation are the quantum corrections terms $Q_{0,4}$ and $Q_{1,4}$ and, similarly to case $N=2,3,$ the terms proportional to $\hbar^4$ in (\ref{Qjl}) vanish identically. 

The only known attempt to solve these equations directly, was given in \cite{PW2011} where a nonseparable superintegrable system was constructed with a third and fourth-order integral. What resulted was the following system, which is a quantum version of a special case of the Drach systems:
\begin{theorem} The operator triplet $( H, X, Y)$ with 
\[ \label{hq} H=\frac12(p_1^2+p_2^2)+\frac{\alpha y}{x^{\frac23}}-\frac{5\hbar^2}{72x^2}\]
 \[\label{xq} X=3p_1^2p_2+2p_2^3 +\lbrace\frac{9\alpha}{2}x^{\frac13}, p_1\rbrace +\lbrace \frac{3\alpha y}{x^{\frac23}}-\frac{5\hbar^2}{24x^2}, p_2\rbrace\]
 \[ \fl \label{yq}  Y=p_1^4 +\left\{\frac{2\alpha y}{x^{\frac23}}-\frac{5\hbar^2}{36 x^2},p_1^2\right\}-\left\{6x^{\frac13}\alpha, p_1p_2\right\}-\frac{2\alpha^2(9x^2-2y^2)}{x^{\frac43}}  -\frac{5\alpha\hbar^2 y}{9x^{\frac83}}+\frac{25\hbar^4}{1296x^4}\]
 constitutes a quantum superintegrable system that does not allow multiplicative separation of variables in the Schr\"odinger equation in any system of coordinates. 
\end{theorem}

%\subsection{Conclusion}
The work on third-order superintegrable systems which admit separation of variables in one coordinate system represents the only successful effort to classify solutions for higher-order superintegrability directly from the structure equations. The study of third-order superintegrability has lead to qualitatively new types of superintegrable quantum potentials, the Painlev\'e transcendent ones. These are solutions of second-order nonlinear ODEs and they do not satisfy any linear ODE. The system of determining equations obtained in Section 5.1 is overdetermined. Hence there will always exist compatibility conditions for the potential, some of them linear. In the case of potentials allowing separation of variables in subgroup type coordinates (Cartesian and polar) the compatibility conditions could be satisfied trivially. This was done by setting some of the leading coefficients in the third-order integral equal to 0 without out annihilating completely the highest order term. For potentials separating in parabolic coordinates this was not possible.  A study of higher-order superintegrability,  in particular the fourth-order case, could lead to interesting results, such as new transcendent functions with the Painlev\'e property. 
 
As demonstrated by the example of fourth-order integrals, the number of determining equations and unknowns grow quickly  and it becomes computationally inefficient to address these determining equations directly.  For a long time research into higher-order superintegrability was stymied by the intractability of the determining equations for such higher-order integrals. However, as we will see in the next sections, new methods of construction and verification of higher-order integrals, beyond brute force analysis of the determining equations, has led to a flourishing of new results.

%% file: higherorderclassicalchapter.tex
\section{Higher order  classical superintegrable systems} \label{higherorderclassicalchapter}

\subsection{The problems and the breakthroughs}

 Until very recently, few examples beyond the anisotropic oscillator were known of superintegrable systems with orders $N>3$ and there was almost no structure theory in either the classical or quantum case.  There were two basic issues: 1) How to construct numerous examples of systems of higher order and in many dimensions,  to develop enough insight to produce a classification  theory. 2) How to compute efficiently the commutators  of symmetry operators of arbitrarily high order, to verify superintegrability and determine the structure equations.
A breakthrough for the first issue came with the publication of two papers by Tremblay, Turbiner and Winternitz, \cite{TTW, TTW2}, the first in 2009.  They started with the 2nd order superintegrable system
\be\label{WSop}H=\partial_x^2+\partial_y^2-\omega^2(x^2+y^2)+\frac{\alpha}{x^2}+\frac{\beta}{y^2},\ee
with a classical analog also superintegrable. (Actually, they used the physical version of the Hamiltonian rather than the one used here, so that the sign of the energy flips.) They wrote the Schr\"odinger equation in polar coordinates $r,\theta$, separable for this system, and replaced $\theta$ by $k\theta$ where $k=p/q$ and $p,q$ are relatively prime integers. After renormalization, the   TTW system and its classical analog become
\be\label{TTWquant} H=\partial ^2_r+\frac{1}{ r}\partial _r+\frac{1}{
r^2}\partial ^2_\theta  -\omega^2 r^2+\frac{1}{
r^2}( 
\frac{\alpha }{ \sin ^2(k\theta )} + \frac {\beta }{ \cos ^2(k\theta )}),\ee
\be\label{TTWHam} {\cal H}=p_r^2+\frac{1}{
r^2}\partial ^2_\theta  +\omega^2 r^2+\frac{1}{
r^2}( 
\frac{\alpha }{ \sin ^2(k\theta )} + \frac {\beta }{ \cos ^2(k\theta )}).\ee
The authors noted that the Hamilton-Jacobi equation ${\cal H}=E$  is still separable in $r,\theta$ coordinates so it admits $\cal H$ and the 2nd order 
symmetry ${\cal L}_2$  responsible for the separation as symmetries. The TTW system is superintegrable if there exists a third constant of the motion, which may be of arbitrarily high order, dependent on $k$.
They noted that for small values of $p,q$ these systems  were already known to be superintegrable (but usually represented in Cartesian coordinates).
 For $k=1$ this is the caged isotropic oscillator,  for $k=1/2$ it is  [E16] in the constant curvature system  list in \cite{KKMP}, for $k=2$ it is a Calogero 
 system on the line, \cite{CALO, CDR2008},  etc. They conjectured and gave strong evidence that the TTW systems were classically and quantum superintegrable for 
{\bf all} rational $k$. This conjecture had broad influence  and led to a flurry of activity to prove the conjectures. 
 It showed how an infinite number of higher order superintegrable systems could be generated from a single 2nd order system. 
It was soon applied to generate numerous other families of higher order systems and in $n>2$ variables, \cite{KKM10, KKM10a, KKM10b}.

The basic problem for proof of the conjectures was exhibiting symmetries of arbitrarily high order. For simple choices such as $k=6$ 
the expression for the  symmetry operator required several printed pages.  The problem was solved for odd $k$ in \cite{CQ10} and first solved in general  in the papers 
\cite{KMPog10,  KKM10}. There have been other proofs of the conjecture using different methods:  \cite{MPY2010} using the methods of differential Galois theory, \cite{CG, hakobyan2012integrable} by direct computation of the action angle variables and \cite{ranada2012new} using factorization of the integrals.   In \cite{KM2012a} a general method was 
introduced that enabled explicit structure equations to be calculated.  We here discuss the present state of the classical theory from this point of view.

\subsubsection{The construction tool for classical systems}  \label{subsection1}
There are far  more verified superintegrable   Hamiltonian systems in 
classical mechanics than was the case in 2009. The principal method for constructing  and verifying these new systems requires the assumption that the root system, such as (\ref{TTWHam}), 
 admits orthogonal separation of variables. For a Hamiltonian system in $2n$-dimensional
 phase space the variable separation gives us $n$ 2nd order constants of the motion in involution. We first review a  general procedure, essentially the construction 
of action angle variables,  which
 yields an additional $n-1$ constants, such that the set of $2n-1$ is functionally independent. This is particularly of interest when it is possible to 
extract $n$ new constants of the motion that are polynomial in the momenta. 
 We  show  how this can be done in many cases, starting with the  construction of action angle variables for nD Hamiltonian systems.

Consider a classical system in $n$ variables on a  Riemannian manifold that admits separation in orthogonal separable coordinates $\bf x$.
Then there is an $n\times n$ nonsingular St\"ackel matrix 
$ S=\left( S_{ij}(x_i)\right)$, \cite{Stackel1891} and  the Hamiltonian is
$$ {\cal H}={\cal L}_1=\sum _{i=1}^n T_{1i}\left(p_i^2+v_i(x_i)\right)=\sum _{i=1}^n T_{1i}\ p_i^2+V(x_1,\cdots, x_n) ,$$
where $V=\sum _{i=1}^n T_{1i}\ v_i(x_i)$, and $T$ is the matrix inverse to $S$: 
\be \label{inversematrix} \sum _{j=1}^nT_{ij}S_{jk}=\sum _{j=1}^n S_{ij}T_{jk}=\delta_{ik}, \quad 1\le i,k\le n.\ee  Here, we must require $ \Pi_{i=1}^n T_{1i}\ne 0$. We define the quadratic constants of the motion ${\cal L}_k,\ k=1,\cdots, n$ by  
\be \fl \label{sepconsts} {\cal L}_k=\sum_{i=1}^nT_{ki}(p_i^2+v_i),\quad k=1,\cdots, n,\quad {\rm
or}\quad
 p_i^2+v_i=\sum_{j=1}^n S_{ij}{\cal L}_j,\quad 1\le i\le n.\ee
As is well known,
$\{{\cal L}_j,{\cal L}_k\}=0,\quad 1\le j,k\le n$.
Furthermore, by differentiating identity (\ref{inversematrix}) with respect to $x_h$  we obtain
$$\partial_h T_{i\ell}=-\sum_{j=1}^n T_{ih}S'_{hj}T_{j\ell},\quad 1\le h,i,\ell\le n,\quad 
S'_{hj}\equiv \partial_hS_{hj}.$$
Now we define nonzero functions $M_{kj}(x_j, {\cal L}_1,\cdots,{\cal L}_n)$ by the requirement
$\{M _{kj},{\cal L}_\ell\}=T_{\ell j}S_{jk},\ 1\le k,j,\ell\le n$. It is straightforward to check that these conditions are equivalent to the differential equations
$ 2p_j\partial_j M_{kj}=S_{jk}$.
Set
$ {\tilde {\cal L}}_q=\sum_{j=1}^n M_{qj},\quad 1\le q\le n$.
Then we have
$ \{{\tilde{\cal  L}}_q,{\cal L}_\ell\}=\sum_{j=1}^n T_{\ell j}S_{jq}=\delta_{\ell q}$.
This shows that the $2n-1$ functions 
${\cal H}={\cal L}_1, {\cal L}_2,\cdots,{\cal L}_n, {\tilde {\cal L}}_2,\cdots, {\tilde{\cal L}}_n$,
are constants of the motion and  they are functionally independent.

In the special case of $n=2$ dimensions we can always assume that  the St\"ackel matrix and its inverse are  of the form
\be\label{Staeckel1} S=\left(\ba{rr} f_1&1\\ f_2&-1\ea\right),\quad T=\frac{1}{f_1+f_2}\left(\ba{rr} 1&1\\ f_2 &-f_1\ea\right),\ee
where $f_j$ is a function of the variable $x_j$ alone, \cite{EIS} . The constants of the motion ${\cal L}_1={\cal H}$ and ${\cal L}_2$ are given to us via variable separation.
 We want to compute a new constant of the motion ${\tilde {\cal L}}_2$ functionally independent of ${\cal L}_1,{\cal L}_2$.  
Setting $M_{21}=M,\ M_{22}=-N$, we see that 
\be \label{MN1} 2p_1\frac{d}{dx_1}M=1,\quad 2p_2\frac{d}{dx_2}N=1.\ee 
from which we  determine $M,N$. Then ${\tilde {\cal L}}_2=M-N$ is the constant  we seek. 

\subsubsection{Application for $n=2$: The TTW system}
As we have seen, for $n=2$ and separable coordinates $x_1=x,x_2=y$ we have
\be\label{sepconst2} {\cal H}={\cal L}_1=\frac{1}{f_1(x)+f_2(y)}(p_x^2+p_y^2+v_1(x)+v_2(y),\ee
$$  {\cal  L}_2 =\frac{f_2(y)}{f_1(x)+f_2(y)}\left(p_x^2+v_1(x)\right)-\frac{f_1(x)}{f_1(x)+f_2(y)}\left(p_y^2+v_2(y)\right).$$
The constant of the motion ${\tilde {\cal L}}_2=M-N$  constructed by solving equations (\ref{MN1}) is usually not a polynomial in the momenta. We describe a procedure for obtaining a polynomial  from $M-N$, based on the observation that 
integrals $$M=\frac12\int \frac{dx_1}{\sqrt{f_1{\cal H}+{\cal L}_2-v_1}},\quad N=\frac12\int \frac{dx_2}{\sqrt{f_2{\cal H}-{\cal L}_2-v_2}},$$
can often be expressed in terms of multiples of the inverse hyperbolic sine or cosine , and the hyperbolic functions satisfy addition formulas. Thus we will search for functions $f_j,v_j$ such that $M$ and $N$ possess this property. There is a class of prototypes 
for this construction, namely the 2nd order superintegrable systems. We  start our construction with one of these 2nd order systems 
and add parameters to get a family of higher order superintegrable systems. We illustrate this by considering the TTW system in detail.

The TTW Hamiltonian  (\ref{TTWHam}) admits a second order constant of the motion 
corresponding to separation of variables in polar coordinates, viz 
\be\label{angmom}
{\cal L}_2=p^2_\theta +\frac{\alpha}{ \cos ^2k\theta } +\frac{\beta} {\sin ^2k\theta }.\ee
To demonstrate superintegrability we need to exhibit a third polynomial constant of the motion 
that is functionally independent of ${\cal H}, {\cal L}_2$. We apply the action angle variable construction above. 
In terms of the new variable $r=e^R$ the Hamiltonian assumes the canonical form 
$${\cal H}=e^{-2R}\left(p^2_R+p^2_\theta +\omega^2 e^{4R}+\frac{\alpha }{ \cos ^2(k\theta )}+ 
\frac{\beta }{ \sin ^2(k\theta )}\right).$$
To find  extra invariants we  first  construct a 
function $M(R)$ such that $\{M,{\cal H}\}=e^{-2R}$, or
$2p_R\partial_R M=1$.
This equation has a solution  
$M=\frac{i}{ 4\sqrt{ {\cal L}_2}} B$
where 
$$
\sinh{ B} =i \frac{(2{\cal L}_2e^{-2R}-{\cal H})}{ \sqrt{ {\cal  H}^2-4\omega^2 {\cal  L}_2}},\quad
\cosh{B }= \frac{2\sqrt{ {\cal L}_2}e^{-2R}p_R}{ \sqrt{ {\cal  H}^2-4\omega^2 {\cal L}_2}},$$
%\be\label{L2}{\cal L}_2=p^2_\theta +\frac{\alpha }{ \cos ^2(k\theta )}+ 
%\frac{\beta }{ \sin ^2(k\theta )},\ee
and we also have the relation 
$p^2_R+{\cal L}_2+\omega^2 e^{4R}-e^{2R}{\cal H}=0$.
Similarly $N(\theta)$ 
 satisfies $\{N,{\cal H}\}=e^{-2R}$, or $2p_\theta \partial _\theta N=1$,
with
solution 
$N= - \frac{i}{ 4\sqrt{{\cal  L}_2}k} A$
where  
$$
\sinh{A }= 
i\frac{-\beta +\alpha -{\cal L}_2\cos (2k\theta )}{ \sqrt{({\cal L}_2-\alpha -\beta )^2-4\alpha\beta }},\quad
\cosh{ A }= 
\frac{\sqrt{ {\cal L}_2}\sin (2k\theta )p_\theta }{\sqrt{ ({\cal L}_2-\alpha -\beta )^2-4\alpha \beta }}.$$
Then $M-N$ is a constant of the motion and, since it satisfies  $\{M-N,{\cal L}_2\}\ne 0$, it is functionally independent of ${\cal L}_2$.

From these expressions for $M$ and $N$  we see that if $k=p/q$ 
 (where $p, q$ are relatively prime integers) then  
$-4ip\sqrt{{\cal  L}_2}[N-M]=qA+pB$ and any function of $qA+pB$ is a constant of the motion (not polynomial in the momenta).
Rather than dealing with  
 $\sinh x$, $\cosh x$, we  make use of  an observation in \cite{CG}  and work with the exponential function.
We have 
\be\label{expA} e^A=\cosh A +\sinh A= {X}/{U},\quad e^{-A}=\cosh A -\sinh A= {\overline{X}}/{U},\ee
\be\label{expB} e^B=\cosh B +\sinh B= {Y}/{S},\quad e^{-B}=\cosh B -\sinh B= {\overline{Y}}/{S},\ee
$$ X= \sqrt{{\cal L}_2}\sin (2k\theta) p_\theta+i(-\beta+\alpha-{\cal L}_2\cos(2k\theta)),\   Y= 2\sqrt{{\cal L}_2} e^{-2R}p_R+i(2{\cal L}_2e^{-2R}-{\cal H}),$$
$${\overline X}= \sqrt{{\cal L}_2}\sin (2k\theta) p_\theta-i(-\beta+\alpha-{\cal L}_2\cos(2k\theta)),\
{\overline Y}= 2\sqrt{{\cal L}_2} e^{-2R}p_R-i(2{\cal L}_2e^{-2R}-{\cal H}),$$
$$ U=\sqrt{({\cal L}_2-\alpha-\beta)^2-4\alpha\beta},\quad S=\sqrt{{\cal H}^2-4\omega^2{\cal L}_2}.$$

Now note that $e^{qA+pB}$ and $e^{-(qA+pB)}$ are constants of the motion, where
\be\fl \label{expAB} e^{qA+pB}=(e^A)^q (e^B)^p=\frac{X^qY^p}{U^qS^p},\quad e^{-(qA+pB)}=(e^{-A})^q (e^{-B})^p=\frac{(\overline{X})^q
(\overline{Y})^p}{U^qS^p}.\ee Moreover, the identity $e^{qA+pB}e^{-(qA+pB)}=1$ can be expressed as
\be\label{fundident} X^q(\overline{X})^qY^p(\overline{Y})^p=U^{2q}S^{2p}=P({\cal L}_2, {\cal H})\ee
where $P$ is a polynomial in ${\cal L}_2$ and ${\cal H}$.

We restrict to the case $p+q$ even; the case $p+q$ odd is very similar, \cite{KM2012a}. Let $a,b,c,d$ be nonzero complex numbers and consider the 
expansion
\be\label{plusbinom}\frac12\left[(\sqrt{{\cal L}_2}a+ib)^q(\sqrt{{\cal L}_2}c+id)^p+(\sqrt{{\cal L}_2}a-ib)^q(\sqrt{{\cal L}_2}c-id)^p\right]
\ee
$$=\sum_{0\le \ell\le q, 0\le s\le p}\left(\ba{cc} q\\ \ell\ea\right) 
\left(\ba{cc} p\\ s\ea\right)b^\ell d^s a^{q-\ell}c^{p-s}\frac{{\cal L}_2^{(q+p-\ell-s)/2}}{2}[i^{\ell+s}+(-i)^{\ell+s}].$$
 If $p+q$ is even then the sum (\ref{plusbinom}) takes the 
form $T_{\rm even}({\cal L}_2) $,   a polynomial in ${\cal L}_2$. 
Similarly, 
expansion 
\be\label{minusbinom}\frac{1}{2i}\left[(\sqrt{{\cal L}_2}a+ib)^q(\sqrt{{\cal L}_2}c+id)^p-
(\sqrt{{\cal L}_2}a-ib)^q(\sqrt{{\cal L}_2}c-id)^p\right] 
\ee
$$=\sum_{0\le \ell\le q, 0\le s\le p}\left(\ba{cc} q\\ \ell\ea\right) 
\left(\ba{cc} p\\ s\ea\right)b^\ell d^s a^{q-\ell}c^{p-s}\frac{{\cal L}_2^{(q+p-\ell-s)/2}}{2i}[i^{\ell+s}-(-i)^{\ell+s}],$$ takes the 
form $\sqrt{{\cal L}_2}V_{\rm even}({\cal L}_2) $ where 
$V_{\rm even}$ is a polynomial in ${\cal L}_2$. 
 Let 
\be \label{evenL34} {\cal L}_4=\frac{1}{i\sqrt{{\cal L}_2}}({\cal L}^- -{\cal L}^+),\ 
{\cal L}_3={\cal L}^+ +{\cal L}^-,\ee
where ${\cal L}^\pm$ are defined as 
$ {\cal L}^+=X^qY^p,\quad {\cal L}^-=(\overline{X})^q(\overline{Y})^p$.
Then we see from (\ref{plusbinom}), (\ref{minusbinom}) that ${\cal L}_3,{\cal L}_4$ are constants of the motion,
 polynomial in the momenta. Moreover, the identity 
${\cal L}^+{\cal L}^-=P({\cal L}_2,{\cal H})$ is  applicable, as are 
$\{{\cal L}_2,{\cal L}^\pm\}=\mp 4ip\sqrt{{\cal L}_2}{\cal L}^\pm$,
Now we obtain
$\{{\cal L}_2,{\cal L}_3\}=-4p{\cal L}_2{\cal L}_4$,
$\{{\cal L}_2,{\cal L}_4\}=4p{\cal L}_3$.
Similarly, since
\[ {\cal L}_4^2=-\frac{1}{{\cal L}_2}\left[ ({\cal L}^+)^2-2{\cal L}^+{\cal L}^-+({\cal L}^-)^2\right],
{\cal L}_3^2=\left[ ({\cal L}^+)^2+2{\cal L}^+{\cal L}^-+({\cal L}^-)^2\right].\]
  we can  derive 
$ {\cal L}_3^2=-{\cal L}_2{\cal L}_4^2+4P({\cal L}_2,{\cal H})$. Using
$\{{\cal L}^+,{\cal L}^-\}=4ip\sqrt{{\cal L}_2}\frac{\partial P}{\partial{\cal L}_2}$,  we  derive 
$\{{\cal L}_3,{\cal L}_4\}=-4p{\cal L}_4^2+8p\frac{\partial P}{\partial{\cal L}_2}$. Hence the structure equations are 
\[\{{\cal L}_2,{\cal L}_4\}={\cal R},\ \{{\cal L}_2,{\cal R}\}=-16p^2{\cal L}_2{\cal L}_4,\ 
\{{\cal L}_4,{\cal R}\}=16p^2{\cal L}_4^2-32p^2\frac{\partial P}{\partial{\cal L}_2},\]
\be\label{structure+} {\cal R}^2=-16p^2{\cal L}_2{\cal L}_4^2+64p^2P({\cal L}_2,{\cal H}).\ee

\subsubsection{The classical constant of the motion ${\cal L}_5$}\label{L5}
As for the classical TTW system with $k=1$, our method doesn't  give the generator of minimal order. In fact, for all rational $k$,  we can extend our 
 algebra by adding a generator of order one less than ${\cal L}_4$. To see this, we  consider   the case
 $p+q$ even, leaving out the similar  odd case.
 Note that ${\cal L}_3$ is  a polynomial in ${\cal L}_2$  with constant term 
$2(-1)^{(q-p)/2} (\alpha-\beta)^q {\cal H}^p$. Thus
\be\label{L54} {\cal L}_5=\frac{{\cal L}_3-2(-1)^{(q-p)/2} (\alpha-\beta)^q {\cal H}^p}{{\cal L}_2}\ee
is a constant, polynomial in the momenta.
For example, if $p=q=1$ then 
$${\cal L}_5=\frac{{\cal L}_3-2(\alpha-\beta){\cal H}}{{\cal L}_2},$$
is a 2nd order constant of the motion. Just as before ${\cal L}_2,{\cal L}_5, {\cal H}$ generate a closed symmetry algebra that properly contains the
original algebra.

\subsection{The extended Kepler-Coulomb system }

Now we apply our construction to an important 3D potential, the 4-parameter extended  Kepler-Coulomb system. 
 We write it in the form
\begin{equation}\label{4Keplerham} {\cal H}=p_x^2+p_y^2+p_z^2+\frac{\alpha}{r}+\frac{\beta}{x^2}+\frac{\gamma}{y^2}+\frac{\delta}{z^2}.\end{equation}
Almost simultaneously Verrier and Evans \cite{Evans2008a}  and Rodr\'iguez,  Tempesta and Winternitz \cite{RTW2009} showed  this system is  4th order superintegrable and,  later, Tanoudis and Daskaloyannis \cite{DASK2011} showed in 
the quantum case that,
 if a second 4th order symmetry is added to the generators, the symmetry algebra  closes polynomially in the sense that all second commutators of the generators can be 
expresses as symmetrized polynomials in the generators. 
 We introduce an analog of the TTW construction  and  consider an infinite class of extended Kepler-Coulomb 
systems indexed by a pair of rational numbers $(k_1,k_2)$. We construct explicitly a set of generators, 
show these systems to be superintegrable and determine the structure of the generated  symmetry algebras. The 
symmetry algebras close rationally; only for systems admitting extra discrete symmetries is polynomial closure achieved. 

The extended  Hamiltonian is
${\cal H}=p_r^2+\frac{\alpha}{r}+\frac{{\cal L}_2}{r^2}$
where
$${\cal L}_2=p^2_{\theta_1}+\frac{{\cal L}_3}{\sin^2(k_1\theta_1)}+\frac{\delta}{\cos^2(k_1\theta_1)},\quad {\cal L}_3=p^2_{\theta_2}+\frac{\beta}{\cos^2(k_2\theta_2)}
+\frac{\gamma}{\sin^2(k_2\theta_2)}.$$
Here ${\cal L}_2,\ {\cal L}_3$ are constants of the motion, in involution. They determine additive separation in the variables $r,\theta,\phi$. Let , $k_1=\frac{p_1}{q_1}, k_2=\frac{p_2}{q_2}$ where $p_1,q_1$, resp.\ $p_2,q_2$, are relatively prime.
Applying the construction to get two new constants of the motion  we note that $x_1=r, x_2=\theta_1,x_3=\theta_2$ and 
$f_1=\frac{1}{r^2}$, $f_2=\frac{1}{\sin^2(k_1\theta_1)}$, $f_3=0$, $V_1=\frac{\alpha}{r}$,
$V_2=\frac{\delta}{\cos^2(k_1\theta_1)}$, $V_3=\frac{\beta}{\cos^2(k_2\theta_2)}+\frac{\gamma}{\sin^2(k_2\theta_2)}$,
to obtain functions $A_j,B_j$, $j=1,2$ such that 
$$N_1=-\frac{iA_1}{4k_1\sqrt{{\cal L}_2}},\ M_1=-\frac{iB_1}{2\sqrt{{\cal L}_2}},\ N_2=-\frac{iA_2}{4k_2\sqrt{{\cal L}_3}},\ M_2=-\frac{iB_2}{4k_1\sqrt{{\cal L}_3}},$$
and 
$p_1q_2A_2-p_2q_1B_2,\ q_1A_1-2p_1B_1$,
are two  constants  such that the full set of five constants of the motion is functionally independent.
We find
\begin{footnotesize}
\[ e^{A_j}=\cosh A_j +\sinh A_j= {X_j}/{U_j},\quad e^{-A_j}=\cosh A_j -\sinh A_j= {\overline{X_j}}/{U_j},\]
\[ e^{B_j}=\cosh B_j +\sinh B_j= {Y_j}/{S_j},\quad e^{-B_j}=\cosh B_j -\sinh B_j= {\overline{Y_j}}/{S_j},\]
\[ X_1= \sqrt{{\cal L}_2}\sin (2k_1\theta_1) p_{\theta_1}+i(-{\cal L}_2\cos(2k_1\theta_1)+\delta-{\cal L}_3),\]
\[ X_2= -\sqrt{{\cal L}_3}\sin (2k_2\theta_2) p_{\theta_2}+i({\cal L}_3\cos(2k_2\theta_2)+\gamma-\beta),\]
\[Y_1=2\sqrt{{\cal L}_2}p_r-i(\alpha+2\frac{{\cal L}_2}{r}),\
Y_2= -2{\cal L}_3 \cot^2(k_1\theta_1)+({\cal L}_2-{\cal L}_3-\delta)-2i\sqrt{{\cal L}_3}\cot(k_1\theta_1)p_{\theta_1}, \]
\[ U_1=\sqrt{{\cal L}^2_3-2{\cal L}_3({\cal L}_2+\delta)+({\cal L}_2-\delta)^2},\quad U_2=\sqrt{(\beta-\gamma-{\cal L}_3)^2-4\gamma {\cal L}_3},\]
\[S_1=\sqrt{\alpha^2+4{\cal H}{\cal L}_2}, \quad S_2=\sqrt{{\cal L}^2_3-2{\cal L}_3({\cal L}_2+\delta)+({\cal L}_2-\delta)^2},\]
\end{footnotesize}
 where $\overline{X_j}$, $\overline{Y_j}$ are obtained from $X_j,Y_j$ by replacing $i$ by $-i$.

Here, $e^{q_1A_1-2p_1B_1}$ and $e^{-q_1A_1+2p_1B_1}$ are constants of the motion, where
\[ e^{q_1A_1-2p_1B_1}=(e^{A_1})^{q_1} (e^{-B_1})^{2p_1}=\frac{X_1^{q_1}\overline{Y_1}^{2p_1}}{U_1^{q_1}S_1^{2p_1}},\]
\[ e^{-q_1A_1+2p_1B_1}=\frac{\overline{X_1}^{q_1}
Y_1^{2p_1}}{U_1^{q_1}S_1^{2p_1}}.\] The identity $e^{q_1A_1-2p_1B_1}e^{-q_1A_1+2p_1B_1}=1$ can be expressed as
\begin{footnotesize}
\[ X_1^{q_1}\overline{X_1}^{q_1}Y_1^{2p_1}\overline{Y_1}^{2p_1}=U_1^{2q_1}S_1^{4p_1}={\cal P}_1
=[{\cal L}^2_3-2{\cal L}_3({\cal L}_2+\delta)+({\cal L}_2-\delta)^2]^{q_1}
[\alpha^2+4{\cal H}{\cal L}_2]^{2p_1}\]
\end{footnotesize}
where 
${\cal P}_1$ is a polynomial in ${\cal H}, {\cal L}_2$ and ${\cal L}_3$. 
Similarly, $e^{p_1q_2A_2-p_2q_1B_2}$,  $e^{-p_1q_2A_2+p_2q_1B_2}$ are constants of the motion, where
\begin{footnotesize}
\[ e^{p_1q_2A_2-p_2q_1B_2}=(e^{A_2})^{p_1q_2} (e^{-B_1})^{p_2q_1}=\frac{X_2^{p_1q_2}\overline{Y_2}^{p_2q_1}}{U_2^{p_1q_2}S_2^{p_2q_1}},\]
\[
e^{-p_1q_2A_2+p_2q_1B_2}=\frac{\overline{X_2}^{p_1q_2}
Y_2^{p_2q_1}}{U_2^{p_1q_2}S_2^{p_2q_1}}.\]
\end{footnotesize}
 The identity $e^{p_1q_2A_2-p_2q_1B_2}e^{-p_1q_2A_2+p_2q_1B_2}=1 $ becomes
\begin{footnotesize}
\[ X_2^{p_1q_2}\overline{X_2}^{p_1q_2}Y_2^{p_2q_1}\overline{Y_2}^{p_2q_1}=U_2^{2p_1q_2}S_2^{2p_2q_1}={\cal P}_2({\cal H}, {\cal L}_2,{\cal L}_3)
\]
$$=[(\beta-\gamma-{\cal L}_3)^2-4\gamma {\cal L}_3]^{p_1q_2}
[{\cal L}^2_3-2{\cal L}_3({\cal L}_2+\delta)+({\cal L}_2-\delta)^2]^{p_2q_1}.$$
\end{footnotesize}

We define basic raising and lowering symmetries 
\begin{footnotesize}
$$ {\cal J}^+=X_1^{q_1}\overline{Y_1}^{2p_1},\ {\cal J}^-=\overline{X_1}^{q_1}Y_1^{2p_1},\  {\cal K}^+=X_2^{p_1q_2}\overline{Y_2}^{p_2q_1},\ {\cal K}^-
=\overline{X_2}^{p_1q_2}Y_2^{p_2q_1}.$$
\end{footnotesize}
and restrict to the case where each of $p_1,q_1,p_2,q_2$ is an odd integer.
Let
\begin{footnotesize}
\[ {\cal J}_1=\frac{1}{\sqrt{{\cal L}_2}}({\cal J}^- +{\cal J}^+),\ 
i{\cal J}_2={\cal J}^- -{\cal J}^+,\
 {\cal K}_1=\frac{1}{\sqrt{{\cal L}_3}}({\cal K}^- +{\cal K}^+),\  i{\cal K}_2={\cal K}^- -{\cal K}^+.
\]\end{footnotesize}
We see from the explicit expressions for the symmetries that ${\cal J}_1,{\cal J}_2, {\cal K}_1,{\cal K}_2$ are constants of the motion,
 polynomial in the momenta. Moreover, the identities 
 ${\cal J}^+{\cal J}^-={\cal P}_1,\quad  {\cal K}^+{\cal K}^-={\cal P}_2$
hold.
 The following relations are straightforward to derive from the definition of the Poisson bracket:
 $$ \{{\cal L}_3,X_1\}=\{{\cal L}_3,Y_1\}= \{{\cal L}_2,Y_1\}= \{{\cal L}_3,Y_2\}= 0,$$
 $$ \{{\cal L}_2,X_1\}= -4ik_1\sqrt{{\cal L}_2}X_1,\  \{{\cal L}_3,X_2\}=- 4ik_2\sqrt{{\cal L}_3}X_2,$$
$$\{{\cal L}_2,X_2\}= -\frac{4ik_2}{\sin^2(k_1\theta_1)}\sqrt{{\cal L}_3}X_2,\  \{{\cal L}_2,Y_2\}= -\frac{4ik_1\sqrt{{\cal L}_3}}{\sin^2(k_1\theta_1)}Y_2,.$$  
$$\{{\cal H},X_1\}=-\frac{4ik_1\sqrt{{\cal L}_2}}{r^2}X_1,\ \{{\cal H},X_2\}= -\frac{4ik_2}{r^2\sin^2(k_1\theta_1)}\sqrt{{\cal L}_3}X_2, $$
 $$ \{{\cal H},Y_1\}= -\frac{2i\sqrt{{\cal L}_2}}{r^2}Y_1,\   \{{\cal H},Y_2\}= -\frac{4ik_1\sqrt{{\cal L}_3}}{r^2\sin^2(k_1\theta_1)}Y_2,$$  with the corresponding results for $\overline{X_j}$, $\overline{Y_j}$  obtained  by replacing $i$ by $-i$.

Commutators relating the $\cal J$ and $\cal K$ follow from
\begin{footnotesize}
$$\{X_1,X_2\}=-\frac{4k_2\sqrt{{\cal L}_3}}{\sqrt{{\cal L}_2}}\left(i\cot(k_1\theta_1)p_{\theta_1}+\sqrt{{\cal L}_2}\cot^2(k_1\theta_1)\right)X_2,$$
$$ \{\overline{Y_1},\overline{Y_2}\}=\frac{4ik_1\sqrt{{\cal L}_3}}{\sin^2(k_1\theta_1)}\left( \frac{p_r}{\sqrt{{\cal L}_2}}+\frac{2i}{r}\right)\overline{Y_2},$$
$$ \{X_1,\overline{Y_2}\}= 4ik_1\sqrt{{\cal L}_2}X_1+ \frac{4ik_1\sqrt{{\cal L}_3}}{\sin^2(k_1\theta_1)}\left(\frac{\sin(2k_1\theta_1)p_{\theta_1}}{2\sqrt{{\cal L}_2}}
-i\cos(2k_1\theta_1)\right)\overline{Y_2}$$
$$+8k_1\cot^2(k_1\theta_1)\left(\frac{i\sqrt{{\cal L}_2{\cal L}_3}\sin(2k_1\theta_1)}{2}p_{\theta_1}-
\sqrt{{\cal L}_3}{\cal L}_2\sin^2(k_1\theta_1)-\sqrt{{\cal L}_2}{\cal L}_3\right),$$
  $$ \{X_2,\overline{Y_1}\} =
\frac{4ik_2\sqrt{{\cal L}_3}}{\sin^2(k_1\theta_1)}\left(\frac{p_r}{\sqrt{{\cal L}_2}}+\frac{2i}{r}\right)X_2.$$
\end{footnotesize}
From these results we find 
$$\frac{\{{\cal J}^+,{\cal K}^+\}}{{\cal J}^+{\cal K}^+}=-\frac{\{{\cal J}^-,{\cal K}^-\}}{{\cal J}^-{\cal K}^-}=
\frac{4iq_1p_1p_2(\sqrt{{\cal L}_2}-\sqrt{{\cal L}_3})({\cal L}_2+2\sqrt{{\cal L}_2{\cal L}_3}+{\cal L}_3
-\delta)}{({\cal L}_3-{\cal L}_2-\delta)^2-4\delta{\cal L}_2}.$$
$$ \frac{\{{\cal J}^+,{\cal K}^-\}}{{\cal J}^+{\cal K}^-}=-\frac{\{{\cal J}^-,{\cal K}^+\}}{{\cal J}^-{\cal K}^+}
=\frac{4iq_1p_1p_2(\sqrt{{\cal L}_2}+\sqrt{{\cal L}_3}) ({\cal L}_2-2\sqrt{{\cal L}_2{\cal L}_3}+{\cal L}_3
-\delta)    }{({\cal L}_3-{\cal L}_2-\delta)^2-4\delta{\cal L}_2}.$$
These relations prove closure of the symmetry algebra in the space of functions polynomial in ${\cal J}^\pm,{\cal K}^\pm$, rational
 in ${\cal L}_2,{\cal L}_3, {\cal H}$ and  linear in $\sqrt{{\cal L}_2},\sqrt{{\cal L}_3}$.

 \subsubsection{Structure relations for polynomial symmetries of the 4-parameter potential}\label{polystructure4}
 From the preceding section we have the structure relations (for $1\le j,k\le 2$)
\be\label{ident1}\{{\cal J}_j,{\cal K}_k\}=\frac{4q_1p_1p_2}{({\cal L}_3-{\cal L}_2-\delta)^2-4\delta{\cal L}_2}\times\ee
$$\left[{\cal J}_{3-j}{\cal K}_k(-{\cal L}_2)^{j-1}({\cal L}_2-{\cal L}_3-\delta)+{\cal J}_j{\cal K}_{3-k}(-{\cal L}_3)^{k-1}
({\cal L}_2+{\cal L}_3+\delta)\right],$$
$$\{{\cal L}_2,{\cal J}_2\}=4p_1{\cal L}_2{\cal J}_1,\quad \{{\cal L}_2,{\cal J}_1\}=-4p_1{\cal J}_2,\
\{{\cal L}_3,{\cal J}_1\}=\{{\cal L}_3,{\cal J}_2\}=0,$$
$$\{{\cal L}_3,{\cal K}_2\}=4p_1p_2{\cal L}_3{\cal K}_1,\ \{{\cal L}_3,{\cal K}_1\}=-4p_1p_2{\cal K}_2,\
\{{\cal L}_2,{\cal K}_2\}=\{{\cal L}_2,{\cal K}_1\}=0,$$
$$ {\cal J}_2^2=-{\cal L}_2{\cal J}_1^2+4{\cal P}_1({{\cal H}, \cal L}_2,{\cal L}_3),
\quad {\cal K}_2^2=-{\cal L}_3{\cal K}_1^2+4{\cal P}_2({{\cal H}, \cal L}_2,{\cal L}_3),$$
$$\{{\cal J}_1,{\cal J}_2\}=-2p_1{\cal J}_1^2+8p_1\frac{\partial {\cal P}_1}{\partial{\cal L}_2},\
\{{\cal K}_1,{\cal K}_2\}=-2p_1p_2{\cal K}_1^2+8p_1p_2\frac{\partial {\cal P}_2}{\partial{\cal L}_3}.$$

The generators for the polynomial symmetry algebra  produced so far are not of minimal order. Here
 ${\cal L}_1, {\cal L}_2, {\cal L}_3$ are of order 2 and the orders of ${\cal J}_1, {\cal K}_1$ are one less than 
the orders of ${\cal J}_2,{\cal K}_2$, respectively. We will construct  symmetries ${\cal J}_0$, ${\cal K}_0$ of order one less 
than ${\cal J}_1$, ${\cal K}_1$, respectively. (In the standard case $k_1=k_2=1$ it is easy to see that ${\cal J}_1$ is of order 5 and ${\cal J}_2$ is of order 6, 
whereas ${\cal K}_1$ is of order 3 and ${\cal K}_2$ is of order 4. Then ${\cal J}_0$, ${\cal K}_0$  will be of orders 4 and 2, respectively, which we know 
corresponds to the minimal generators of the symmetry algebra in this case, \cite{Evans2008a}.) 
The symmetry ${\cal J}_2$ is a polynomial in ${\cal L}_2$ with constant term 
${\cal D}_1=2(-1)^{(q_1-1)/2}(\delta-{\cal L}_3)^{q_1}\alpha^{2p_1}$,
 itself a constant of the motion. 
 Thus  
$ {\cal J}_0=\frac{{\cal J}_2-{\cal D}_1}{{\cal L}_2}$
is a polynomial symmetry of order two less than ${\cal J}_2$. We have the identity
$ {\cal J}_2={\cal L}_2{\cal J}_0+{\cal D}_1$.
{}From this,  $\{{\cal L}_2,{\cal J}_2\}={\cal L}_2 \{{\cal L}_2,{\cal J}_0\}$. We already know that $\{{\cal L}_2,{\cal J}_2\}=4p_1{\cal L}_2{\cal J}_1$
so 
$ \{{\cal L}_2,{\cal J}_0\}=4p_1{\cal J}_1$, $ \{{\cal L}_3,{\cal J}_0\}=0$.

The same construction works for ${\cal K}_2$. It is a polynomial in ${\cal L}_3$, with constant term 
${\cal D}_2=2(-1)^{(p_1q_2+1)/2} (\gamma-\beta)^{p_1q_2}({\cal L}_2-\delta)^{p_2q_1}$.
Thus  
$ {\cal K}_0=\frac{{\cal K}_2-{\cal D}_2}{{\cal L}_3}$
is a polynomial symmetry of order two less than ${\cal K}_2$ and  we have the identity
$ {\cal K}_2={\cal L}_3{\cal K}_0+{\cal D}_2$.
Further,
$  \{{\cal L}_3,{\cal K}_0\}=4p_1p_2{\cal K}_1$, $  \{{\cal L}_2,{\cal K}_0\}=0$.

Now we choose ${\cal L}_1,{\cal L}_2, {\cal L}_3,{\cal J}_0,{\cal K}_0$ as the generators of our algebra. We define the basic nonzero commutators as
 $${\cal R}_1=\{{\cal L}_2,{\cal J}_0\}=4p_1{\cal J}_1,\quad {\cal R}_2=\{{\cal L}_3,{\cal K}_0\}=4p_1p_2{\cal K}_1, \quad {\cal R}_3=\{{\cal J}_0,{\cal K}_0\}.$$

Then we have
\be\label{R1^2} \frac{{\cal R}_1^2}{16p_1^2}={\cal J}_1^2=-{\cal L}_2{\cal J}_0^2-2{\cal D}_1{\cal J}_0+\frac{4{\cal P}_1-{\cal D}_1^2}{{\cal L}_2},\ee
where the  right-hand term is polynomial in  generators ${\cal H},{\cal L}_2,{\cal L}_3$.
 Further, 
\be\label{R2^2}\frac{{\cal R}_2^2}{16p_1^2p_2^2}={\cal K}_1^2=-{\cal L}_3{\cal K}_0^2-2{\cal D}_2{\cal K}_0+\frac{4{\cal P}_2-{\cal D}_2^2}{{\cal L}_3},\ee
which again can be verified to be a polynomial in the generators.
Note that this symmetry algebra cannot close polynomially in the usual sense. If it did close then the product ${\cal R}_1{\cal R}_2$ 
would be expressible as a polynomial in the generators. The preceding two equations show that ${\cal R}_1^2{\cal R}_2^2$  is so expressible, but that the resulting polynomial is 
not a perfect square. Thus the only possibility to obtain closure is to add new generators to the algebra, necessarily functionally dependent on the original set. Continuing in this way, see \cite{KM2012b} for details,  one finds that that ${\cal L}_1, {\cal L}_2,{\cal L}_3, {\cal K}_0, {\cal J}_0$ generate a symmetry algebra that closes rationally.
In particular, each of the commutators ${\cal R}_1,{\cal R}_2,{\cal R}_3$ satisfies an explicit polynomial equation in the generators.

\subsubsection{The special case $k_1=k_2=1$}
In the case $k_1=k_2=1$  we are in Euclidean space and our system has additional symmetry. In terms of Cartesian coordinates
$x=r\sin\theta_1\cos\theta_2,\ y=r\sin\theta_1\sin\theta_2,\ z=r\cos\theta_1$,
the Hamiltonian is
\be\label{cartham} {\cal H}=p_x^2+p_y^2+p_z^2+\frac{\alpha}{r}+\frac{\beta}{x^2}+\frac{\gamma}{y^2}+\frac{\delta}{z^2}.\ee
Note that any permutation of the ordered pairs $(x,\beta),(y,\gamma),(z,\delta)$  leaves the Hamiltonian unchanged. This leads to 
additional structure in the symmetry algebra.
The basic symmetries are
\be\label{L3cart} {\cal L}_3={\cal I}_{xy}=(xp_y-yp_x)^2+\frac{\beta(x^2+y^2)}{x^2}+\frac{\gamma(x^2+y^2)}{y^2},\ee
$${\cal L}_2= {\cal I}_{xy}+{\cal I}_{xz}+{\cal I}_{yz}-(\beta+\gamma+\delta).$$
Permutation symmetry of the Hamiltonian shows  ${\cal I}_{xz},{\cal I}_{yz}$ are  constants, and 
$${\cal K}_0=4{\cal I}_{yz}+2{\cal L}_3-2({\cal L}_2+\beta+\gamma+\delta)=2({\cal I}_{yz}-{\cal I}_{xz}).$$
Here ${\cal J}_0$ is 4th order:
\[\fl  {\cal J}_0=-16\left({\cal M}_3^2+\frac{\delta(xp_x+yp_y+zp_z)^2}{z^2}\right)+8{\cal H}({\cal I}_{xz}+{\cal I}_{yz}-\beta-\gamma-\delta)+2\alpha^2,\]
\be\label{J0cart} {\cal M}_3=(yp_z-zp_y)p_y-(zp_x-xp_z)p_x-z\left(\frac{\alpha}{2r}+\frac{\beta}{x^2}+\frac{\gamma}{y^2}+\frac{\delta}{z^2}\right).\ee
If $\delta=0$ then ${\cal M}_3$  is itself a constant of the motion.

 The symmetries ${\cal H}, {\cal L}_2,{\cal L}_3,{\cal J}_0,{\cal K}_0$ form a generating (rational) basis for the constants of the motion.
Under the transposition $(x,\beta)\leftrightarrow(z,\delta)$ this basis is mapped to an alternate basis ${\cal H}, {\cal L}'_2,{\cal L}'_3,{\cal J}'_0,{\cal K}'_0$
where
\be\label{R'idents} {\cal L}'_2 = {\cal L}_2,\quad  {\cal L}'_3=\frac14{\cal K}_0+\frac12{\cal L}_2-\frac12{\cal L}_3+\frac{\beta+\gamma+\delta}{2},\ee
\[\fl  {\cal K}'_0=\frac12{\cal K}_0-{\cal L}_2+3{\cal L}_3-(\beta+\gamma+\delta),\ {\cal R}_1'=\{{\cal L}_2,{\cal J}_0'\},\ {\cal R}_2'= \{{\cal L}_3',{\cal K}_0'\}=-\frac54{\cal R}_2,\]
$$ {\cal R}_3'=\{{\cal J}_0',{\cal K}_0'\}=2{\cal R}_1'-2\{{\cal L}_3,{\cal J}_0'\}.$$

All the identities in Section \ref{polystructure4} hold for the primed symmetries. The ${\cal K}'$ symmetries are simple polynomials in the
 ${\cal L},{\cal K}$ symmetries already constructed, e.g., ${\cal K}'_1=\frac14\{{\cal L}'_3,{\cal K}'_0\}=-\frac54{\cal K}_1$.
However, the ${\cal J}'$ symmetries are new. 
\be\label{newJ0} \fl {\cal J}'_0=-16\left({\cal M}_1^2+\frac{\beta(xp_x+yp_y+zp_z)^2}{x^2}\right)+8{\cal H}({\cal I}_{xy}+{\cal I}_{xz}-\beta-\gamma-\delta)+2\alpha^2,\ee
\[{\cal M}_1=(yp_x-xp_y)p_y-(xp_z-zp_x)p_z-x\left(\frac{\alpha}{2r}+\frac{\beta}{x^2}+\frac{\gamma}{y^2}+\frac{\delta}{z^2}\right).\]
Note that the transposition $(y,\gamma)\leftrightarrow(z,\delta)$ does not lead to anything new. Indeed, under the symmetry we would obtain a constant of the motion
$ {\cal J}_0''$
but it is straightforward to check that
${\cal J}_0+{\cal J}'_0+{\cal J}''_0 =2\alpha^2$,
so that the new constant depends linearly on the previous constants.

In the paper \cite{DASK2011}, Tanoudis and Daskaloyannis show that the quantum symmetry algebra generated by the 6 functionally dependent symmetries 
${\cal H}, {\cal L}_2,{\cal L}_3,{\cal J}_0,{\cal K}_0$ and ${\cal J}'_0$  closes polynomially, in the sense that all double commutators of the generators are 
again expressible as polynomials in the generators, strong evidence that the classical analog also closes polynomially. The complicated details in proving this can be found in \cite{KM2012b} where it is shown that the 6 generators obey a functional identity of order 12.

\subsubsection{Further results}  Using the methods described above, many more classical systems have been shown to be higher order superintegrable, and in some cases the structure of 
the symmetry algebras has been worked out. In \cite{KM2012b} a extended version of the 3-parameter Kepler-Coulomb system was studied for all rational $k_1,k_2$ and the
 structure of the symmetry algebra exposed. This is the restriction of the 4-parameter potential to $\delta=0$. It is of interest because the symmetry algebra of the restriction is larger 
than the original symmetry algebra; the system is 2nd order superintegrable in the $k_1=k_2=1$ case. Always the symmetry algebra closes rationally, but not polynomially.

In \cite{KKM10b} the scope of the construction given here is explored for $n=2$ to find all superintegrable systems with rational $k$, where the $k=1$ case is 2nd order 
superintegrable on a constant curvature space, separating in spherical coordinates.
In \cite{KKM10} the construction of extended superintegrable systems for rational $k_i$ is considered in $n$ dimensions for systems that admit separation in subgroup coordinates,
 i.e., polyspherical coordinates. For $n=4$ the first known examples of superintegrable systems on non-conformally flat spaces are presented.

The method of construction of higher order superintegrable systems presented here always yields  separable systems. However, recently examples of nonseparable superintegrable systems 
have been produced: \cite{MPT}, using differential Galois theory, and \cite{PW2011}, which also has a quantum version.
 A general method of construction of such nonseparable systems and 
determination of their structure algebras has not yet been achieved.

%% file: higherorderquantumchapter.tex
\section{Higher order  quantum superintegrable systems} \label{higherorderquantum}

The problem of finding and verifying higher order quantum superintegrability is analogous to that of classical superintegrability, only harder. In addition to the problems of finding  numerous examples  of these  systems and  working out ways to compute the commutators  of symmetry operators of arbitrarily high order, there is the quantization problem.
The relationship between classical and quantum superintegrability is not $1-1$, the difference shows up with greater frequency in superintegrable systems 
in dimensions $>2$, or order $>2$.

As for the classical case, a  breakthrough  came with the publication of the Tremblay, Turbiner and Winternitz papers, \cite{TTW, TTW2}.
They provided strong evidence for the  conjecture  that the system, which we call the TTW system,  (\ref{TTWquant}), was  quantum superintegrable for 
all rational $k$. There was a flurry of activity to prove the conjectures;
the first results were obtained in \cite{CQ10} where Dunkl operators were used to prove quantum superintegrability for odd integer $k$.
 The  problem for all rational $k$ was solved first in \cite{KKM10a} by a very general approach that verified superintegrability but didn't give
 information about 
the structure equations, see also \cite{ Marquette20103}. Then in \cite{KKM2011Recurr} a method was introduced that enabled 
superintegrability to be proved and explicit structure equations to be calculated. This is the recurrence approach. It applies to large families 
of quantum systems that permit
 separation in coordinates yielding hypergeometric eigenfunctions
and works for all $n$. It has been conjectured that all maximally superintegrable Euclidean systems are exactly solvable, thus  it should always apply for these systems \cite{TempTW}. 
 
Here is the procedure for $n=2$ where one new symmetry  is needed: 
 1) Require that the system admit  a 2nd order symmetry that determines a separation of variables.
 2) The formal eigenspaces of the
Hamiltonian are invariant under action of 
any  symmetry, so the operator must induce recurrence relations for the
basis of separated  eigenfunctions.
3) Suppose  the separated eigenfunctions are of hypergeometric type. Use the known recurrence relations 
for hypergeometric functions to reverse this process and determine a symmetry
operator from the
 recurrences. 
 4) Compute the symmetry operators and structure
equations by restricting  
to a formal ``basis'' of separated eigenfunctions.
5) Appeal to a general theory
of  canonical forms for symmetry 
operators to show that results 
obtained on  a formal eigenbasis  hold as true identities
for purely differential operators 
defined independent of basis.

We sketch how this method works for the TTW system,  (\ref{TTWquant}),
$H\Psi =E\Psi $, $H=L_1$. A formal eigenbasis takes the form 
$$\Psi  = e^{-\frac{\omega }{ 2}r^2}r^{k(2n+a+b+1)} 
L^{k(2n+a+b+1)}_m(\omega r^2)
(\sin (k\theta ))^{a+\frac{1}{ 2}}
(\cos (k\theta ))^{b+\frac{1}{ 2}} 
P^{a,b}_n(\cos (2k\theta )),$$
where $\alpha =k^2(\frac{1}{ 4} -a^2)$ and $\beta =k^2(\frac{1}{
4} 
-b^2)$. The $L^A_B(z)$ are associated Laguerre functions and the $P^{a,b}_n(w)$ are Jacobi functions, \cite{AAR}.
We express the eigenfunctions in the form $\Psi=(\sin (k\theta ))^{a+\frac{1}{ 2}}
(\cos (k\theta ))^{b+\frac{1}{ 2}} \Pi $ with 
\[\Pi  = e^{-\frac{\omega }{2}r^2}r^{k(2n+a+b+1)} 
L^{k(2n+a+b+1)}_m(\omega r^2)P^{a,b}_n(x)
=Y^{ k\mu}_m(r)X^{a,b}_n(x)\]
where $x=\cos (2k\theta )$, $\mu=2n+a+b+1$. 
The energy eigenvalue is 
$$ E=-2\omega \left[2(m+nk)+1+(a+b+1)k\right]$$
and the symmetry operator responsible for the separation of variables is 
 $$
{ L}_2\Psi=(\partial ^2_\theta + \frac{\alpha }{ \sin ^2(k\theta )} + 
\frac{\beta }{\cos ^2(k\theta )})\Psi
=-k^2\mu^2 \Psi$$ where $ \mu=2n+a+b+1$, $k=p/q$. Recall that the separation constant is just the eigenvalue of the symmetry operator $L_2$. The energy values for the Hamiltonian are negative because here we have the non-physical convention of setting $-\hbar^2/2m=1$. The transformation to the un-scaled Hamiltonian is given by (\ref{scaledseqn}).  

The strategy is to look for recurrences that change $m,n$ but fix
 $u=m+nk$. This will map eigenfunctions of $H$ with eigenvalue $E$ to eigenfunctions with the same $E$.   Two transformations 
achieving this are 
\[ (1)\quad
n\rightarrow n+q,\quad m\rightarrow m-p,\quad
(2)\quad n\rightarrow n-q,\quad m\rightarrow m+p.\]

First, consider $X^{a,b}_n(x)$. Raise or lower
 $n$ with 
$$(1):\quad J^+_n X^{a,b}_n=
(2n+a+b+2)(1-x^2)\partial
_xX^{a,b}_n$$
$$+(n+a+b+1)(-(2n+a+b+2)x-(a-b))X^{a,b}_n
=2(n+1)(n+a+b+1)X^{a,b}_{n+1},$$
$$(2):\quad J^-_n X^{a,b}_n=
-(2n+a+b)(1-x^2)\partial
_xX^{a,b}_n$$
$$-n((2n+a+b)x-(a-b))X^{a,b}_n
=2(n+a)(n+b)X^{a,b}_{n-1}.$$

For ${\cal Y}^{k\mu}_m(R)=\omega^{k\mu/2}Y^{k\mu}_m(r)$ where
$R= r^2$, change $m$ by
\[\quad K^+_{k\mu,m}{\cal Y}^{k\mu}_m=\left\{(k\mu+1)\partial _R-\frac{E}{ 4 }-\frac{1}{
2R}k\mu(k\mu+1)\right\}{\cal Y}^{k\mu}_m
= -\omega{\cal Y}^{k\mu+2}_{m-1},\]
$$\quad K^-_{k\mu,m}{\cal Y}^{k\mu}_m=\left\{(-k\mu+1) 
\partial _R-\frac{E}{ 4}+\frac{1}{ 2R}k\mu(1-k\mu)\right\}{\cal Y}^{k\mu}_m$$
$$=-\omega(m+1)(m+k\mu) {\cal Y}^{k\mu-2}_{m+1}.$$

From these recurrences we  construct the two operators
$$\Xi _\pm=K^\pm_{k\mu\pm 2(p-1),m\mp (p-1)}\cdots
K^\pm_{k\mu,m}J^\pm _{n\pm q-1}\cdots J^\pm_n.$$
For fixed
$u=m+kn$, we have 
$$\Xi _+\Psi_n=2^q(-1)^p \omega^p(n+1)_q(n+a+b+1)_q\Psi_{n+q},$$
$$ \Xi _-\Psi_n=2^q \omega^p
(-n-a)_q(-n-b)_q(u-kn+1)_p
(-u-k[n+a+b+1])_p\Psi_{n-q}.$$
These are basis dependent operators. However, under the transformation $n \rightarrow  -n-a-b-1$, i.e., $\mu\to -\mu$,  we have ${\Xi_+} \rightarrow
\Xi_-$ and ${\Xi_-} \rightarrow \Xi_+$. 
Thus $\Xi=\Xi_++\Xi_-$  is a polynomial in $\mu^2$. 
 Therefore we can replace $(2n+a+b+1)^2=\mu^2$
by $L_2/k^2$ and $E$ by $H$ everywhere they occur, and express $\Xi$ as a
pure differential symmetry operator. 
Note also that  under the transformation $n \rightarrow  -n-a-b-1$, i.e., $\mu\to-\mu$ 
the operator $\Xi_+ -\Xi_-$ changes sign, hence 
 ${\tilde \Xi}=(\Xi_+ -\Xi_-)/{\mu}\equiv (\Xi_+ -\Xi_-)(1/\mu)$ is unchanged.  
This defines $\tilde \Xi$ as a symmetry operator.
We set 
\be L_3= \Xi, \quad L_4={\tilde \Xi}\label{defL3L}. \ee

 We have shown that $L_3,L_4$ commute with $H$ on any formal eigenbasis.   In fact, we have constructed  pure differential operators which commute with $H$, independent of  basis.
 This takes  proof, which is provided in  \cite{KKM10a} via a canonical form for higher order differential operators.
Thus operator relations verified on formal eigenbases  must hold identically.

\subsection{The structure of the TTW  algebra}\label{TTWstructure}
Taking a product of  raising and lowering operators we obtain 
\[ \Xi_-\Xi_+\Psi_n
=(-1)^p4^q\omega^{2p}(n+1)_q(n+a+1)_q(n+b+1)_q\]
\[\times (n+a+b+1)_q(-u+kn)_p(u+k[n+a+b+1]
+1)_p\Psi_n
=\xi_n\Psi_n,\]
\[\Xi_+\Xi_-\Psi_n=(-1)^p4^q\omega^{2p}(-n)_q(-n-a)_q(-n-b)_q\]
\[\times(-n-a-b)_q(u-kn+1)_p(-u-k[n+a+b+1])_p\Psi_n
 =\eta_n\Psi_n.\]
Thus  $\Xi^{(+)}=\Xi_-\Xi_++ \Xi_+\Xi_-$ multiplies any
basis function   by 
$\xi_n+\eta_n$.
 However, the transformation  $\mu\to-\mu$  maps
$\Xi_-\Xi_+\leftrightarrow \Xi_+\Xi_-$ and   
  $\xi_n\leftrightarrow \eta_n$.
  Thus $\Xi^{(+)}$ is an even polynomial operator in $\mu$, 
polynomial in $u$, and 
 $\xi_n+\eta_n$ is an even polynomial function in $\mu$, polynomial in
$u$. 
 Furthermore, each of $\Xi_-\Xi_+$ and $\Xi_+\Xi_-$ is unchanged under 
$u\longrightarrow -u-(a+b+1)-1$, hence  a polynomial of order $p$ in 
$[2u+(a+b+1)k+1]^2$.
 We conclude 
$$ \Xi^{(+)}=P^{(+)}(H^2,L_2,\omega^2,a,b).$$
Similarly 
$$\Xi^{(-)}=(\Xi_-\Xi_+-
\Xi_+\Xi_-)/\mu
=P^{(-)}(H^2,L_2,\omega^2,a,b).$$

Setting $R=-4k^2qL_3-4k^2q^2L_4$, we find the structure equations
\be\label{structureTTW} [L_2,L_4]=R,\quad
 [L_2,R]=-8k^2q^2\{L_2,L_4\}-16k^4q^4L_4,\ee
\[ [L_4,R]=8k^2q^2L_4^2-8k^2qP^{(-)}(H^2,L_2,\omega^2,a,b),\]
\[\fl \frac{3}{8k^2q^2}R^2=-22k^2q^2L_4^2+\{L_2,L_4,L_4\}
+4k^2qP^{(-)}(H^2,L_2,\omega^2,a,b)
+12k^2P^{
(+)}(H^2,L_2,\omega^2,a,b).\]
Thus $H,L_2,L_4$ generate a closed
 algebra.
Note that the symmetries $P^{(-)},P^{
(+)}$ can be of arbitrarily high order, depending on $p$ and $q$, but we have explicit compact expressions for these symmetries.

There is a complication: 
Examples show that $L_4$ is not the lowest order generator. To find the lowest order generator 
we look for a symmetry operator $L_5$ such that $[L_2,L_5]=L_4$.
 Applying this condition to a formal eigenbasis of functions $\Psi_n$ we obtain the 
 general solution 
$$ L_5=-\frac{1}{4qk^2}\left(\frac{\Xi_+}{(\mu+q)\mu}+\frac{\Xi_-}{(\mu-q)\mu}\right)+\frac{\beta_n}{\mu^2-q^2}$$
where $\beta_n$ is a polynomial function of $H$.
Simple algebra gives 
$$ L_5(-\mu^2+q^2)\equiv L_5(\frac{L_2}{k^2}+q^2)=\frac{1}{4qk^2}(L_4-qL_3)-\beta_n.$$
To find $\beta_n$ we take the limit as $\mu\to -q$. 
  There are 3 similar cases, depending on the relative parities of $p$ and $q$.
For $p$, $q$ both odd we have
$$\beta_n\equiv Q(H)=
 -\frac{H(a^2-b^2)}{4}\Pi_{\ell=1}^{(p-1)/2}\left[(-\omega^2)(\frac{H}{4\omega}-\ell)(\frac{H}{4\omega}+\ell)\right]\times $$
$$\Pi_{s=1}^{(q-1)/2}\left[\frac14(-a-b+2s)(a+b+2s)(a-b+2s)
(-a+b+2s)\right].$$

Note that the raising and lowering operators $\Xi_\pm$ are key to the success of this method. See also \cite{Marquette20103}. 
They themselves are not symmetries,  not even pure differential operators. However, from them the true symmetries and  structure equations can be constructed explicitly. In \cite{Celeghini2013} the authors give a unified ladder operator, basis dependent, construction of the dynamical symmetries of the TTW system that includes both classical and quantum systems.

\subsection{The 3D  Kepler-Coulomb system}
Now we extend the recurrence method to an important class of systems in 3 dimensions, the quantum analog of the classical extended Kepler problem. 
This is a St\"ackel transform of the extended oscillator system that is the natural 3D analog of the TTW system. An interesting feature of this system is that the quantum potential now differs from the classical potential on nonflat spaces.
The extended Kepler-Coulomb system is
\[ { H}=\partial_r^2+\frac{2}{r}\partial_r+\frac{\alpha}{r}+\frac{1-k_1^2}{4r^2}+\frac{{ L}_2}{r^2},\]
\[
 { L}_3= \partial_{\theta_2}^2+\frac{\beta}{\cos^2(k_2\theta_2)}+\frac{\gamma}{\sin^2(k_2\theta_2)}.\]
\be\label{Hq2}{ L}_2 =\partial_{\theta_1}^2+k_1\cot (k_1\theta_1)\partial_{\theta_1}+\frac{{ L}_3}{\sin^2(k_1\theta_1)}+\frac{\delta}{\cos^2(k_1\theta_1)}.
\ee
Here, ${ L}_2$, ${ L}_3$ are symmetry operators that determine multiplicative separation of the Schr\"odinger eigenvalue  equation $H\Psi=E\Psi$, and $k_j=p_j/q_j$ where $p_j,q_j$ are nonzero relatively prime positive integers for $j=1,2$, respectively. Note that  ${ L}_2$ and ${ L}_3$ commute:
$[{ L}_2,{ L}_3]=0$. The scalar potential is 
$${\tilde V}=\frac{\alpha}{r}+\frac{1-k_1^2}{4r^2}+ \frac{1}{r^2}\left(\frac{\beta}{\sin^2(k_1\theta_1)\cos^2(k_2\theta_2)}+
\frac{\gamma}{\sin^2(k_1\theta_1)\sin^2(k_2\theta_2)}+\frac{\delta}{\cos^2(k_1\theta_2)}\right).$$
It differs from the classical potential $V$ by the term $\frac{1-k_1^2}{4r^2}$, proportional to the scalar curvature. It is relatively easy to prove superintegrability of the system, but the computations to determine the structure algebra are quite involved. 
The main message that we want to commucate is that it is indeed possible to determine this structure.

The separation equations for the equations, $H\Psi=E\Psi$,  $L_3\Psi=-\mu^2\Psi$,  $L_2\Psi=\lambda\Psi$ with $\Psi=R(r)\Theta(\theta_1)\Phi(\theta_2)$, are:
\be \label{quant1} (\partial_{\theta_2}^2+\frac{k_2^2(\frac14- b^2)}{\cos^2(k_2\theta_2)}+\frac{k_2^2(\frac14- c^2)}{\sin^2(k_2\theta_2)})\Phi(\theta_2)=-\mu^2\Phi(\theta_2),\ee
where we have taken $\beta =k_2^2(\frac14- b^2), \gamma= k_2^2(\frac14- c^2)$, and written the separation constant $\mu=k_2(2m+b+c+1)$.
If we look for solutions
$\Theta(\theta_1)=\frac{\Psi(\theta_1)}{\sqrt{\sin(k_1\theta_1)}}$, $R({\rm R})=S(r)/{\rm R}$ with ${\rm R}=2\sqrt{E}r$
 the equations satisfied by $\Psi$ and $S$ are
\be\label{quant2}  (\partial_{\theta_1}^2+\frac{\frac{k_1^2}{4}-\mu^2}{\sin^2(k_1\theta_1)}+\frac{\delta}{\cos^2(k_1\theta_1)}+\frac{k_1^2}{4}-
\lambda)\Psi(\theta_1)=0,\ee
\be \label{quant3}  (\partial _{{\rm R}}^2 + \frac{1-k_1^2\rho^2}{4{\rm R}^2}+\frac{\alpha}{2\sqrt{E}{\rm R}}-\frac14)S({\rm R})=0.\ee
In (\ref{quant3}) we have set $\lambda=\frac{k_1^2}{4}(1-\rho^2)$, $\delta=k_1^2(\frac14-d^2)$.
The separated solutions are 
$$\Xi_{p,m,n}=R^{k_1\rho}_p(r)\Phi_m^{(c,b)}(\cos(2k_2\theta_2)) \frac{\Psi^{(\mu/k_1,d)}_n(\cos k_1\theta_1)}{\sqrt{\sin(k_1\theta_1)}},$$
$$\Phi_m^{(c,b)}(\cos(2k_2\theta_2))=\sin^{c+1/2}(k_2\theta_2)\cos^{b+1/2}(k_2\theta_2)P_m^{(c,b)}(\cos(2k_2\theta_2)),$$
where the $P_m^{(c,b)}(\cos(2k_2\theta_2))$ are Jacobi functions, \cite{AAR};
$$ \Psi^{(\mu/k_1,d)}_n(\cos(k_1\theta_1))=\sin^{\mu/k_1+1/2}(k_1\theta_1)\cos^{d+1/2}(k_1\theta_1)P^{(\mu/k_1,d)}_n(\cos(2k_1\theta_1)),$$
where the $P^{(\mu/k_1,d)}_n(\cos(2k_1\theta_1))$ are Jacobi  functions; 
$$ R^{k_1\rho}_p(r)=\frac{S(r)}{2\sqrt{E}r},\quad S(r)=e^{-\sqrt{E}r}r^{(k_1\rho+1)/2}L_p^{k_1\rho}(2\sqrt{E}r),$$
and $\rho=2(2n+\frac{\mu}{k_1}+d+1)$, where the $L_p^{k_1\rho}(2\sqrt{E}r)$ are associated Laguerre functions, \cite{AAR}, and the relation between $E$, $\rho$,  $p$ is the quantization condition 
\be\fl \label{E_p} E=\frac{\alpha^2}{(2p+k_1\rho+1)^2}=\frac{\alpha^2}{(2[p+2k_1n+2k_2m]+2k_1[d+1]+2k_2[b+c+1]+1)^2}.\ee

Note: As in the computations with the TTW potential, we  are only interested in the space of generalized eigenfunctions, not the normalization of any individual eigenfunction. Thus the  relations to follow are valid on generalized eigenspaces and don't necessarily agree with the normalization of common polynomial eigenfunctions.

There are transformations that preserve $E$ and imply quantum superintegrability. Indeed for $k_1=p_1/q_1, k_2=p_2/q_2$ the transformations
\[ \fl 1):\ p\to p+2p_1,\ m\to m,\ n\to n-q_1,\quad 
2):\  p\to p-2p_1,\ m\to m,\ n\to n+q_1,\]
\[\fl 3):\ p\to p,\ m\to m-p_1q_2,\ n\to n+q_1p_2,\quad 
4):\ p\to p\ m\to m+p_1q_2,\ n\to n-q_1p_2,\]
will accomplish this.

To effect the $r$-dependent transformations 1) and 2) we use $Y(1)^{ p}_\pm$: 
\be\label{Dnpmt1} {Y(1)}^p_\mp R^{k_1\rho}_p(r)=[2(\pm k_1\rho+1)\partial_r+(2\alpha+\frac{1-k_1^2\rho^2}{r})]R_p^{k_1\rho}(r)\ee
\[ {Y(1)}^p_\mp R^{k_1\rho}_p(r)=-\frac{2\alpha}{2p+k_1\rho+1}[(1\pm 1)+(1\mp 1)(p+1)(p+k_1\rho)]R^{k_1\rho\pm2}_{p\mp 1}(r).\]

To  incorporate the $\theta_1$-dependent parts of 1) and 2)  we use the following recurrence formulas for the
functions $\Psi_n^{\mu/k_1,d}(z)$ where $z=\cos(2k_1\theta_1)$:
\[ Z(1)^n_-\frac{\Psi_n^{\mu/k_1,d}(z)}{{\sqrt{\sin(k_1\theta_1)}}}=-2(\frac{\mu}{k_1}+n)(d+n)\frac{\Psi_{n-1}^{\mu/k_1,d}(z)}{{\sqrt{\sin(k_1\theta_1)}}}\equiv\]
\be\label{Dnpmt2}\left((1-z^2)(\frac{\rho}{2}-1)\partial_z+\frac12(\frac{\rho}{2}-1)(\frac{\rho}{2})z+
\frac{\mu^2}{k_1^2}-d^2)\right)\frac{\Psi_n^{\mu/k_1,d}(z)}{{\sqrt{\sin(k_1\theta_1)}}},\ee
\[ Z(1)^n_+\frac{\Psi_n^{\mu/k_1,d}(z)}{{\sqrt{\sin(k_1\theta_1)}}}=-2(\frac{\mu}{k_1}+d+n+1)(n+1)\frac{\Psi_{n+1}^{\mu/k_1,d}(z)}{{\sqrt{\sin(k_1\theta_1)}}}\equiv\]
\[ \left(-(1-z^2)(\frac{\rho}{2}+1)\partial_z+\frac12(\frac{\rho}{2}-1)(\frac{\rho}{2})z+
\frac{\mu^2}{k_1^2}-d^2)\right)\frac{\Psi_n^{\mu/k_1,d}(z)}{{\sqrt{\sin(k_1\theta_1)}}}.\]
With the identification  $\rho=2(2n+\frac{\mu}{k_1}+d+1)$ we see that the operators $Z(1)_\pm$ depend 
on $\mu^2$ (which can be interpreted as a differential operator) and are polynomial  in $\rho$. We now
form the two operators 
$$J^\pm  =\left(Y(1)^{p\pm 2p_1\mp 1}_\pm Y(1)^{p\pm 2p_1\mp 2}_\pm \cdots Y(1)^{p\pm 1}_\pm Y(1)^p_\pm \right)\left(Z(1)^{n\mp (q_1-1)}_\mp\cdots Z(1)^n_\mp\right).$$ 
Since $J^+$ and $J^-$ switch places under the reflection $\rho\to -\rho$  we see that 
\be\label{J1J2} J_2=J^+ +J^-,\quad J_1=(J^--J^+)/\rho\ee
 are even functions in
 both $\rho$ and $\mu$, hence,  pure differential operators. 
 
 To implement the $\theta_1$-dependent parts of 3) and 4) we set $w=\sin^2(k_1\theta_1)$ and consider $\Psi_n^{\mu/k_1,d}$ as a function of $w$. The relevant
 recurrences are
 $$W^n_-(1)\frac{\Psi_n^{\mu/k_1,d}}{w^{1/4}}\equiv \left[(1+\frac{\mu}{k_1})\left((w-1)\frac{d}{dw}-\frac{\mu}{4k_1}(1-\frac{2}{w})\right)
+\frac{d^2-\frac{\rho^2}{4}}{4}\right]\frac{\Psi_n^{\mu/k_1,d}}{w^{1/4}}$$ 
\be\label{R2a}=\frac{(\frac{\mu}{k_1}+\frac{\rho}{2}+d+1)(-\frac{\mu}{k_1}+\frac{\rho}{2}+d-1)}{4(\frac{\mu}{k_1}+1)(\frac{\mu}{k_1}+2)}\frac
{\Psi_{n-1}^{\mu/k_1+2,d}}{w^{1/4}},\ee
 $$W^n_+(1)\frac{\Psi_n^{\mu/k_1,d}}{w^{1/4}}\equiv \left[(1-\frac{\mu}{k_1})\left((w-1)\frac{d}{dw}+\frac{\mu}{4k_1}(1-\frac{2}{w})\right)
+\frac{d^2-\frac{\rho^2}{4}}{4}\right]\frac{\Psi_n^{\mu/k_1,d}}{w^{1/4}}$$ 
$$=\frac14(-\frac{\mu}{k_1}+\frac{\rho}{2}-d+1)(\frac{\mu}{k_1}+\frac{\rho}{2}-d-1)\frac{\mu}{k_1}(\frac{\mu}{k_1}-1)\frac{\Psi_{n+1}^{\mu/k_1-2,d}}{w^{1/4}}.$$
  To implement the $\theta_2$-dependent parts of 3) and 4) we use the recurrences $X(2)^m_\pm$:
$$ X(2)_+^m \Phi^{(c,b)}_m(z)=2(m+1)(m+c+b+1)\Phi^{(c,b)}_{m+1}(z)\equiv$$
\be\label{R2b}\left[(M+1)(1-z^2)\frac{d}{dz}-\frac12 M(M+1)z-\frac12(c^2-b^2)\right]\Phi^{(c,b)}_m(z),\ee
$$ X(2)_-^m \Phi^{(c,b)}_m(z)=2(m+c)(m+b)\Phi^{(c,b)}_{m-1}(z)\equiv$$
$$\left[ -(M-1)(1-z^2)\frac{d}{dz}-\frac12(M-1)Mz-\frac12(c^2-b^2)\right]\Phi^{(c,b)}_m(z).$$
Here $z=\cos(2k_2\theta_2)$, $M=2m+b+c+1$.
We define 
$$K^\pm =\left(W(1)^{n\pm (q_1p_2-1)}_\pm \cdots W(1)^n_\pm \right)\left(X(2)^{m\mp (p_1q_2-1)}_\mp\cdots X(2)^m_\mp\right).$$ 
From the form of these operators we see that they are even functions of $\rho^2$ and they switch places under the reflection $\mu\to -\mu$. Thus
$$ K_2= K^++K^-,\quad K_1=(K^--K^+)/\mu$$
 are even polynomial functions in
 both $\rho$ and $\mu$, hence, pure differential operators.

We have now constructed  partial differential operators $J_1,J_2,K_1,K_2$, each of which commutes with the 
Hamiltonian $H$ on its 8-dimensional formal eigenspaces. 
However, to prove that they are true symmetry operators  we must show that they commute 
with $H$ when acting on {\it any} analytic functions, not just separated eigenfunctions. This is established in \cite{KKM2012}
via a  canonical form, so that the  $J_i,K_i$ are 
true symmetry operators. We will work out the structure of the algebra generated by $H,L_2,L_3,J_1,K_1$ and from this it will be clear  the system is superintegrable.

To determine the structure relations it is sufficient to establish them on the generalized eigenbases. Then a canonical form  argument shows that the relations hold for general analytic functions. We start by using the 
definitions (\ref{J1J2}) and computing on a generalized eigenbasis:
 \bea
 \fl J^-\Xi_{p,m,n}=\frac{(2)^{4p_1+q_1}(-1)^{q_1}\alpha^{2p_1}(n+1)_{q_1}(\mu/k_1+d+n+1)_{q_1}}{(2p+k_1\rho+1)^{2p_1}}\Xi_{p-2p_1,m,n+q_1},\\
 \fl \label{Jpm2} (-1)^{q_1}\alpha^{-2p_1}(2p+k_1\rho+1)^{2p_1}J^+\Xi_{p,m,n}=\\
(2)^{4p_1+q_1}(-\mu/k_1-n)_{q_1}(-d-n)_{q_1}(p+1)_{2p_1}(-p-k_1\rho)_{2p_1}\Xi_{p+2p_1,m,n-q_1},\nn
\fl \label{Kpm2}(-1)^{q_1p_2}2^{-p_1q_2} K^+\Xi_{p,m,n}=\\
(n+1)_{q_1p_2}(-\frac{\mu}{k_1}-n)_{q_1p_2}(-\frac{\mu}{k_1})_{2q_1p_2}(-m-c)_{p_1q_2}(-m-b)_{p_1q_2}\Xi_{p,m-p_1q_2,n+q_1p_2},\nn
\fl (-1)^{q_1p_2}2^{-p_1q_2}(\frac{\mu}{k_1}+1)_{2q_1p_2}K^-\Xi_{p,m,n}=\\
(\frac{\mu}{k_1}+d+n+1)_{q_1p_2}(-d-n)_{q_1p_2}(m+1)_{p_1q_2}(m+c+b+1)_{p_1q_2}\Xi_{p,m+p_1q_2,n-q_1p_2}.\nonumber\eea

From these 
definitions  we  obtain: 
$$[J_1,L_2]=2k_1^2q_1J_2+4p_1^2J_1,\
[J_2,L_2]=-4q_1\{J_1,L_2\}-4q_1^2k_1^2J_2+2k_1^2q_1(1-8q_1^2)J_1,$$
$$[J_1,L_3]=[J_2,L_3]=0,\ [K_1,L_2]=[K_2,L_2]=0,$$ 
$$[K_1,L_3]=4p_1p_2K_2+4p_1^2p_2^2K_1,\ 
[K_2,L_3]=-2p_1p_2\{L_3,K_1\}-4p_1^2p_2^2K_2-8p_1^3p_2^3K_1.$$
Further we find 
\[\fl 4^{-4p_1-q_1}E^{-2p_1}J^+J^-\Xi_{p,m,n}=\]
$$(\frac{\rho/2-\frac{\mu}{k_1}-d+1}{2})_{q_1}(\frac{\rho/2+\frac{\mu}{k_1}+d+1}{2})_{q_1}(\frac{\rho/2+\frac{\mu}{k_1}-d+1}{2})_{q_1}
\times$$
$$(\frac{\rho/2-\frac{\mu}{k_1}+d+1}{2})_{q_1}(\frac{k_1\rho-\frac{\alpha}{\sqrt{E}}+1}{2})_{2p_1}(\frac{k_1\rho+\frac{\alpha}{\sqrt{E}}+1}{2})_{2p_1}\Xi_{p,m,n}\ ,$$
\[\fl 4^{-4p_1-q_1}E^{-2p_1}J^-J^+\Xi_{p,m,n}=\]
$$(\frac{-\rho/2-\frac{\mu}{k_1}-d+1}{2})_{q_1}(\frac{-\rho/2+\frac{\mu}{k_1}+d+1}{2})_{q_1}(\frac{-\rho/2+\frac{\mu}{k_1}-d+1}{2})_{q_1}
\times$$
$$(\frac{-\rho/2-\frac{\mu}{k_1}+d+1}{2})_{q_1}(\frac{-k_1\rho-\frac{\alpha}{\sqrt{E}}+1}{2})_{2p_1}(\frac{-k_1\rho+\frac{\alpha}{\sqrt{E}}+1}{2})_{2p_1}\Xi_{p,m,n}\ ,$$
\[\fl 2^{-2p_1q_2}K^+K^-\Xi_{p,m,n}=\]
$$I(p,m,n) (m+1)_{p_1q_2}
 (m+c+b+1)_{p_1q_2}
 (m+c+1)_{p_1q_2}(m+b+1)_{p_1q_2}\Xi_{p,m,n}\ ,$$
 \[\fl 2^{-2p_1q_2}K^-K^+\Xi_{p.m.n}=\]
 $$I (p,m,n)(-m)_{p_1q_2}(-m-c-b)_{p_1q_2}(-m-c)_{p_1q_2}(-m-b)_{p_1q_2}\Xi_{p,m,n}\ ,$$
\[I(p,m,n)=\Pi_{\epsilon=\pm 1}(\frac{\frac{\epsilon\rho}{2}+\frac{\mu}{k_1}+d+1}{2})_{q_1p_2}(\frac{\frac{\epsilon\rho}{2}+\frac{\mu}{k_1}-d+1}{2})_{q_1p_2}.\]
 From these expressions it is easy to see that each of $J^+J^-,J^-J^+$ is a polynomial in $\mu^2$ and $E$ and that these operators switch places under the reflection $\rho\to-\rho$. Thus $P_1(H,L_2,L_3)=J^+J^-+J^-J^+$ and $P_2(H,L_2,L_3)=(J^+J^--J^-J^+)/\rho$ are each polynomials in $H,L_2,L_3$. Similarly, each of $K^+K^-,K^-K^+$ is a polynomial in $\rho^2$ and in $\mu$ and  these operators switch places under the reflection $\mu\to -\mu$. Thus $P_3(L_2,L_3)=K^+K^-+K^-K^+$ and $P_4(L_2,L_3)=(K^+K^--K^-K^+)/\mu$ are each polynomials in $L_2,L_3$.
 
 Straightforward consequences of these formulas are the structure relations
 \[[J_1,J_2]=-2q_1J_1^2-2P_2,\ J_1^2(\frac14-k_1^{-2}L_2)+2q_1J_1J_2=J_2^2-2P_1,\]
 \[\ K_1^2+K_2^2=2P_3+2p_1p_2K_1K_2,\  [K_1,K_2]=-2p_1p_2K_1^2-2P_4.\]

\subsubsection{Lowering the orders of the generators}
Just as for the classical analogs, we can find generators that are of order one less than $J_1$ and $K_1$. First we look for a symmetry operator $J_0$ such that $[L_2,J_0]=J_1$ and $[L_3,J_0]=0$. A straightforward computation yields the solution
$$J_0=-\frac{1}{2k_1^2q_1}\left(\frac{J^-}{\rho(\rho+2q_1)}+\frac{J^+}{\rho(\rho-2q_1)}\right)+\frac{S_1(H,L_3)}{\rho^2-4q_1^2}.$$
 {}From this it is easy to show that 
$2k_1^2q_1(\rho^2-4q_1^2)J_0=-J_2+2q_1J_1+2k_1^2q_1S_1(H,L_3).$
To determine $S_1$ we evaluate both sides of this equation for $\rho=-2q_1$:
$S_1(H, L_3)=\frac{1}{k_1^2q_1}J^-_{\rho=-2q_1}$.
The detailed computation can be found in \cite{KKM2012}: 
$$2k_1^2q_1(-4)^{(1-q_1)/2}4^{-p_1}S_1(H,L_3)=$$
$$(\frac{L_3}{k_1^2}+d^2)\ \Pi_{s=0}^{p_1-1}\left(\alpha^2-(1+2s)^2H\right)\Pi_{s=1}^{(q_1-1)/2}((s-\frac{d}{2})^2+\frac{L_3}{4k_1^2})((s+\frac{d}{2})^2+\frac{L_3}{4k_1^2}),$$
$$[L_2,J_0]=J_1,\  [L_3,J_0]=0,\ 2k_1^2q_1J_0\ (1-\frac{4L_2}{k_1^2}-4q_1^2)=-J_2+2q_1J_1+2k_1^2q_1S_1.$$

Next we look for a symmetry operator $K_0$ such that $[L_3,K_0]=K_1$ and $[L_2,K_0]=0$. A straightforward computation yields the solution
$$K_0=-\frac{1}{4p_1p_2}\left(\frac{K^-}{\mu(\mu+p_1p_2)}+\frac{K^+}{\mu(\mu-p_1p_2)}\right)+\frac{S_2(L_2)}{\mu^2-p_1^2p_2^2},$$
where the symmetry operator $S_2$ is to be determined. {}From this it is easy to show that 
$4p_1p_2K_0(\mu^2-p_1^2p_2^2)=-K_2+p_1p_2K_1+4p_1p_2S_2(L_2).$
To determine $S_2$ we evaluate both sides of this equation for $\mu=-p_1p_2$:
$S_2(L_2)=\frac{1}{2p_1p_2}K^-_{\mu=-p_1p_2}$. The result is \cite{KKM2012}: 
\bea\fl S_2(L_2)=(d^2-\frac{\rho^2}{4})2^{p_1q_2-5}\frac{(b^2-c^2)}{p_1p_2}\Pi_{s=1}^{(q_1p_2-1)/2}\left((\frac{d}{2}+\frac{\rho}{4})^2-s^2\right)\left((\frac{d}{2}-\frac{\rho}{4})^2-s^2\right)\nn
\times \Pi_{s=1}^{(p_1q_2-1)/2}(s+\frac{c+b}{2})(s-\frac{c+b}{2})(s-1-\frac{b-c}{2})(s-1+\frac{b-c}{2}), \label{S2L2} \eea
$$ [L_3,K_0]=K_1,\ [L_2,K_0]=0,\ 4p_1p_2K_0(L_3+p_1^2p_2^2)=K_2-p_1p_2K_1-4p_1p_2S_2.$$

\subsubsection{The structure equations}

Now we determine the  the $J,K$-operator commutators.
We write 
$$ J^\pm\Xi_{p,m,n}={\cal J}^\pm(p,m,n)\ \Xi_{p\pm 2p_1,m,n\mp q_1},\ 
 K^\pm \Xi_{p,m,n}={\cal K}^\pm (m,n)\ \Xi_{p,m\mp p_1q_2,n\pm q_1p_2},$$
 where ${\cal J}^\pm,{\cal K}^\pm$ are defined by the righthand sides of  (\ref{Jpm2}),  (\ref{Kpm2}).
We find
$$[J^\pm,K^+]\Xi_{p,m,n}=\frac{1-A(\mp\rho,\mu,)}{1+A(\mp\rho,\mu,)}\{J^\pm,K^+\}\Xi_{p,m,n}\equiv C(\mp\rho,\mu)\{J^\pm,K^+\}\Xi_{p,m,n},$$
$$[J^\pm,K^-]\Xi_{p,m,n}= C(\mp\rho,-\mu)\{J^\pm,K^-\}\Xi_{p,m,n},$$
$$A(\rho,\mu)=
\frac{(\frac{\rho}{4}+\frac{\mu}{2k_1}+\frac12-\frac{d}{2})_{q_1}(\frac{\rho}{4}+\frac{\mu}{2k_1}+\frac12+\frac{d}{2})_{q_1}}
{(\frac{\rho}{4}+\frac{\mu}{2k_1}+\frac12-q_1p_2-\frac{d}{2})_{q_1}(\frac{\rho}{4}+\frac{\mu}{2k_1}+\frac12-q_1p_2+\frac{d}{2})_{q_1}}.$$
From this we can compute the relations
$$[J_j,K_k]=Q^{jk}_{11}\{J_1,K_1\}+Q^{jk}_{12}\{J_1,K_2\}+Q^{jk}_{21}\{J_2,K_1\}+Q^{jk}_{22}\{J_2,K_2\},$$
where $1\le j,k\le2$ and the $Q^{jk}_{i\ell}$ are rational in $\rho^2$, $\mu^2$: $4\rho^{i-j}\mu^{\ell-k} Q^{jk}_{i\ell}=$
$$ (-1)^{j+k+i+\ell}C(-\rho,\mu)+(-1)^{k+\ell}C(\rho,\mu)+(-1)^{j+i}C(-\rho,-\mu)+C(\rho,-\mu).$$
 The relations can be cast into the pure operator form 
\be\label{ratstructure}[K_\ell, J_h] Q=\{J_1,K_1\}P^{h\ell}_{11}+\{J_1,K_2\}P^{h\ell}_{12}+\{J_2,K_1\}P^{h\ell}_{21}+\{J_2,K_2\}P^{h\ell}_{22},\quad\ee
where $h,\ell=1,2$, and $Q,P^{h\ell}_{jk}$ are polynomials in $L_2,L_3$. In particular, 
$$Q=B(\rho,\mu)B(-\rho,\mu)B(\rho,-\mu)B(-\rho,-\mu),\quad B(\rho,\mu)=$$
$$
\Sigma_{\epsilon=0,1}(\frac{\rho}{4}+\frac{\mu}{2k_1}+\frac12-\epsilon q_1p_2-\frac{d}{2})_{q_1}(\frac{\rho}{4}+\frac{\mu}{2k_1}+\frac12-\epsilon q_1p_2+\frac{d}{2})_{q_1} $$
on a generalized eigenbasis.
Thus for general $k_1,k_2$ the symmetry algebra closes algebraically but not polynomially. The basis generators are $H,L_2,L_3,J_0,K_0$ and the commutators $J_1,K_1$  are appended to the algebra.

\subsubsection{The special case $k_1=k_2=1$}
In the case $k_1=k_2=1$  we are in Euclidean space and just as for the classsical system we have  additional permutation symmetry. The extra symmetry gives rise to a new 4th order symmetry 
operator $J_0'$ and the 6 functionally dependent symmetries 
${ H}, { L}_2,{ L}_3, { K}_0$ (2nd order)  and $J_0, { J}'_0$ (4th order) generate a symmetry algebra that  closes polynomially. The complicated details can be found in 
\cite{DASK2011, KKM2012}. The algebraic relation obeyed by the 6 generators is of order 12, \cite{ KKM2012}.

%% file: Stackeltransformchapter.tex
\section{Generalized St\"ackel transform}\label{Stackeltransform}

In Chapter \ref{Chapter2ndorder} we introduced the St\"ackel transform on 2nd order  systems. Here we review the  fundamentals 
of coupling constant metamorphosis (CCM) and  St\"ackel transform, and 
apply them to map superintegrable systems of all orders into other
such  systems on different manifolds. In general, CCM does not preserve structure
but we study specializations  which do preserve
polynomials and  symmetry structures in both the classical and quantum cases.  Details of the proofs can be found in \cite{KMP10}.

\subsection{Coupling constant metamorphosis for classical systems }
The basic tool that we employ follows from ``coupling constant
metamorphosis" (CCM), a general fact about Hamiltonian systems, 
\cite{HGDR}.
Let ${\cal H}({\bf x},{\bf p})+\alpha U({\bf x})$ define a Hamiltonian system
in 2$n$ dimensional phase space. Thus the
Hamilton-Jacobi equation is  ${\cal H}({\bf x},{\bf p})+\alpha
U({\bf x})=E$.
\begin{theorem}\label{theorem0} (CCM)  Assume that the system
admits a constant of the motion ${\cal K}(\alpha)$, locally analytic in parameter $\alpha$. The
Hamiltonian ${\cal H}'=({\cal H}-E)/U$ admits the constant of the motion
${\cal K}'={\cal K}(-{\cal H}')$, where now $E$ is a parameter. \end{theorem}

\medskip\noindent PROOF: If $F,G$ are functions on phase space of the
form $G({\bf x}, {\bf p})$, $F=F(a)=F(a,{\bf x},{\bf p})$ where $a=\alpha({\bf
x},{\bf p})$ then
$$\{F,G\}= \{F(a),G\}|_{a=\alpha({\bf x},{\bf p})}+\partial_a F(a)
|_{a=\alpha({\bf x},{\bf p})}\{\alpha,G\}.$$
By assumption, $\{{\cal K}(\alpha),{\cal H}\}=-\alpha\{{\cal K}(\alpha),U\}$
for any value of the parameter $\alpha$. Thus
$\{{\cal K}(\alpha),{\cal H}'\}=\frac{\{U,{\cal K}(\alpha)\}}{U}({\cal
H}'+\alpha)$,
$$\{{\cal K}({-\cal H}'),{\cal H}'\}=\left[\partial_{\alpha} {\cal
K}(\alpha)\{{\cal H}',{\cal H}'\}+\frac{\{U,{\cal K}(\alpha)\}}{U}({\cal
H}'+\alpha)\right]_{\alpha=-{\cal H}'}=0.$$
$\Box$

\begin{corollary} Let ${\cal K}_1(\alpha), {\cal K}_2(\alpha)$ be constants of
the motion for the system  ${\cal H}({\bf x},{\bf p})+\alpha U({\bf x})$. Then 
$\{{\cal K}_1, {\cal K}_2\}(\alpha)\equiv \{{\cal K}_1(\alpha), {\cal
K}_2(\alpha)\}$ is also a constant of the motion and
$ \{{\cal K}_1(-{\cal H}'), {\cal K}_2(-{\cal H}')\}=\{{\cal K}_1, {\cal
K}_2\}(-{\cal H}')$. 
\end{corollary}

Clearly CCM takes  superintegrable
systems to superintegrable systems. We are concerned with the case where $\cal K$ is
 polynomial in the momenta and
${\cal H}=\sum_{i,j=1}^ng^{ij}p_ip_j+V({\bf x})+\alpha U({\bf x})\equiv {\cal
H}_0+V+\alpha U$.  For 2nd order
constants of the motion  there is special structure: The 2nd order constants
 are typically linear in $\alpha$, so they transform to 2nd
order symmetries again. Then CCM agrees with the St\"ackel transform.
 However, in general the order of
constants of the motion is not preserved by CCM.
\begin{example}\label{example1} The system
$ {\cal H}= p_1^2+p_2^2 + b_1\sqrt{x_1}+b_2x_2$
admits the 2nd order constant  ${\cal K}^{(2)}=p_2^2+b_2x_2$ and
the 3rd order constant  ${\cal K}^{(3)}=p_1^3+\frac32
b_1\sqrt{x_1}p_1-\frac{3b_1^2}{4b_2}p_2$, (\cite{Gravel, MW2008, MW2007} and references
contained therein).
If we choose  $\alpha U=\alpha\sqrt{x_1}$ then the transform of ${\cal K}^{(3)}$
will be 5th order. If we choose $\alpha U=\alpha x_2$ then the transform of
${\cal K}^{(3)}$ will be   nonpolynomial. To obtain useful
structure results from CCM we need to restrict the
generality of the transform action.
\end{example}

\subsection{The Jacobi transform}
Here we study a specialization of CCM  to the
 case $V=0$.  This special version   takes
$N$th order constants of the motion for Hamiltonian systems to $N$th order
constants.
An $N$th order constant  ${\cal K}({\bf x},{\bf p})$ for the
system
\be\label{hamiltonian1} {\cal H}=\sum_{i,j=1}^ng^{ij}p_ip_j+U({\bf x})={\cal
H}_0+U\ee
 is a function on the phase space such that
$\{{\cal K},{\cal H}\}=0$ where
\[\fl {\cal K}={\cal K}_N+{\cal K}_{N-2}+{\cal K}_{N-4}+\cdots+{\cal K}_0,\ 
N\ {\rm even},\quad
 {\cal K}={\cal K}_N+{\cal K}_{N-2}+\cdots+{\cal K}_1,\
N\ {\rm odd}.\]
Here, ${\cal K}_N\ne 0$ and ${\cal K}_j$ is homogeneous in $\bf p$ of order
$j$.
This implies
\[ \{{\cal K}_N,{\cal H}_0\}=0,\quad
\{{\cal K}_{N-2k},U\}+\{{\cal K}_{N-2k-2},{\cal H}_0\}=0,
\quad k=0,1,\cdots,[N/2]-1,\]
and, for $N$ odd, $ \{{\cal K}_1,U\}=0$.

The case $N=1$ is very special. Then ${\cal K}={\cal K}_1$ and the conditions
are
$\{ {\cal K}, {\cal H}_0\}=0,\quad \{{\cal K},U\}=0,$
so $\cal K$ is a Killing vector and $U$ is invariant under the local group
action generated by the Killing vector.

For $N=2$, ${\cal K}={\cal K}_2+{\cal K}_0$ and the conditions are
\be\label{N=2}\{ {\cal K}_2, {\cal H}_0\}=0,\quad \{{\cal K}_2,U\}+\{{\cal
K}_0,{\cal H}_0\}=0,\ee
so ${\cal K}_2$ is a 2nd order Killing tensor and $U$ satisfies 
Bertrand-Darboux  conditions.

For $N=3$, ${\cal K}={\cal K}_3+{\cal K}_1$ and the conditions are
$$\{ {\cal K}_3, {\cal H}_0\}=0,\quad \{{\cal K}_3,U\}+\{{\cal K}_1,{\cal
H}_0\}=0,\quad \{{\cal K}_1,U\}=0. $$
 Integrability conditions for
the last 2 eqns.\  lead to nonlinear PDEs for $U$.

\begin{theorem}\label{theorem1} Suppose  the system (\ref{hamiltonian1}) admits
an $N$th order constant of the motion ${\cal K}$ where $N\ge 1$.  Then
$${\hat{\cal K}}=\sum_{j=0}^{[N/2]}\left(-\frac{{\cal H}_0-E}{U}\right)^j{\cal
K}_{N-2j}$$
 is an $N$th order constant of the motion for the  system $({\cal
H}_0-E)/U$.\end{theorem}

\begin{corollary}\label{cor2} Suppose the system ${\cal H}_0+U$ is $N$th  order
superintegrable. Then the free system ${\cal H}_0/U$ is also $N$th order
superintegrable.\end{corollary}

We call  ${\hat {\cal K}}$ a {\it Jacobi transform} of ${\cal K}$, since 
 it is related to the Jacobi metric, \cite{LW} page 172.
 Note that it is  invertible.

\begin{corollary}\label{cor3} The Jacobi transform satisfies the properties
 $\widehat{\{{\cal K}, {\cal L}\}}=\{\hat{\cal K},\hat{\cal L}\}$,\break $ \widehat{{\cal
K}{\cal L}}=\hat{\cal K}\hat{\cal L}$,
and, if ${\cal K},\ {\cal L}$ are of the same order, $\widehat{a{\cal K}+b{\cal L}}=a\hat{\cal K}+b\hat{\cal L}.$
Thus it defines a homomorphism from the graded symmetry algebra of the system ${\cal
H}_0+U$ to the graded symmetry algebra of the  system $({\cal H}_0-E)/U$.
\end{corollary}
\begin{example}\label{example2}
Consider the  system of Example \ref{example1}:
$ {\cal H}= p_1^2+p_2^2 + b_1\sqrt{x_1}+b_2x_2$, and let $U=
b_1\sqrt{x_1}+b_2x_2+b_3$ for some fixed $b_1,b_2,b_3$ with $b_1b_2\ne 0$. 
The Jacobi transforms of ${\cal H}$, ${\cal K}^{(2)}$, ${\cal K}^{(3)}$ are
$$\hat{\cal H}= \frac{p_1^2+p_2^2-E}{b_1\sqrt{x_1}+b_2x_2+b_3},\
\hat{{\cal
K}}^{(2)}=p_2^2-b_2x_2\left(\frac{p_1^2+p_2^2-E}{b_1\sqrt{x_1}+b_2x_2+b_3}\right),$$
$$ \hat{{\cal K}}^{(3)}=p_1^3-(\frac32
b_1\sqrt{x_1}p_1-\frac{3b_1^2}{4b_2}p_2)\left(\frac{p_1^2+p_2^2-E}{b_1\sqrt{x_1}+b_2x_2+b_3}\right).$$
\end{example}

\subsection{The St\"ackel transform}
We use  the same notation as in the previous section, and a particular nonzero
potential $U=V({\bf x},{\bf b}_0)$.
 The St\"ackel transform  for a 2nd order  system was treated in Section \ref{4.1.3}. It is not a special case of
CCM, although the two transforms are closely
related. However in the situation where the potential functions $V({\bf x},{\bf
b})$ form a finite dimensional vector space,    usual in the study of
2nd order superintegrability,  the
transforms coincide. In this case, by redefining parameters if necessary, we
can assume $V$ is linear in  $\bf b$.

Now we  investigate extensions of the St\"ackel transform to higher order
constants, assuming  $V({\bf x},{\bf b})$  is
linear in ${\bf b}=(b_0,b_1,\cdots b_M)$, $U$ is of the form $U({\bf x}) =
V({\bf x},{\bf b}^0)$ and the potentials $V({\bf x}, {\bf b})$ span a space of
dimension $M+1$:
\be \label{Spot}V({\bf x},{\bf b})=b_0+\sum_{i=1}^M U^{(i)}({\bf x})b_i\ee
where the set of functions $\{1,U^{(1)}({\bf x}),\cdots,U^{(M)}({\bf x})\}$ is
linearly independent.
In the study of 2nd order superintegrability, typically the 2nd order
constants are linear in the $\bf b$ and the algebra generated by
these symmetries via products and commutators has the property that a constant
 of order $N$ depends polynomially on the parameters with order
$\le [N/2]$.
Thus we  consider only those higher order constants of the motion of order $N$
of the form
\be\label{Ssymm}{\cal K}=\sum_{j=0}^{[N/2]} {\cal K}_{N-2j}({\bf p}, {\bf
b})\ee
where ${\cal K}_{N-2j}(a{\bf p},{\bf b})=a^{N-2j}{\cal K}_{N-2j}({\bf p},{\bf
b})$ and ${\cal K}_{N-2j}({\bf p},a{\bf b})=a^{j}{\cal K}_{N-2j}({\bf p},{\bf
b})$ for any parameter $a$.
Let ${\cal K}({\bf b})$ be such an $N$th order constant.
Then
\be\label{Ssym1}{\cal K}(\alpha)\equiv {\cal K}({\bf p},{\bf b} +\alpha {\bf
b}^{(0)})\ee
 is an $N$th order constant  for 
${\cal H}_0+V({\bf x},{\bf b})+\alpha U({\bf x})$.
 Applying Theorem \ref{theorem0}:
 \begin{theorem} Let ${\cal K}$ be an $N$th order constant of the motion for
the system ${\cal H}_0+V({\bf x},{\bf b})$ where $V$ is of the form
(\ref{Spot}) and $\cal K$ is of the form (\ref{Ssymm}). Let ${\cal K}(\alpha)$ be defined by 
(\ref{Ssym1}). Then
$$\tilde{\cal K}={\cal K}\left(-\frac{{\cal H}_0 +V({\bf x},{\bf b})}{U({\bf
x})}\right)=\sum_{j=0}^{[N/2]}\tilde{{\cal K}}_{N-2j}({\bf p},{\bf b}) $$
is an $N$th order constant for the system $({\cal H}_0 +V({\bf
x},{\bf b}))/{U({\bf x})}$, whenever
\be\label{Ssym2}\tilde{{\cal K}}_{N-2j}(a{\bf p},{\bf b})=a^{N-2j}\tilde{{\cal
K}}_{N-2j}({\bf p},{\bf b}),\quad \tilde{{\cal K}}_{N-2j}({\bf p},a{\bf
b})=a^{j}\tilde{{\cal K}}_{N-2j}({\bf p},{\bf b}).\quad \ee
\end{theorem}

\begin{example}\label{example3}
Let, \cite{TTW},
$$ {\cal H}=p^2_1+p^2_2+a(x_1^2+x_2^2)+b \frac{(x_1^2+x_2^2)}{ (x_1^2-x_2^2)^2} +c 
\frac{(x_1^2+x_2^2)}{ x_1^2x_2^2}.$$
There are  two basic constants of the motion, 
$${\cal K}_2=(x_1p_2-x_2p_1)^2+4b \frac{x_1^2x_2^2}{(x_1^2-x_2^2)^2} + c \frac{(x_1^4+x_2^4)}{ x_1^2x_2^2},$$
$${\cal K}_4=(p^2_1-p^2_2)^2+[2ax_1^2+ 2b \frac{(x_1^2+x_2^2)}{(x_1^2-x_2^2)^2} -2c 
\frac{(x_1^2-x_2^2)}{ x_1^2x_2^2} ]p^2_1$$
$$+[-4ax_1x_2+ 8b \frac{x_1x_2}{ (x_1^2-x_2^2)^2} ]p_1p_2+[2ax_2^2+ 2b 
\frac{(x_1^2+x_2^2)}{ (x_1^2-x_2^2)^2} +2c \frac {(x_1^2-x_2^2)}{ x_1^2x_2^2} ]p^2_2$$
$$+a^2(x_1^2-x_2^2)^2+ \frac{b^2}{ (x_1^2-x_2^2)^2} + c^2 \frac{(x_1^2-x_2^2)^2}{ x_1^4x_2^4}+8ab 
\frac{x_1^2x_2^2}{ (x_1^2-x_2^2)} +2 \frac{bc}{ x_1^2x_2^2}.$$
Then the  transformed system  also has  2nd and 4th order  constants:
$${\tilde {\cal H}}=\frac{p^2_1+p^2_2+a(x_1^2+x_2^2)+b \frac{(x_1^2+x_2^2)}{ (x_1^2-x_2^2)^2} +c 
\frac{(x_1^2+x_2^2)}{ x_1^2x_2^2}+d}{(x_1^2+x_2^2)+B \frac{(x_1^2+x_2^2)}{ (x_1^2-x_2^2)^2} +C 
\frac{(x_1^2+x_2^2)}{ x_1^2x_2^2}+D }.$$
\end{example}

\begin{example} In \cite{PW2010} it is shown that the TTW system is St\"ackel equivalent to the caged isotropic oscillator. We sketch the  corresponding construction in the 3D case.
Consider the  caged isotropic oscillator 
\be\label{isotropic}{\cal H}'=p_R^2+\alpha'R^2+\frac{{\cal L}'_2}{R^2},\ee
$${\cal L}'_2=p^2_{\phi_1}+\frac{{\cal L}'_3}{\sin^2(j_1\phi_1)}+\frac{\delta'}{\cos^2(j_1\phi_1)},\quad {\cal L}'_3=p^2_{\phi_2}+\frac{\beta'}{\cos^2(j_2\phi_2)}
+\frac{\gamma'}{\sin^2(j_2\phi_2)}.$$
Here ${\cal L}'_2,\ {\cal L}'_3$ are constants of the motion  that determine additive separation in the spherical coordinates  $R,\phi_1,\phi_2$. 
Also, $j_1,j_2$ are nonzero rational numbers. If $j_1=j_2=1$, then in  Cartesian coordinates 
we have ${\cal H}'=p_x^2+p_y^2+p_z^2+ \alpha'R^2 +\beta'/x^2+\gamma'/y^2+\delta'/z^2$,  and (\ref{isotropic})  can be considered as a 
3-variable analog of the TTW system, flat space  only if $j_1=1$.  Consider the Hamilton-Jacobi 
equation ${\cal H}'=E'$ and take the St\"ackel transform  corresponding  to division by $R^2$. Then, for
 $r=R^2$, $2\phi_1=\theta_1$, $2\phi_2=\theta_2$, we obtain the new Hamilton-Jacobi equation 
${\cal H}=E$ where 
$ {\cal H}= p_r^2+\frac{\alpha}{r} +\frac{{\cal L}_2}{r^2}$
with
$${\cal L}_2=p^2_{\theta_1}+\frac{{\cal L}_3}{\sin^2(k_1\theta_1)}+\frac{\delta}{\cos^2(k_1\theta_1)},\quad {\cal L}_3=p^2_{\theta_2}+\frac{\beta}{\cos^2(k_2\theta_2)}
+\frac{\gamma}{\sin^2(k_2\theta_2)},$$
$$E=-\alpha'/4,\ \alpha=-E'/4, \beta=\beta'/4,\ \gamma=\gamma'/4,\ \delta=\delta'/4,\ k_1=j_1/2,\ k_2=j_2/2.$$
This is  the extended Kepler-Coulomb system.  St\"ackel transforms preserve structure so all  structure results apply to the caged 
  oscillator. Note  that $k_1=k_2=1$ for
 Kepler-Coulomb corresponds to  $j_1=j_2=2$ for the oscillator.\end{example}

\subsection{Coupling Constant Metamorphosis for  quantum symmetries}
Unlike the case of classical Hamiltonian, CCM for quantum systems is not guaranteed to preserve integrals of the motion. In general, it is not clear how to replace the coupling constant by the corresponding operator as you would for the function in classical mechanics. However, we can isolate the cases where it is possible to preserve some of the symmetries. In particular, for second-order superintegrable systems in 2D, in is known that all potential and the integrals of motion depend linearly on at most 4 constants. Thus, CCM is well defined on these systems, see \cite{KKM20061}. 

For higher-order integrals of the motion, we require that the Hamiltonian admit a part of the potential parameterized by a coupling constant and that the integrals of the motion be polynomial in this coupling constant. In this case, the coupling constant metamorphosis preserves integrals of the motion. The results are given in the following theorem from \cite{KMP10}. 

\begin{theorem}\label{Nthorderthm} Let 
$$H(\alpha) \equiv H(0)+\alpha U$$ be a Hamiltonian operator
with integral of motion  of the form 
$$K(\alpha)=\sum_{j=0}^{[N/2]} K_{N-2j}\alpha^j, \qquad [H(\alpha), K(\alpha)]=0$$
where $H(0)$, $U$  and $K_{N-j}$ are independent of the coupling constant $\alpha$. Then the operator ${\tilde K} =\sum_{h=0}^{[N/2]}(-1)^h K_{N-2h} (U^{-1}(H+b))^h$ is a well-defined finite-order differential operator which commutes with the Hamiltonian  ${\tilde H}=U^{-1}(H+b)$.
\end{theorem}
Furthermore, as with many known cases, if each $K_{N-2j}$ is a differential operator of order $N-2j$ then the transformed operator ${\tilde K}$ will remain $N$th order. This is the case in the following example. 
\begin{example} \label{91os} (The 3-1 anisotropic oscillator)  Let $H(\alpha)=\partial_{11}+\partial_{22}+\alpha(9x_1^2+x_2^2)$. This is a superintegrable system with generating 2nd and 3rd order symmetries
 $$ L=\partial_{22}+\alpha x_2^2,\quad K=\{x_1\partial_2-x_2\partial_1,\partial_{22}\} +\frac{\alpha}{3}(\{x_2^3,\partial_1\}-9\{x_1x_2^2,\partial_2\}),$$ where $\{S_1,S_2\}\equiv S_1S_2+S_2S_1$.
Let $U=(9x_1^2+x_2^2) +c$. It follows that  system 
$ {\tilde H}= \frac{1}{(9x_1^2+x_2^2) +c}\left(\partial_{11}+\partial_{22}+b\right)$
has  a 2nd and a 3rd order symmetry.
\end{example}
We also mention that as in the classical case, CCM preserves the structure of the symmetry algebras as shown in the following corollary. 
\begin{corollary} Let $K(\alpha)$ $L(\alpha)$ be $N$th and $M$th order operator symmetries, respectively,  of $H(\alpha)$, each satisfying 
the conditions of Theorem \ref{Nthorderthm}. Then 
$[{\tilde L},{\tilde K}]=\widetilde {[L,K]},\quad  {\tilde L}{\tilde K}=\widetilde {LK}$.
\end{corollary}

Note that Theorem \ref{Nthorderthm} does not require that the quantum system go to a classical system, only that a scalable potential term can be split off. Thus it applies to ``hybrid'' quantum systems that have a classical part which depends linearly on an arbitrary (scalable) parameter and a quantum part depending on $\hbar^2.$ 
\begin{example} \label{91qos} (The hybrid  3-1 anisotropic oscillator)  Let $H=-\hbar^2(\partial_{11}+\partial_{22})+a(9x_1^2+x_2^2)+2\hbar^2/x_2^2$. This is a superintegrable system with generating 2nd and 3rd order symmetries,
 $$ L=\partial_{22}+ax_2^2,\quad K=\{x_1\partial_2-x_2\partial_1,\partial_{22}\} +\{\frac{a}{3}x_2^3+\frac{\hbar^2}{x_2},\partial_1\}-\{3x_1(ax_2^2+\frac{\hbar^2}{x_2^2}),\partial_2\}.$$  
 Let\  $U=(9x_1^2+x_2^2) +c$. Then the system also has   one 2nd and one 3rd order symmetry:
$$ {\tilde H}= \frac{-\hbar^2}{(9x_1^2+x_2^2) +c}\left(\partial_{11}+\partial_{22}+a(9x_1^2+x_2^2)+\frac{2\hbar^2}{x_2^2}+b\right).$$
\end{example}

\begin{example} \label{91qost} (A translated hybrid  3-1 anisotropic oscillator)  This is a slight modification of Example \ref{91qos}. Let $H=-\hbar^2(\partial_{11}+\partial_{22})+a(9x_1^2+x_2^2)+cx_1+2\hbar^2/x_2^2$. This is a superintegrable system with generating 2nd and 3rd order symmetries.
 Let\  $U=x_1$. It follows that the system 
$$ {\tilde H}= \frac{-\hbar^2}{x_1}\left(\partial_{11}+\partial_{22}+a(9x_1^2+x_2^2)+cx_1+\frac{2\hbar^2}{x_2^2}+b\right)$$
is superintegrable with one 2nd and one 3rd order symmetry. 
\end{example}
Note that the general  anisotropic oscillator with singular  terms 
\be H=-\hbar^2(\partial_{11}+\partial_{22})+a(k^2x_1^2+x_2^2) +b/x_1^2+c/x_2^2,\ee
is superintegrable for all integers $k$ \cite{RTW2008}, however it is only with the specific choices of constants as in Examples \ref{91qos} and \ref{91qost} that the additional higher-order integral of motion is of third-order. 
In this sense, the systems in in Examples \ref{91qos} and \ref{91qost} are truly quantum since in the classical limit the potential reduces to the 3-1 anisotropic oscillator and the classical Hamiltonian with the potentials as in the quantum case ($\hbar$ is then considered an arbitrary constant) has different symmetry generators from the quantum case. Thus, we see that CCM can be extended to quantum systems which differ from their classical analog as long as the potential has a term proportional to an arbitrary constant.

%\begin{example}  Let $H(0)=-\hbar^2(\partial_{11}+\partial_{22})+a/x_1^2+2\hbar^2/x_2^2$. This is a superintegrable system with two linearly independent  2nd and three linearly independent 3rd order symmetries, \cite{Gravel}.
 %Let\  $U=1/x_1^2 +c$. Then the system has  two   2nd and three  3rd order symmetries:
%$$ {\tilde H}= -\frac{\hbar^2x_1^2}{1 +cx_1^2}\left(\partial_{11}+\partial_{22}+\frac{a}{x_1^2}+\frac{2\hbar^2}{x_2^2}+b\right).$$
 %\end{example} 

For simplicity, we have restricted our quantum constructions  to 2D manifolds though some partial results hold in $n$ dimensions. 
There appears to be no insurmountable barrier to extending these results to 3D and higher conformally flat manifolds, but the details have not yet 
been worked out. Clearly gauge transformations are required and the gauge will depend on the  curvature of the manifold. In the classical case a multiple parameter
extension of the St\"ackel transform has been studied \cite{SB2008}.

%% file: Askeyschemechapter.tex
\section{Polynomial algebras and their irreducible representations}
\label{algebrareps}
Just as the representation theory for Lie algebras has seen wide applications in mathematical physics and in special functions, it is natural that the representation theory for the algebras generated by superintegrable systems would have interesting analysis and applications. Indeed, the study of polynomial algebras and their representations continues to be a rich field of current inquiry, see e.g. \cite{BDK, Dask2001, GZLU, Zhedanov1992a,Zhedanov1992b,Quesne1994, VILE, Quesne2007, Marquette20103, Marquette2012}. In particular, the irreducible representations give important information about the spectrum of the Hamiltonian and other symmetry operators, as well as the multiplicities of the bound state energy levels. In this chapter, we survey results concerning only quadratic algebras with 3 and 5 functionally independent operators. In the former case, discussed in section \ref{Askeyscheme},  the representations of 2nd order superintegrable systems in 2D is directly connected with the Askey scheme of orthogonal polynomials and the limits are obtained from limits of the physical systems which in turn give contractions of the algebras. The case of quadratic algebras with 5 generators is less well understood: we  discuss only one model in section \ref{generic3sphere} as a generalization of the 2D results and an indication of possible areas of future research.

\subsection{Quadratic algebras: Contractions  \& Askey Scheme} \label{Askeyscheme}
Special functions  arise as solutions of exactly solvable problems, so it shouldn't be  surprising that the Askey scheme for organizing hypergeometric 
orthogonal polynomials is a consequence of  contractions of  superintegrable systems.
We describe how  all 2nd order superintegrable systems in 2 dimensions  are limiting cases of a single system: 
the generic 3-parameter potential on the 2-sphere,  $S9$ in our listing. This implies that the quadratic symmetry algebras of these systems are 
contractions of that of $S9$. The irreducible representations of $S9$ have a realization  in terms of difference operators in 1 variable,  the 
structure algebra for  the Wilson and Racah polynomials: the Askey-Wilson algebra for $q=1$. Recently, Genest, Vinet and Zhedanov \cite{genest2013superintegrability}  have given an elegant proof of the equivalence between the symmetry algebra for S9 and the Racah problem of $\mathfrak{su}(1,1)$.   By 
contracting the  representations of S9 we can  obtain  representations of the quadratic symmetry algebras of  other   systems we 
obtain the full Askey scheme, \cite{Koorn, KLS}. This 
 ties the  scheme directly to physical phenomena. It is more general: 
it applies to all special functions that arise from these systems via separation of variables, not just those of hypergeometric type, and it extends to
 higher dimensions.

The special functions of mathematical physics are associated with realizations of the irreducible representations of the quadratic 
symmetry algebras of 2nd order superintegrable systems, \cite{BDK, Dask2001,GZLU,Zhedanov1992a,Zhedanov1992b,Quesne1994}. Since the structures of
 these algebras are essentially preserved under the St\"ackel transform it is sufficient to study only one system in each equivalence class of transforms. 
 There are 13 St\"ackel equivalence classes of 2D systems with non-constant potentials but one is an isolated Euclidean singleton unrelated to the Askey scheme. 
 Since every 
2nd order 2D superintegrable system is St\"ackel equivalent to a constant curvature system,  we choose our examples in flat space and on complex  2-spheres
 There are exactly 6 St\"ackel equivalence classes of nondegenerate potentials and 6 of degenerate ones, listed in Section \ref{thelist}. 
 
Each of the  12 superintegrable systems related to the Askey scheme restricts to a free 1st order superintegrable system when the potential is set to 0:\break
\noindent 
{\bf 1)}\ {\bf  The complex 2-sphere}:
Here
$s_1^2+s_2^2+s_3^2=1$ is the embedding of the   2-sphere in complex Euclidean space, and  
the Hamiltonian is $H=J_1^2+J_2^2+J_3^2$,
where      $J_3=s_1\partial_{s_2}-s_2\partial_{s_1}$  and $J_2,J_3$
are obtained by cyclic permutations of  $1,2,3$.  
The basis symmetries are $J_1,J_2,J_3$.
They generate the Lie algebra $so(3)$ with relations 
 $[J_1,J_2]=-J_3$,
 $[J_2,J_3]=-J_1$, $[J_3,J_1]=-J_2$ and Casimir $H$.

\medskip\noindent
{\bf 2)}\ {\bf The complex Euclidean plane}:\label{freeplane}
Here $H= \partial_x^2+\partial_y^2$ with 
 basis symmetries  $P_1=\partial_x$, $P_2=\partial_y$ and $M=x\partial_y-y\partial_x$.
The symmetry Lie algebra is $e(2)$ with  relations 
$ [P_1,P_2]=0$, $[P_1,M]=P_2$, $[P_2,M]=-P_1$ and Casimir $H$.
As will be shown in a forthcoming article \cite{KMSH}, all of the contractions of the quadratic algebras are in fact generated by the contractions of these Lie algebras, which have been classified \cite{WW}.

\subsubsection{Contractions of superintegrable systems}\label{contractions}  A detailed treatment of contractions will appear elsewhere, \cite{KMSH}.
 Here we just describe ``natural" contractions. Suppose we have a nondegenerate superintegrable system with generators $H,L_1,L_2$ and structure
 equations (\ref{Casimir1}), defining a quadratic algebra $Q$.
If we make a change of basis to new generators ${\tilde H},{\tilde L_1}, {\tilde L_2}$ and parameters ${\tilde a_1},{\tilde a_2}, {\tilde a_3}$ such that 
\[
\left(\begin{array}{c}
{\tilde L_1}\\
{\tilde L_2} \\
{\tilde H}
\end{array}\right)
=\left(\begin{array}{ccc}
A_{1,1} & A_{1,2}&A_{1,3} \\
A_{2,1}&A_{2,2} &A_{2,3}  \\
0 &0 &A_{3,3}
\end{array}\right)
\left(\begin{array}{c}
L_1\\
L_2 \\
H
\end{array}\right)+
\left(\begin{array}{ccc}
B_{1,1} & B_{1,2}&B_{1,3} \\
B_{2,1}&B_{2,2} &B_{2,3}  \\
B_{3,1} &B_{3,2} &B_{3,3}
\end{array}\right)
\left(\begin{array}{c}
a_1\\
a_2 \\
a_3
\end{array}\right),\]
\[ \left(\begin{array}{c}
{\tilde a_1}\\
{\tilde a_2} \\
{\tilde a_3}
\end{array}\right)
=\left(\begin{array}{ccc}
C_{1,1} & C_{1,2}&C_{1,3} \\
C_{2,1}&C_{2,2} &C_{2,3}  \\
C_{3,1} &C_{3,2} &C_{3,3}
\end{array}\right)
\left(\begin{array}{c}
a_1\\
a_2 \\
a_3
\end{array}\right)
\]
for some $3\times 3$ constant matrices $A=(A_{i,j}),B,C$ such that $\det A \cdot \det C\ne 0$,  we will have the same 
system with new structure equations  of the form (\ref{Casimir1}) for ${\tilde R}=[{\tilde L_1},{\tilde L_2}]$, $[{\tilde L_j},
{\tilde R}]$,  ${\tilde R}^2$,  but with transformed structure constants.
 We choose a continuous 1-parameter family of basis transformation matrices $A(\epsilon),B(\epsilon),C(\epsilon)$, $0<\epsilon\le 1$ such 
that $A(1)=C(1)$ is the identity matrix, $B(1)=0$ and  $\det A(\epsilon)\ne 0$, $\det C(\epsilon)\ne 0$. Now 
suppose as $\epsilon\to 0$ the basis change becomes singular, (i.e., the limits of $A,B,C$ either do not exist or, if they exist
 do not satisfy $\det A(0)\det C(0)\ne 0$) but the structure equations involving $A(\epsilon),B(\epsilon),C(\epsilon)$, go to a limit,
  defining a new quadratic algebra $Q'$. We call $Q'$ a {\it contraction} of $Q$ in analogy with Lie algebra contractions \cite{Wigner}.

For a degenerate superintegrable system with generators $H,X, L_1,L_2$ and structure equations (\ref{structure2}),(\ref{Casimir2}), 
defining a quadratic algebra $Q$,
 a change of basis to new generators ${\tilde H},{\tilde X}{\tilde L_1}, {\tilde L_2}$ and parameter ${\tilde a}$ such that ${\tilde a} = Ca$, and 
\[
\left(\begin{array}{c}
{\tilde L_1}\\
{\tilde L_2} \\
{\tilde H}\\
{\tilde X}
\end{array}\right)
=\left(\begin{array}{cccc}
A_{1,1} & A_{1,2}&A_{1,3}&0 \\
A_{2,1}&A_{2,2} &A_{2,3}&0  \\
0 &0 &A_{3,3}&0\\
0&0&0&A_{4,4}
\end{array}\right)
\left(\begin{array}{c}
L_1\\
L_2 \\
H\\
X
\end{array}\right)+
\left(\begin{array}{c}
B_1 \\
B_2  \\
B_3\\
0
\end{array}\right)
a
\]
for some $4\times 4$   matrix $A=(A_{i,j})$ with $\det A\ne 0$, complex 4-vector $B$ and constant $C\ne 0$ yields the same superintegrable 
system with new structure equations  of the form (\ref{structure2}),(\ref{Casimir2}) for $[{\tilde X},{\tilde L_j}]$,  
$[{\tilde L_1},{\tilde L_2}]$, and $\tilde G=0$,  but with transformed structure constants. 
 Suppose we choose a continuous 1-parameter family of basis transformation 
matrices $A(\epsilon),B(\epsilon),C(\epsilon)$, $0<\epsilon\le 1$ such that $A(1)$ is the identity matrix, $B(1)=0$, $C(1)=1$, and  
$\det A(\epsilon)\ne 0$, $ C(\epsilon)\ne 0$. Now suppose as $\epsilon\to 0$ the basis change becomes singular, (i.e., 
the limits of $A,B,C$ either do not exist or, exist and do not satisfy $ C(0)\det A(0)\ne 0$), but that the structure equations 
involving $A(\epsilon),B(\epsilon),C(\epsilon)$, go to a finite limit, thus defining a new quadratic algebra $Q'$. We call $Q'$ a {\it contraction} of $Q$.

It has been established that all 2nd order 2D superintegrable systems 
can be obtained from system $S9$ by limiting processes in the coordinates and/or  a St\"ackel transformation, e.g. \cite{KMWP, KKM20041, KKM20042, KKWMPOG}.
All systems listed in Section \ref{thelist} are limits of $S9$.  It follows that the
quadratic algebras generated by each system are contractions of the algebra of $S9$. (However,  an abstract quadratic algebra may 
not be associated with a superintegrable system, and a contraction of a quadratic algebra associated with one superintegrable system to a quadratic algebra 
associated with another superintegrable system does not necessarily imply that this  is associated with a coordinate  limit process.)

\subsubsection{Models of superintegrable systems}
A representation of a quadratic algebra is a homomorphism of the algebra into the associative  algebra of linear operators on some vector space. 
 In this paper a {\it model} is  a
 faithful  representation in which the vector space is a space of polynomials in one complex variable and the action is via   differential/difference 
operators acting on that space. We will  study classes of irreducible representations realized by these models.
 Suppose a superintegrable system with quadratic algebra $Q$ contracts to a superintegrable system with quadratic algebra $Q'$ via a 
continuous family of transformations indexed by the parameter $\epsilon$. If we have a model of a representation of $Q$ we can 
try to ``save" this representation, as did Wigner for Lie algebra representations \cite{Wigner}, by passing through a continuous family of  representations of $Q(\epsilon)$ in the model to 
obtain a representation of $Q'$ in the  limit. We will show that as a byproduct of contractions  from $S9$ for which we save representations in the  limit, we obtain the Askey Scheme for hypergeometric 
orthogonal polynomials. (The full details can be found in \cite{KMPost13}.) In all the models to follow the polynomials we classify are eigenfunctions of formally self-adjoint or formally 
skew-adjoint operators.  The weight functions for the orthogonality
 can be found in \cite{koekoek1996askey}. They can be derived by requiring that the 2nd order operators $H,L_1,L_2$ are formally self-adjoint 
and the 1st order operator $X$ is formally skew-adjoint. See \cite{KMPost11}  for examples.

\subsubsection{The $S9$ model} 
There is no differential model for $S9$ but  a difference operator model yielding structure equations for the Racah and Wilson polynomials \cite{KMPost1}, defined as  
\begin{equation}\label{Wilson} w_n(t^2)\equiv w_n(t^2,\alpha,\beta,\gamma,\delta)=(\alpha+\beta)_n(\alpha+\gamma)_n(\alpha+\delta)_n\times\end{equation}
$$ {}_4F_{3}\left(\begin{array} {llll}-n,&a+b+c+d+n-1,&a-t,&a+t \\ a+b,&a+c,& a+d\end{array};1\right)$$
$$ = (a+b)_n(a+c)_n(a+d)_n\Phi^{(a,b,c,d)}_{n}(t^2),$$
where $(a)_n$ is the Pochhammer symbol and ${}_4F_3(1)$ is a  hypergeometric function of unit argument \cite{AAR}. The polynomial $w_n(t^2)$ is symmetric in 
$\alpha,\beta,\gamma,\delta$.
For the finite dimensional representations the spectrum of $t^2$ is $\{(a+k)^2,\ k=0,1,\cdots,m\}$ and the
 orthogonal basis eigenfunctions are Racah polynomials. In the infinite dimensional case they are Wilson polynomials. They are eigenfunctions for the difference 
operator $\tau^*\tau$ defined via $E_t^AF(t)=F(t+A)$ and
 \bea \tau&=&\frac{1}{2t}(E_t^{1/2}-E_t^{-1/2}),\label{tau}\\
 \tau^*&=&\frac{1}{2t}\left[(a+t)(b+t)(c+t)(d+t)E_t^{1/2}-(a-t)(b-t)(c-t)(d-t)E_t^{-1/2}\right].\label{taustar}\eea

A finite or infinite dimensional bounded below  representation  is defined by $H=E$, $a_i=\frac14-\alpha_i^2$ and 
\medskip\noindent
\be L_1=-4\tau^*\tau -2(\alpha_2+1)(\alpha_3+1)+\frac12,\ 
        L_2=-4t^2+\alpha_1^2+\alpha_3^2-\frac12,\ee
 \be E=-4(m+1)(m+1+\alpha_1+\alpha_2+\alpha_3)+2(\alpha_1\alpha_2+\alpha_1\alpha_3+\alpha_2\alpha_3)+ \alpha_1^2+\alpha_2^2+\alpha_3^2-\frac14 \nonumber,\ee 
 and the  constants of the Wilson polynomials are chosen as
\be \ba{ll} a=-\frac12(\alpha_1+\alpha_3+1)-m,& d=\alpha_2+m+1+\frac12(\alpha_1+\alpha_3+1),\\
                  b=\frac12(\alpha_1+\alpha_3+1), & c=\frac12(-\alpha_1+\alpha_3+1).\ea \nonumber\ee
Here $n=0,1,\cdots,m$ if $m$ is a nonnegative integer and $n=0,1,\cdots$ otherwise. 

 For the basis  $f_{n,m}\equiv \Phi^{(a,b,c,d)}_{n}(t^2)$,
 the model action  is $Hf_{m,n}=Ef_{n,m}$,
 \bea L_1f_{n,m}&=&-\left(4n^2+4n[\alpha_2+\alpha_3+1]+2[\alpha_2+1][\alpha_3+1]-\frac12\right)f_{n,m},\nonumber\\
         L_2f_{n,m}&=&K_{n+1,n}f_{n+1,m}+K_{n-1,n}f_{n-1,m}+\left( K_{n,n}+ \alpha_1^2+\alpha_3^2-\frac12   \right)f_{n,m},\nonumber \eea
\[K_{n+1,n}={\frac { \left( \alpha_{{3}}+1+\alpha_{{2}}+n \right)  \left( m-n
 \right)  \left( m-n+\alpha_{{1}}\right)  \left( 1+\alpha_{{2}}+n
 \right) }{ \left( \alpha_{{3}}+1+\alpha_{{2}}+2\,n \right)  \left( 
\alpha_{{3}}+2+\alpha_{{2}}+2\,n \right) }}
,\]
\[ K_{n-1,n}={\frac {n \left( \alpha_{{3}}+n \right)  \left( \alpha_{{1}}+\alpha_{{
3}}+1+\alpha_{{2}}+m+n \right)  \left( 1+\alpha_{{3}}+\alpha_{{2}}+m+n
 \right) }{ \left( \alpha_{{3}}+1+\alpha_{{2}}+2\,n \right)  \left( 
\alpha_{{3}}+\alpha_{{2}}+2\,n \right) }},\]
\[ K_{n,n}=\left(\frac12(\alpha_1+\alpha_3+1)-m\right)^2-K_{n+1,n} -K_{n-1,n}.\]

Note that these models give the possible energy eigenvalues of the quantum Hamiltonians and their multiplicities.
\subsubsection{Some  $S9 \to E1$ contractions}

There are at least two ways to take this contraction; it is possible to contract the sphere about the point $(0,1,0)$ which gives  continuous Dual Hahn polynomials
 as limits of  Wilson polynomials. Contracting about the point $(1,0,0)$ leads to continuous Hahn polynomials or Jacobi polynomials. 
 The dual Hahn and continuous dual Hahn polynomials correspond to the same superintegrable system but they are eigenfunctions of different generators. 
For the finite dimensional restrictions ($m$ a positive integer) we have the restrictions of Racah polynomials to dual Hahn and Hahn respectively.

\noindent {\bf 1)}\ {\bf Wilson $\to$ continuous dual Hahn  }
For the first limit, in the quantum system, we contract about  $(0,1,0)$ so that the points of our two dimensional space lie in the plane $(x,1, y)$.
We set $s_1=\sqrt{\epsilon}x, s_2=\sqrt{1-s_1^2-s_3^2}\approx 1-\frac{\sqrt{\epsilon}}{2}(x^2+y^2) $, $ s_3=\sqrt{\epsilon }y$, 
for small $\epsilon$. The coupling constants  transform as 
\be  \left(\begin{array}{c}
{\tilde a_1}\\
{\tilde a_2} \\
{\tilde a_3}
\end{array}\right)
=\left(\begin{array}{ccc}
0 & \epsilon^2 &0\\
1 &0 & 0\\
0 &0 & 1
\end{array}\right)
\left(\begin{array}{c}
a_1\\
a_2 \\
a_3
\end{array}\right),\ee
 and we get $E1$ as $\epsilon\to 0$. This gives the quadratic algebra contraction 
\be\label{FirstS9-E1contraction}
\left(\begin{array}{c}
{\tilde L_1}\\
{\tilde L_2} \\
{\tilde H}
\end{array}\right)
=\left(\begin{array}{ccc}
\epsilon &0  &0 \\
0 & 1 &0 \\
0 &0 &\epsilon
\end{array}\right)
\left(\begin{array}{c}
L_1\\
L_2 \\
H
\end{array}\right)+
\left(\begin{array}{ccc}
0 & 0 &0  \\
0 & 0  &0   \\
0 & -\epsilon & 0
\end{array}\right)
\left(\begin{array}{c}
a_1\\
a_2 \\
a_3
\end{array}\right).
\ee
 As in S9, it is advantageous in the model to express the 3 coupling constants as quadratic functions of other parameters, so that with $\alpha_2\rightarrow \infty$,
\be\label{betas} \left(\begin{array}{c}
{\tilde a_1}\\
{\tilde a_2} \\
{\tilde a_3}
\end{array}\right)=\left(\begin{array}{c}
-\beta_1^2\\
\frac14-\beta_2^2\\
\frac14-\beta_3^2
\end{array}\right)=\left(\begin{array}{c}
\frac{\epsilon^2}4-\epsilon^2 \alpha_2^2\\
\frac14-\alpha_1^2\\
\frac14-\alpha_3^2
\end{array}\right).\ee
In the contraction limit the operators tend to $ H'=\lim_{\epsilon\rightarrow 0}\widetilde{H}=E'$ and 
\[  L_1'=\lim_{\epsilon\rightarrow 0}\widetilde{L}_1=-4\tau'^*\tau' -2\beta_1(\beta_3+1),\
        L_2'=\lim_{\epsilon\rightarrow 0}\widetilde{L}_2=-4t^2+\beta_2^2+\beta_3^2-\frac12,\]
  \[ E'=-2\,\beta_{{1}} \left( 2\,m+2+\beta_{{2}}+\beta_{{3}} \right) \]
  The eigenfunctions of $L_1$, the Wilson polynomials, transform in the contraction limit to the eigenfunctions of $ L_1'$,  the dual Hahn polynomials $S_n$,  
 $$ S_n(-t^2,a',b',c') =(a'+b')_n(a'+c')_n {}_3 F_2\left(\ba{lll} -n,&a'+t,&a'-t\\ a'+b',& a'+c'& \ea;1\right),$$
\be a'=-\frac12(\beta_2+\beta_3+1)-m,\quad   b'=\frac12(\beta_2+\beta_3+1), \quad  c'=\frac12(-\beta_2+\beta_3+1).\nonumber\ee
Again, $n=0,1,\dots,m$ if $m$ is a nonnegative integer and $n=0,1,\dots$ otherwise. The operators $\tau'^*$ and $\tau'$ are given by
 \bea \tau'&=&\tau=\frac{1}{2t}(E_t^{1/2}-E_t^{-1/2}),\nonumber\\
 \tau^*&=&\frac{\beta_1}{2t}\left[(a'+t)(b'+t)(c'+t)E_t^{1/2}-(a'-t)(b'-t)(c'-t)E_t^{-1/2}\right],\nonumber\eea 
 
 The action  on the basis $f'_{n,m}\equiv S_{n}(-t^2,a',b',c')/(a'+b')_n(a'+c')_n$ is 
 \[L_1'f'_{n,m}=-2\beta_1\left(2n+\beta_3+1\right)f'_{n,m},\   H'f_{m,n}=E'f_{n,m}.\]
      \[   L_2'f'_{n,m}=K'_{n+1,n}f'_{n+1,m}+K'_{n-1,n}f'_{n-1,m}+\left( K'_{n,n}+ \beta_2^2+\beta_3^2-\frac12   \right)f_{n,m},\]
\[K'_{n+1,n}=\left( m-n \right)  \left(m-n+\beta_2 \right) , \qquad 
 K'_{n-1,n}=n(n+\beta_3)\]
\[ K'_{n,n}=\left(\frac12(\beta_2+\beta_3+1)-m\right)^2-K'_{n+1,n} -K'_{n-1,n}.\]

\noindent {\bf 2)}\ {\bf Wilson $\to$ continuous Hahn  }
Next,  we contract about  $(1,0,0)$ so that the points of our two dimensional space lie in the plane $(1,x, y).$ 
We set $s_1=\sqrt{1-s_1^2-s_3^2}\approx 1-\frac{\sqrt{\epsilon}}{2}(x^2+y^2) , s_2=\sqrt{\epsilon} x$, $ s_3=\sqrt{\epsilon }y$, for small $\epsilon$. 
\be  \left(\begin{array}{c}
{\tilde a_1}\\
{\tilde a_2} \\
{\tilde a_3}
\end{array}\right)
=\left(\begin{array}{ccc}
\epsilon^2 & 0 &0\\
0 &1 & 0\\
0 &0 & 1
\end{array}\right)
\left(\begin{array}{c}
a_1\\
a_2 \\
a_3
\end{array}\right),\ee
 \be\label{SecondS9-E1contraction}
\left(\begin{array}{c}
{\tilde L_1}\\
{\tilde L_2} \\
{\tilde H}
\end{array}\right)
=\left(\begin{array}{ccc}
0 &\epsilon  &0 \\
1 & 0 &0 \\
0 &0 &\epsilon
\end{array}\right)
\left(\begin{array}{c}
L_1\\
L_2 \\
H
\end{array}\right)+
\left(\begin{array}{ccc}
0 & 0 &0  \\
0 &0  &0   \\
 -\epsilon & 0 & 0
\end{array}\right)
\left(\begin{array}{c}
a_1\\
a_2 \\
a_3
\end{array}\right).
\ee
 In terms of the constants of (\ref{betas}) the transformation gives, with $\alpha_1\rightarrow \infty$,
\be\left(\begin{array}{c}
{\tilde a_1}\\
{\tilde a_2} \\
{\tilde a_3}
\end{array}\right)=\left(\begin{array}{c}
-\beta_1^2\\
\frac14-\beta_2^2\\
\frac14-\beta_3^2
\end{array}\right)=\left(\begin{array}{c}
\frac{\epsilon^2}4-\epsilon^2 \alpha_1^2\\
\frac14-\alpha_2^2\\
\frac14-\alpha_3^2
\end{array}\right).\ee
Saving a representation: We set
$t=-x+{\frac {\beta_1}{2\epsilon}}+m+\frac12(\beta_{{3}}+1)$.
 In the contraction limit, the operators go as $L_i'=\lim_{\epsilon\rightarrow 0}\widetilde{L}_i$, 
\bea L_1'&=&2\beta_{{1}} \left( 2\,x-2m-\beta_{{3}}-1 \right),\   H'=-2\beta_1(2m+2+\beta_2+\beta_3),\nonumber \\
        L_2'&=&-4\left( B(x)E_x+C(x)E_x^{-1}-B(x)-C(x)\right)-2(\beta_2+1)(\beta_3+1)+\frac12\nonumber 
      \eea
with $B(x)=(x-m)(x+\beta_2+1), \, C(x)=x(x-m-1-\beta_3).$
The operators $L_1', L_2' $ and $H'$ satisfy the algebra relations in (\ref{E1Structure}). 
The eigenfunction of $L_1$, the Wilson polynomials, transform in the contraction limit to the eigenfunctions of $ L_2'$, which are the Hahn
 polynomials, ${f'}_{m,n}=Q_n$, 
$$Q_n(x;\beta_2,\beta_3,m)= {}_3 F_2\left(\ba{lll} -n,&\beta_2+\beta_3+n+1,&-x\\ -m,& \beta_2+1& \ea;1\right).$$
The action of the operators on this basis is given by
\bea L_1' f'_{n,m}&=&K'_{n+1,n}f'_{n+1,m}+K'_{n,n}f'_{n,m}+K'_{n-1,n}f'_{n-1,n},\nonumber\\
  L'_2f'_{n,m}&=&-\left(4n^2+4n[\beta_2+\beta_3+1]+2[\beta_2+1][\beta_3+1]-\frac12\right)f'_{n,m}.\nonumber\\
 K'_{n+1, n}&=&-4\beta_1\frac{(m-n)(n+\beta_2+\beta_3+1)(n+\beta_2+1)}{(2n+\beta_2+\beta_3+1)(2n+\beta_2+\beta_3+2)}\nonumber \\
K'_{n-1, n}&=&-4\beta_1\frac{n(n+\beta_3)(m+n+\beta_2+\beta_3+1)}{(2n+\beta_2+\beta_3+1)(2n+\beta_2+\beta_3)}\nonumber\\
K'{n,n}&=&-2\beta_1(2m+\beta_3+1)-K'_{n+1,n}-K_{n-1,n}
.\nonumber\eea
\medskip\noindent
{\bf 3)}\ {\bf Wilson $\to$ Jacobi:} The previous operator contraction degenerates when $\alpha_1=\beta_1=0$. However, we can save this representation by setting 
\[ m=\frac{\sqrt{-E'}}{2\sqrt{\epsilon}}-1+\frac{\beta_2+\beta_3}{2},\qquad t=\frac{\sqrt{-E'}}{2\sqrt{\epsilon}}\sqrt{\frac{1+x}{2}}\]
 for $E'$ a constant and letting 
$m\to\infty$.  Then (\ref{SecondS9-E1contraction}) gives a contraction of the model for S9  to a differential operator model for $E1$ with $\beta_1=0$:
\begin{footnotesize}
\bea \label{E1Model3}\ba{rl} 
 L_1'=&\!\!\!\!\frac{E'}{2}(x+1),\quad  H'= E',\\
        L_2'=&\!\!\!\!4(1-x^2)\partial_x^2+4 \left[\beta_3-\beta_2-(\beta_2+\beta_3+2)x\right]\partial_x -2(\beta_2+1)(\beta_3+1)+\frac12.
      \ea\nonumber\eea
        The eigenfunctions for $L_1$, the Wilson polynomials,  tend in the limit to eigenfunction of $L_2'$,  the Jacobi polynomials: 
\be \label{Jacobia}P_n^{\beta_2, \beta_3}(x)=\frac{(\beta_2+1)_n}{n!} {}_2 F_1\left(\ba{ll} -n,&\beta_2+\beta_3+n+1\\  \beta_2+1& \ea;\frac{x-1}{2}\right).\ee
\end{footnotesize}
In terms of the basis $f_{n}=\frac{n!}{(\beta_2+1)_n}P_n^{\beta_2, \beta_3}(x)$, the action of the operators is 
\begin{footnotesize}
\bea  L_1'f'_{n}=K'_{n+1,n}f'_{n+1}+K'_{n-1,n}f_{n-1}+K'_{n,n}f_{n}\nonumber \\
         L_2'f'_n=-4n(n+\beta_2+\beta_3+1)-2(\beta_2+1)(\beta_3+1)+\frac12,\nonumber\eea
       \bea K'_{n+1,n}&=& \frac{E'(\beta_2+\beta_3+n+1)(\beta_2+n+1)}{(\beta_2+\beta_3+2n+1)(\beta_2+\beta_3+2n+2)},\nonumber\\
 K'_{n-1,n}&=& \frac{E'n(n+\beta_3)}{(\beta_2+\beta_3+2n)(\beta_2+\beta_3+2n+1)},\
 K'_{n,n}=E'-K'_{n+1,n}-K'_{n-1,n}.\nonumber\eea 
 \end{footnotesize}

\subsubsection{A nondegenerate $\to$ degenerate limit}
This  appears initially  a  mere restriction of the 3-parameter potential to 1-parameter. However, after 
 restriction one 
2nd order generator $L_i$ becomes a perfect square $L_i=X^2$. The spectrum of  $L_i$ is nonnegative  but that of $X$
 can 
take both positive/negative values. This results in a virtual doubling of the support of the  measure in the finite case. Also, 
the commutator of $X$ and the remaining 2nd order symmetry 
leads to a new 2nd order symmetry.  
In \cite{KMSH} we will show how this follows directly as a contraction from  $R^2$.

\medskip\noindent
{\bf 1)}\ {\bf Wilson $\to$ special dual Hahn (1st model):}
The quantum system $E3$ (\ref{S3algebra}) is given in the singular limit from system $S9$, (\ref{Wilson}) by 
\[ a_2=a_3=\epsilon\to 0 , \ a_1=a_1,\ X'^2= \lim_{\epsilon\rightarrow 0} L_1, \ L_2'=\lim_{\epsilon\rightarrow 0} L_2, \ L_1'=[X',L_2'].\]
The operators in this contraction differ from those given in subsection \ref{thelist} by a cyclic permutation of the coordinates $s_i\rightarrow s_{i+1}$.   

Now we investigate  how the difference operator realization of    $S9$ contracts to irreducible representations of the $S3$ algebra. This is more 
complicated since the original restricted algebra is now  a proper subalgebra of the contracted algebra.  
The contraction  is realized in the model by setting $\alpha_2=\alpha_3=-1/2$ and $\alpha_1=\alpha$ (the subscript is dropped since there
is now a sole $\alpha$). The restricted operators  become $H'=E'$ with $E'=-4(m+1)(m+\alpha)-(\alpha-1)^2+\frac14$ and 
$ X'^2=-4\tau^*\tau$, $L_2'=-4t^2+\alpha^2-\frac14$.
 The eigenfunctions for $X'^2$, the Wilson polynomials, become         
    \begin{equation}\label{Wilson1} \Phi_{\pm n}(t^2)=
{}_4F_{3}\left(\begin{array} {l}-n,\quad n,\quad-\frac{4m+2 \alpha+1}{4}-t,\quad-\frac{4m+2 \alpha+1}{4}+t \\ -m,\quad-m-\alpha,\quad \frac12\end{array};1\right).\ee    
Here $n=0,1,\dots,m$ if $m$ is a nonnegative integer and $n=0,1,\dots$ otherwise.
For finite dimensional representations, the spectrum of $t$ is the set $\{ \frac{\alpha}{2}+\frac14+m-k,\ k=0,1,\dots,m\}$.
The restricted polynomial functions (\ref{Wilson1}) are no longer the correct basis functions for the contracted superintegrable system:
 \[ L_2'f_{n,m}=K_{n+1,n}f_{n+1,m}+K_{n-1,n}f_{n-1,m}+\left( K_{n,n}+ \alpha_1^2-\frac14   \right)f_{n,m}\nonumber \nonumber\]
\[ K_{n+1,n}=\frac14 \left( m-n+\alpha \right)  \left( m-n \right),\ 
K_{n-1,n}=\frac14 \left( m+n+\alpha \right)  \left( m+n \right),\]
\[ K_{n,n}=\left(\frac12(\alpha+\frac12)-m\right)^2-K_{n+1,n} -K_{n-1,n}.
\]
 Now $K_{n-1,n}$ no longer vanishes for $n=0$, so $f_{0,m}$ is no longer the lowest weight eigenfunction; note $f_{-1,m}=f_{1,m}$ is still a polynomial in $t^2.$
 To understand the contraction 
we set $n=N-M/2$ where $N$ is a nonnegative integer and $M=2m$. Then the equations for the $K$'s become
\[ K(N+1,N)=\frac14 \left( M-N+\alpha \right)  \left( M-N \right), \
K(N-1,N)=\frac14 N \left( N+\alpha \right),\]
\[K(N,N)=\frac14\left(\alpha+\frac12-M\right)^2-K(N+1,N) -K(N-1,N).
\]  The three term recurrence relation gives a new set of basis orthogonal polynomials for 
 representations of $S3$. The lowest eigenfunction occurs for $N=0$; if $M$ is a nonnegative integer the representation is $2m+1$-dimensional with highest eigenfunction for $N=M$. 
The new basis functions  are
\be\label{dualHahna1}
f_{N,M}(t^2)=\frac{(\alpha+1)_N}{(-\alpha-M)_N}{}_3F_2\left(\ba{lll}-N,&-s,&s+2\alpha +1\\-M,&1+\alpha \ea;1\right).\ee
Here $f_N$ is a polynomial of  order $2N$ in $s$ and of order $n$ in $\lambda(s)=s(s+2\alpha+1)$,
a special case of dual Hahn polynomials. These dual Hahn polynomials admit 
\be\label{XS3dualHahn} X=i\left(B(s)E_s+C(s)E_s^{-1}\right),\
B(s)+C(s)=M,\ee
\[B(s)=\frac{(s+2\alpha+1)(M-s)}{2s+2\alpha+1}, \qquad C(s)=\frac{s(s+M+2\alpha+1)}{2s+2\alpha+1}.\]
The operators which form a model for the algebra (\ref{S3algebra}) are  $X$ (\ref{XS3dualHahn}),
\[ L_1=-\left(s+\alpha+\frac12\right)^2+\alpha^2-\frac14, \qquad L_2=[L_1,X] .\]
For  finite dimensional representations  the spectrum of $s$ is  $\{0,1,\dots,M\}$. From the operator $L_1$ we can determine the relation  $s=2t-\alpha-1/2$.

What is the relation between the functions (\ref{Wilson1}) and the proper basis functions (\ref{dualHahna1})?  
Note this model $X, L_1, L_2$ can be obtained from the contracted model $X', L_1', L_2'$ by conjugating by the ``ground state" of the
 contracted model $\Phi_{-\frac{M}{2}}(t^2).$ We find explicitly the gauge function
$ \Phi_{-\frac{M}{2}}(t^2) =\frac{(\frac12-\frac{M}{2})_k(-\alpha-M)_k}{(\frac12)_k(-\alpha-\frac{M}{2})_k}$,
when $t$ is evaluated at the weights $t=\frac{\alpha}{2}+\frac14+\frac{M}{2}-k,\ k=0,1,\dots \frac{M}{2}$,
so the operator $X'$ is related to $X$ via conjugation by $\Phi_{-\frac{M}{2}}(t^2)$.

Note that the functions $\Phi_n(t^2)$ are only defined for discrete values of $t$. However, on this 
restricted set the functions 
$\Phi_{-\frac{M}{2}+N}$ and $f_{N,M}$ satisfy exactly the same three term recurrence formula under multiplication by $-4t^2-a$, with 
the bottom of the weight ladder at $N=0$. From this we find 
\be\label{except} \Phi_{-\frac{M}{2}}(t^2) f_{N,M}(t^2) =\Phi_{-\frac{M}{2}+N}(t^2), \qquad t=\frac{\alpha}{2}+\frac14+\frac{M}{2}-k.\ee
Since $\Phi_{-\frac{M}{2}}(t^2) =\Phi_{\frac{M}{2}}(t^2)$,  $ f_{M,M}(t^2)=1$ restricted to the spectrum of $t$.

\medskip\noindent
{\bf 2)}\ {\bf Wilson $\to$ special  Hahn (2nd model):}
The quantum system $S3$ (\ref{S3Structure}) can also be obtained from  system $S9$, (\ref{Wilson}) by 
\[ a_1=a_3=\epsilon\to 0 , \ a_2=\frac14 -\alpha^2,\
L_1'=\lim_{\epsilon\rightarrow 0} L_1, \ X'^2= \lim_{\epsilon\rightarrow 0} L_2, \ L_2'=[X',L_1'].\]
Again, the physical model obtained by this contraction is related to the that given in subsection \ref{thelist} by a cyclic permutation of
 the coordinates $s_i\rightarrow s_{i-1}$.   

In this limit, the operator $X^2$ can be immediately factorized to obtain the skew-adjoint operator $X=2it$.  Taking $ x=t+m$, we find
\[
 L_1'=-\left[B(x)E_x+C(x)E_x^{-1}-B(x)-C(x)\right] -\alpha-\frac12,\ X'=2i(x-m),\]
\[B(x)=(x-2m)(x+\alpha+1), \ C(x)=x(x-2m-\alpha-1),\ 
 L_2'=[X',L_1],\]
which is diagonalized by  Hahn polynomials
\[\widehat{f}_{k,m}= {}_3F_2\left(\ba{c} -k, k+2\alpha+1, -x \\ \alpha+1, -2m\ea ; 1\right)=Q_k(x;B_2,B_2,2m),\]
\[ L_1'\widehat{f}_{k,m}=\left(-\left(k+\alpha+\frac12\right)^2+\alpha^2-\frac14\right)\widehat{f}_{k,m},\quad   k=0,1,\dots,2m. \]
These polynomials satisfy special  relations not obeyed by general  Hahn polynomials.
The dimension of the  space has jumped from $m+1$ to $2m+1.$ Comparing these eigenfunctions with the limit of the Wilson polynomials,  
\[ \lim_{\epsilon\rightarrow 0}L_1'f_{n,m}=\left(-\left(2n+\alpha+\frac12\right)^2+\alpha-\frac14\right)f_{n,m},\ n=0,1,\dots,m,\]
$$f_{n,m}(t)={}_4F_3\left(\ba{cccc} -n,&n+\alpha+\frac12,&-m-t,&-m+t\\ -m,&\frac12 -m,&\alpha+1\ea;1\right),$$
we see that in the limit only about half of the spectrum is uncovered. Here, the functions $f_{n,m}$ are even functions of $t$ whereas
 $\widehat{f}_{k,m}(-t)=(-1)^k\widehat{f}_{k,m}(t)$.  

 The recurrences for mutiplication by $2it$ and $-4t^2$ are compatible, so we obtain the following identity obeyed by special Wilson polynomials:
\[{}_4F_3\left(\ba{cccc} -n,&n+\alpha+\frac12,&-m-t,&-m+t\\ -m,&\frac12 -m,&\alpha+1\ea;1\right)= \]
$${}_3F_2\left(\ba{c}
 -2n, 2n+2\alpha+1, -t-m \\\alpha+1, -2m\ea ; 1\right),\ n=0,1,\dots, m.$$

\subsubsection{The  scheme and final comments}
The full set of contractions leading to the Askey Scheme can be found in \cite{KMPost13}. 
The top half of Figure \ref{fig1} shows the standard Askey Scheme indicating which orthogonal polynomials can be obtained by pointwise limits from other polynomials and, ultimately,
from the Wilson or Racah polynomials.  The bottom half of Figure \ref{fig1} shows how each of the superintegrable systems can be obtained by a series of contractions from the generic system $S9$.
Not all possible contractions are listed, partly due to complexity and partly to keep the graph from being too cluttered. 
(For example, {\it all} nondegenerate and degenerate superintegrable systems contract to the Euclidean system $H=\partial_{xx} +\partial_{yy}$.)
 The {\it singular systems} are superintegrable in the sense that they have 3 algebraically independent generators, but the coefficient matrix of the 2nd 
order terms in the Hamiltonian is singular. They follow naturally as contractions of nonsingular systems. Figure \ref{fig2} shows which orthogonal polynomials are associated with models of which quantum 
superintegrable system and how contractions enable us to reach all of these functions from $S9$. Again not all contractions have been exhibited, but 
enough to demonstrate that the Askey Scheme is a consequence of the contraction structure linking 2nd order quantum superintegrable systems in 2D.
Forthcoming papers will simplify considerably the compexity of this approach, \cite{KMSH}. Indeed  the structure equations for
nondegenerate superintegrable systems can be derived directly from the expression for $R^2$ alone, and the structure equations for degenerate superintegrable systems can be 
derived, up to a multiplicative factor, from the Casimir alone. It will also be demonstrated that all of the contractions of quadratic algebras in the Askey scheme can be induced 
by natural contractions of the Lie algebras $e(2,\C)$ and $o(3,\C)$.

There is a close association between  the models and the the symmetry operators in the original physical quantum systems that describe bases 
of eigenfunctions via separation of variables. All of the quantum systems are multiseparable. Some separable systems are exactly solvable in the sense that 
the physical solutions are products of hypergeometric 
functions.  
The special functions arising in the models we consider can  be described as the coefficients in the expansion of a separable eigenbasis for the 
original quantum system in terms of another separable eigenbasis.
The functions in the Askey Scheme are all hypergeometric polynomials that arise as the expansion coefficients relating two separable eigenbases 
that are {\it both} of hypergeometric type. 
Special polynomials in the Bannai-Ito classification have been associated with superintegrability, see \cite{PostVinet2011,PostVinet2012}.
The method obviously extends to 2nd order systems in more variables; a start  can be found in \cite{KMPost11}.
Examples of models for higher order superintegrable systems can be found in \cite{KKM2012}. The 
eigenfunctions are now rational in general, rather than polynomial.

To extend the method to Askey-Wilson  polynomials  we would need to find  appropriate  $q$-quantum mechanical systems with $q$-symmetry algebras
and, so far, this hasn't been done.

\begin{figure}[p]
\centerline{\includegraphics[scale=1]{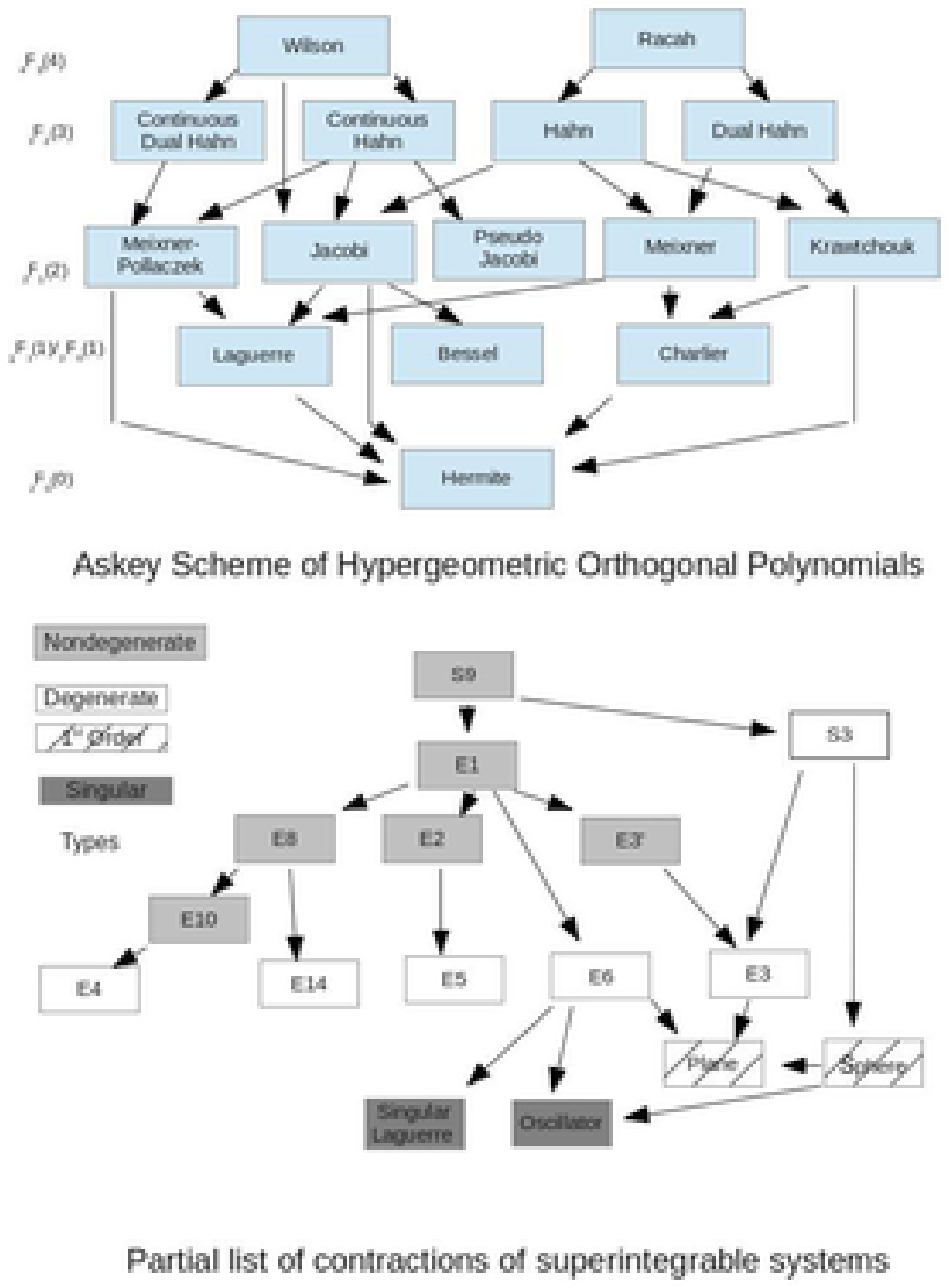}}
\caption{The Askey scheme and contractions of superintegrable systems} \label{fig1}
\end{figure}
\begin{figure}[p]
\centerline{\includegraphics[scale=1]{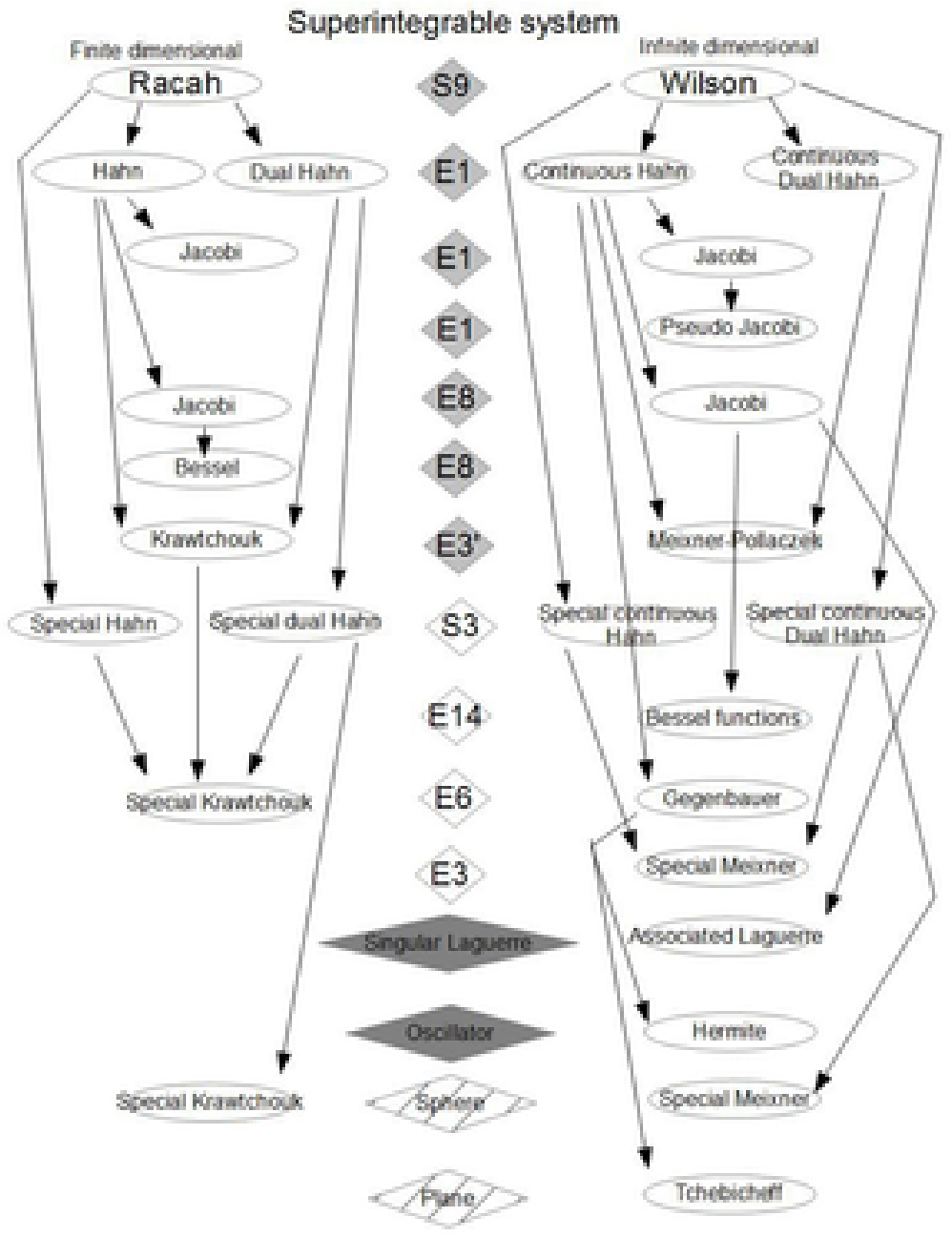}}
\caption{The Askey contraction scheme} \label{fig2}
\end{figure}

\subsection{Quadratic algebras of 3D systems}
As described in the previous section \ref{Askeyscheme}, the representation theory for quadratic operators with two functionally independent generators (i.e. those associated with second-order superintegrable systems in 2D) is now well understood. Although the analysis is not complete for 3D, there has been some work in this area, namely models of the singular isotropic oscillator \cite{KMPost10}, and models for the generic system on the 3-sphere \cite{KMPost11}. It is this latter model that we present here. 

\subsubsection{The system on the 3-sphere}\label{generic3sphere}
The Hamiltonian operator is defined via the  the embedding of the unit 3-sphere in 4D  flat space, $x_1^2+x_2^2+x_3^2+x_4^2=1,$
\be\label{qham} H=\sum_{1\le i<j\le 4}(x_i\partial_j-x_j\partial_i)^2+\sum_{k=1}^4\frac{a_k}{x_k^2}, \quad
\partial_i\equiv\partial_{x_i}.\ee
 A basis for the 2nd order integrals of the motion is given by 
\be\label{q2ordsyma}
L_{ij}\equiv L_{ji}=(x_i\partial_j-x_j\partial_i)^2+\frac{a_i x_j^2}{x_i^2}+\frac{a_j x_i^2}{x_j^2},\ee
for $1\le i<j\le 4$.
While there are 6 such operators which commute with the Hamiltonian (\ref{qham}), there is a functional relations which appears at 8th order as well as the linear relation, 
$$H =\sum_{1\le i< j\le 4} L_{ij}+\sum_{k=1}^4a_k.$$
Thus, there are indeed $6$ linearly  independent integrals and 5 functionally independent, in agreement with the $5$ implies 6 theorem. 

Let us know consider the algebra  generated by these operators. In the following $i,j,k,\ell$ are pairwise distinct integers such that $1\le i,j,k,\ell\le 4$, and  $\epsilon_{ijk}$ is the completely skew-symmetric tensor such that $\epsilon_{ijk}=1$ if  $i<j<k$.
There are 4 linearly independent commutators of the 2nd order
symmetries (no sum on repeated indices): \be\label{q3ordsyma} R_\ell=\epsilon_{ijk}[L_{ij},L_{jk}]\ee
This implies, for example, that 
$$R_1=[L_{23},L_{34}]=-[L_{24},L_{34}]=-[L_{23},L_{24}].$$
Also, 
$$[L_{ij},L_{k\ell}]=0.$$
Here we define the commutator of linear operators $F,G$ by
$[F,G]=FG-GF$. 

 The 4th order structure equations are
\be\label{q4ordsym1}{} [L_{ij},R_j]=4\epsilon_{i\ell k}(\{L_{ik},L_{j\ell}\}-\{L_{i\ell},L_{jk}\}+L_{i\ell}-L_{ik}+L_{jk}-L_{j\ell})\ee
\be\label{q4ordsym2}[L_{ij},R_k]=4\epsilon_{ij\ell}(\{L_{ij},L_{i\ell}-L_{j\ell}\}+(2+4a_j)L_{i\ell}-(2+4a_i)L_{j\ell}+2a_i-2a_j).\ee
Here, $\{F,G\}=FG+GF$.
The 5th order structure equations are obtainable directly from the
fourth order equations and the Jacobi identity. 
%\be\label{q5ordsym} [R_\ell,R_k]=4\epsilon_{ik\ell}(R_i-\{L_{ij},R_i\}) +4\epsilon_{jk\ell}(R_j-\{L_{ij},R_j\}).\ee

The 6th order structure equations are 
\be\label{q6ordsym1}
R_\ell^2=\frac83 \{L_{ij},L_{ik},L_{jk}\}-(12+16a_k)L_{ij}^2-(12+16a_i)L_{jk}^2-(12+16a_j)L_{ik}^2\ee
$$+\frac{52}{3}(\{L_{ij},L_{ik}+L_{jk}\}+\{L_{ik},L_{jk}\})+
(\frac{16}{3}+\frac{176}{3}a_k)L_{ij}+(\frac{16}{3}+\frac{176}{3}a_i)L_{jk}$$
$$+(\frac{16}{3}+\frac{176}{3}a_j)L_{ij}
+64a_ia_ja_k +48(a_ia_j+a_ja_k+a_ka_i)+\frac{32}{3}(a_i+a_j+a_k),
$$
\be\label{q6ordsym2}
\frac{\epsilon_{ik\ell}\epsilon_{jk\ell}}2\{R_i,R_j\}=\frac43(\{L_{i\ell},L_{jk},L_{k\ell}\}+\{L_{ik}, L_{j\ell}, L_{k\ell}\}-\{L_{ij}, L_{k\ell}, L_{k\ell}\})\ee
$$+\frac{26}{3}\{L_{ik},L_{j\ell}\}+\frac{26}{3}\{L_{i\ell},L_{jk}\}+\frac{44}{3}\{L_{ij},L_{k\ell}\}+4L_{k\ell}^2
$$
$$-2\{L_{j\ell}+L_{jk}+L_{i\ell}+L_{ik},L_{k\ell}\}
-(6+8a_\ell)\{L_{ik},L_{jk}\}-(6+8a_k)\{L_{i\ell},L_{j\ell}\}$$
$$-\frac{32}{3}L_{k\ell}-(\frac83-8a_\ell)(L_{jk}+L_{ik})-(\frac83-8a_k)(L_{jl}+L_{i\ell})$$
$$+(\frac{16}{3}+24a_k+24a_\ell+32a_ka_\ell)L_{ij}
-16(a_ka_\ell+a_k+a_\ell).$$
Here, $\{A,B,C\}=ABC+ACB+BAC+BCA+CAB+CBA$.

The 8th order functional relation is \be\label{q8ordsym}
\sum_{i,j,k,l}\left[ \frac1{8} L_{ij}^2 L_{kl}^2
-\frac1{92} \lbrace L_{ik}, L_{il}, L_{jk}, L_{jl}\rbrace  
-\frac{1}{36} \lbrace L_{ij}, L_{ik}, L_{kl}\rbrace \right.\ee
$$-\frac{7}{62} \lbrace L_{ij},L_{ij},L_{kl}\rbrace 
+ \frac1{6}(\frac12+\frac23 a_l) \lbrace L_{ij}L_{ik}L_{jk}\rbrace$$
$$
+\frac 2 3  L_{ij}L_{kl}
-(\frac13-\frac{3}{4}a_k-\frac{3}{4} a_l-a_ka_l)L_{ij}^2
+(\frac 13 +\frac16 a_l) \lbrace L_{ik}, L_{jk} \rbrace
+(\frac{4} 3 a_k+\frac{4}3 a_l +\frac{7}{3}a_ka_l)L_{ij}$$
$$\left.
+\frac{2}{3}a_{i}a_{j}a_{k}a_{l}+2a_{i}a_{j}a_{k}+\frac{4}{3}a_{i}a_{j}\right]=0
$$
Here, $\{A,B,C,D\}$ is the 24 term symmetrizer of 4 operators and the sum is taken over all pairwise distinct $i,j,k,\ell$.

We note here that the algebra described above contains several copies of the algebra generated by the corresponding potential on the two-sphere. Namely, if $\cal A$ is defined to be the algebra generated by all the operators $\{L_{ij}, {\cal I} \}$ for all $i,j=1,..,4$ where $\cal I$ is the identity. Then, the subalgebras ${\cal A}_{k}$ generated by $\{L_{ij}, {\cal I} \}$ for $i,j \ne k $  are exactly those associated with the 2D analog of this system. For example, the algebra ${\cal A}_4$ generated by $\lbrace L_{12}, L_{13}, L_{23}\rbrace$ admits the following operator $\tilde{H}=L_{12}+L_{13}+L_{23}+(3/4-b_1^2-b_2^2-b_3^2){\cal I}$, which is the Hamiltonian for the associated system on the two sphere and which is in the center of ${\cal A}_4.$ Thus, the representation of ${\cal A}_4$ will be used as a basis for the representation of $\cal A$. 

\subsubsection{The model}
As described above,  a representation of  $\cal A$ can be obtained by extending the representations for the subalgebras ${\cal A}_k$, namely the representation of the system S9 in terms of Wilson polynomials. 
 In order to extend this representation, we note that the operator  $\widetilde{ H}=L_{12}+L_{13}+L_{23}+3/4-(b_1^2+b_2^2+b_3^2)$ is in the center of ${\cal A}_4$ but not $\cal A$ and so it is no longer the energy of the system but instead a variable. Thus, if we begin with a representation of ${\cal A}_4$ in terms of a variable $t$, we adjoin a new variable $s$ associated with the operator $\widetilde{H}$. Thus, the model for the algebra $\cal A$ will consist of difference operators in two variables.
 
 The model is given by 
 \bea
   H&=& -\left((2M+\sum_{j=1}^4 b_j+3)^2+1\right){\cal I}, \
 {\tilde H}=\frac14-4s^2, \\
  L_{13}&=&-4t^2-\frac12+b_1^2+b_3^2,\
  L_{12}=-4\tau_t^*\tau_t-2(b_1+1)(b_2+1)+1/2, \nonumber \\
  L_{24}&=& -4{\tilde \tau}^*_s{\tilde \tau}_s-2(b_2+1)(b_4+1)+\frac12,\nonumber\\ 
  L_{34}&=& A(s)S (\sigma_{\alpha\beta}\sigma_{\alpha\gamma})_t+B(s)S^{-1}(\sigma_{\gamma\delta}\sigma_{\beta\delta})_t+C(s)4\tau_t^*\tau_t+D(s),\nonumber\eea
  where capital letters $S^k$ and $T^k$ are the shift operators in $S$ and $T$ respectively. 
 The operators $\tau$ and $\tau^*$ are correspond to the factorization of the eigenvalue equation for Wilson polynomials (\ref{tau}, \ref{taustar})  in the variable indicated by the subscript, with parameters for $\tau^*$ given by 
\be \label{para1} \alpha=\frac{b_2+1}{2}+s,\ \beta =\frac{b_1+b_3+1}{2},\ \gamma=\frac{b_1-b_3+1}{2},\ \delta=\frac{b_2+1}{2}-s\ee
and parameters for $\tilde{\tau}^*$ given by
\be\label{tildeparameters}  {\tilde \alpha}=t+\frac{b_2+1}{2},\ {\tilde \beta}=-M-\frac{b_1+b_2+b_3}{2}-1,\ee
$$ {\tilde \gamma}=M+b_4+\frac{b_1+b_2+b_3}{2}+2,\ {\tilde \delta}=-t+\frac{b_2+1}{2}.$$

 The coefficient functions in $L_{34}$ are 
 \bea  \label{As} A(s)&=&-\frac{(2 M+b_1+b_2+b_3-2 s+2) (2 M+b_1+b_2+b_3+2 b_4+2 s+4)}{2s (2 s+1)},\nonumber \\
        B(s)&=&-\frac{(2M+b_1+b_2+b_3+2 s+2) (2 M+b_1+b_2+b_3+2 b_4-2 s+4)}{2s (2 s-1)},\nonumber\\
        C(s)&=&-2+\frac{2(2M+b_1+b_2+b_3+3)(2M+b_1+b_2+b_3+2b_4+3)}{4s^2-1},\nonumber\\
        D(s)&=&2s^2-2(\frac{2M+b_1+b_2+b_3+b_4+4}{2})^2-\frac{(b_1+b_2)^2}2+\frac{b_3^2+b_4^2}{2}+b_3+b_4+2M+3\nn
      \label{Ds}  &+&\frac{((b_1+b_2+1)^2-b_3^2)(2M+b_1+b_2+b_3+3)(2M+b_1+b_2+b_3+2b_4+3)}{2(4s^2-1)},\nonumber\eea
  and the operators 
 \be \sigma_{\mu, \nu}=\frac{-1}{2t}\left[(\mu-\frac12+t)(\nu-\frac12+t)T^{\frac12}-(\mu-\frac12-t)(\nu-\frac12-t)T^{-\frac12}\right], \ee
 shift the parameters of Wilson polynomials, see \cite{KMPost11} for the action of these operators on the basis. 
 
There are several bases forming finite-dimensional irreducible representations for this model, namely 
\bea \label{dbasis}d_{\ell,m}(s,t)&=&\delta(t-t_\ell )\delta(s-s_m),\quad 0\le \ell \le m\le M,\\
\label{fbasis}f_{n,m}(s,t)&=&w_n(t^2,\alpha,\beta,\gamma,\delta)\delta(s-s_m),\quad 0\le n\le m\le M,\\
\label{gbasis}g_{\ell,k}(s,t)&=&w_k(s^2,\tilde{\alpha},\tilde{\beta},\tilde{\gamma},\tilde{\delta})\delta(t-t_\ell),\quad 0\le \ell \le k+\ell \le M,\eea
the latter two being given in terms of Wilson polynomials in one-variable. Another basis, for which the operators $L_{12}$ and  $L_{12} + L_{14} + L_{24}$ are diagonal is given in terms of two-variable Wilson polynomials constructed by Tratnik \cite{Trat1991, GI2010}.

The relation between this system and this two-variable generalization of the Wilson polynomials leads directly to at least two open question. First, whether there is an analogous ``Askey-tableau" for two variable orthogonal polynomials. Second, though related, whether the $n-$variable version of these polynomials gives representations for the associated system on the $n$-sphere and to consider their contractions. 

%% file: Conclusions.tex
\section{Conclusions}
In this review we have concentrated on certain aspects of superintegrability. First of all, we have restricted ourselves to finite-dimensional nonrelativistic classical and quantum Hamiltonian systems. The Hamiltonian always has the form $H=T+V$ where $T$ is the kinetic energy and $V$ a scalar potential. Physically this corresponds to a scalar particle in a scalar potential field, or the interaction between two scalar particles. 

We first consider the case of quadratic integrability where all integrals of motion are at most second-order polynomials in the momenta (Sections \ref{Chapter1} and \ref{Chapter2ndorder}). In 2D, the classification of such systems is complete and the results are presented.  This is combined with a review of the structure of the quadratic algebras of their integrals of motion. This classification is performed for 2D Riemannian, pseudo-Riemannian or complex Riemannian spaces of constant or non-constant curvature that allow at least two Killing tensors (Darboux spaces) \cite{Koenigs}. An important role in the study of superintegrability is played by the St\"ackel transform which relates physically different systems, often even in different spaces, to each other. 

Sections 5, 6 and 7 are devoted to higher-order superintegrability, mainly in 2D real Euclidean spaces. In section 5, we present the determining equations for the existence of integrals of finite order $N\geq 2$ in the Euclidean space $E_2.$ We review the case of $N=2$ and verify the results of the previous section that the classical and quantum integrable (and superintegrable) potentials coincide and the integrals are the same, up to symmetrization. Starting from $N=3,$ the determining equations acquire non-vanishing quantum corrections and hence the classical and quantum potentials no longer necessarily coincide. The general solution to the determining equations is not known for $N>2$. However, we present all third-order superintegrable systems that also allow a first-order integral or a second-order one leading to separation of variables in Cartesian, polar or parabolic coordinates. (The case of separation in elliptic coordinates has not yet been studied.) We also present the determining equations for fourth-order integrals and a new fourth-order superintegrable potential which does not admit separation of variables. 

For higher-order integrals of motion, the determining equations quickly become intractable. However, a break-through in this area was in the discovery of a family (since named the TTW system) of exactly-solvable systems which was conjectured to be superintegrable with integrals of arbitrarily high order. The proof of this conjecture has led to novel approaches to construction and analysis of superintegrable systems. In Section \ref{higherorderclassicalchapter}, the system is introduced and  a method for constructing additional integrals of motion for classical Hamiltonians admitting separation of variables is introduced. This method is then applied to prove the superintegrability of the TTW system as well as to construct an infinite family of superintegrable systems containing the extended Kepler-Coulomb systems in 3D. Section \ref{higherorderquantum} presents a method for proving the superintegrability of systems with higher-order integrals via recurrence relations for the separable solutions. This method makes use of the conjectured exact solvability of superintegrable systems to construct differential operators which fix energy eigenstates of the Hamiltonian. This method is used to prove the superintegrability of the TTW system and an extension of the 3D Kepler-Coulomb system  and to construct their symmetry algebras. 

Section \ref{Stackeltransform} reviews the St\"ackel transform introduced in Section \ref{Chapter2ndorder} and its relation to coupling constant metamorphosis (CCM). While CCM is well defined for Hamiltonians in classical mechanics, for quantum systems there are additional requirements on the form of the integral in order for it to be preserved as a symmetry under the transformation, namely it be a polynomial in the coupling constant. Section \ref{algebrareps} is devoted to a more mathematical aspect of superintegrability, namely the relation between the representation theory of quadratic algebras and special function theory. Just as Lie algebra
relations were exploited to determine energy levels of the hydrogen atom, so  can the representation of the polynomial algebras give energy levels and expansion coefficients from one separable basis to another. These representations give a direct connection between second-order superintegrability systems in 2D and the Askey scheme of orthogonal polynomials. Contractions of the algebras correspond to the limits between the families of classical orthogonal polynomials. 

Some aspects of superintegrability theory are not included in this review because of (self-imposed) restrictions on space and time. They include the study of systems not of the form $H=\Delta+V(\vec{x})$, for example velocity dependent potentials
\be\label{Hvdep} H=(\vec{p})^2+V(\vec{r})+\left(\vec{A}(\vec{r}), \vec{p}\right)+\left(\vec{p},\vec{A}(\vec{r})\right),\ee
where $\vec{A}(\vec{r})$ is a vector potential. The Hamiltonian \eref{Hvdep} describes the motion of a spin zero particle in a magnetic field, for instance. Integrable and superintegrable systems of this type in $E_2$ have been studied systematically with a restriction to first- and second-order integrals of the motion \cite{benenti2001variable, berube2004integrable, dorizzi1985integrable, mcsween2000integrable, pucacco2005integrable, charest2007quasiseparation, escobar2013two}. 

Another class of Hamiltonians that have been investigated from the point of view of integrability and superintegrability is of the form  
\be H=(\vec{p})^2+V_0(r)+V_1(r)(\vec{\sigma}\vec{L})\ee
and describes the interaction of a spin $1/2$ particle with a spin 0 one (e.g. pion-nucleon interaction) \cite{desilets2012superintegrable, winternitz2006integrable, winternitz2008integrable, winternitz2009integrable}. For further relevant articles on superintegrability for particles with spin see e.g. \cite{niederle2006relativistic, nikitin2012matrix, nikitin2012new, nikitin2013laplace, pronko2007quantum, pron1977new}. Finally, we mention a few other superintegrable systems where the Hamiltonian is of the different form that those studied here including those related to Dunkl oscillators and quantum spin lattices \cite{genest2013dunkl,genest2013singular, PostVinet2011, MPVZ, Miki2012quantum} as well as those associate with super-symmetric systems \cite{brink1998hidden, quesne2010chiral, PVZ2011}. 

%% file: bib.tex
% this is the bibliography 
%\chapter{Bibliography}
 \bibliography{bib}
 \bibliographystyle{jphysa}

\subsection*{Acknowledgement}
This work was partially supported by a grant from the Simons Foundation (\# 208754 to W. M.). We thank Jared Aurentz and  Bjorn Berntson for helping to work out 
the examples and Libor Snobl for helpful comments on the manuscript. The research of P.W. was partially
supported by a grant from NSERC of Canada.